\documentclass[11pt,a4paper,francais]{book}

\usepackage{amsmath,amsfonts,amssymb,amsthm,cite,babel}
\usepackage{fancyheadings}
\usepackage{makeidx}

\evensidemargin=\oddsidemargin
\advance\topmargin by -\headheight
\advance\topmargin by -\headsep

\newtheorem{thm}{Th\'eor\`eme}[section]
\newtheorem{lem}[thm]{Lemme}
\newtheorem{cly}[thm]{Corollaire}
\newtheorem{prop}[thm]{Proposition}

\newtheorem{defn}[thm]{D\'efinition}
\newtheorem{rem}[thm]{Remarque}
\numberwithin{equation}{section}             

\newcommand{\A}{\mathcal{A}}                 
\renewcommand{\a}{\alpha}                    
\DeclareMathOperator{\ad}{ad}                
\DeclareMathOperator{\Aut}{Aut}
\newcommand{\Aun}{\widetilde{\mathcal{A}}}   
\newcommand{\B}{\mathcal{B}}                 
\renewcommand{\b}{\beta}                     

\newcommand{\braket}[2]{\langle#1\mathbin|#2\rangle} 
\newcommand{\C}{\mathbb{C}}                  
\newcommand{\CC}{\mathcal{C}}                
\newcommand{\cc}{\mathbf{c}}                 
\DeclareMathOperator{\ch}{ch}                
\DeclareMathOperator{\Cl}{C\ell}             
\newcommand{\Coo}{C^\infty}                  
\newcommand{\D}{\mathcal{D}}                 
\newcommand{\dl}{\delta}                     
\newcommand{\del}{\partial}                  
\DeclareMathOperator{\Dom}{Dom}              
\DeclareMathOperator{\Diff}{Diff}
\newcommand{\Dslash}{{D\mkern-11.5mu/\,}}    
\newcommand{\dslash}{{\pa\mkern-10mu/\,}}
\newcommand{\E}{\mathcal{E}}                 
\newcommand{\ep}{\varepsilon}                
\newcommand{\eps}{\varepsilon}               
\newcommand{\F}{\mathcal{F}}                 
\newcommand{\G}{\mathcal{G}}                 
\newcommand{\Ga}{\Gamma}                     
\newcommand{\ga}{\gamma}                     
\renewcommand{\H}{\mathcal{H}}               
\newcommand{\HH}{\mathbb{H}}
\newcommand{\half}{\tfrac{1}{2}}             
\newcommand{\hookto}{\hookrightarrow}        
\newcommand{\Ht}{{\widetilde{\mathcal{H}}}}  
\newcommand{\I}{\mathcal{I}}                 
\DeclareMathOperator{\im}{Im}
\DeclareMathOperator{\Int}{Int}
\DeclareMathOperator{\Isom}{Isom}
\DeclareMathOperator{\Junk}{Junk}            
\newcommand{\K}{\mathcal{K}}                 
\DeclareMathOperator{\Ker}{Ker}
\renewcommand{\L}{\mathcal{L}}               
\newcommand{\La}{\Lambda}                    
\newcommand{\la}{\lambda}                    
\DeclareMathOperator{\Lie}{Lie}
\newcommand{\M}{\mathcal{M}}                 

\newcommand{\mop}{{\mathchoice{\mathbin{\star_{_\theta}}}
             {\mathbin{\star_{_\theta}}}           
             {{\star_\theta}}{{\star_\theta}}}}    
\newcommand{\Mop}{{\mathchoice{\mathbin{\star_{_\Theta}}}
             {\mathbin{\star_{_\Theta}}}           
             {{\star_\Theta}}{{\star_\Theta}}}}    

\DeclareMathOperator{\mood}{mod}
\newcommand{\N}{\mathbb{N}}                  
\newcommand{\NN}{\mathcal{N}}                
\newcommand{\nb}{\nabla}                     
\DeclareMathOperator{\Out}{Out}
\newcommand{\Oh}{\mathcal{O}}                
\newcommand{\opp}{{\mathrm{op}}}             
\newcommand{\ox}{\otimes}                    
\newcommand{\oxyox}{\otimes\cdots\otimes}    
\newcommand{\pa}{\partial}                   
\newcommand{\pd}[2]{\frac{\partial#1}{\partial#2}}
\newcommand{\piso}[1]{\lfloor#1\rfloor}      
\newcommand{\PsiDO}{\Psi\mathrm{DO}}         
\newcommand{\Q}{\mathbb{Q}}                  
\newcommand{\R}{\mathbb{R}}                  

\newcommand{\roundbraket}[2]{(#1\mathbin|#2)} 
\DeclareMathOperator{\Ricci}{Ricci}         
\newcommand{\sepword}[1]{\quad\mbox{#1}\quad} 
\newcommand{\set}[1]{\{\,#1\,\}}             
\newcommand{\Sf}{\mathbb{S}}                 
\DeclareMathOperator{\spec}{sp}              
\renewcommand{\SS}{\mathcal{S}}              
\DeclareMathOperator{\supp}{supp}            
\newcommand{\T}{\mathbb{T}}                  
\DeclareMathOperator{\Trw}{\Tr_\omega}
\newcommand{\Th}{\Theta}                     
\renewcommand{\th}{\theta}                   
\newcommand{\thalf}{\tfrac{1}{2}}            
\newcommand{\tihalf}{\tfrac{i}{2}}           
\newcommand{\sihalf}{{\scriptstyle\frac{i}{2}}}
\newcommand{\tpi}{{\tilde\pi}}               
\DeclareMathOperator{\Tr}{Tr}                
\DeclareMathOperator{\tr}{tr}                
\newcommand{\tri}{\Delta}                    
\DeclareMathOperator{\tsum}{{\textstyle\sum}} 
\newcommand{\V}{\mathcal{V}}                 
\newcommand{\vf}{\varphi}                    
\newcommand{\x}{\times}                      
\newcommand{\Z}{\mathbb{Z}}                  
\newcommand{\8}{\bullet}                     
\renewcommand{\.}{\cdot}                     
\renewcommand{\:}{\colon}                    

\def\<#1,#2>{\langle#1,#2\rangle}            
\def\ee_#1{e_{{\scriptscriptstyle#1}}}       
\def\wick:#1:{\mathopen:#1\mathclose:}       

\newcommand{\hideqed}{\renewcommand{\qed}{}} 

\numberwithin{equation}{section}        

\newcommand{\slim}{\mathop{\mathrm{s\mbox{-}lim}}} 

\makeindex

\begin{document}

\thispagestyle{empty}

\begin{center}
{\large \bf  Universit\'e  de Provence,  Aix-Marseille I}\\

\vspace{1.3cm}

{\Large \bf TH\`ESE}\\

\vspace{1.3cm}

Pr\'esent\'ee par\\

\bigskip

{\huge \bf  Victor GAYRAL}\\

\vspace{1.8cm}

pour obtenir le grade de\\

\medskip

{\large \bf  Docteur de l'Universit\'e de Provence}\\

\bigskip

{\it Sp\'ecialit\'e}: Physique des Particules, Physique Math\'ematique et Mod\'elisation\\

\bigskip

{\large \bf \'Ecole Doctorale}: Physique et Science de la Mati\`ere\\

\vspace{1.3cm}

{\bf  \large Titre:}

\bigskip

{\huge \bf {D\'eformations Isospectrales Non Compactes }}

\medskip

{\huge \bf { et Th\'eorie Quantique des Champs}} \\

\vspace{1.3cm}

Soutenue publiquement le 20 Mai 2005 devant le jury compos\'e de\\
\end{center}
\bigskip

\hspace{4cm}{\large T. Fack (Pr\'esident)}\\

\vspace{-0.4cm}
\hspace{4cm}{\large B. Iochum (Directeur de th\`ese)}\\

\vspace{-0.4cm}
\hspace{4cm}{\large G. Landi (Rapporteur)}\\

\vspace{-0.4cm}
\hspace{4cm}{\large T. Sch\"ucker}\\

\vspace{-0.4cm}
\hspace{4cm}{\large J. C. V\'arilly (Rapporteur)}\\

\vspace{1.6cm}

\begin{center}
{\large \bf Centre de Physique Th\'eorique\\
CNRS--UMR  6207}\\
\end{center}

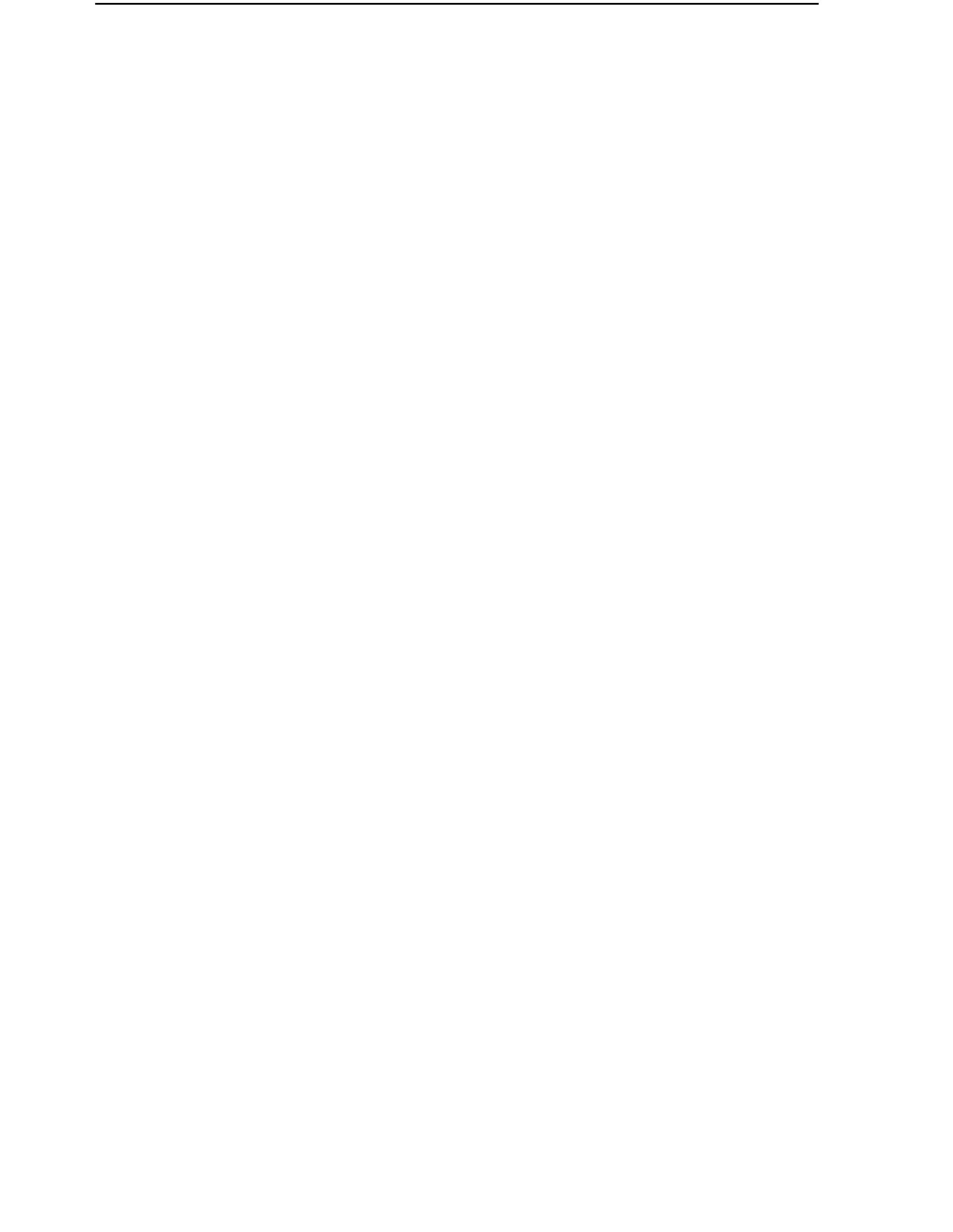
\chapter*{Remerciements}

Un directeur de th\`ese est rarement une unique personne. Je tiens tout
d'abord \`a remercier Bruno Iochum, mon directeur officiel, qui m'a
beaucoup appris, tr\`es justement conseill\'e et sans qui tout cela n'aurait
\'et\'e possible.\\
Viennent ensuite Jos\'e Gracia-Bond\'ia et Joseph V\'arilly, qui ont \'et\'e bien plus que de merveilleux
collaborateurs.\\
Tous trois ont su me faire confiance et ont \'et\'e d'un grand soutien. Je tiens \`a leur exprimer
ma plus profonde gratitude.

\medskip

La formation de chercheur n'est \'epanouissante qu'au sein d'une \'equipe. \\
Je voudrai
remercier `les jeunes' qui m'ont entour\'e et avec qui j'ai eu de nombreuses discussions fertiles:
Thomas Krajewski, Raimar Wulkenhaar, Micha\"el Grasseau, Rapha\"el Zentner, Christoff
Stephan, Jan Jureit, Ya\"el Fr\'egier et Pierre Martinetti.\\
Je tiens aussi \`a remercier tous les chercheurs et enseignants chercheurs du CPT et de l'IML
qui m'ont accord\'e beaucoup de leur temps \`a tenter d'apporter des r\'eponses
aux innombrables questions que je me suis pos\'ees durant ces trois ann\'ees
de th\`ese, en particulier \`a Thomas Sch\"ucker, Serge Lazzarini,
Christian Duval, Michel Rouleux, Valentin Zagrebnov, Pierre Duclos, Eric Soccorsi,
Claude-Alain Pillet, Micha\"el Puschnigg, Antony Wassermann, Jean-Michel Combes.\\
Un grand merci \`a tout le `staff' technique du CPT, en particulier \`a Dolly Roche,
Isabelle Morganti, Vincent Bayle, Denis Patrat, Corinne Roux, Annette Elbaz, Sylvie Retourna.

\medskip

Un grand merci aussi \`a Fernando Ruiz Ruiz pour m'avoir invit\'e trois mois au
Departemento de F\'isica Te\'orica I de la Universidad Complutense de Madrid,
pendant lesquels une partie de ce travail a \'et\'e fait et \`a
Carmelo Mart\'in pour m'avoir permis de partager son bureau pendant
cette p\'eriode.

\medskip
 
Je remercie vivement Thierry Fack, Gianni Landi, Thomas Sch\"ucker et
Joseph V\'arilly pour avoir accept\'e de faire partie du jury.
\newpage
\tableofcontents
\newpage
{\renewcommand{\thechapter}{}\renewcommand{\chaptername}{}
\addtocounter{chapter}{-1}
\chapter{Introduction}\markboth{\sl INTRODUCTION}{\sl INTRODUCTION}}

Cette th\`ese constitue un recueil des travaux de recherche que j'ai effectu\'es ces trois
derni\`eres ann\'ees en collaboration avec J. M. Gracia-Bond\'ia,
B. Iochum, F. Ruiz Ruiz, T. Sch\"ucker et J. C. V\'arilly.

\bigskip

La g\'eom\'etrie non commutative (GNC) est une g\'en\'eralisation
(alg\'ebrique et op\'eratorielle) des concepts principaux de la g\'eom\'etrie
diff\'erentielle ordinaire; elle permet ainsi d'\'elargir le
cadre math\'ematique qui d\'ecrit la physique fondamentale au niveau classique.
Les pr\'emisses de la GNC datent de la premi\`ere moiti\'e du si\`ecle dernier,
de l'av\`enement de la m\'ecanique quantique, \`a partir duquel il fut de plus en
plus \'evident que les notions g\'eom\'etriques s\'eculaires et intuitives devraient
\^etre repens\'ees pour permettre d'y int\'egrer des situations (g\'eom\'e\-tri\-ques)
plus g\'en\'erales.

Plus sp\'ecifiquement, les \'etudes que nous avons men\'ees et qui sont report\'ees
dans cette th\`ese, s'inscrivent dans le courant de pens\'ee qui s'est construit autour
des id\'ees \'elabor\'ees par Alain Connes. L'une des motivations principales
de son approche, fut la description d'espaces `pathologiques' pour lesquels
les outils g\'eom\'etriques et analytiques classiques perdent leur pertinence.\\
La n\'ecessit\'e d'\'etendre les notions g\'eom\'etriques ordinaires est aussi primordiale
en physique fondamentale, en l'occurrence si on
d\'ecide de s'affranchir de l'hyphoth\`ese du continu pour mod\'eliser et
comprendre l'espace-temps ainsi que les `objets vivant dessus', tels
les champs de mati\`ere et les champs de jauge. En effet, une description continue
de l'espace-temps est contradictoire avec le principe d'incertitude spatio-temporel  \cite{FDR},
analogue \`a celui d'Heisenberg sur la position et l'impulsion en m\'ecanique quantique,
et r\'esultant de la nature dynamique de l'espace-temps induite par les \'equations d'Einstein.
Il se produit \'evidemment dans le r\'egime des petites distances, typiquement \`a l'\'echelle
de Planck $l_p\sim 1.6\,\,10^{-33}$cm.
En effet, vouloir localiser une particule dans l'espace-temps
avec une pr\'ecision arbitraire, n'est pas envisageable d\'ej\`a dans le
cadre de la relativit\'e g\'en\'erale. L'\'energie
n\'ecessaire \`a une telle localisation (par exemple celle d'une particule incidente
si on veut proc\'eder par exp\'erience de diffusion) peut provoquer une singularit\'e,
dont l'horizon rendrait
impossible toute d\'etection. L'espace-temps ne peut  avoir la structure d'une vari\'et\'e
diff\'erentiable qu'en premi\`ere approximation.\\
La probl\'ematique ultime de la physique fondamentale, consiste en la description unifi\'ee
des interactions r\'egissant les particules \'el\'ementaires au niveau quantique.
En dehors des difficult\'es d'ordres techniques et/ou conceptuelles persistantes
en th\'eorie quantique des champs (en parti\-cu\-lier dans sa formulation perturbative),
une partie de ce programme a \'et\'e r\'ealis\'ee: la quantification du mod\`ele standard,
i.e. des interactions \'electromagn\'etique, faible et forte. L'\'etape
suivante est la quantification de la gravitation. Elle n'a \'et\'e abord\'ee que dans
le contexte de la quantification canonique (gravit\'e quantique \`a boucle) et dans celui
de la th\'eorie des cordes. La question, qui pour l'heure n'a aucun embryon de r\'eponse,
est de savoir si les id\'ees de la g\'eom\'etrie non commutative sont suffisamment robustes
pour fournir un cadre appropri\'e \`a la description de la gravitation au niveau
quantique. Nous verrons qu'il existe plusieurs arguments conceptuels en faveur de
la GNC, mais que de nombreuses obstructions techniques rendent pour l'heure
probl\'ematique toute tentative de quantification de la gravitation dans le cadre de la GNC.

Les fondements de la g\'eom\'etrie non commutative reposent en quelque sorte \`a la fois sur le
th\'eor\`eme de dualit\'e de Gelfand--Naimark et sur la m\'ecanique quantique elle-m\^eme. En effet,
ce th\'eor\`eme affirme que toute $C^*$-alg\`ebre commutative est isomorphe
\`a l'alg\`ebre des fonctions continues et s'annulant \`a l'infini sur un espace
localement compact. Tout espace topologique localement compact $X$ peut ainsi
\^etre  d\'ecrit de mani\`ere duale en terme de la $C^*$-alg\`ebre $\Coo_0(X)$.
Par ailleurs, la m\'ecanique quantique nous indique, en particulier pour
comprendre la stabilit\'e des atomes et d\'ecrire leurs spectres en \'energie,
que les coordonn\'ees de l'espace des phases ne commutent
plus. Ainsi, d'une part les espaces ordinaires peuvent \^etre caract\'eris\'es en termes
purement alg\'ebrique et d'autre part il existe des ph\'enom\`enes physiques ne
pouvant \^etre compris qu'en termes de coordonn\'ees non commutatives.\\
Un espace non commutatif est alors d\'efini par dualit\'e, et consiste en la donn\'ee
d'une $C^*$-alg\`ebre non commutative. Dans ce cadre, on peut aussi donner
du sens \`a la notion de mesure \`a travers celle d'alg\`ebre de von Neumann,
de topologie via des th\'eories (co)homologiques telles la K-th\'eorie et la
(co)homologie cyclique mais aussi de structure diff\'erentielle, g\'en\'eralement
impl\'ement\'ee par un op\'erateur externe (un op\'erateur de Dirac abstrait).
Les notions usuelles de fibr\'es vectoriels et de connections
sont aussi (par dualit\'e) directement transposable au monde non commutatif.\\
Finalement, l'objet central dans l'interface entre la GNC et la physique fondamentale
est celui de triplet spectral, g\'en\'eralisation non commutative de la notion
de vari\'et\'e Riemannienne \`a spin; le point de d\'epart naturel pour l'\'elaboration de
th\'eories physiques.

\medskip

En physique des particules, la GNC a permis deux avanc\'ees conceptuelles majeures:
une interpr\'etation g\'eom\'etrique du m\'ecanisme de Higgs et l'unification
au niveau classique de toutes les interactions connues.

Un des avantages de la GNC est de pouvoir traiter dans un m\^eme cadre espaces
continus et espaces discrets. C'est en comprenant l'espace-temps en termes du
`produit' (rigoureusement du produit de triplets spectraux) d'un espace continu
(une quatre-vari\'et\'e) par un espace \`a deux points, produisant
un espace \`a deux couches chacune correspondant heuristiquement au monde
des fermions de chiralit\'e donn\'ee, que l'on peut comprendre
g\'eom\'etriquement le m\'ecanisme de Higgs. En consid\'erant la fonctionnelle
d'action de Connes--Lott (qui n'est rien d'autre qu'une transposition directe dans
le langage de la GNC de celle de Yang--Mills), cette particule scalaire appara\^it alors
(avec son fameux potentiel quartique) comme la composante (dans la direction discr\`ete)
d'une connection \cite{ConnesL1,ConnesL2}.

En plus de cette m\^eme interpr\'etation du champs de Higgs, l'action spectrale permet
d'unifier au niveau classique les interactions \'electro-faible, forte et gravitationnelle.
Cette fonctionnelle d'action est d\'efinie \`a partir du spectre d'un op\'erateur de Dirac
$\D$: $S=\Tr\chi(\D^2/\Lambda^2)$ (nombre de ses valeurs propres inf\'erieures ou
\'egales \`a une \'echelle de masse $\Lambda$). On obtient par fluctuation de la m\'etrique,
c'est-\`a-dire par transformations de jauge g\'en\'eralis\'ees au groupe des unitaires de l'alg\`ebre, et en
d\'eveloppant cette action en puissances de $\Lambda$, le Lagrangien
du mod\`ele standard coupl\'e \`a la gravitation d'Einstein--Weyl \cite{CIKS, AliAlain, AliAlain2}.

\medskip

Alors que sur le plan conceptuel les id\'ees de la GNC sont en parfaite ad\'equation avec
celles de la gravitation quantique, en particulier sur la structure microscopique de
l'espace-temps, sur le plan technique de nombreuses difficult\'es existent.
Pour ne citer que la plus importante, il n'est pas clair comment impl\'ementer
(sur l'hypoth\'etique alg\`ebre non commutative d\'ecrivant notre espace-temps
au niveau quantique) l'invariance sous diff\'eomorphismes. Ce probl\`eme est bien plus
s\'erieux qu'il n'y para\^it, car pour l'heure toutes les constructions d'espaces non
commutatifs utilisent le point de vue `syst\`eme de coordonn\'ees'.

Les th\'eories quantiques des champs sur espace non commutatif (NCQFT)
ne souffrent pas de telles obstructions. Elles constituent plut\^ot une g\'en\'eralisation
directe de celles sur espace commutatif. Les champs classiques sont interpr\'et\'es
comme les \'el\'ements d'une alg\`ebre non commutative dans le cas scalaire,
ou comme les \'el\'ements d'un module projectif de type fini (la notion non
commutative de fibr\'e vectoriel) dans le cas vectoriel. Le seul autre ingr\'edient
dont on a besoin (en dehors d'un op\'erateur d\'eterminant la cin\'etique) est une trace
sur l'alg\`ebre pour construire une action classique.\\
Ces th\'eories peuvent \^etre interpr\'et\'ees comme la limite \`a basse \'energie
d'une th\'eorie quantique unifi\'ee. La motivation principale pour les introduire,
tomb\'ee en d\'esu\'etude aujourd'hui \cite{CMarco3,CMarco1,CMarco2},
\'etait de faire dispara\^itre les divergences ultraviolettes (aux grandes \'energies
ou aux petites distances) dont sont affubl\'ees les th\'eories quantiques des champs
sur espace ordinaire. En effet, lorsque l'on substitue un espace quantique \`a un espace
ordinaire, on s'attend a voir dispara\^itre ces divergences car la notion de point,
donc celle de petite distance, n'existe plus sur les espaces non commutatifs.\\
Ce ne fut pas le cas pour les premiers exemples \'etudi\'es, plans de Moyal
et tores non commutatifs, ou en plus des divergences UV ordinaires,
des divergences m\'elangeant courtes et grandes distances apparurent:
le ph\'enom\`ene de m\'elange UV/IR.\\
C'est pour mieux comprendre l'origine de tels ph\'enom\`enes, mais aussi
pour diff\'erencier les aspects g\'en\'eriques de ceux sp\'ecifiques
\`a un mod\`ele, qu'il est primordial
d'\'etudier la th\'eorie quantique des champs sur d'autres espaces non commutatifs.

\bigskip

Dans cette th\`ese, nous nous sommes int\'eress\'e aux
aspects math\'ematiques ainsi qu'\`a certaines applications physiques,
d'une classe importante d'espaces non commutatifs: les d\'eformations
isospectrales. Ce sont des g\'en\'eralisations en espace courbe
des sans doute plus anciens espaces quantiques connus: les tores non commutatifs
et les plans de Moyal.

En cherchant \`a construire et \`a classifier, \`a partir de consid\'erations homologiques,
les vari\'et\'es non commutatives sph\'eriques de dimension trois,
Connes, Landi et Dubois-Violette ont donn\'e dans les articles \cite{CL, CDV} une
m\'ethode pour g\'en\'erer \`a partir d'un espace Riemannien classique, une famille
d'espaces quantiques fond\'es sur le paradigme du tore non commutatif.
L'ingr\'edient de base est une vari\'et\'e Riemannienne compacte munie d'une structure spin et
dont le groupe d'isom\'etrie est de rang (dimension du sous-groupe commutatif
maximal) sup\'erieur ou \'egal \`a deux. On construit alors par dualit\'e une famille
de vari\'et\'es  (d'alg\`ebres) non commutatives $\Coo(M_\Th)$, en d\'eformant l'alg\`ebre
commutative $\Coo(M)$ \`a partir de l'action sur $M$ du sous-groupe Ab\'elien maximal
(isomorphe \`a un tore car la vari\'et\'e est compacte) du groupe d'isom\'etrie de la
vari\'et\'e. Dans la construction de Connes--Dubois-Violette \cite{CDV} (celle
de Connes--Landi \cite{CL} est sensiblement diff\'erente), $\Coo(M_\Th)$ est d\'efinie
en termes d'alg\`ebre des points fixes pour l'action d'un groupe:
\begin{equation}
\label{fixedpoint}
\Coo(M_\Th):=\Big(\Coo(M)\widehat{\otimes}
\Coo(\T^l_\Th)\Big)^{\a\widehat{\otimes}\tau^{-1}},
\end{equation}
o\`u $\Coo(\T^l_\Th)$ d\'esigne l'alg\`ebre d'un $l$-tore non commutatif de matrice
de d\'eformation $\Th$, $\tau$ est l'action sur $\Coo(\T^l_\Th)$ du $l$-tore ordinaire
et $\a$ est l'action par automorphismes sur $\Coo(M)$ donn\'ee par le sous-groupe
Ab\'elien maximal de $\Isom(M,g)$.\\
La terminologie `d\'eformation isospectrale' vient du fait que le triplet spectral sous-jacent,
i.e. l'objet dual $(\Coo(M_\Th),L^2(M,S),\Dslash)$
codant les structures topologique, diff\'erentielle, m\'etrique et spin de la vari\'et\'e
de d\'epart $M$, poss\`ede le m\^eme espace de repr\'esentation, l'espace de Hilbert des
sections de carr\'e sommable du fibr\'e des spineurs, et le m\^eme op\'erateur de Dirac
que celui non d\'eform\'e; seule l'alg\`ebre est modifi\'ee.

Quelques temps plus tard, V\'arilly \cite{Larissa} et Sitarz~\cite{Sitarz} remarqu\`erent
ind\'ependamment, que cette construction s'inscrit dans la th\'eorie de Rieffel de la
d\'eformation/quantification par action de $\R^l$ \cite{RieffelDefQ}. Se donnant une alg\`ebre
de Fr\'echet $A$ munie d'une famille de semi-normes $\{p_i\}_{i\in I}$ ainsi que d'une action
du groupe Ab\'elien $\R^l$,
fortement continue et isom\'etrique pour chaque semi-norme, on peut d\'eformer la
sous-alg\`ebre $A^\infty$ consistant en les \'el\'ements de $A$ qui sont lisses (pour la
topologie Fr\'echet) par rapport \`a l'action des g\'en\'erateurs infinit\'esimaux
$X^k$, ${k\in\{1,\ldots,l\}}$ de l'action $\a$. L'alg\`ebre $A^\infty$ peut alors \^etre
munie canoniquement d'une nouvelle famille de semi-normes
$\tilde{p}_{i,m}(.):=\sup_{j\leq i}\sum_{|\beta|\leq m} p_j(X^\beta.)$, $\beta\in\N^l$.
Elles ont la propri\'et\'e d'\^etre compatibles avec le produit d\'eform\'e, d\'efini par:
\begin{equation}
\label{Rieffelprod}
a\Mop b:=(2\pi)^{-l}\int_{\R^{2l}}\,d^ly\,d^lz\;e^{-i<y,z>}\;
\a_{\thalf\Th y}(a)\a_{-z}(b),\;a,b\in A^\infty.
\end{equation}
Ici $\Th$ est une matrice r\'eelle et antisym\'etrique; la matrice de d\'eformation.
Les int\'egrales intervenant dans la d\'efinition de ce produit sont des int\'egrales
oscillantes \`a valeur dans $A^\infty$. La propri\'et\'e principale de ce processus de d\'eformation
est d'\^etre r\'eversible: en notant $A^\infty_\Th$ l'alg\`ebre $(A^\infty,\Mop)$,
on a $(A^\infty_\Th)^\infty_{\Th'}=A^\infty_{\Th+\Th'}$.

En utilisant ce cadre plus g\'en\'eral, la construction de Connes--Landi--Dubois-Violette
peut \^etre \'etendue aux vari\'et\'es Riemanniennes non compactes munies
d'une action de $\R^l$ par isom\'etrie.

\smallskip

Dans cette th\`ese, on s'int\'eressera
principalement aux espaces non commutatifs non compacts ainsi obtenus,
qui sont qualifi\'es de d\'eformations isospectrales non p\'eriodiques non compactes.
Ils ont pour paradigmes les plans de Moyal, avec lesquels ils co\"incident lorsque
$M=\R^l$ et lorsque $\R^l$ agit sur lui-m\^eme par translation.

\bigskip

Dans la premi\`ere partie de cette th\`ese, nous commencerons par revoir les motivations
homologiques qui ont abouti \`a la construction (et \`a la classification) des vari\'et\'es
non commutatives tri-dimensionnelles de type sph\'erique. Nous reverrons ensuite les constructions
respectives de Connes--Landi et de Connes--Dubois-Violette des d\'eformations isospectrales
p\'eriodiques compactes. Les d\'eformations isospectrales non p\'eriodiques non compactes
seront alors introduites en utilisant la th\'eorie de Rieffel.
Leurs propri\'et\'es alg\'ebriques seront \'etablies et les
aspects analytiques g\'en\'eriques, en particulier Hilbertien, seront \'etudi\'es.

La deuxi\`eme partie consiste en la construction de triplet spectraux sans unit\'e.
On commencera par le cas unif\`ere, puis on
motivera les modifications n\'ecessaires pour formuler les axiomes des
`espaces Riemanniens \`a spin quantiques non compacts'. Apr\`es avoir revu
un certain nombre de faits concernant l'analyse fonctionnelle sous-jacente
au produit de Moyal, on v\'erifiera que ces axiomes modifi\'es sont satisfait pour les d\'eformations
non compactes plates. Enfin, nous donnerons les points clefs pour construire
des triplets spectraux sans unit\'e \`a partir des d\'eformations isospectrales
non compactes courbes.

Dans le chapitre suivant, nous calculerons pour les plans de Moyal,
deux types de `fonctionnelle d'action pour champs de jauge non commutatifs':
l'action de Connes--Lott et l'action spectrale.

Dans le dernier chapitre, nous nous int\'eresserons aux th\'eories quantiques
des champs sur d\'eformations isospectrales courbes. Nous nous focaliserons sur le
ph\'enom\`ene de m\'elange des divergences ultraviolette et infrarouge
(m\'elange UV/IR); ph\'enom\`ene
tr\`es important car rendant probl\'ematique la renormalisation de la th\'eorie et
qui jusque-l\`a n'\'etait connu que pour les d\'eformations plates (plans de Moyal
et tores non commutatifs).\\
Au travers de l'\'etude d'une th\'eorie scalaire en dimension quatre, nous allons
montrer le caract\`ere intrins\`eque de ce ph\'enom\`ene et \'etudier ses
cons\'equences sur la renormalisabilit\'e. Nous verrons appara\^itre des manifestations
nouvelles et/ou plus fines du m\'elange UV/IR, en relations avec les propri\'et\'es
g\'eom\'etriques de ces espaces quantiques et arithm\'etiques des param\`etres de d\'eformation.
D'autres perspectives de recherches seront propos\'ees dans la conclusion, et deux appendices
r\'esumeront les propri\'et\'es des traces de Dixmier et de la base de Wigner de
l'oscillateur harmonique.

\chapter{D\'eformations isospectrales}
\section{D\'eformations isospectrales p\'eriodiques}
\label{defperio}
\subsection{Sph\`eres de Connes--Landi--Dubois-Violette}
\label{sec:CLDV-basics}

Nous avons d\'ej\`a stipul\'e dans l'introduction que
l'origine des d\'eformations isospectrales est
de nature homologique. Ce point de vue homologique ne sera
pas vraiment utilis\'e dans ce m\'emoire; nous nous concentrerons sur les
aspects analytiques. Il nous a sembl\'e cependant instructif, de rappeler la construction
homologique des sph\`eres de Connes--Landi--Dubois-Violette, en particulier pour
motiver l'introduction des d\'eformations isospectrales.
En effet, une sous-famille de ces espaces non commutatifs
constitue l'arch\'etype des d\'eformations isospectrales courbes.

Le point de d\'epart de leur construction
est l'\'etude et  la classification des vari\'et\'es non commutatives
tridimensionnelles sph\'eriques. Ces derni\`eres sont d\'efinies alg\'ebriquement
\`a travers des contraintes sur les composantes de leur caract\`ere de Chern
en homologie cyclique, i.e. un cycle cyclique local sur l'alg\`ebre $\A$
d\'efinissant la vari\'et\'e non commutative et dont le couplage
avec la K-homologie du triplet spectral $(\A,\H,\D)$ (voir chapitre 2)
donne un indice de Fredholm (voir \cite{ConnesBook,Polaris}).

On cherche ainsi \`a construire une alg\`ebre $\A$,  de dimension $n$,
par g\'en\'erateurs et relations.
Les g\'en\'erateurs sont dans le cas pair les \'el\'ements de matrice
d'un projecteur $e\in M_q(\A)$, $e^2=e^*=e$ et sont dans le cas impair ceux
d'un unitaire $U\in M_q(\A)$, $U^*U=UU^*=1$. Les relations sont
impl\'ement\'ees par l'annulation de toutes les composantes de degr\'e non maximal
du caract\`ere de Chern. Ces derni\`eres conditions
correspondent \`a la condition de sph\'ericit\'e impos\'ee et s'expriment en
dimension paire $n=2m$ par
\begin{equation}
\label{idem}
\ch_k(e)=0,\,\forall\, k=0,\cdots,m-1,
\index{chern@$\ch_k$}
\end{equation}
et en dimension impaire $n=2m+1$ par
\begin{equation}
\label{unit}
\ch_{k+1/2}(U)=0,\,\forall\, k=0,\cdots,m-1.
\end{equation}
La composante maximal ($\ch_m$ ou $\ch_{m+1/2}$) du caract\`ere de
Chern d\'efinit, quant \`a elle, un cycle de Hochschild jouant le r\^ole de la forme
volume sur l'espace non commutatif en question.
A des constantes  $\lambda_k$, $\lambda_k'$ pr\`es,
les composantes du caract\`ere de Chern sont donn\'ees dans le cas pair par
\begin{equation*}
\ch_k(e)=\lambda_k\sum_{i_0,\cdots,i_{2k}=1}^q\big(e^{i_0}_{i_1}-\thalf\delta^{i_0}_{i_1}\big)\otimes
e^{i_1}_{i_2}\otimes e^{i_2}_{i_3}\otimes\dots\otimes e^{i_{2k}}_{i_0}
\in \A\otimes(\widetilde{\A})^{\otimes 2k}\label{chernpaire},
\end{equation*}
et dans le cas impair par
\begin{align*}
\label{chernimpaire}
\ch_{k+1/2}(U)=&\lambda_k'\sum_{i_0,\cdots,i_{2k+1}=1}^q U^{i_0}_{i_1}\otimes
U^{*i_1}_{i_2}\otimes U^{i_2}_{i_3}\otimes\dots\otimes U^{*i_{2k+1}}_{i_0}\\
&\hspace{2cm}-U^{*i_0}_{i_1}\otimes
U^{i_1}_{i_2}\otimes U^{*i_2}_{i_3}\otimes\dots\otimes U^{i_{2k+1}}_{i_0}
\in\A\otimes(\widetilde{\A})^{\otimes 2k+1},
\end{align*}
o\`u $\widetilde{\A}:=\A/\C$. Il existe de plus, une condition de
consistance entre la dimension $n$ et le rang $q$ des matrices $e$ et
$U$, \`a savoir $q=2^m$ pour $n=2m$ ou $n=2m+1$.

A. Connes a montr\'e dans \cite{ConnesSurvey} qu'une
solution des \'equations (\ref{idem}) et (\ref{unit}) est donn\'ee par
le g\'en\'erateur de Bott dans le cas des sph\`eres ordinaires $\Sf^n$
pour $n=1,2,3,4$, et a trouv\'e avec G. Landi \cite{CL} une famille \`a
un param\`etre de solutions non commutatives dans le cas $n=3,4$.
L'ensemble des solutions de (\ref{unit}) pour $n=3$, i.e. l'ensemble
des vari\'et\'es non commutatives tri-dimensionnelles de type
sph\'erique, a ensuite \'et\'e classifi\'e par A. Connes et M. Dubois-Violette
dans l'article \cite{CDV}. Il en r\'esulte une famille de solutions \`a trois param\`etres,
dont les alg\`ebres homog\`enes associ\'ees (pour lesquelles la condition d'unitarit\'e
de la matrice $U\in M_2(\A)$ a \'et\'e remplac\'ee par $U^*U=UU^*\subset {\bf 1}_2
\ox\A$) sont isomorphes, pour
des valeurs g\'en\'eriques des param\`etres, aux alg\`ebres de Sklyanin,
introduites en connexion avec l'\'equation de Yang-Baxter. C'est cette
construction que nous revoyons maintenant, en suivant de pr\`es l'article
\cite{CDV}. Notons finalement que l'ensemble de solutions pour $n=4$, $q=4$, constitue
quant \`a lui une famille \`a sept param\`etres de vari\'et\'es
non commutatives \cite{DV}.

Soient $n=3$, $q=2$. Nous cherchons une alg\`ebre complexe,
unitale et involutive $\A$ dont les g\'en\'erateurs sont les \'el\'ements d'une
matrice $U\in M_2(\C)\ox\A\simeq M_2(\A)$ et les relations sont exprim\'ees
au travers des \'equations
\begin{equation}
\label{unit2}
\ch_{1/2}(U) = \sum_{i,j=1,2}U_{ij} \ox U^*_{ij} - U^*_{ji} \ox U_{ji}=0,
\end{equation}
et
\begin{equation}
\label{unit3}
UU^*=U^*U={\bf 1}_2
\end{equation}
o\`u
$$
U = \begin{pmatrix} U_{11} & U_{12} \\ U_{21} & U_{22} \end{pmatrix}
\sepword{et} U^* = \begin{pmatrix} U^*_{11} & U^*_{21} \\
U^*_{12} & U^*_{22} \end{pmatrix},\sepword{avec} U_{ij}\in\A.
$$
Pour reparam\'etrer cette derni\`ere \'equation, il est commode
d'introduire les matrices de Pauli
$$
\sigma := (1_2, -\vec\sigma), \qquad \bar\sigma := (1_2, \vec\sigma),
$$
satisfaisant \`a
\begin{equation}
\sigma^l\sigma^j = \dl^{lj} + i\eps^{ljm}\sigma^m.
\label{eq:more-trivialities}
\end{equation}
Dans  ce param\'etrage, $U$ s'\'ecrit
$$
U = (\bar\sigma\tilde z) = z_0 + i\vec\sigma\cdot\vec z,
$$
o\`u $(\tilde z):=(z_0,iz_1,iz_2,iz_3)\in\A^4$ et $\vec z:=(z_1,z_2,z_3)$,
c'est-\`a-dire
\begin{equation}
U = \begin{pmatrix} z_0 + iz_3 & iz_1 + z_2 \\ iz_1 - z_2 & z_0 - iz_3
\end{pmatrix}.
\label{eq:so-what}
\end{equation}
L'\'equation (\ref{unit2}) devient alors:
\begin{equation}
\ch_{1/2}(U) = 2\,\bigl(\tsum_{\mu=0}^3 z_\mu \ox z^*_\mu -
z^*_\mu \ox z_\mu\bigr) = 0.
\label{eq:cond-top}
\end{equation}
Cette condition est satisfaite si et seulement si (voir \cite{CDV} pour la preuve)
il existe une matrice $\La\in M_4(\C)$ unitaire et sym\'etrique telle que
$$
z_\nu^* = \La^\mu_\nu z_\mu.
$$
Finalement, l'ensemble des relations (\ref{unit2}) et  (\ref{unit3}) devient:
\begin{align}
\tsum_{\nu=0}^3 z_\nu z^*_\nu = \tsum_{\nu=0}^3 z^*_\nu z_\nu &= 1,
\nonumber \\
z_k z_0^* - z_0 z^*_k + \eps^{lkm} z_lz^*_m &= 0,
\label{eq:cond-nonsense} \\
z_0^* z_k - z^*_k z_0 + \eps^{lkm} z_l^*z_m &= 0.
\nonumber
\end{align}
Pour param\'etrer l'ensemble des solutions, il faut remarquer que les relations $UU^*=U^*U={\bf 1}_2$ et
$\ch_{1/2}(U) = 0$ ne changent pas sous la transformation
\begin{equation}
U \mapsto uV_1UV_2,
\label{eq:less-freedom}
\end{equation}
o\`u $u\in U(1)$ et $V_1,V_2\in SU(2)$. Utilisant ce degr\'e de libert\'e,
$\La$ peut \^etre diagonalis\'ee et r\'e\'ecrite comme:
$$
\La = \begin{pmatrix} 1 & & & \\ & e^{-2i\vf_1} & & \\ & & e^{-2i\vf_2}
& \\ & & & e^{-2i\vf_3} \end{pmatrix}.
$$
C'est-\`a-dire, nous pouvons supposer que des g\'en\'erateurs $\{z_\mu\}$
s'\'ecrivent
$$
z_0 = x_0; \qquad z_j = e^{i\vf_j}x_j,
$$
o\`u les \'el\'ements  $x_\nu$ sont maintenant hermitiens.
Pour ces nouveaux g\'en\'erateurs, les relations de commutation
(\ref{eq:cond-nonsense}) deviennent
\begin{align*}
[x_0,x_1]_- &=  i\frac{\sin(\vf_2-\vf_3)}{\cos\vf_1}[x_2,x_3]_+
\\
[x_2,x_3]_- &= -i\frac{\sin\vf_1}{\cos(\vf_2-\vf_3)}[x_0,x_1]_+,
\end{align*}
plus celles obtenues par permutation circulaire de $\{x_1,x_2,x_3\}$.
Il est aussi \`a noter qu'une
rotation de $SO(3)$ permute les $\vf_l$
et que l'on peut remplacer $\vf_l$ par $\vf_l-\pi$ en
\'echangeant $x_l$ par $-x_l$, sans changer les relations de l'alg\`ebre;
tous les cas de figure sont donc couverts par
\begin{equation}
\set{(e^{i\vf_1},e^{i\vf_2},e^{i\vf_3})\in\T^3 \big| \pi>\vf_1\ge\vf_2\ge\vf_3\ge0}.
\label{eq:the-orange}
\end{equation}

Il est aussi possible de v\'erifier, que la composante maximale du
caract\`ere de Chern satisfait
\begin{align*}
\ch_{3/2}(U) =& -\sum_{0\leq\a,\b,\ga,\delta\leq3}\ep_{\a\b\ga\delta}
\cos(\vf_\a-\vf_\b+\vf_\ga-\vf_\delta)x_\a\ox x_\b\ox x_\ga \ox x_\delta\\
&+i\sum_{0\leq\mu,\nu\leq3}\sin(2(\vf_\mu-\vf_\nu))x_\mu\ox x_\nu\ox
x_\mu\ox x_\nu.
\end{align*}
Le caract\`ere de Chern est par construction un cycle de Hochschild sur l'alg\`ebre
$\A$, dont la composante de degr\'e maximale est l'\'equivalent
non commutatif de la forme volume. Pour des valeurs g\'en\'eriques des
param\`etres, $\ch_{3/2}(U)$ est non trivial et correspond \`a la forme volume
associ\'ee \`a la m\'etrique ronde de $\Sf^3$ dans le cas commutatif limite o\`u
$\vf_1=\vf_2=\vf_3=0$.

Dans la suite, nous n'allons nous int\'eresser qu'\`a une sous famille
particuli\`ere \`a un param\`etre des sph\`eres de Connes--Dubois-Violette,
donn\'ee par $\vf_1=\vf_2=\th/2$, $\vf_3=0$. Ces solutions correspondent
aux 3-sph\`eres de Connes--Landi \cite{CL}. Les relations de commutation
deviennent pour ces valeurs:
\begin{equation}
[x_0,x_1]_- = i\tan(\th/2)[x_2,x_3]_+; \quad
[x_0,x_2]_- = -i\tan(\th/2)[x_1,x_3]_+,
\label{eq:commut-one}
\end{equation}

\begin{equation}
[x_1,x_3]_- = i\tan(\th/2)[x_0,x_2]_+; \quad
[x_2,x_3]_- = -i\tan(\th/2)[x_0,x_1]_+,
\label{eq:commut-two}
\end{equation}
et
\begin{equation}
[x_0,x_3]_- =[x_1,x_2]_- =0.
\end{equation}
En d\'efinissant les \'el\'ements complexes
$$
\zeta_1:=x_0+ix_3; \qquad \zeta_2:=x_1+ix_2,
$$
ces relations deviennent:
\begin{align*}
\zeta_1\,\zeta_2&=e^{-i\th}\,\zeta_2\,\zeta_1,\\
\zeta_1^*\,\zeta_2^*&=e^{-i\th}\,\zeta_2^*\,\zeta_1^*,\\
\zeta_1^*\,\zeta_2&=e^{i\th}\,\zeta_2\,\zeta_1^*,\\
\zeta_1\,\zeta_2^*&=e^{i\th}\,\zeta_2^*\,\zeta_1,\\
\zeta_1\,\zeta_1^*&=\zeta_1^*\,\zeta_1,\\
\zeta_2\,\zeta_2^*&=\zeta_2^*\,\zeta_2.
\end{align*}
Ces derni\`eres \'egalit\'es rappellent les relations canoniques
du tore non commutatif:
c'est ce point de vue qui va \^etre dor\'enavant exploit\'e.\\
En posant finalement,
\begin{equation}
\label{TNC}
\zeta_1=u\,\cos\phi, \,\,\zeta_2=v\,\sin\phi,
\end{equation}
o\`u $u$ et $v$ sont des unitaires et $\phi$ un \'el\'ement central,
on obtient exactement les relations du tore non commutatif pour $u$ et $v$,
c'est-\`a-dire:
\begin{equation}
u\,v=e^{-i\th}v\,u.
\end{equation}
Cette construction donne en r\'ealit\'e le premier exemple de d\'eformation
isospectrale p\'eriodique, d\'eformations que nous d\'efinirons dans le paragraphe
suivant. En effet, et au vu de cette derni\`ere param\'etrisation, on peut d\'eformer
la structure commutative de l'alg\`ebre des fonctions sur la trois-sph\`ere
ordinaire, en utilisant l'action isom\'etrique de $\T^2$ sur $\Sf^3$,
donn\'ee dans le syst\`eme de coordonn\'ees sous-jacent \`a
la param\'etrisation (\ref{TNC}) par
$$
\T^2\ni z: (u,v,\phi)\mapsto (z_1.u,z_2.v,\phi).
$$
Il suffit ensuite d'introduire le 2-cocycle du groupe additif $\Z^2$
$$
\sigma(r,s):=\exp\{-\tihalf\th(r_1s_2-r_2s_1)\},
$$
pour obtenir une alg\`ebre non commutative, en sp\'ecifiant le nouveau produit sur les
`modes de Fourier' $U_r:=u^{r_1}v^{r_2}$, par
$$
U_r\,U_s=\sigma(r,s)\,U_s\,U_r.
$$
Autrement dit, l'alg\`ebre du tore non commutatif a \'et\'e
plong\'ee dans l'alg\`ebre commutative des fonctions lisses sur la trois-sph\`ere,
de mani\`ere compatible avec sa structure Riemannienne standard (h\'erit\'ee
de celle de $\R^4$): La compatibilit\'e s'exprime par le fait que cette action de
groupe est isom\'etrique pour la m\'etrique ronde.
Nous verrons d\`es le paragraphe suivant que cette
construction se g\'en\'eralise facilement pour une grande classe de
vari\'et\'e Riemannienne.

\subsection{La construction de Connes--Landi}
\label{constructionCL}
Nous revoyons dans ce paragraphe la construction des d\'eformations
isospectrales p\'eriodiques introduite dans \cite{CL}.

{\it Soit $(M,g)$ une vari\'et\'e Riemannienne compacte, sans bord,
g\'eod\'esiquement compl\`ete, connexe, orient\'ee, de dimension $n$ et
munie d'une structure spin. Soit ensuite, $\tilde{\a}$ une action lisse et effective
du groupe compact et Ab\'elien $\T^l$, $2\leq l\leq n$, par isom\'etrie
$$
\tilde{\a}:\T^l\longrightarrow \Isom(M,g)\subset\Diff(M),
$$
o\`u $l$ est inf\'erieur ou \'egal au rang du groupe d'isom\'etrie
de $(M,g)$. }\\
Il est \`a noter que la classe des vari\'et\'es dont le rang du
groupe d'isom\'etrie est sup\'erieur ou \'egal \`a deux
est loin d'\^etre vide, puisque les isom\'etries pr\'eservant l'orientation
(ce que nous supposeront toujours)
forment un sous-groupe de $SO(n)$,
d'apr\`es le th\'eor\`eme de Myers--Steenrod.
En anticipant sur le chapitre 3, notons aussi $\A:=\Coo(M)$, l'alg\`ebre
commutative des fonctions ind\'efiniment diff\'erentiables sur $M$,
$\H:=L^2(M,S)$ l'espace de Hilbert des sections de carr\'e sommable
du fibr\'e spinoriel et $\Dslash$ l'op\'erateur de Dirac associ\'e \`a la
m\'etrique $g$.\medskip \\
Pour fixer les notations, notons que l'op\'erateur de Dirac
s'\'ecrit localement $\Dslash=-i(\pa_\mu
+\omega_\mu)\ox\gamma^\mu$, o\`u $\omega_\mu$ d\'esigne la connection
de spin, i.e. la connection de Levi--Civita relev\'ee au fibr\'e des spineurs.
Lorsque l'espace sera plat, c'est-\`a-dire lorsque $M=\R^n$ ou $M=\T^l$,
on notera $\dslash$ l'op\'erateur de Dirac plat, i.e. $\dslash=-i\pa_\mu
\ox\gamma^\mu$.\medskip\\
L'action $\tilde{\a}$ de $\T^l$ sur $M$ induit sur
$\A$ une action de $\T^l$ par
automorphisme, que nous continuerons \`a d\'esigner par $\tilde{\a}$:
$$
(\tilde{\a}_zf)(p):=f(\tilde{\a}_{-z}(p)),\,z\in\T^l,\,f\in\A,\,p\in M.
$$
Sur le fibr\'e des spineurs, l'action $\tilde{\a}$ se rel\`eve mais seulement
modulo $\pm 1$. On peut construire (voir \cite{GIV} pour une construction
particuli\`ere) un recouvrement double $p:\widetilde{\T^l}\to\T^l$,
 o\`u $\widetilde{\T^l}$ est isomorphe \`a $\T^l$, de telle sorte
que l'on puisse trouver un groupe d'op\'erateurs unitaires
$\{V_{\tilde z}:\,\tilde z\in\widetilde{\T^l}\}\subset\mathcal{U}
(\H)$ couvrant le groupe d'isom\'etries $\{\tilde{\a}_z:\,z\in\T^l\}$.
En terme de sections lisses, ce recouvrement satisfait \`a
$$
V_{\tilde z}(f\cdot\psi)=\tilde{\a}_z(f)\cdot V_{\tilde z}(\psi),
$$
quel que soit $\psi\in\H$, $f\in\Coo(M)$ et avec $p(\tilde z)=z$.
En g\'en\'eral ce rel\`evement au fibr\'e des spineurs n'est pas
trivial, i.e. si $p(\tilde z)=p(\tilde z')$ alors $V_{\tilde z}=\pm
V_{\tilde z'}$, mais le signe ne peut \^etre choisi globalement.
De plus, en notant $(\cdot,\cdot)$ la structure Hermitienne du fibr\'e des
spineurs, on a
$$
(V_{\tilde z}\psi,V_{\tilde z}\psi')=\tilde{\a}_z(\psi,\psi'),
$$
pour $\psi,\psi'\in\H$. L'action de $\T^l$ \'etant suppos\'ee
isom\'etrique, le pull-back de l'action laisse invariante
la m\'etrique:
$$
\tilde{\a}_z^*\,g=g.
$$
De m\^eme,
il est facile de montrer que l'op\'erateur
de Dirac, le Laplacien (scalaire et  son rel\`evement
au fibr\'e des spineurs) ainsi que les champs de vecteurs $X_j$,
$j=1,\cdots,l$,
associ\'es \`a l'action infinit\'esimale de $\T^l$,
commutent avec les unitaires $V_{\tilde z}$.
Pour les m\^emes raisons, la forme volume Riemannienne
$\mu_g$ est aussi laiss\'ee invariante par (le pull-back de)
l'action.
\begin{rem}
Dans le cas des d\'eformations p\'eriodiques, il n'est en fait
pas n\'ecessaire de supposer l'action isom\'etrique. En
effet, \`a partir d'une action lisse de $\T^l$ et d'une
structure Riemannienne quelconque $g$, on peut fabriquer une
m\'etrique $g_{inv}$ invariante sous l'action de $\T^l$ en
moyennant cette derni\`ere par l'action du tore
$$
g_{inv}:=\int_{\T^l}d^lz\,\tilde{\a}_z^*(g),
$$
avec $d^lz$ la mesure de Haar normalis\'ee du $l$-tore.
Cependant, pour les d\'eformations isospectrales
non p\'eriodiques (pour lesquelles l'action de $\T^l$ est
remplac\'ee par une action de $\R^l$, voir
section~\ref{defnonperio}) ce processus de moyenne n'est
pas forc\'ement bien d\'efini, et il est commode, dans notre
contexte, de conserver l'hypoth\`ese d'isom\'etrie pour l'action.
\end{rem}

La construction de Connes--Landi \cite{CL} des d\'eformations
isospectrales p\'eriodiques est tout \`a fait analogue \`a celle des
sph\`eres de Connes--Landi utilisant le point de vue
`syst\`eme de coordonn\'ees versus s\'erie de Fourier'.\\
Nous allons exploiter ici le fait que tout $f\in\A=\Coo(M)$
poss\`ede une d\'ecomposition en sous-espaces spectraux
(ou d\'ecomposition de Peter--Weyl) $f=\sum_{r\in\Z^l}f_r$,
indic\'ee par les \'el\'ements de $\Z^l$ le
groupe dual de $\T^l$, et pour laquelle chaque $f_r$ satisfait \`a la relation
$$
\tilde{\a}_z(f_r)=e^{-i(r_1z_1+\cdots+r_lz_l)}f_r.
$$
Pour \'eviter les probl\`emes li\'es au rel\`evement de l'action
sur le fibr\'e spinoriel, dans ce qui suit,
nous allons pour simplifier travailler uniquement au niveau
scalaire, c'est-\`a-dire au niveau de l'espace de Hilbert r\'eduit
$\H_r:=L^2(M,\mu_g)$ des fonctions de carr\'e sommable
par rapport \`a l'espace mesur\'e $(M,\mu_g)$ o\`u $\mu_g$
est la forme volume Riemannienne. Notons encore $V_z$ la
repr\'esentation de $\T^l$ induite sur $\H_r$ par des op\'erateurs
unitaires, d\'efinie par $V_z(\psi)(p):=\psi(\tilde{\a}_{-z}(p))$,
pour tout $\psi\in\H_r$.\\
Chaque op\'erateur $A$ born\'e sur
$\H_r$ et lisse en norme
relativement \`a l'action du tore, i.e. qui soit tel que l'application
$\T^l\ni z \mapsto V_zAV_ {-z}$
soit lisse pour la topologie normique de $\L(\H)$,
poss\`ede une unique d\'ecomposition en \'el\'ements
homog\`enes
$$
A=\sum_{r\in\Z^l}A_r,
$$
convergente en norme. Cette d\'ecomposition d\'efinit ainsi une
$l$-graduation; chaque $A_r$ est de $l$-degr\'e
$r=(r_1,\cdots,r_l)$. Explicitement, chaque composante
$A_r$ satisfait \`a la relation
$$
V_z\,A_r\,V_{-z}=e^{-ir.z}\,A_r.
$$
Pour tout $f\in\Coo(M)$, on obtient alors une
d\'ecomposition en `s\'erie de Fourier'
\begin{equation}
\label{eq:PW}
f=\sum_{r\in\Z^l}f_r,
\end{equation}
avec $\tilde{\a}_z(f_r)=e^{-ir.z}f_r$. Cette assertion
provient du fait que $M_f$, l'op\'erateur
de multiplication point \`a point par $f$, est lisse relativement \`a l'action car
il appartient \`a l'intersection des domaines des puissances des d\'erivations
$\delta_j(.):=[X_j,.]$, o\`u les $X_j$, $j=1,\cdots,l$, sont les g\'en\'erateurs infinit\'esimaux
de l'action $\tilde\a$. On obtient alors le r\'esultat en it\'erant la relation
$$
\|[X_j,M_f]\|=\|M_{X_j(f)}\|=\|X_jf\|_\infty,
$$
qui est finie car $f\in\Coo(M)$ et donc $X_j(f)\in\Coo(M)$ aussi.

On d\'efinit alors $M_\Th$, la d\'eformation isospectrale de $M$
de param\`etre $\Th$, par dualit\'e comme \'etant l'alg\`ebre $\Coo(M_\Th)$
des fonctions lisses sur $M$ munie du produit $\Mop$, que l'on d\'efinit sur
les \'el\'ements homog\`enes par
$$
f_r\Mop g_s:=e^{-\tihalf\Th(r,s)}\,f_r\,g_s=:\sigma_\Th(r,s)\,f_r\,g_s,
$$
avec $\Th(r,s):=\sum_{i,j=1}^lr^i\Th^{ij}s^j$. Ici, $\Th^t=-\Th$ est la
matrice ($l\times l$ \`a entr\'ees r\'eelles) de d\'eformation.
Utilisant la d\'ecomposition en sous-espaces spectraux, ce produit s'\'etend par
lin\'earit\'e \`a tous les \'el\'ements de  $\Coo(M)$. L'associativit\'e du produit ainsi d\'efini,
est alors garantie par le fait que $\sigma_\Th$ est un 2-cocycle sur le groupe
additif $\Z^l$:
$$
\sigma_\Th(r,s)\sigma_\Th(r+s,t)=\sigma_\Th(s,t)\sigma_\Th(r,s+t).
$$

Il est alors justifi\'e de consid\'erer les d\'eformations isospectrales
p\'eriodiques comme g\'en\'eralisa-tions en espace courbe des paradigmes
que sont les $l$-tores non commutatifs. Notons au passage que ces derniers peuvent \^etre
alternativement construits en utilisant cette proc\'edure.
Notons aussi que les plans
de Moyal jouiront du m\^eme statut vis-\`a-vis des d\'eformations isospectrales
non p\'eriodiques.

Concernant la topologie, avec la pr\'esente d\'efinition,
il est seulement possible de compl\'eter,  par rapport \`a la norme
op\'eratorielle, $\Coo(M_\Th)$ en une $C^*$-alg\`ebre.
C'est-\`a-dire, avec un abus \'evident de notation
\begin{equation}
\label{C*}
C^*(M_\Th):=\overline{{L^\Th(\Coo(M))}}^{\|.\|},
\end{equation}
o\`u $L^\Th(f)\equiv L^\Th_f$ est l'op\'erateur
de multiplication twist\'ee \`a gauche
par $f$ sur $\H_r$, i.e. $L_f\psi:=f\Mop \psi$ pour tout $\psi\in\H_r$. Il est ensuite
imm\'ediat de r\'ealiser que l'op\'erateur $L^\Th_f$ ainsi d\'efini, est born\'e sur
$\H_r$ pour tout $f\in\Coo(M)$. En effet,
$$
L^\Th_f\psi=\sum_{r\in\Z^l}f_r\Mop\psi
=\sum_{r\in\Z^l}f_r.V_{-\thalf\Th r}(\psi),
$$
ainsi
$$
\|L^\Th_f\|\leq \sum_{r\in\Z^l}\|M_{f_r}V_{-\thalf\Th r}\|\leq
 \sum_{r\in\Z^l}\|M_{f_r}\|\leq \sum_{r\in\Z^l}\|f_r\|_\infty<\infty,
$$
car $f\in\Coo(M)$ et car sa d\'ecomposition de Peter--Weyl est convergente
en norme $\|.\|_\infty$.

En revanche, nous allons voir que la construction de Connes--Dubois-Violette
des d\'eforma\-tions isospectrales p\'eriodiques, permet de munir
canoniquement $\Coo(M_\Th)$ d'une topologie plus fine,
rendant $\Coo(M_\Th)$ localement convexe.

Nous aurions aussi bien pu utiliser les op\'erateurs de multiplication
twist\'ee \`a droite, d\'efinis par $R^\Th_f\psi:=\psi\mop f$
et satisfaisant aux m\^emes propri\'et\'es que ceux de
multiplication \`a gauche, pour d\'efinir la $C^*$-compl\'etion (\ref{C*}),
\'etant donn\'e que $L^\Th(\Coo(M))$ et $R^\Th(\Coo(M))$ sont
isomorphes. De plus,
par associativit\'e du produit d\'eform\'e, on obtient directement que les
repr\'esentations r\'eguli\`eres droite et gauche commutent:
$$
[L^\Th_f,R^\Th_g]=0,\,\forall\,f,g\in\Coo(M).
$$

A la diff\'erence des cas non d\'eg\'en\'er\'es
(plans de Moyal et tores non commutatifs irrationnels, avec matrice de d\'eformation
symplectique), les alg\`ebres de von Neumann associ\'ees ne sont pas
\`a priori des facteurs. En effet, m\^eme pour des valeurs
irrationnelles des param\`etres de d\'eformation, leurs centres contient
l'alg\`ebre des fonctions invariantes sous l'action du tore, c'est-\`a-dire les
fonctions constantes sur les orbites de l'action. Cette alg\`ebre
se r\'eduit au corps des scalaires seulement lorsque $M=\T^l$
muni d'une action libre.

Cette construction permet aussi de d\'efinir, pour n'importe quel op\'erateur $A$
lisse en norme relativement \`a l'action du tore, ses twists gauche $L^\Th_A$
et droit $R^\Th_A$ par
\begin{align*}
L^\Th_A:&=\sum_{r\in\Z^l}A_r\,V_{-\thalf\Th r},\\
R^\Th_A:&=\sum_{r\in\Z^l}A_r\,V_{\thalf\Th r}.
\end{align*}
Ici, $A_r$ r\'ef\`ere \`a la d\'ecomposition de $A$ en \'el\'ements
homog\`enes. Cette d\'ecomposition \'etant convergente
en norme et $V_z$ \'etant un unitaire, il est clair que les op\'erateurs
ainsi d\'efinis sont born\'es. De plus on a
\begin{align*}
[L^\Th_A,R^\Th_B]=&
\sum_{r,s}[A_r\,V_{-\thalf\Th r},B_s\,V_{\thalf\Th s}]\\
=&\sum_{r,s}[A_r,B_s]V_{\thalf\Th(s- r)}+
B_s[A_r,V_{\thalf\Th s}]V_{-\thalf\Th r}+
A_r[V_{-\thalf\Th r},B_s]V_{\thalf\Th s}\\
=&\sum_{r,s}[A_r,B_s]\,e^{-i\thalf \Th(r,s)}\,V_{\thalf\Th(s- r)},
\end{align*}
o\`u la commutation des $\{V_z\}$ et les relations
$[A_r,V_z]=(1-e^{-irz})A_rV_z$ ont \'et\'e utilis\'ees.
Ainsi, si les composantes homog\`enes de $A$ et de $B$
commutent entre elles, le twist gauche de $A$ commutera avec le twist
droit de $B$.

Pour terminer, notons d\`es \`a pr\'esent qu'il n'y a pas d'obstruction
\`a \'etendre cette construction au cas o\`u $M$ ne serait que
localement compacte. Cependant, les difficult\'es techniques \'etant
plus importantes, nous ne traiterons ce cas de figure que lors du paragraphe
\ref{defnonperio}.

\subsection{La construction de Connes--Dubois-Violette}

Nous revoyons maintenant la construction de Connes--Dubois-Violette
des d\'eformations isospectrales p\'eriodiques~\cite{CDV}. A la diff\'erence de
celle de Connes--Landi, qui comme nous le verrons au paragraphe \ref{defnonperio}
s'inscrit dans une th\'eorie des d\'eformations plus
g\'en\'erale due \`a Rieffel
\cite{RieffelDefQ}, l'approche de Connes--Dubois-Violette adopte un point
de vue totalement diff\'erent. Cette approche a non seulement la vertu de
permettre de munir canoniquement les alg\`ebres
d'une topologie plus fine que celle donn\'ee par la norme op\'eratorielle
(n\'ecessaire pour la construction de triplets spectraux), mais aussi
d'\^etre enti\`erement compatible avec les techniques standards
de calcul cohomologique. Elle donne un acc\`es quasi direct
\`a la cohomologie de Hochschild de ces d\'eformations, qui se trouve
\^etre identique \`a celle des alg\`ebres non d\'eform\'ees \cite{CDV}. Finalement,
cette nouvelle caract\'erisation va donner une image heuristique (mais
correcte) de la situation, \`a savoir que l'on transf\`ere la structure du tore
non commutatif dans $\Coo(M)$ d'une mani\`ere compatible avec
la structure Riemannienne de la vari\'et\'e.

Soit $(M,g,\tilde{\a})$, satisfaisant aux m\^emes hypoth\`eses que celles du paragraphe
pr\'ec\'edent. On construit la vari\'et\'e non commutative $M_\Th$
par dualit\'e, en d\'efinissant $\Coo(M_\Th)$ en terme d'alg\`ebre de points fixes
sous l'action d'un groupe.\\
Pour fixer les notations, d\'esignons par
$\Coo(\T^l_\Th)$,  l'alg\`ebre du $l$-tore non commutatif de matrice
de d\'eformation $\Th\in M_l(\R)$, $\Th^t=-\Th$,
$$
\Coo(\T^l_\Th):=\left\{c_r\,U^r:\{c_r\}\in\SS(\Z^l)\right\},
$$
avec $U^r:=u_1^{r_1}\cdots u_l^{r_l}$, o\`u les unitaires $\{u_{i}\}_{i=1,\cdots,l}$
satisfont aux relations
$$
u_i\,u_j=e^{-i\Th_{ij}}u_j\,u_i,
$$
qui \'equivaut \`a
$$
U^r\,U^s=e^{-ir.\Th s}U^s\,U^r.
$$
Rappelons que muni de la famille de semi-normes canoniques
$$
p_k(a):=\sup_{r\in\Z^l}\left|(1+|r|^2)^k a_r\right|,\sepword{avec}
\Coo(\T^l_\Th)\ni a=\sum_r a_r\,U^r,
$$
$\Coo(\T^l_\Th)$ devient un espace de Fr\'echet. Enfin, notons $\tau$ l'action
usuelle du $l$-tore ordinaire sur le $l$-tore non commutatif, donn\'ee sur les
g\'en\'erateurs par
$$
\tau:\T^l\ni z\mapsto \tau_z(U^r)=z^rU^r=z_1^{r_1}\cdots z_l^{r_l} U^r.
$$
On d\'efinit alors
\begin{equation}
\label{carac-point-fixe}
\Coo(M_\Th):=\left(\Coo(M)\widehat{\ox}\Coo(\T^l_\Th)\right)
^{\tilde{\a}\widehat{\ox}\tau^{-1}},
\end{equation}
o\`u $\widehat{\ox}$ est la compl\'etion
du produit tensoriel alg\'ebrique par rapport \`a la topologie
inductive ou projective, qui co\"incident ici puisque
$\Coo(\T^l_\Th)$ est un espace de Fr\'echet nucl\'eaire et
$\Coo(M)$ munie de sa topologie ordinaire est localement convexe.

Il n'est pas non plus n\'ecessaire ici de se restreindre au cas
o\`u $M$ est compacte. Ainsi, il suffit de remplacer $\Coo(M)$ par
$\Coo_c(M)$ dans le cas o\`u $M$ ne serait que localement compacte.
Une fois encore, les d\'etails du cas non compact ne seront donn\'es
que lors de la description du cas non p\'eriodique.

\section{D\'eformations isospectrales non p\'eriodiques}
\label{defnonperio}
Nous allons voir une troisi\`eme construction, permettant
de g\'en\'eraliser les d\'eformations isospectrales
dans le cas d'une action de $\R^l$ par isom\'etrie.
Ces espaces courbes non commutatifs auront alors pour
paradigme le plan de Moyal. Cette approche type `star-produit'
est bas\'ee sur la th\'eorie de Rieffel des d\'eformations \cite{RieffelDefQ},
faisant appel \`a une formule int\'egrale pour le produit d\'eform\'e. Elle
permettra un traitement unifi\'e avec le cas p\'eriodique et
fournira un cadre adapt\'e \`a l'analyse Hilbertienne.

\subsection{Le produit d\'eform\'e, d\'efinition et premi\`eres propri\'et\'es}

Dans la suite, nous allons faire les hypoth\`eses (sensiblement
diff\'erentes de celles du paragraphe \ref{constructionCL}) suivantes:

{\it Soit $(M,g)$ une vari\'et\'e Riemannienne sans bord,
g\'eod\'esiquement compl\`ete, connexe, orien\-t\'ee, de dimension $n$ et
munie d'une structure spin. Soit ensuite $\a$ une action lisse
du grou\-pe  Ab\'elien $\R^l$, $2\leq l\leq n$, par isom\'etrie
$$
\a:\R^l\longrightarrow \Isom(M,g)\subset\Diff(M),
$$
o\`u $l$ est inf\'erieur ou \'egal au rang du groupe d'isom\'etrie
de $(M,g)$.}

Pour simplifier les notations, nous utiliserons de mani\`ere
\'equivalente $\a_z(p)$ et $z.p$ pour d\'esigner l'action d'un
\'el\'ement du groupe sur un point de la vari\'et\'e et nous noterons encore
$\a$ l'action induite par automorphisme sur $\Coo_c(M)$:
$(\a_zf)(p)=f(\a_{-z}(p))$. Soient aussi
$\{X_j\}_{j=1,\cdots,l}$ les champs de vecteurs associ\'es \`a
l'action infinit\'esimale:
$$
\left.X_j(f):=\frac{\pa}{\pa z^j}\a_z(f)\right|_{z=0},\hspace{0.5cm}
f\in\Coo(M).
$$

Pour des raisons qui deviendront plus claires dans la suite,
nous devons supposer dans le cas d'une action effective ($\ker(\a)=\{0\}$),
qu'en plus d'\^etre isom\'etrique, l'action soit propre. Nous nous
restreindrons donc \`a des actions pour lesquelles l'application
$$
\R^l\times M\ni(z,p)\mapsto (p,\a_z(p))\in M\times M,
$$
est propre. Rappelons qu'une application entre deux espaces topologiques
$f:E\to F$ est propre si et seulement si la pr\'e-image de tout sous-ensemble
compact $K$ de $F$ est un compact de $E$. Il convient cependant de noter
que cette hypoth\`ese est tout-\`a-fait naturelle car \'etant automatiquement
satisfaite lorsque
$\{\a_z:z\in\R^l\}$ est ferm\'e dans $\Isom(M,g)$ pour la topologie des
ouverts compacts \cite[lemme 5.5]{Michor}.

\begin{defn}
Soit $\Theta \in M_l(\R)$ une matrice r\'eelle et antisym\'etrique.
Pour $f,h\in \Coo_c(M)$, l'espace des fonctions ind\'efiniment
diff\'erentiables \`a support compact, le produit ordinaire de $f$ avec
$g$ peut \^etre d\'eform\'e par l'action de groupe $\a$,
de la mani\`ere suivante~\cite{RieffelDefQ}:
\begin{equation}
\label{product}
f \Mop h := (2\pi)^{-l} \int_{\R^l}\int_{\R^l}\,d^ly \,d^lz\,
e^{-i<y,z>} \, \a_{\half\Th y}(f) \, \a_{-z}(h) .
\end{equation}
\end{defn}
Ici, $<y,z>=\sum_{j=1}^ly^iz^i$ est vu comme le couplage entre
$\R^l$ et son groupe dual. Nous utiliserons sans distinction
$<y,z>$ et $y.z$ pour d\'esigner ce couplage.

\begin{defn}
\label{MP}
Lorsque $M=\R^l$ et lorsque l'espace Euclidien agit sur
lui-m\^eme par translation, pour une matrice de d\'eformation non d\'eg\'en\'er\'ee
(ce qui implique que la dimension soit paire $l=2N$), le produit twist\'e \eqref{product} co\"incide avec
le produit de Moyal \cite{Moyal}:
\begin{equation*}
f \mop h(x) := (2\pi)^{-2N} \int_{\R^{2N}}\int_{\R^{2N}}\,d^{2N}y \,d^{2N}z\,
e^{-i<y,z>} \, f(x-\half\Th y) \, h(x+z).
\end{equation*}
\end{defn}
Ce produit a \'et\'e introduit dans la premi\`ere moiti\'e
du si\`ecle pr\'ec\'edent pour formuler la m\'ecani\-que quantique
sur l'espace de phase: munie d'un tel produit, l'alg\`ebre des fonctions sur
l'espace de phase $\R^{2N}$ (l'espace cotangent de $\R^N$
avec sa structure symplectique canonique) est isomorphe \`a
l'alg\`ebre des op\'erateurs sur $L^2(\R^N)$. La forme asymptotique
de ce produit (cf. paragraphe \ref{casMoyal}) a donn\'e naissance \`a une th\'eorie
g\'eom\'etrique de quantification, pour les vari\'et\'es de
Poisson, appel\'ee quantification par d\'eformation \cite{Bayenetall}.

\begin{rem}
\label{casdege}
Notons d\`es \`a pr\'esent que l'on peut toujours supposer
$\Th$ inversible. En effet, lorsque $\Th$ ne l'est pas, le produit
$\Mop$ se r\'eduit au produit d\'eform\'e associ\'e \`a
l'action restreinte $\sigma:=\a|_{V^\bot}$, o\`u $V$ est le noyau
de $\Th$ vu comme endomorphisme sur $\R^l$
\cite[proposition 2.7]{RieffelDefQ}.
\end{rem}

En d\'epit des apparences, cette formule est
`sym\'etrique'; m\^eme avec une matrice de
d\'eforma\-tion d\'eg\'en\'er\'ee, le produit twist\'e
peut se r\'e\'ecrire comme
$$
f \Mop h := (2\pi)^{-l} \int_{\R^l}\int_{\R^l}\,d^ly \,d^lz\,
e^{i<y,z>} \, \a_{-z}(f) \, \a_{\half\Th y}(h) .
$$
Cette `sym\'etrie' montre en particulier que la conjugaison complexe
est une involution (cf. \'equation \eqref{pro1}).

Il y a fondamentalement deux situations distinctes \`a consid\'erer.
Lorsque l'action $\a$ est effective, nous parlerons de d\'eformations
non p\'eriodiques. Les d\'eformations p\'eriodiques seront quant \`a elles,
caract\'eris\'ees par $\ker(\a)\simeq \Z^l$. Dans ce dernier cas,
l'action $\a$ se factorise au travers de l'action d'un tore
$\tilde{\a}:\R^l/\Z^l\to \Isom(M,g)$, qui est automatiquement propre.
Bien que pouvant \^etre  trait\'e dans ce cadre, nous ne
ferons pas mention explicite du
cas mixte $\a:\R^d\times\T^{l-d}\to\Isom(M,g)$.
D'\'evidentes modifications de nos arguments permettent de
g\'en\'eraliser nos r\'esultats dans cette situation.

Alors que dans le cas d'une vari\'et\'e non compacte les deux situations
(p\'eriodique ou non) existent,
lorsque la vari\'et\'e $M$ est compacte, l'action $\a$ doit \^etre
p\'eriodique pour donner lieu \`a une action propre.

Finalement, dans le cas non p\'eriodique uniquement,
que l'action soit propre implique qu'elle est aussi libre. Rappelons d'abord
que l'action d'un groupe $G$ sur un espace $X$ est dite libre
si tous les groupes d'isotropie ou stabilisateurs,
$H_x:=\{g\in G:g.x=x, \,x\in X\}$, sont r\'eduits \`a l'identit\'e du groupe.
Aussi, une action est propre si et seulement si (voir
\cite{Michor} par exemple) $\{g\in G:\,g.U\cap V\ne\emptyset\}$ est compact
pour tout compact $U,V\subset X$.\\
En prenant alors $U=V=\{p_0\}$ pour n'importe quel point $p_0\in M$,
son groupe d'isotropie $H_{p_0}=
\{z\in \R^l:z.p_0=p_0\}=\{z\in \R^l:z.\{p_0\}\cap \{p_0\}
\ne \emptyset\}$ doit \^etre compact. Le seul sous-groupe compact
de $\R^l$ \'etant $\{0\}$, l'action est alors automatiquement libre.
Cela implique en particulier que l'application quotient
$$
\pi:M \longrightarrow M/\R^l,
$$
d\'efinit une projection de $\R^l$-fibr\'e principal, qui sont en d\'efinitive les vari\'et\'es
que nous consid\'erons dans le cas non p\'eriodique.

Dans le cas p\'eriodique, l'action n'est \'evidemment pas automatiquement
libre. L'ensemble $M_{sing}$ des points de $M$ ayant un
stabilisateur non trivial donne lieu \`a des difficult\'es. Nous verrons
dans le chapitre \ref{NCQFT}
que cet ensemble est en particulier responsable d'un nouveau type
de m\'elange des divergences ultraviolettes et infrarouges,
pour les th\'eories quantiques
des champs sur d\'eformations isospectrales.

\medskip

Pour \'etudier
les propri\'et\'es g\'en\'erales du produit d\'eform\'e
dans le cas non p\'eriodique, il est
suffisant d'\'etablir le r\'esultat suivant.

\begin{lem}
\label{lm:sup}
Soient $\a$ une action propre
de $\R^l$ par isom\'etrie et
$\tri_\a:=-\sum_{j=1}^l {X_j}^2$, o\`u les $X_j$ sont les champs
de vecteurs associ\'es \`a l'action infinit\'esimale. Pour tout
$k\in\N$ et tout $f\in\Coo_c(M)$, on a
$$
\sup_{p\in M} \int_{\R^l}\,d^ly\, |\a_y (\Delta_\a^kf)(p)|  < \infty.
$$
\end{lem}

\begin{proof}[Preuve]
Puisque $f\in\Coo_c(M)$, $\Delta_\a^kf$ est aussi lisse
et \`a support compact. Il est donc suffisant de traiter le cas $k=0$.
L'action \'etant propre, l'application
$p\mapsto \tilde{f}(p):=\int_{\R^l} d^ly\,|\a_y( f)(p)| $
est bien d\'efinie puisque  $\set{y \in \R^l : \a_y(p) \in \supp f}$ est
compact pour chaque $p \in M$ \cite[p.~41]{Michor} car $f$ est \`a
support compact.
Ainsi, la fonction $\tilde f(p)$ est finie
et est constante sur chaque orbite de l'action.
Soit $\pi \: M \to M/\R^l$ la projection sur l'espace des orbites.
Alors $\tilde{f}$ se factorise \`a travers~$\pi$
pour donner une application $\bar{f}$ d\'efinie par
$\bar{f}(\pi(p)) := \tilde{f}(p)$.
On obtient alors le r\'esultat car $\bar{f} \in \Coo_c(M/{\R^l},\C)$.
En effet, si $p \notin \a_{\R^l}(\supp f)$, de telle
mani\`ere que  $\pi(p)$ n'appartienne pas \`a l'ensemble compact
$\pi(\supp f)$, alors $\bar{f}(\pi(p)) = 0$.
\end{proof}

La non-localit\'e de ce produit, ayant pour cons\'equence premi\`ere
la non-pr\'eservation des supports (le produit de deux fonctions
\`a supports disjoints n'est pas nul a priori), implique que
$\big(\Coo_c(M),\Mop\big)$ ne poss\`ede pas une structure
d'alg\`ebre. En revanche, il est ais\'e de montrer
dans le cas non p\'eriodique, que $\Mop$ est un produit bilin\'eaire sur
$\Coo_c(M)$ \`a valeur dans $L^\infty(M,\mu_g)$.

\begin{lem}
\label{lem:borne}
Dans le cas d'une action lisse, propre et isom\'etrique de $\R^l$,
$\Mop$ est un produit bilin\'eaire sur $\Coo_c(M)$
\`a valeur dans $L^\infty(M,\mu_g)$.
\end{lem}
\begin{proof}[Preuve]
En effet,
avec $\th:=\det(\Th)^{1/l}$ (rappelons que l'on peut toujours supposer
$\Th$ inversible) on a
$$
\sup_{p\in M} |f\mop h(p)|\leq
(\pi\th)^{-l} \sup_{p\in M}\int d^ly|\a_y(f)(p)|\,
\sup_{p\in M}\int d^lz|\a_z(h)(p)|,
$$
qui est fini d'apr\`es le lemme \ref{lm:sup}.
\end{proof}
Simplement \`a partir de sa d\'efinition, on peut montrer que $\Mop$
est associatif dans tous les cas de figures (action p\'eriodique ou
action effective) et qu'il satisfait \`a toutes les propri\'et\'es du produit de
Moyal ordinaire.
\begin{lem}
Le produit $\Mop$ est formellement associatif. De plus, la conjugaison
complexe est une involution:
\begin{equation}
\label{pro1}
(f\Mop h)^*=h^*\Mop f^*.
\end{equation}
L'action $\a$ est toujours un automorphisme pour le produit d\'eform\'e:
\begin{equation}
\label{pro2}
\a_z(f\Mop h)=\a_z(f)\Mop\a_z(h).
\end{equation}
La r\`egle de Leibniz est satisfaite pour les g\'en\'erateurs infinit\'esimaux
de l'action:
\begin{equation}
\label{pro3}
X_j(f\Mop h)=X_j(f)\Mop h+f\Mop X_j(h),\,\,j=1,\cdots,l.
\end{equation}
\label{lem:proo}
\end{lem}
\begin{proof}[Preuve]
Pour l'associativit\'e du produit twist\'e, nous avons
d'une part:
$$
((f\Mop g)\Mop h)=(2\pi)^{-2l}
\int d^ly\,d^lz\,d^ly'\,d^lz'\,
e^{-i(<y,z>+<y',z'>)}
\a_{\thalf\Th(y+y')}(f)\,
\a_{\thalf\Th y-z'}(g)\,\a_{-z}(h).
$$
En effectuant les changements de variables $y'\to y'-y$,
$z'\to z'+\thalf\Th y$ on obtient apr\`es une int\'egration
d'ondes planes:
$$
(2\pi)^{-l}\int d^ly\,d^lz\,
e^{-i<y,z>}\,\a_{\thalf\Th y}(f)\,
\a_{-z}(g)\,\a_{-z-\thalf\Th y}(h).
$$
D'autre part nous avons:
$$
(f\Mop (g\Mop h))=(2\pi)^{-2l}
\int d^ly\,d^lz\,d^ly'\,d^lz'\,
e^{-i(<y,z>+<y',z'>)}
\a_{\thalf\Th y}(f)\,
\a_{-z+\thalf\Th y'}(g)\,\a_{-z-z'}(h),
$$
qui, apr\`es avoir effectu\'e les translations $z'\to z'-z$, $y'\to y'+2\Th^{-1}z$,
devient
$$
(2\pi)^{-l}\int d^ly\,d^lz\,
e^{-i<y,z>}\a_{\thalf\Th y+z}(f)\,
\a_{\thalf\Th y}(g)\,\a_{-z}(h).
$$
En effectuant finalement $y\to y-2\Th^{-1}z$,
$z\to z+\thalf\Th y$, on obtient le r\'esultat.

Que la conjugaison complexe soit une involution,
est une cons\'equence directe du fait qu'elle
commute avec l'action $\a$ et de la d\'efinition produit d\'eform\'e.
Pour obtenir la troisi\`eme assertion, il suffit d'utiliser
le fait que  $\{\a_z\}$ forme un groupe Ab\'elien. Etant la version
infinit\'esimale de la pr\'ec\`edente, la derni\`ere affirmation
en est une cons\'equence.
\end{proof}
Plus g\'en\'eralement, un op\'erateur diff\'erentiel
d'ordre un satisfait \`a la r\`egle de Leibniz
pour le produit d\'eform\'e si et seulement si il commute avec l'action $\a$.
Ce sera en particulier le cas pour l'op\'erateur de Dirac.

Comme dans le cas plat, l'int\'egrale avec la forme volume
Riemannienne $\mu_g$ est une trace pour le produit
twist\'e:

\begin{lem}
\label{trace}
Pour  $f,h\in\Coo_c(M)$, on a
$$
\int_M\,\mu_g\, f \Mop h  =
\int_M\,\mu_g\,h \Mop f  = \int_M \mu_g\,fh .
$$
\end{lem}

\begin{proof}[Preuve]
Il suffit de remarquer que
\begin{align*}
\int_M f \Mop h(p) \,\mu_g(p)
&=
(2\pi)^{-l} \int_M \mu_g(p)\int_{\R^{2l}} \,d^ly \,d^lz \,
e^{-i<y,z>} f((-\half\Th y)\.p)\,h(z\.p)
\\
&= (2\pi)^{-l} \int_M \mu_g(p)\int_{\R^{2l}}\,d^ly \,d^lz \,
 e^{-i<y,z>}f((-\half\Th y - z)\.p) \,h(p)
\\
&= (2\pi)^{-l} \int_M \mu_g(p)\int_{\R^{2l}}\,d^ly \,d^lz \,
 e^{-i<y,z>} f((-z)\.p) \,h(p)
\\
&= \int_M\,\mu_g(p)\, f(p) \,h(p) ,
\end{align*}
en utilisant l'isom\'etrie $p \mapsto (-z)\.p$ et la translation
$z \mapsto z - \half\Th y$.
\end{proof}

Nous reviendrons au cours de ce manuscrit sur les diff\'erentes
alg\`ebres que l'on peut construire avec le produit d\'eform\'e
dans sa forme g\'en\'erique, mais surtout avec le produit de Moyal
sur l'espace Euclidien $\R^l$. Cependant, nous allons d'ores et d\'ej\`a
mentionner l'existence de
l'alg\`ebre $\big(\B(M,\a),\Mop\big)$, car la preuve de la stabilit\'e des
\'el\'ements de $\B(M,\a)$ sous $\Mop$, utilise de
mani\`ere fondamentale ce qui doit \^etre consid\'er\'e comme
la d\'efinition correcte des int\'egrales intervenant dans la d\'efinition du produit
d\'eform\'e, \`a savoir des int\'egrales oscillantes.
\begin{defn}
Soit $\B(M,\a)$ l'espace de Fr\'echet des fonctions essentiellement
born\'ees sur $M$ et ayant toutes leurs d\'eriv\'ees, par rapport
aux g\'en\'erateurs $\{X_j\}_{j=1,\cdots,l}$ de l'action, born\'ees
$$
\B(M,\a):=\left\{f\in\Coo(M)\cap L^\infty(M,\mu_g):X^\b f\in L^\infty(M,\mu_g),
\,\forall \b\in\N^l\right\},
$$
o\`u $X^\beta:=X_1^{\b_1}\cdots X_l^{\b_l}$. L'espace $\B(M,\a)$
est muni de la topologie donn\'ee par la famille de semi-normes
$$
q_k(f):=\sup_{|\b|\leq k}\sup_{p\in M}
\big|(X^\b f)(p)\big|.
$$
\end{defn}
\begin{lem}
\label{lem:B}
$\Big(\B(M,\a),\Mop\Big)$ est une alg\`ebre de Fr\'echet
associative et involutive avec un produit jointement continu.
\end{lem}
\begin{proof}[Preuve]
En d\'efinissant le produit d\'eform\'e en terme d'int\'egrales oscillantes
\cite{RieffelDefQ, Hormander} on a avec $k\in\N$ arbitraire
\begin{align*}
\big(X^\b f\big)\Mop \big(X^\gamma h\big)&=
(2\pi)^{-l}\int d^ly\,d^lz\,\frac{\a_{\thalf\Th y}\big(X^\b f\big)}
{(1+|y|^2)^k}
\frac{\a_{-z}\big(X^\gamma h\big)}
{(1+|z|^2)^k}\,(1+|y|^2)^k\,(1+|z|^2)^k\,e^{-i<y,z>}\\
&=
(2\pi)^{-l}\int d^ly\,d^lz\,\frac{\a_{\thalf\Th y}\big(X^\b f\big)}
{(1+|y|^2)^k}
\frac{\a_{-z}\big(X^\gamma h\big)}
{(1+|z|^2)^k}\,[P_k(\pa_y,\pa_z)e^{-i<y,z>}],
\end{align*}
o\`u $P_k$ est un polyn\^ome de degr\'e $2k$ dans ses deux variables.
En utilisant ensuite
$$
|\pa^\b_y (1+|y|^2)^{-k}|\leq C_{\b,k}(1+|y|^2)^{-k},
$$
on obtient d'apr\`es la r\`egle de Leibniz
\begin{align*}
\left|\big(X^\b f\big)\Mop \big(X^\gamma h\big)\right|\leq&
\sum_{|\delta|,|\delta'|\leq2k}C_{\delta,\delta'}'\int d^ly\,d^lz\,
\left|\a_{\thalf\Th y}\big(X^{\b+\delta} f\big)\right|\,
\left|\a_{-z}\big(X^{\ga+\delta'} h\big)\right|\\
&\hspace{5cm}\x (1+|y|^2)^{-k}\,(1+|z|^2)^{-k}\\
\leq&\sum_{|\delta|,|\delta'|\leq2k}C_{\delta,\delta'}'
q_{_{|\b+\delta|}}(f)\,q_{_{|\ga+\delta'|}}(h)\left(\int d^ly\,(1+|y|^2)^{-k}\right)^2,
\end{align*}
qui est finie pour $k>l/2$. Ces estimations montrent que le produit est
s\'epar\'ement continu, donc jointement continu car $\B(M,\a)$
est un espace de Fr\'echet.
\end{proof}

\begin{rem}
Dans le cas p\'eriodique ($\ker(\a)=\Z^l$), compact ou non,
nous avons vu au paragraphe \ref{constructionCL} que
chaque fonction born\'ee et ind\'efiniment diff\'erentiable
poss\`ede une d\'ecompo\-si\-tion en \'el\'ements homog\`enes (\ref{eq:PW}).
On obtient alors pour $f,h\in\Coo(M)\cap L^\infty(M,\mu_g)$
\begin{align*}
f\Mop h&=\sum_{r,s\in\Z^l}
(2\pi)^{-l} \int_{\R^l}\int_{\R^l}d^ly \,d^lz\,
e^{-i<y,z>} \, \a_{\half\Th y}(f_r) \, \a_{-z}(h_s) \\
&=\sum_{r,s\in\Z^l}
(2\pi)^{-l} \int_{\R^l}\int_{\R^l}d^ly \,d^lz\,
e^{-i<y,z>} \, e^{-\tihalf r\Th y}\,e^{isz} \,f_r\,h_s\\
&=\sum_{r,s\in\Z^l}e^{-\tihalf \Th(r,s)}\,f_r\,h_s.
\end{align*}
La construction par produit twist\'e des d\'eformations isospectrales
p\'eriodiques, co\"incide donc avec celle de Connes--Landi.
Ce petit calcul montre aussi que dans le cas p\'eriodique non compact,
$\Coo_c(M)$ est stable par multiplication twist\'ee.
En effet, bien que non local sur les orbites de l'action, le produit
twist\'e p\'eriodique de deux fonctions
\`a support compact est encore une fonction \`a support compact, puisque les orbites
sont compactes (elles sont isomorphes \`a $\T^l$) et les directions transverses
n'interviennent pas dans la proc\'edure de d\'eformation. En d'autres termes, puisque
$\supp(f_r)\subset \T^l.\supp(f)$, on a
$$
\supp(f\mop h)\subset\bigcup_{r,s\in\Z^l}\supp(f_r)\cap\supp(h_s)
\subset\T^l.\supp(f)\cap\T^l.\supp(h).
$$
\end{rem}

\subsection{Analyse Hilbertienne du produit d\'eform\'e}
\subsubsection{1.2.2.1 Propri\'et\'es du noyau du produit twist\'e}
\label{parahil}

Dans cette partie, nous allons commencer par montrer que
l'op\'erateur $L_f$ (resp. $R_f$) de multiplication twist\'ee
\`a gauche (resp. \`a droite) pour $f\in\Coo_c(M)$ est un
op\'erateur \`a noyau par rapport \`a l'espace mesur\'e
$(M,\mu_g)$ et qu'il est born\'e. Lorsque nous voudrons
insister sur sa d\'ependance en $\Th$, nous \'ecrirons
$L^\Th_f$ (resp. $R^\Th_f$) \`a la place de $L_f$ (resp. $R_f$).
Nous adopterons aussi la notation $M_f$ pour d\'esigner l'op\'erateur
de multiplication ordinaire par $f$, correspondant au cas
limite $\Th=0$.

Soit, comme au paragraphe~\ref{constructionCL},
$\H=L^2(M,S)$ l'espace de Hilbert des sections de carr\'e
sommable du fibr\'e des spineurs et soit $\H_r=L^2(M,\mu_g)$
l'espace de Hilbert r\'eduit. Sous l'hypoth\`ese de compl\'etude
g\'eod\'esique de la vari\'et\'e, l'op\'erateur de Dirac $\Dslash$,
restreint au domaine dense $\Gamma^\infty_c(M,S)$ des sections
lisses et \`a support compact, est essentiellement auto-adjoint ~\cite{Wolf}.
Nous utiliserons alors la m\^eme notation pour d\'esigner l'op\'erateur
densement d\'efini et son unique fermeture. Il en va de m\^eme pour
le Laplacien scalaire $\tri$ (ainsi que son rel\`evement au fibr\'e des spineurs)
de domaine $\Coo_c(M)\subset\H_r$. Dans nos conventions,
$\tri=(d+\delta)^2$ est positif et restreint aux 0-formes,
$\tri=\delta d=\ast_{_H}d\ast_{_H}d$, o\`u $\ast_{_H}$ est l'\'etoile de Hodge.

Le groupe $\R^l$ \'etant simplement connexe,
\`a la diff\'erence de l'action d'un tore,
l'action $\a$ de $\R^l$ se rel\`eve directement au fibr\'e des spineurs.
Nous noterons simplement $\{V_ z:  z\in\R^l\}$
le groupe d'unitaires associ\'e au groupe d'isom\'etries
$\{\a_z:z\in\R^l\}$. On a encore $V_z(f.\psi)=
\a_z(f).V_z(\psi)$ et $\big(V_z\psi,V_z\psi'\big)=
\a_z\big(\psi,\psi'\big)$, pour tout $\psi,\psi'\in\H$, $f\in\Coo_c(M)$
et avec $\big(.,.\big)$ la structure Hermitienne du fibr\'e des spineurs.
Notons aussi que $\Dslash$, $\tri$, $X_k$ commutent avec $V_z$,
cons\'equence de l'isom\'etrie de l'action.

\begin{defn}
\label{df:delta-g}
Pour tout $p \in M$, soit $\delta_p^g \in \Coo_c(M)'$ la
distribution d\'efinie pour tout $\phi \in \Coo_c(M)$ par
$$
\langle \delta_p^g, \phi \rangle
= \int_M \delta_p^g(p') \,\phi(p') \,\mu_g(p') := \phi(p).
$$
La distribution $\delta_p^g$, repr\'esent\'ee dans un syst\`eme
de coordonn\'ees locales par
$$
(\det g(x))^{-1/2} \,\delta(x - x'),
$$
peut aussi \^etre pens\'ee comme un $n$-courant de de~Rham
\cite{Schwartz} en la multipliant par la forme volume Riemannienne; elle
est surtout le noyau, au sens de l'espace mesur\'e $(M,\mu_g)$, de
l'identit\'e:
$$
\delta_p^g(p')=K_{\bf 1}(p,p').
$$
\end{defn}
\noindent Une cons\'equence de la propri\'et\'e d'isom\'etrie de l'action,
que nous utiliserons largement, est l'identit\'e
$$
\delta^g_{z\cdot p}(p')=\delta^g_p(-z\cdot p').
$$

\begin{defn}
\label{df:left-multn}
Pour $f \in \Coo_c(M)$, les op\'erateurs de multiplication twist\'ee
\`a gauche $L_f$ et \`a droite $R_f$ sur $\H$,
sont d\'efinis par
\begin{equation}
\label{eq:left-multn}
L_f \psi(p) :=(2\pi)^{-l} \int_{\R^l} \int_{\R^l} \,d^ly \,d^l z\,
e^{-iy z}
\a_{\half\Th y}(f)(p) \,V_{-z}\psi(p) ,
\end{equation}
et
\begin{equation}
\label{eq:right-multn}
R_f \psi(p) :=(2\pi)^{-l} \int_{\R^l} \int_{\R^l} \,d^ly \,d^lz\,
e^{-iz y}
\a_{-z}(f)(p) \,V_{\thalf\Th y}\psi(p) .
\end{equation}
\end{defn}

Nous allons aussi tirer profit du fait que $L_f$ et $R_f$
peuvent \^etre alternativement d\'efinis en terme d'int\'egrales
\`a valeur op\'erateur:
\begin{equation}
\label{Lfint}
L_f=(2\pi)^{-l}\int_{\R^{2l}}\,d^ly\,d^lz\;
e^{-i<y,z>}\;V_{\thalf \Th y}\,M_f\,V_{-z},
\end{equation}
\begin{equation}
\label{Rfint}
R_f=(2\pi)^{-l}\int_{\R^{2l}}\,d^ly\,d^lz\;
e^{-i<y,z>}\;V_{-z}\,M_f\,V_{\thalf \Th y}.
\end{equation}
Les formules (\ref{Lfint}) et (\ref{Rfint}) peuvent \^etre facilement
obtenues \`a partir de (\ref{eq:left-multn}) et (\ref{eq:right-multn})
en utilisant
$$
V_zM_fV_{-z}=M_{\a_z(f)},
$$
ainsi que la translation $z\to z-\thalf\Th y$ qui laisse la phase inchang\'ee
car $\Th$ est antisym\'etrique.

N\'eanmoins, ces int\'egrales ne sont pas d\'efinies au sens de Bochner, mais
sont plut\^ot des int\'egrales oscillantes \`a valeur dans $\L(\H)$. En effet,
il est facile de voir que la norme (op\'eratorielle) de l'int\'egrant n'est pas
une fonction sommable sur $\R^{2l}$, car ne d\'ependant de $y,z$ qu'au
travers d'op\'erateurs unitaires.

Nous allons voir que ces
derni\`eres formules permettent de d\'efinir les twists gauche et droit
pour une classe d'op\'erateurs born\'es beaucoup plus importante.\\
Avec cette pr\'esentation, il est aussi quasiment direct de
v\'erifier que $L$ et $R$ sont deux repr\'esenta\-tions qui commutent
(en fait $R$ est une anti-repr\'esentation):
$$
[L_f,R_h]=0,\;\;\forall f,h\in C^\infty_c(M).
$$
Les formules (\ref{Lfint}) et (\ref{Rfint}) fournissent alors un autre moyen
de v\'erifier l'associativit\'e du produit d\'eform\'e, l'associativit\'e \'etant
\'equivalente \`a la commutation de $L$ avec $R$.

En utilisant la propri\'et\'e de trace (lemme \ref{trace}), on peut aussi montrer
que l'adjoint de $L_f$ (resp. $R_f$) est \'egal \`a $L_{f^*}$ (resp. $R_{f^*}$).
Ce r\'esultat peut \^etre encore plus simplement \'etabli en
utilisant les formules~(\ref{Lfint})
et~(\ref{Rfint}). Pour $L_f$ on obtient
\begin{align*}
(L_f)^*&=(2\pi)^{-l}\int_{\R^{2l}}\,d^ly\,d^lz\;
e^{i<y,z>}\;V_{z}\,M_{f^*}\,V_{-\thalf\Th y}\\
&=(2\pi)^{-l}\int_{\R^{2l}}\,d^ly\,d^lz\;
e^{-i<y,z>}\;V_{\thalf \Th z}\,M_{f^*}\,V_{-y}=L_{f^*},
\end{align*}
o\`u les changements de variables $z\to \thalf\Th z$, $y\to 2\Th^{-1}y$
et la relation $<\Th^{-1}y,\Th z>=-<y,z>$ ont \'et\'e utilis\'es.

Nous allons \'etablir une estimation pour la norme
des op\'erateurs $L_f$, $R_f$ dans le cas d'une action propre de $\R^l$.
Au chapitre \ref{Moyal}, nous verrons que dans le cas non d\'eg\'en\'er\'e
(caract\'eris\'e par $n=l$, $\Th$ inversible), on peut obtenir de plus fines estimations sur $\|L_f\|$,
en utilisant une caract\'erisation matricielle du produit d\'eform\'e.

\begin{prop}
\label{pr:kernel}
Lorsque $f \in \Coo_c(M)$, alors $L_f$ et $R_f$, dans le cas d'une
action propre et isom\'etrique de $\R^l$, sont des op\'erateurs \`a
noyau par rapport \`a l'espace mesur\'e $(M,\mu_g)$ et sont
born\'es sur $L^2(M,\mu_g)$. De plus, on a l'expression suivante
pour leur noyau distributionnel:
\begin{equation}
\label{eq:noyauxL}
K_{L_f}(p,p') = (2\pi)^{-l} \int_{\R^l\x\R^l} \,d^ly \,d^lz\,e^{-i<y,z>}
f((-\half\Th y)\.p) \,\delta_{z\.p}^g(p') .
\end{equation}
\begin{equation}
\label{eq:noyauxR}
K_{R_f}(p,p') = (2\pi)^{-l} \int_{\R^l\x\R^l}\,d^ly \,d^lz\, e^{-i<y,z>}
f(z\.p) \,\delta_{(-\half\Th y)\.p}^g(p') .
\end{equation}
\end{prop}
\begin{proof}[Preuve]
Le cas $R_f$ \'etant tout \`a fait similaire, nous ne traiterons que
celui de $L_f$.
Soit $\psi \in \H$; par d\'efinition
\eqref{eq:left-multn}, nous avons
$$
L_f \psi(p) = (2\pi)^{-l} \int_{\R^l} \int_{\R^l}\,d^ly \,d^lz\, e^{-i<y,z>}\,
\a_{\half\Th y}(f)(p) \int_M \,\mu_g(p')\,\delta^g_{z\.p}(p') \,\psi(p').
$$
L'expression du noyau (\ref{eq:noyauxL}) est alors obtenue
par permutation des int\'egrales.\\
Que $L_f$ soit born\'e est une cons\'equence
de sa d\'efinition en terme d'int\'egrales oscillantes
\cite{RieffelDefQ,Hormander,Amalthea} \`a valeur op\'erateur:
\begin{align*}
L_f \psi(p)
&= (2\pi)^{-l} \int_{\R^l} \int_{\R^l} \,d^ly \,d^lz\,
e^{-i<y,z>} \,\a_{\half\Th y}(f)
\,V_{-z}\psi(p)
\\
&= (2\pi)^{-l} \int_{\R^l} \,d^lz\,(1+|z|^2)^{-r} \int_{\R^l}d^ly \, (1+|z|^2)^r
\,e^{-i<y,z>} \,\a_{\half\Th y}(f) \,V_{-z}\psi(p)
\\
&= (2\pi)^{-l} \int_{\R^l}\,d^lz\, (1+|z|^2)^{-r} \int_{\R^l}\,d^ly\,
((1+\Delta_\a)^r \,e^{-i<y,z>}) \,\a_{\half\Th y}(f)
\,V_{-z}\psi(p)
\\
&= (2\pi)^{-l} \int_{\R^l} \,d^lz\,(1+|z|^2)^{-r} \int_{\R^l}\,d^ly \, e^{-i<y,z>}\,
((1+\Delta_\a)^r \a_{\half\Th y}(f)) \,V_{-z}\psi(p) ,
\end{align*}
o\`u $\Delta_\a := -\sum_{j=1}^l X_j^2$. Nous avons utilis\'e ici
le fait que les d\'eriv\'ees totales ne contribuent pas \`a l'expression
pr\'ec\'edente car $f$ est \`a support compact. Alors,
\begin{equation}
\label{eq:borne}
\|L_f \psi\| \leq (2\pi)^{-l} \|\psi\|\,
\biggl( \int_{\R^l}\,d^lz\, (1+|z|^2)^{-r}  \biggr) \sup_{p\in M}
\int_{\R^l} \,d^ly\,|(1+\Delta_\a)^r \a_{\half\Th y}(f)(p)| .
\end{equation}
Cette expression est finie d'apr\`es le lemme~\ref{lm:sup} et pour $r > l/2$.
\end{proof}

\begin{rem}
Dans le cas p\'eriodique (compact ou non), que $L_f$ soit
born\'e pour une fonction $f\in\Coo_c(M)$, est une cons\'equence de la
d\'ecomposition de $f$ en \'el\'ements homog\`enes. D'une part on a $L_{f_r}=
M_{f_r}V_{-\thalf\Th r}$ et d'autre part:
$$
\|L_f\|\leq \sum_{r\in\Z^l}\|M_{f_r}V_{-\thalf\Th r}\|
\leq \sum_{r\in\Z^l}\|f_r\|_\infty<\infty.
$$
On obtient dans ce cas pour le noyau de Schwartz
de l'op\'erateur $L_f$:
$$
K_{L_f}(p,p')=\sum_{r\in\Z^l}f_r(p)\,\delta^g_{\thalf\Th r.p}(p').
$$
Evidemment, des expressions analogues pour les op\'erateurs de multiplication
twist\'ee \`a droite sont valables.
\end{rem}

\subsubsection{1.2.2.2 Invariance de la norme de Hilbert--Schmidt}

Dans ce paragraphe, nous allons \'etablir un r\'esultat central:
{\it la norme de Hilbert--Schmidt est un invariant des d\'eformations
isospectrales}.

Tout d'abord, nous avons besoin d'obtenir des propri\'et\'es
d'invariance pour les noyaux des op\'erateurs du type $h(\Dslash)$,
avec  $h$ une fonction r\'eelle, lisse et born\'ee.

\begin{lem}
\label{lm:invkernel}
Pour tous $z \in \R^l$, $p,p' \in M$, $h$ fonction r\'eelle,
lisse et born\'ee, le noyau $K_{h(\Dslash)}$
est invariant sous $\a$:
$$
K_{h(\Dslash)}(z\.p, z\.p') = K_{h(\Dslash)}(p,p'),
$$
presque partout sur $(M \x M,\mu_g\x\mu_g)$.
\end{lem}

\begin{proof}[Preuve]
C'est une cons\'equence directe de la propri\'et\'e d'isom\'etrie
de l'action. En effet,
l'invarian\-ce de la connection de Levi-Civita associ\'ee \`a~$g$, implique
l'invariance de la connection de spin sous le rel\`evement de l'action au
fibr\'e des spineurs. Ainsi $V_z \Dslash V_{-z} = \Dslash$ pour tout $z$.
Si $h$ est une fonction r\'eelle et born\'ee sur~$\R$, on obtient alors
par calcul fonctionnel
$[V_z, h(\Dslash)] = 0$ pour $z \in \R^l$. De plus, pour
$\psi \in \H$, l'invariance de la forme volume Riemannienne
sous le diff\'eomorphisme $p'\mapsto \a_{-z}(p')$ donne
$$
\int_M \,\mu_g(p')\,K_{h(\Dslash)}(z\.p, z\.p') \,\psi(p')
= \int_M \,\mu_g(p')\,K_{h(\Dslash)}(z\.p, p') \,\psi((-z)\.p') .
$$
Le terme de droite est \'egal \`a
$(h(\Dslash) V_z\psi) (z\.p) = (V_{-z} h(\Dslash) V_z\psi)(p)
= (h(\Dslash)\psi)(p)$.
Alors, $K_{h(\Dslash)}(\a_z(\.),\a_z(\.))$ et $K_{h(\Dslash)}(\.,\.)$
repr\'esentent le m\^eme op\'erateur sur~$\H$.
\end{proof}

Le th\'eor\`eme suivant
montre que la norme de Hilbert--Schmidt de $L_f\, h(\Dslash)$
est ind\'ependante des param\`etres de d\'eformation.

\begin{thm}
\label{th:HS-norm}
Soient $f \in \Coo_c(M)$ et $h$ une fonction lisse, r\'eelle et born\'ee sur~$\R$,
telle que $M_f\, h(\Dslash)$ soit un op\'erateur de Hilbert--Schmidt. Alors, lorsque
l'action $\a$ est lisse, propre et isom\'etrique
les op\'erateurs $L_f^\Th\, h(\Dslash)$, $R_f^\Th\, h(\Dslash)$
sont aussi de Hilbert--Schmidt, avec
$$
\|L_f^\Th\, h(\Dslash)\|_2 =\|R_f^\Th\, h(\Dslash)\|_2 = \|M_f\, h(\Dslash)\|_2.
$$
\end{thm}

\begin{proof}[Preuve]
Les arguments des cas $L_f$ et $R_f$ \'etant tout \`a fait semblables, nous ne
traiterons que le cas de la multiplication twist\'ee \`a gauche.
Tout d'abord et d'apr\`es la proposition \ref{pr:kernel}, on peut repr\'esenter
le noyau de $L_f\, h(\Dslash)$ en terme du noyau $K_{h(\Dslash)}$:
\begin{align*}
K_{L_f h(\Dslash)}(p,p')
&= \int_M\,\mu_g(q)\, K_{L_f }(p,q) K_{h(\Dslash)}(q,p')
\\
&= (2\pi)^{-l} \int_M \mu_g(q)\int_{\R^{2l}} \,d^ly \,d^lz \,e^{-i<y,z>}
f((-\half\Th y)\.p)\, \delta^g_{z\.p}(q)\, K_{h(\Dslash)}(q,p')
\\
&= (2\pi)^{-l} \int_{\R^{2l}}\,d^ly \,d^lz\, e^{-i<y,z>} f((-\half\Th y)\.p)\,
K_{h(\Dslash)}( z\.p,p') .
\end{align*}
Ainsi,
\begin{align*}
 \|L_f\, h(\Dslash)\|_2^2
&= \int_{M\x M}\,\mu_g(p)\,\mu_g(p')\, |K_{L_f h(\Dslash)}(p,p')|^2
\\
&= (2\pi)^{-2l} \int_{M\x M} \,\mu_g(p)\,\mu_g(p')\int_{\R^{4l}}
\,d^ly_1 \,d^lz_1 \,d^ly_2 \,d^lz_2\,
e^{i(y_1 z_1 - y_2 z_2)}\\
&\qquad \x\bar{f}((-\half\Th y_1)\.p) \, f((-\half\Th y_2)\.p)
\overline{K_{h(\Dslash)}}(z_1\.p, p') \, K_{h(\Dslash)}(z_2\.p, p')
\\
&= (2\pi)^{-2l} \int_{M\x M} \,\mu_g(p)\,\mu_g(p')\int_{\R^{4l}}
\,d^ly_1 \,d^lz_1 \,d^ly_2 \,d^lz_2\,
e^{i(y_1 z_1 - y_2 z_2)}
\bar{f}((-\half\Th y_1 - z_2)\.p) \\
&\qquad \x f((-\half\Th y_2 - z_2)\.p)
\overline{K_{h(\Dslash)}}((z_1-z_2)\.p, (z_1-z_2)\.p') \,
K_{h(\Dslash)}(p, (z_1-z_2)\.p') ,
\end{align*}
o\`u l'invariance de~$\mu_g$ sous les isom\'etries
$p \mapsto (-z_2)\.p$ et $p' \mapsto (z_1 - z_2)\.p'$ a \'et\'e
utilis\'ee. En\-suite, d'apr\`es le
lemme~\ref{lm:invkernel}, en effectuant la translation
$z_1 \mapsto z_1 + z_2$, la derni\`ere expression devient
\begin{align*}
&(2\pi)^{-2l} \int_{M\x M}\,\mu_g(p)\,\mu_g(p') \int_{\R^{4l}}
\,d^ly_1 \,d^lz_1 \,d^ly_2 \,d^lz_2\,
e^{i(y_1 (z_1+z_2) - y_2 z_2)} \\
&\qquad\qquad \x\, \bar{f}((-\half\Th y_1 - z_2)\.p)
\, f((-\half\Th y_2 - z_2)\.p)
\overline{K_{h(\Dslash)}}(p, p') \, K_{h(\Dslash)}(p, z_1\.p'),
\\
&= (2\pi)^{-2l} \int_{M\x M} \,\mu_g(p)\,\mu_g(p')\int_{\R^{4l}}
\,d^ly_1 \,d^lz_1 \,d^ly_2 \,d^lz_2
e^{i((y_1-2\Th^{-1}z_2) (z_1+z_2) - y_2 z_2)} \\
&\qquad\qquad \x\,
\bar{f}((-\half\Th y_1)\.p) \, f((-\half\Th y_2)\.p)
\overline{K_{h(\Dslash)}}(p, p') \, K_{h(\Dslash)}(p, z_1\.p') ,
\end{align*}
en faisant les changements $y_1 \mapsto y_1 - 2\Th^{-1}z_2$ et
$y_2 \mapsto y_2 - 2\Th^{-1}z_2$. On obtient alors,
\begin{align*}
& (2\pi)^{-l} \int_{M\x M} \mu_g(p)\,\mu_g(p')\int_{\R^{2l}}
\,d^ly \,d^lz \,e^{i<y,z>} \,\\
&\qquad\qquad \x\bar{f}((-\half\Th y)\.p) \, f((-\half\Th y - z)\.p)
\overline{K_{h(\Dslash)}}(p, p') \, K_{h(\Dslash)}(p, z\.p')
\\
&= (2\pi)^{-l} \int_{M\x M} \mu_g(p)\,\mu_g(p')\int_{\R^{2l}}
\,d^ly \,d^lz \,e^{i<y,z>} \,
\bar{f}(p) \, f((-z)\.p) \,
\overline{K_{h(\Dslash)}}(p, p') \, K_{h(\Dslash)}(p, z\.p')\\
&= \int_{M\x M} \,\mu_g(p)\,\mu_g(p')
|f(p)|^2 |K_{h(\Dslash)}(p,p')|^2
= \|M_f\, h(\Dslash)\|_2^2.
\end{align*}
Dans la deuxi\`eme \'egalit\'e, nous avons utilis\'e l'invariance de la forme volume
ainsi que celle du noyau $K_{h(\Dslash)}$, sous les isom\'etries
$p \mapsto (\half\Th y)\.p$ et $p' \mapsto (\half\Th y)\.p'$.
\end{proof}

\begin{rem}
Nous avons d\'emontr\'e ce r\'esultat dans le cas spinoriel, mais
il est \'evidem\-ment aussi valable dans le cas scalaire, i.e. lorsque $L_fh(\tri)$
agit sur l'espace de Hilbert r\'eduit $\H_r$. On obtient alors:
$$
\|L_fh(\tri)\|_2=\|R_fh(\tri)\|_2=\|M_fh(\tri)\|_2.
$$
\end{rem}

Nous avons eu besoin de supposer ici que
$M_f\, h(\Dslash)$ appartienne \`a l'id\'eal
des op\'erateurs de Hilbert--Schmidt. Nous verrons au paragraphe
\ref{section26} des conditions suffisantes sur la fonction $h$, pour
r\'ealiser cette hyphoth\`ese.

Dans la suite, nous allons utiliser la pr\'esentation de $L_f$ et $R_f$ en
termes d'int\'egrales \`a valeur op\'erateur (\ref{Lfint})
et (\ref{Rfint}) pour obtenir une deuxi\`eme d\'emonstration du th\'eor\`eme
pr\'ec\'edent, avec une formulation l\'eg\`erement diff\'erente:
$h(\Dslash) \, L_f \, h(\Dslash)$ est \`a trace si et seulement si l'op\'erateur `non d\'eform\'e'
$h(\Dslash) \, M_f $ $ h(\Dslash)$ l'est et leurs traces co\"incident.
Avant de donner cette deuxi\`eme  preuve, il est int\'eressant de comprendre
l'origine de ce r\'esultat au travers du calcul heuristique suivant:
\begin{align}
\label{heuristique}
\Tr(h(\Dslash)\, L_f \, h(\Dslash))
&= (2\pi)^{-l}\Tr\biggl( \int_{\R^{2l}} \,d^ly \,d^lz\,e^{-i<y,z>}\,
h(\Dslash) V_{\half\Th y}\, M_f \,V_{-\half\Th y-z} h(\Dslash)
 \biggr)\nonumber\\
&= (2\pi)^{-l} \int_{\R^{2l}} \,d^ly \,d^lz\,e^{-i<y,z>} \Tr\bigl(
V_{\half\Th y} h(\Dslash)\, M_f \,h(\Dslash) V_{-\half\Th y-z} \bigr)
\nonumber\\
&= (2\pi)^{-l} \int_{\R^{2l}} \,d^ly \,d^lz\,e^{-i<y,z>} \Tr\bigl(
h(\Dslash)\, M_f \,h(\Dslash)\, V_{-z} \bigr)
\nonumber\\
&= \Tr\biggl( h(\Dslash)\, M_f \,h(\Dslash) \int_{\R^l}\,d^lz\,
\delta_0(z)\, V_{-z}  \biggr)
\nonumber\\
&= \Tr(h(\Dslash)\, M_f \,h(\Dslash)).
\end{align}

\begin{rem}
Nous allons voir que ces manipulations
(inversion de la trace et de l'int\'egra\-le)
peuvent \^etre justifi\'ees en introduisant un r\'egularisateur
fortement convergent. Cependant, avec la trace de Dixmier
\`a la place de la trace ordinaire (avec des contraintes bien pr\'ecises
sur la fonction $h$), ces manipulations restent formelles.
En effet, tous les r\'egularisateurs
naturels donnent lieu \`a des op\'erateurs \`a trace, et donc de
trace de Dixmier nulle.
\end{rem}

{\it Deuxi\`eme preuve du th\'eor\`eme~\ref{th:HS-norm}.}
Ici, nous rempla\c cons
l'hypoth\`ese $M_f \,h(\Dslash)$ est Hilbert--Schmidt par $h(\Dslash)\,
M_f \,h(\Dslash)$ est \`a trace, pour $f=g^*\Mop g$ et
$g\in\Coo_c(M)$.

Soit $\{u_k\}_{k\in\N}$ un syst\`eme uniform\'ement born\'e d'unit\'es
approch\'ees pour $\Coo_c(M)$, c'est-\`a-dire,
une famille croissante de fonctions positives \`a support compact
telles que $u_k \uparrow 1$ point \`a point sur $M$. En particulier, ceci implique que
$\slim M_{u_k}= 1$. De plus, $M_{u_k}\,h(\Dslash)$ est Hilbert--Schmidt
pour tout~$k$ car $u_k$ est \`a support compact (cf. proposition \ref{pr:HiSc}).
Consid\'erons les op\'erateurs positifs r\'egularis\'es
$$
0\leq A_k := M_{u_k} \, h(\Dslash)\, L_f \,h(\Dslash)\, M_{u_k} .
$$
Etant donn\'e que $M_{u_k}\,h(\Dslash)$ est Hilbert--Schmidt, $A_k$ est
\`a trace et
\begin{align*}
A_k &= (2\pi)^{-l}\int_{\R^{2l}}\,d^ly \,d^lz\, e^{-i<y,z>} \, M_{u_k} \, h(\Dslash)
V_{\half\Th y} \, M_f \, V_{-\half\Th y-z} \, h(\Dslash) \, M_{u_k}
\\
&=:\int_{\R^{2l}} \,d^ly \,d^lz\,A_k(y,z)
\end{align*}
est bien d\'efinie comme une int\'egrale d'op\'erateurs \`a trace.
De plus, comme $A_k$ et
$A_k(y,z)$ sont \`a trace, on peut \'echanger la trace et l'int\'egrale,
pour obtenir en utilisant $[h(\Dslash),V_z]=0$
$$
\Tr (A_k) = (2\pi)^{-l}\int_{\R^{2l}}\,d^ly \,d^lz\, e^{-i<y,z>} \Tr\bigl(
h(\Dslash) \,M_f\, h(\Dslash) \, V_{-\half\Th y-z} \, M_{u_k^2} \,
V_{\half\Th y}
\bigr) .
$$
De m\^eme,
$$
\int_{\R^{2l}} \, d^ly \,d^lz\,e^{-i<y,z>} \, h(\Dslash) \, M_f \,h(\Dslash) \,
V_{-\half\Th y-z} \, M_{u_k^2} \, V_{\half\Th y}
$$
est une int\'egrale o\`u l'int\'egrant est \`a trace, car $h(\Dslash)\, M_f \, h(\Dslash)$
est \`a trace par hypoth\`ese et
\begin{equation}
\label{eq:int-rmult}
\int_{\R^{2l}} \, d^ly \,d^lz\,e^{-iy.z}\, V_{-\half\Th y-z} \, M_{u_k^2} \, V_{\half\Th y}
 = R_{u_k^2}
\end{equation}
est born\'e d'apr\`es la proposition~\ref{pr:kernel}.
Ainsi, $\Tr(A_k)
= \Tr\bigl( h(\Dslash)\, M_f \, h(\Dslash) \, R_{u_k^2} \bigr)$,
et donc
\begin{equation}
\label{eq:cont}
\|A_k\|_1
\leq \|R_{u_k^2}\| \, \|h(\Dslash)\, M_f \,h(\Dslash)\|_1 .
\end{equation}
Etant donn\'e que $\{M_{u_k^2}\}_{k\in\N}$ est un
syst\`eme d'unit\'es approch\'ees uniform\'ement born\'e, par
la proposition 2.18 \cite{RieffelDefQ}, $\{R_{u_k^2}\}_{k\in\N}$ est aussi
un syst\`eme d'unit\'es approch\'ees uniform\'ement born\'e. Alors
$\|R_{u_k^2}\| \leq C$, d'o\`u $\|A_k\|_1\leq C \, \|h(\Dslash) \,
M_f \, h(\Dslash)\|_1$ pour tout $k$. Nous avons ensuite besoin de montrer
que $\slim A_k=A:=h(\Dslash) \, L_f \, h(\Dslash)$, pour garantir la tra\c cabilit\'e
de $A$ car d'apr\`es la proposition 2 de \cite{DeSi}, la limite d'une famille fortement
continue d'op\'erateurs uniform\'ement born\'es en norme trace, est aussi
\`a trace. Nous avons
$$
A-A_k=A-M_{u_k}AM_{u_k}=(1-M_{u_k})A+M_{u_k}A(1-M_{u_k}).
$$
Puisque $\|M_{u_k}\|\leq1$, on obtient pour tout $\psi\in\H$
$$
\|(A-A_k)\psi\|\leq \|(1-M_{u_k})A\psi\|+
\|A\|\|(1-M_{u_k})\psi\|.
$$
Cela conclut la preuve, car $\slim M_{u_k}=1$.
\qed

La propri\'et\'e d'invariance de la norme de Hilbert--Schmidt peut \^etre
g\'en\'eralis\'ee de la mani\`ere suivante. Pour un op\'erateur born\'e $A$,
on peut d\'efinir formellement ses twists gauche et droit par
$$
L_A^\Th:= (2\pi)^{-l} \int_{\R^{2l}} \,d^ly \,d^lz \,e^{-i<y,z>} \,
V_{\half\Th y} \,A\, V_{-z}  ,
$$
$$
R_A^\Th:= (2\pi)^{-l} \int_{\R^{2l}} \,d^ly \,d^lz \,e^{-i<y,z>} \,
V_{-z} \,A\, V_{\half\Th y}  .
$$
Nous allons montrer que ces expressions peuvent \^etre bien d\'efinies,
par exemple pour les op\'erateurs de Hilbert--Schmidt.
\begin{thm}
Soit $A$ un op\'erateur de Hilbert--Schmidt, alors $L_A^\Th$ et $R_A^\Th$
sont aussi Hilbert--Schmidt avec
$$
\|L_A^\Th\|_2=\|R_A^\Th\|_2=\|A\|_2 \, .
$$
\end{thm}
\begin{proof}[Preuve]
La preuve \'etant similaire \`a celle du th\'eor\`eme~\ref{th:HS-norm},
nous esquisserons uniquement le cas de $L_A$.
D'apr\`es \cite[th\'eor\`eme 2.11]{SimonTrace}, le noyau $K_A$ de $A$ est
une fonction appartenant \`a $L^2(M\times
M,\mu_g\times\mu_g)$. De plus, nous obtenons l'expression du noyau
de $L_A$ en terme de celui de $A$:
$$
K_{L_A}(p,p')=(2\pi)^{-l}\int_{\R^{2l}}\, d^ly\,d^lz \,e^{-iy.z}\,
K_A(\thalf\Th y.p,z.p')  .
$$
Un calcul tout \`a fait semblable \`a celui de la preuve
du th\'eor\`eme~\ref{th:HS-norm}, montre alors que
l'application $K_A\mapsto K_{L_A}$
est une isom\'etrie sur $L^2(M\times
M,\mu_g\times\mu_g)$, c'est-\`a-dire:
\begin{align}
\|L_A\|_2&=\int_{M\times M} \,\mu_g(p)\,\mu_g(p')\,|K_{L_A}(p,p')|^2\\
&=\int_{M\times M} \,\mu_g(p)\,\mu_g(p')\,|K_A(p,p')|^2=\|A\|_2.
\tag*{\qed}
\end{align}
\hideqed
\end{proof}

\chapter{Triplets spectraux sans unit\'e}
\label{Moyal}
\section{G\'eom\'etries spinorielles non commutatives}
\subsection{La notion d'espace Riemannien \`a spin non commutatif}

Une des notions centrales en g\'eom\'etrie non commutative est celle
de triplet spectral; {\it de vari\'et\'e Riemannienne
\`a spin non commutative}.

Pour d\'efinir et construire des espaces m\'etriques
\`a spin non commutatifs, Connes a reformul\'e en terme
alg\'ebrique et op\'eratoriel \cite{ConnesReal, ConnesCollege}
la notion de vari\'et\'e ordinaire. Cette
reformulation va ensuite permettre une g\'en\'eralisation (quasi
directe) au monde non commutatif.\\
Dans ce chapitre, nous allons tout d'abord
revoir le cadre axiomatique qui a \'et\'e
d\'evelopp\'e pour reconstruire \`a partir de donn\'ees op\'eratorielles
les structures topologique, diff\'erentielle, m\'etrique et spin.

Une vari\'et\'e non commutative compacte \`a spin,
consiste en un triplet $(\A,\H,\D)$ (ainsi que deux autres
op\'erateurs externes $\chi$ et $J$), sujet aux sept conditions
propos\'ees dans \cite{ConnesReal}. Ici, $\A$ est une
$*$-alg\`ebre \`a \'el\'ement unit\'e fid\`element repr\'esent\'ee
sur un espace de Hilbert $\H$ (nous noterons
$\pi$ la repr\'esentation) et $\D$
est un op\'erateur non born\'e et essentiellement auto-adjoint,
jouant le r\^ole d'un op\'erateur de Dirac abstrait.
De plus, le triplet doit \^etre tel que chacun des commutateurs
$[\D,\pi(a)]$, avec $a\in\A$,
s'\'etende\ en un op\'erateur born\'e. Il est \'evidemment entendu
que $\pi(a)\Dom(\D)\subset\Dom(\D)$. Le triplet ainsi d\'efini
est dit pair, s'il existe de plus une $\Z_2$-graduation $\chi$ de l'espace de
Hilbert, commutant avec les \'el\'ements de l'alg\`ebre repr\'esent\'ee
et anticommutant avec l'op\'erateur $\D$. Le dernier ingr\'edient est un op\'erateur
antilin\'eaire $J$, l'involution de Tomita--Takesaki.

Dans le cas commutatif, c'est-\`a-dire lorsque l'on veut reconstruire
une vari\'et\'e Riemannienne \`a spin ordinaire $(M,g,S)$, c'est l'alg\`ebre
$\Coo(M)$ qui va permettre de reconstruire l'espace topologique
sous-jacent \`a la vari\'et\'e $M$. En effet, l'alg\`ebre $C(M)$ des
fonctions continues, la $C^*$-compl\'etion de $\Coo(M)$
par rapport \`a la norme $\|.\|_\infty$, est en dualit\'e via le
th\'eor\`eme de Gelfand--Naimark avec l'espace de ses caract\`eres
($*$-homomorphismes d'alg\`ebres entre $\Coo(M)$ et $\C$),
qui se trouve \^etre (en tant qu'espace localement compact) isomorphe \`a $M$.\\
La structure Riemannienne va pouvoir quant \`a elle, \^etre reconstruite
via la formule des distances;
en prenant pour $\D$ l'op\'erateur de Dirac $\Dslash$ associ\'e \`a la
m\'etrique $g$, on obtient la distance g\'eod\'esique via la formule
\begin{equation}
\label{distance}
d_g(p,p')=\sup_{f\in\Coo(M)}\left\{|f(p)-f(p')|:
\|[\Dslash,M_f]\|\leq 1\right\}.
\end{equation}
Ici $\Dslash$ est vu comme op\'erant sur $L^2(M,S)$, l'espace de Hilbert
des sections de carr\'e sommable du fibr\'e des spineurs, sur lequel
$\Coo(M)$ est aussi repr\'esent\'ee par op\'erateur de multiplication
point \`a point.
Rappelons que la distance g\'eod\'esique $d_g(.,.)$ est usuellement
d\'efinie par
\begin{equation}
\label{distance2}
d_g(p,p'):=\inf_\ga\left\{l_g(\ga),\,\ga(0)=p,\,\ga(1)=p'
\right\},
\end{equation}
o\`u $\ga:[0,1]\to M$ est une courbe lisse par morceaux et $l_g(\ga)$ est
la longueur (Riemannienne) de cette courbe:
$$
l_g(\ga):=\int_0^1|\dot{\ga}(t)|dt
:=\int_0^1\sqrt{g\big(\dot{\ga}(t),\dot{\ga}(t)\big)}dt.
$$
Il est important de remarquer qu'\`a la diff\'erence de la formule
(\ref{distance2}), la formule des distances de Connes (\ref{distance})
se g\'en\'eralise pour des espaces non commutatifs.
On peut en effet d\'efinir une notion de distance, associ\'ee \`a la
structure Riemannienne abstraite donn\'ee par l'op\'erateur $\D$, sur l'espace
des \'etats purs (points extr\'emaux de l'espace convexe des fonctionnelles
positives et normalis\'ees) d'une $C^*$-alg\`ebre $A$ par
\begin{equation}
\label{distance3}
d(\psi,\phi)=\sup_{a\in A}\left\{|\psi(a)-\phi(a)|:
\|[\D,a]\|\leq 1\right\},
\end{equation}
pour tout $\psi,\phi$, \'etats purs de $A$.\\
Finalement, lorsque la dimension $n$ de $M$ est paire $n=2m$,
la graduation $\chi$ est l'\'equivalent $n$-dimensionnelle de la matrice $\ga_5$ en
dimension $4$
$$
\chi=(-i)^m\ga^1\cdots\ga^{2m},
$$
et $J$ est la conjugaison de charge pour les spineurs.\\
Les autres structures peuvent \^etre aussi reconstruites,
dans le cas compact et sans bord \cite{ConnesReal,ConnesCollege,
Polaris, RennieSpin}, \`a condition que le
triplet $\big(\Coo(M),L^2(M,S),\Dslash,
J,\chi\big)$ satisfasse \`a des contraintes suppl\'ementaires.
Nous allons revoir ces conditions dans le cas g\'en\'eral compact (\`a unit\'e) et nous
discuterons les points \`a modifier pour qu'elles soient applicables
aussi dans le cas non compact (sans unit\'e).

Soit $\K(\H)$ l'id\'eal des op\'erateurs compacts sur $\H$, et
soit $\L^p(\H)$, $p\geq 1$, les classes de Schatten ordinaires,
i.e.
$$
\L^p(\H):=\{T\in\K(\H): \|T\|_p:=\big(\Tr(|T|^p)\big)^{1/p}
<\infty\}.
$$
L'op\'erateur $\D$ doit de plus
\^etre \`a r\'esolvante compacte, c'est-\`a-dire
$(\D-\lambda)^{-1}\in\K(\H)$ pour tout $\lambda$ en dehors
du spectre de $\D$. Cette premi\`ere condition oblige \`a
se restreindre \`a des op\'erateurs ayant un spectre purement ponctuel,
i.e. consistant de valeurs propres discr\`etes de multiplicit\'es finies.
Cette condition caract\'erise, avec le fait que l'alg\`ebre $\A$
poss\`ede une unit\'e,
les espaces non commutatifs ``compacts'' et assure aussi que la classe de
K-homologie du triplet est bien d\'efinie\cite{HigsonR}.

La condition de sommabilit\'e, ou encore axiome de dimension, stipule
que $(\D-\lambda)^{-1}$ soit non seulement compact mais qu'il
appartienne \`a la $k$-i\`eme classe de Schatten faible $\L^{(k,\infty)}(\H)$,
pour un certain $k\in\N$.
Rappelons que ces id\'eaux d'op\'erateurs compacts sont d\'efinis par
$$
\L^{(k,\infty)}(\H):=\{T\in\K(\H):\mu_m(T)=O(m^{-1/k})\},\hspace{1cm}k\geq 1,
$$
o\`u $\mu_m(T)$ d\'esigne la $m$-i\`eme valeur singuli\`ere de $T$, i.e. la $m$-i\`eme
valeur propre de son module $|T|:=\sqrt{T^*T}$. Aussi, $\L^{(1,\infty)}(\H)$ est le
domaine naturel des traces de Dixmier $\Tr_\omega$
(associ\'ees \`a la divergence logarithmique, cf. appendice \ref{Dixmier}).\\
Autrement dit, il doit exister un unique entier naturel $k$, la dimension
spectrale du triplet, qui soit tel que $(\D-\lambda)^{-k}$ soit
Dixmier-tra\c cable et avec une trace de Dixmier non identiquement nulle.\\
La trace de Dixmier, dont les propri\'et\'es fondamentales sont
donn\'ees dans l'appendice \ref{Dixmier}, jouit d'un statut tr\`es
particulier en g\'eom\'etrie non commutative.
Elle permet non seulement de d\'efinir une notion
purement analytique de dimension, mais aussi de construire
abstraitement une int\'egrale sur l'alg\`ebre, via l'application:
$$
\A\ni a\mapsto\Tr_\omega\Big(\pi(a)|\D|^{-k}\Big).
$$
Par construction, cette
fonctionnelle poss\`ede aussi la propri\'et\'e d'hypertrace
\cite{ConnesBook,Polaris}, i.e. s'annule sur les commutateurs.\\
Nous verrons au chapitre \ref{action}, que cette int\'egrale
non commutative permet de construire des fonctionnelles d'action
g\'en\'eralisant celle de Yang--Mills ordinaire.\\
Lorsque l'op\'erateur $\D$ n'est pas inversible,
on pourra tout de m\^eme donner du sens \`a son inverse.
En effet, que l'op\'erateur $\D$
soit \`a r\'esolvante compacte, implique en particulier que son noyau
est de dimension finie. Son inverse partielle, $\D^{-1}$ est alors d\'efini
en lui assignant la valeur $0$ sur $\ker(\D)$.\\
Ces conditions doivent
\^etre modifi\'ees dans le cas non compact. En effet, pour
une vari\'et\'e non compacte ordinaire,
non seulement le noyau de l'op\'erateur de Dirac n'est pas de dimension
finie mais en plus, son spectre essentiel n'est pas vide.

La condition suivante, dite de r\'egularit\'e, constitue une notion de diff\'erentiabilit\'e
pour les \'el\'ements de l'alg\`ebre. Soit $\delta$ la d\'erivation d\'efinie par
$\delta(T):=[|\D|,T]$, pour $T\in\L(\H)$. La r\'egularit\'e signifie que
$a$ et $[\D,\pi(a)]$ appartiennent \`a l'intersection des domaines des
puissances de $\delta$, pour tout $a\in\A$. Cette condition
ne sera que l\'eg\`erement modifi\'ee dans le cas sans unit\'e.

L'axiome de finitude stipule que l'espace des spineurs lisses $\H^\infty$,
l'intersection des domaines
des puissances de $\D$, $\H^\infty:=\bigcap_{n}\Dom(\D^n)$, soit
un module projectif de type fini. Rappelons qu'un module projectif de
type fini, le pendant non commutatif d'un fibr\'e vectoriel, est un
module \`a gauche sur $\A$ de la forme $\A^m p$,
pour un certain projecteur $p=p^*=p^2\in M_m(\A)$ et un certain
$m\in\N$, le `rang du fibr\'e'. Le projecteur en question devant pouvoir \^etre
l'unit\'e de $M_m(\A)$ (par exemple dans le cas commutatif avec
un fibr\'e des spineurs trivial), il est clair que cet axiome devra \^etre
modifi\'e dans le cas non compact.

La condition de r\'ealit\'e consiste en l'existence d'une involution antilin\'eaire $J$,
d\'efinissant une repr\'esentation de l'alg\`ebre oppos\'ee sur $\H$
par $a\mapsto J\pi(a^*)J^{-1}$. De plus, cette deuxi\`eme repr\'esentation doit
commuter avec la premi\`ere
$$
[\pi(a),J\pi(b^*)J^{-1}]=0,\,\forall\,a,b\in\A,
$$
et doit satisfaire $J^2=\pm1$, $J\D=\pm\D J$, $J\chi=\pm\chi J$,
o\`u les signes ne d\'ependent que de $k$ mod $8$ (voir \cite{ConnesReal,
Polaris} pour la table des signes). Cette condition ainsi que la suivante,
sont de nature purement alg\'ebrique et ne n\'ecessitent pas d'\^etre modifi\'ees
dans le cas sans unit\'e.

L'axiome suivant, dit de premier ordre, caract\'erise le fait que dans le
cas commutatif, l'op\'erateur de Dirac est un op\'erateur diff\'erentiel
d'ordre un. Cette condition est alors alg\'ebrique\-ment traduite par le
fait que les op\'erateurs born\'es $[\D,\pi(a)]$ commutent aussi avec
la repr\'esenta\-tion de l'alg\`ebre oppos\'ee; cela \'equivaut aussi \`a dire
que l'espace $[\D,\pi(\A)]$ est affili\'e \`a $\pi(\A)$.

La condition d'orientabilit\'e correspond \`a l'existence d'une
"forme volume" sur l'espace non commutatif. Il doit ainsi
exister un $k$-cycle de Hochschild (\`a valeur dans le bimodule
$\A\otimes\A^{op}$) ${\bf c}=(a_0\otimes b_0)\otimes a_1
\otimes \cdots \otimes a_k$, $a_0,\cdots, a_k\in\A$, $b_0\in\A^{op}$
o\`u $k$ est la dimension spectrale du triplet, tel que $\pi_\D(
{\bf c})=\chi$. Ici, $\pi_\D$ est la repr\'esentation de
$(\A\otimes\A^{op})\otimes\A^{\otimes k}$ donn\'ee par:
\begin{equation}
\label{eq:cycle-rep}
\pi_\D\big((a_0\otimes b_0)\otimes a_1
\otimes \cdots \otimes a_l\big)=
\pi(a_0)J\pi(b_0^*)J^{-1}[\D,\pi(a_1)]\cdots[\D,\pi(a_l)].
\end{equation}
Il y aura encore lieu de modifier cette condition, car
la valeur ${\bf 1}\in\L(\H)$ doit pouvoir \^etre prise par
$\pi_\D({\bf c})$ lorsque $k$ est impair.

La derni\`ere condition, la dualit\'e de Poincar\'e, est de nature topologique.
Dans le cas compact, elle s'exprime par le fait que la forme d'intersection
$\bigcap:K_*(\A)\x K_*(\A)\to\Z$, donn\'ee par la composition du produit
de Kasparov avec l'application indice de Fredholm, soit non d\'eg\'en\'er\'ee.
Nous laisserons \`a part cette derni\`ere condition, car n'\'etant pas centrale dans
le th\'eor\`eme de reconstruction \cite{ConnesCollege, Polaris} et
faisant aujourd'hui encore l'objet d'investigations dans le cas commutatif
non compact (voir en particulier les travaux d'A. Rennie \cite{RennieProj}).

\subsection{G\'en\'eralisation des axiomes au cas sans unit\'e}
\subsubsection{2.1.2.1 Plongement dans une alg\`ebre avec unit\'e}

En se basant sur la dualit\'e de Gelfand--Naimark dans le cas non compact
(toute $C^*$-alg\`ebre commutative sans unit\'e $A$ est isomorphe \`a l'alg\`ebre
$C_0(\chi(A))$ des fonctions continues et s'annulant \`a l'infini sur l'espace
de ses caract\`eres $\chi(A)$), le point de d\'epart pour la construction de
triplets spectraux non unitaux est la donn\'ee d'une alg\`ebre sans unit\'e $\bar{\A}$.

A l'exception de la condition $(\D-\lambda)^{-1}\in\K(\H)$, qui n'est clairement pas
satisfaite dans le cas commutatif non compact et dont la g\'en\'eralisation
naturelle (d\'ej\`a pr\'esente dans \cite{ConnesReal}) est
$\pi(a)(\D-\lambda)^{-1}\in\K(\H)$ pour tout $a\in\bar{\A}$, toutes les autres
modifications n\'ecessitent de consid\'erer un plongement dans une alg\`ebre
\`a \'el\'ement unit\'e: $\A\hookrightarrow\Aun$. Evidemment, $\Aun$ doit elle aussi \^etre
repr\'esent\'ee par des op\'erateurs born\'es et doit \^etre telle que les
commutateurs $[\D,\pi(a)]$, pour tout $a\in\Aun$, s'\'etendent en des op\'erateurs
born\'es. De plus, on demandera que les conditions de r\'egularit\'e, de r\'ealit\'e
et de premier ordre, soient aussi satisfaites pour les \'el\'ements
du plongement unif\`ere choisie.

Cette adjonction d'unit\'e va jouer un r\^ole pr\'epond\'erant dans les axiomes
d'orientation et de finitude. \\
Le point clef pour l'axiome d'orientation est de pouvoir laisser la possibilit\'e
au cycle de Hoch\-schild, jouant le r\^ole de la forme volume, de prendre
la valeur ${\bf 1}\in\L(\H)$. Il est donc clair que dans le cas sans unit\'e,
le cycle ${\bf c}$ doit \^etre un cycle de Hochschild sur $\Aun$ et non
pas sur $\bar{\A}$.\\
Concernant l'axiome de finitude, le projecteur $p$ devra lui aussi \^etre
choisi dans $M_m(\Aun)$.
Pour donner lieu \`a un module projectif de type fini \`a gauche sur $\bar{\A}$
(et non pas sur $\Aun$), il faudra alors se restreindre \`a un plongement
unif\`ere
pour lequel $\bar{\A}$ soit un id\'eal (\`a droite). Il est alors utile d'introduire
la notion de pull-back de module, dans la terminologie de Rennie
\cite{RennieProj}.

\begin{defn}
Soit $\E:=A^mp$, $p\in M_m(A)$ un module projectif de type fini \`a gauche
et soit $A_1$ un id\'eal \`a droite de $A$ (i.e. $A_1A\subset A_1$). Le
pull-back de $\E$ sur $A_1$ est le module \`a gauche $\E_1$ sur $A_1$,
d\'efini par $\E_1:=A_1\E$.
\end{defn}

Finalement, l'axiome de dimension doit \^etre lui aussi l\'eg\`erement
modifi\'e. Ainsi, nous verrons dans l'exemple du plan de Moyal,
qu'il est trop restrictif de demander que
$$
\pi(a)(\D-\lambda)^{-1}\in\L^{(k,\infty)}(\H),
$$
soit satisfait, pour tout $a\in\bar{\A}$ (et pour un certain $k\in\N$).
En effet, l'alg\`ebre $\bar{\A}$ est quasiment (modulo le projecteur $p$)
uniquement d\'etermin\'ee par l'espace des ``spineurs lisses'' $\H^\infty$;
bien que sous-ensemble des op\'erateurs compact, l'ensemble
$\{\pi(a)(\D-\lambda)^{-1}\}_{a\in\bar{\A}}$ n'est pas a priori inclus
dans $\L^{(k,\infty)}(\H)$.
Cette subtilit\'e \'emanant des propri\'et\'es des espaces
$\L^{(p,\infty)}(\H)$ n\'ecessite l'introduction d'une alg\`ebre
interm\'ediare $\A$, dense dans $\bar{\A}$, qui soit aussi
un id\'eal bilat\`ere de $\Aun$ et
qui soit telle que
$$
\{\pi(a)(\D-\lambda)^{-1}\}_{a\in\A}\subset\L^{(k,\infty)}(\H),
$$
pour un certain $k\in\N$.

\subsubsection{2.1.2.2 Les axiomes modifi\'es}
\label{axiomes}
\begin{defn}
Un triplet spectral sans unit\'e r\'eel de dimension~$k$,
consiste en la donn\'ee de
$$
(\A, \bar{\A},\Aun, \H, D, J, \chi),
$$
o\`u $\A$ est une sous-alg\`ebre dense d'une alg\`ebre (a priori sans unit\'e) $\bar{\A}$,
toutes deux fid\`element repr\'esent\'ees
(via une repr\'esentation $\pi$) sur l'espace de  Hilbert $\H$ et
$\Aun$ est un plongement unif\`ere de~$\bar{\A}$, agissant par op\'erateurs born\'es
sur le m\^eme espace de Hilbert.
$D$ est un op\'erateur auto-adjoint non born\'e sur $\H$ tel que
$[D,\pi(a)]$, pour tout $a$ dans $\Aun$, s'\'etende en un op\'erateur born\'e sur~$\H$.
De plus, $J$ et $\chi$ sont des op\'erateurs respectivement anti-unitaire et
auto-adjoint. Le triplet est dit impair lorsque $\chi = 1$ et pair sinon. Dans le cas paire,
$J$ et $\chi$ doivent satisfaire \`a
$\chi^2 = 1$, $\chi \pi(a) = \pi(a) \chi$ pour $a \in \Aun$ et
$D \chi = -\chi D$.
Finalement, les conditions suivantes doivent \^etre v\'erifi\'ees:
\end{defn}

\begin{enumerate}
\addtocounter{enumi}{-1}

\item\textit{Compacit\'e}:

L'op\'erateur $\pi(a)(D - \la)^{-1}$ est compact pour $a \in \bar{\A}$ et
$\la \notin \spec D$.

\item\textit{Dimension spectrale}:

Il existe un unique entier positif $k$, la dimension spectrale
du triplet, pour lequel l'op\'erateur $\pi(a)(D^2 + \eps^2)^{-1/2}$
appartient \`a la classe de Schatten faible $\L^{(k,\infty)}(\H)$ pour tout
$a \in \A$. De plus, $\Tr_\omega(\pi(a)(|D| + \eps)^{-k})$ doit \^etre non
identiquement nulle. L'entier $k$ est paire si et seulement si le triplet l'est.

\item\textit{R\'egularit\'e}:

Les op\'erateurs born\'es $\pi(a)$ et $[D,\pi(a)]$, pour tout $a \in \Aun$,
appartiennent \`a l'intersection des domaines des puissances de la d\'erivation
$\delta : T \mapsto [|D|,T]$.

\item\textit{Finitude}:

Les alg\`ebres $\bar{\A}$ et $\Aun$ sont des
pr\'e-$C^*$-alg\`ebres (stables par calcul fonctionnel
holomorphe dans leur $C^*$-compl\'etion) et sont telles que l'espace des spineurs lisses
$$
\H^\infty := \bigcap_{k \in \N} \Dom(D^k)
$$
soit le $\bar{\A}$-pullback d'un module projectif de type fini sur $\Aun$.
De plus $\A$ est aussi un id\'eal bilat\`ere de
$\Aun$ et est une pr\'e-$C^*$-alg\`ebre avec la m\^eme
$C^*$-compl\'etion que $\bar{\A}$.\\
Aussi, une structure Hermitienne $\roundbraket{\.}{\.}$ \`a valeur dans $\bar{\A}$
est implicitement d\'efinie sur $\H^\infty$ par l'int\'egrale non commutative:

\begin{equation}
\Tr_\omega\bigl( \roundbraket{\pi(a)\xi}{\eta}(|D| + \eps)^{-k} \bigr)
= \braket{\eta}{\pi(a)\xi},
\label{eq:abs-cont}
\end{equation}
o\`u $a \in \Aun$ et $\braket{\.}{\.}$ d\'esigne le produit scalaire
de~$\H$.

\item\textit{R\'ealit\'e}:

Il existe un op\'erateur antiunitaire $J$ sur~$\H$, qui soit tel que
$[\pi(a), J\pi(b)^*J^{-1}] = 0$ pour tout $a,b\in\Aun$ (donc $b \mapsto J\pi(b)^*J^{-1}$
est une repr\'esentation sur $\H$ de l'alg\`ebre oppos\'ee $\A^\opp$ commutant avec
la repr\'esentation originelle). De plus, $J^2 = \pm 1$, $JD = \pm DJ$ et aussi
$J\chi = \pm \chi J$ dans le cas pair, o\`u les signes ne d\'ependent que de
$k\bmod 8$ (voir par exemple \cite[p.~405]{Polaris} pour la table des signes).

\item\textit{Premier ordre}:

Les op\'erateurs $[D,\pi(a)]$  commutent aussi avec la repr\'esentation de
l'alg\`ebre oppos\'ee: \\
$[[D,\pi(a)], J\pi(b)^*J^{-1}] = 0$ pour tous $a,b \in \Aun$.

\item\textit{Orientation}:

Il existe un \textit{$k$-cycle de Hochschild} $\cc$ sur~$\Aun$, \`a valeur dans
$\Aun \ox \Aun^\opp$, consistant en une somme finie de termes
$(a \ox b^\opp) \ox a_1 \oxyox a_k$, dont la repr\'esentation naturelle
sur~$\H$ est donn\'ee par $\pi_D(.)$ \eqref{eq:cycle-rep};
la ``forme volume'' $\pi_D(\cc)$ doit
r\'esoudre l'\'equation
\begin{equation}
\pi_D(\cc) = \chi \sepword{(cas pair),\quad ou}
\pi_D(\cc) = {\bf 1}    \quad\text{(cas impair)}.
\label{eq:vol-form}
\end{equation}

\end{enumerate}

Finalement, une telle g\'eom\'etrie est dite \textit{connexe}
si les seuls op\'erateurs commutant avec $\Aun$ et $\D$ sont les scalaires.

\section{Les plans de Moyal vus comme triplets spectraux}
\label{casMoyal}

Les r\'esultats que nous allons mentionner dans cette partie, ont \'et\'e obtenus
dans \cite{Himalia} en collaboration avec J. M. Gracia-Bond\'{\i}a,
B. Iochum, T. Sch\"ucker et J.~C. V\'arilly. Ils visent \`a r\'ehabiliter
les d\'eformations du type Moyal dans le cadre de la GNC, en
mettant fin \`a une pol\'emique bas\'ee sur des inexactitudes.

Les critiques faites \`a l'encontre de ces d\'eformations sont
li\'ees \`a deux notions diff\'erentes de dimension.
Premi\`erement, en se basant sur l'isomorphisme de Wigner (voir
la remarque \ref{wigner}), il a \'et\'e argument\'e que la dimension
spectrale du plan de Moyal est nulle: \'etant isomorphe \`a des op\'erateurs
compacts, les \'el\'ements de l'alg\`ebre, multipli\'es par l'inverse de
l'op\'erateur de Dirac \`a la puissance de la dimension de l'hyperplan
consid\'er\'e, donnent lieu \`a des op\'erateurs \`a trace et donc
\`a trace de Dixmier nulle.\\
Comme nous le verrons lors du paragraphe suivant, cette assertion
n'est vraie que lorsque l'on consid\`ere la repr\'esentation irr\'eductible
de Schr\"odinger. La repr\'esentation que nous allons utiliser,
la repr\'esentation r\'eguli\`ere gauche, ne souffre quant \`a elle
d'aucune affection de cette sorte.

La deuxi\`eme critique visait la trivialit\'e de l'homologie
de Hochschild de l'alg\`ebre sous-jacente au plan de Moyal.
Pour cette deuxi\`eme notion non commutative de
dimension (qui est ind\'ependante de la repr\'esentation,
du moment qu'elle soit fid\`ele) les caract\'eristiques topologiques sont
primordiales. Il en r\'esulte que cette critique n'est en fait
que partiellement correcte, i.e. n'est vraie que pour certaines
alg\`ebres construites avec le produit de Moyal.

\subsection{Quelques notions de Moyalologie}
\subsubsection{2.2.1.1 Espace de Schwartz et produit de Moyal}

Soient $l\in\N$ et $\Theta$ une matrice  $l\x l$ r\'eelle
antisym\'etrique. Soit aussi $\SS(\R^l)$ l'espace de  Schwartz
(fonctions lisses ayant toutes leurs d\'eriv\'ees \`a d\'ecroissance rapide)
sur $\R^l$. Pour $f,h \in \SS(\R^l)$, le produit de Moyal a \'et\'e d\'efini (d\'efinition
\ref{MP}) par:
\begin{equation}
f \star_\Theta h(x) := (2\pi)^{-l} \iint_{\R^{2l}} d^ly\,d^lz \, e^{-iyz}
f(x - \half\Theta y) \, h(x + z).
\label{eq:moyal-prod-slick}
\end{equation}
Le produit de Moyal correspond \'evidemment
au produit de Rieffel (\ref{product})
associ\'e \`a l'action de $\R^l$ sur lui-m\^eme par translation.

Au vu de la remarque (\ref{casdege}),
nous supposerons que la matrice de d\'eformation
$\Theta$ soit non d\'eg\'en\'er\'ee, c'est-\`a-dire
que la forme bilin\'eaire $\sigma(y,z) := y\.\Theta z$ soit symplectique.
Cela implique que la dimension soit paire, $l = 2N$.
On d\'efinit aussi $\th > 0$ par $\th^{2N} := \det\Theta$.
La formule~\eqref{eq:moyal-prod-slick} peut \^etre alors r\'e\'ecrite comme
\begin{equation}
f \star_\Theta h(x) = (\pi\th)^{-2N} \iint \,d^{2N}y \,d^{2N}z\,
f(x + y)\, h(x + z)\, e^{-2iy\.\Theta^{-1}z} .
\label{eq:moyal-prod-gen}
\end{equation}

Cette derni\`ere expression est tr\`es famili\`ere dans la formulation
de la m\'ecanique quantique sur l'espace des phases,
o\`u $\R^{2N}$ est param\'etr\'e par $N$ paires de variables de
positions et d'impulsions conjugu\'ees. En s\'electionnant
\begin{equation}
\label{matS}
\Theta = \hbar S
:= \hbar \begin{pmatrix} 0 & 1_N \\ -1_N & 0 \end{pmatrix},
\end{equation}
le star-produit associ\'e (ou plut\^ot son commutateur) a \'et\'e
introduit dans ce contexte par Moyal \cite{Moyal}, en utilisant
un d\'eveloppement en puissance de $\hbar$, dont le premier terme
non trivial donne le crochet de Poisson:
\begin{equation}
f\star_\hbar g(x)
= \sum_{\a\in\N^{2N}} \Bigl(\frac{i\hbar}{2}\Bigr)^{|\a|}
\frac{1}{\a!}\, \pd{f}{x^\a}(x) \, \pd{g}{(Sx)^\a}(x).
\label{eq:moyal-asymp}
\end{equation}
Le d\'eveloppement~\eqref{eq:moyal-asymp} de la forme int\'egrale
~\eqref{eq:moyal-prod-gen}
existe et quelquefois m\^eme devient exact, sous certaines conditions
donn\'ees dans~\cite{Nereid}.

Bien que particuli\`erement simplificatrice, en
l'occurrence lorsque
l'une des deux fonctions est un polyn\^ome, la forme diff\'erentielle
(ou asymptotique)~\eqref{eq:moyal-asymp} du produit de Moyal
n'est pas utili\-sable pour la construction  de triplets spectraux.
En effet, l'op\'erateur de Moyal-multiplication associ\'e \'etant
$$
L^\Th_f=\sum_{\a\in\N^{2N}} \Bigl(\frac{i\th}{2}\Bigr)^{|\a|}
\frac{1}{\a!}\, M_{\pa^\a f} \, \pd{}{(\Th x)^\a},
$$
ne peut en aucun cas d\'efinir un op\'erateur born\'e sur $L^2(\R^{2N})$.

Etant donn\'e que nos pr\'eoccupations sont de nature analytique,
il n'y a pas vraiment d'avan\-ta\-ge \`a travailler avec la matrice de
d\'eformation $\Th$ la plus g\'en\'erale qui soit; on prendra alors pour simplifier
(les notations principalement) $\Theta = \th S$ avec $\th$ r\'eel.
Avec ce choix de $\Th$, le produit de Moyal peut \^etre r\'e\'ecrit
comme
\begin{equation}
f \mop g(x) := (\pi\th)^{-2N} \iint_{\R^{4N}} \,d^{2N}y \,d^{2N}z\,f(y) g(z)\,
e^{\frac{2i}{\th}(x-y)\,\.\,S(x-z)} .
\label{eq:moyal-prod}
\end{equation}

En plus des propri\'et\'es g\'en\'eriques
du produit twist\'e~\eqref{pro1}, \eqref{pro2}, \eqref{pro3},
on peut montrer par int\'egration par partie que le produit de Moyal
satisfait \`a la ``r\`egle de multiplication mixte'',
\begin{equation}
\label{pro4}
x_j. (f \mop g)
= f \mop (x_j. g) + \frac{i\,\th}{2} \pd{f}{(Sx)_j} \mop g
= (x_j .f) \mop g - \frac{i\,\th}{2} f \mop \pd{g}{(Sx)_j},
\end{equation}
qui permettra en particulier d'\'ecrire les op\'erateurs de diff\'erentiation
partielle en termes de Moyal-commutateurs.

Pour simplifier les notations, on posera aussi $\SS := \SS(\R^{2N})$
et $\SS' := \SS'(\R^{2N})$ l'espace des distributions temp\'er\'ees.

\begin{thm} {\rm\cite{Phobos}}
$\A_\th := (\SS,\mop)$ est une alg\`ebre de Fr\'echet
sans unit\'e, associative, involutive avec un produit jointement
continu et une trace fid\`ele.
\end{thm}
La preuve de l'associativit\'e a d\'ej\`a \'et\'e donn\'ee dans le lemme
\ref{lem:proo}. Que l'int\'egrale soit une trace pour le
produit de Moyal, a aussi \'et\'e montr\'e
dans le lemme \ref{trace}. La continuit\'e pour $\mop$, dans la topologie produit
de l'espace de Schwartz s'obtient en remarquant que
$$
\|f\mop g\|_\infty\leq (\pi\th)^{-2N}\|f\|_1\,\|g\|_1.
$$
Des estimations similaires pour $x^\a\,\pa^\beta(f\mop g)$,
$\forall\a,\beta\in\N^{2N}$ sont ensuite obtenues par induction, en
utilisant la r\`egle de multiplication mixte~\eqref{pro4} ainsi que celle de Leibniz.
Ces in\'egalit\'es impliquent alors que le produit est s\'epar\'ement continu,
donc jointement continu car $\SS$ est un espace de Fr\'echet.

\subsubsection{2.2.1.2 La base de l'oscillateur harmonique}

\begin{defn}
\label{df:basis-fns}
L'alg\`ebre $\A_\th$ poss\`ede une base naturelle,
constitu\'ee des ``transitions propres'' $f_{mn}$ de l'oscillateur
harmonique, indic\'ees par $m,n \in \N^N$. Avec comme d'habitude,
la notation multi-indicielle $m = (m_1,\dots,m_N) \in \N^N$,  $|m| := m_1 +\cdots+ m_N$,
$m! := m_1!\dots m_N!$. Soit
$$
H_l := \half(x_{_l}^2 + x_{_{l+N}}^2) \sepword{pour $l=1,\dots,N$\quad et}
H := H_1 + H_2 +\cdots+ H_N,
$$
les~$f_{mn}$ diagonalisent l'Hamiltonien de l'oscillateur harmonique:
\begin{align}
H_l \mop f_{mn} &= \th(m_l+\half) f_{mn},
\nonumber \\
f_{mn} \mop H_l &= \th(n_l+\half) f_{mn}.
\label{eq:moyal-haml}
\end{align}
Ces fonctions peuvent \^etre d\'efinies par
\begin{equation}
f_{mn}
:= \frac{1}{\sqrt{\th^{|m|+|n|}\,m!n!}}\,(a^*)^m \mop f_{00} \mop a^n,
\label{eq:basis}
\end{equation}
o\`u $f_{00}$ est la Gaussienne $f_{00}(x) := 2^N e^{-2H/\th}$,
les fonctions de cr\'eation et d'annihilation sont
\begin{equation}
a_l := \frac{1}{\sqrt{2}} (x_{_l} + i\,x_{_{l+N}})  \sepword{et}
a_l^* := \frac{1}{\sqrt{2}} (x_{_l} - i\,x_{_{l+N}}),
\label{eq:crea-annl}
\end{equation}
et $a^n := a_1^{n_1} \dots a_N^{n_N} =
a_1^{\mop n_1} \mop\cdots\mop a_N^{\mop n_N}$.
\end{defn}

Les $f_{mn}$ peuvent aussi \^etre
exprim\'ees en termes de polyn\^omes de Laguerre et sont \'evidem\-ment
tr\`es singuli\`eres dans la limite $\th\to 0$ (voir l'appendice \ref{fmn}).
Le lemme suivant r\'esume leurs propri\'et\'es principales.

\begin{lem} {\rm\cite{Phobos}}
\label{lm:osc-basis}
Soient $m,n,k,l \in \N^N$. Alors $f_{mn} \mop f_{kl} = \delta_{nk}f_{ml}$
et $f_{mn}^* = f_{nm}$. Donc $f_{nn}$ est un projecteur orthogonal et
$f_{mn}$ est nilpotent pour $m \neq n$. De plus,
$\<f_{mn}, f_{kl}> = 2^N\,\delta_{mk}\,\delta_{nl}$. La famille
$\set{f_{mn} : m,n\in\N^N} \subset \SS \subset L^2(\R^{2N})$ est
une base orthogonale.
\end{lem}

Il est clair que $\ee_K := \sum_{|n|\leq K} f_{nn}$, pour $K \in \N$,
d\'efinit un syst\`eme d'unit\'es approch\'ees $\{\ee_K\}$
pour~$\A_\th$. Qui plus est, ce syst\`eme est uniform\'ement born\'e
par rapport \`a la norme op\'eratorielle sur $L^2(\R^{2N})$ car
$$
0\leq\ee_K \leq \sum_{n=0}^\infty f_{nn}=1.
$$

Une cons\'equence du lemme~\ref{lm:osc-basis} est la caract\'erisation
matricielle du produit de Moyal.

\begin{prop} {\rm\cite{Phobos}}
\label{rien}
Soit $N = 1$. Alors $\A_\th$ est isomorphe (en tant qu'alg\`ebre
de Fr\'echet) \`a l'alg\`ebre matricielle des suites doublement indic\'ees
\`a d\'ecroissance rapide $c = \{c_{mn}\}$, i.e. qui sont telles que pour chaque $k \in \N$,
les semi-normes
$$
r_k(c) := \biggl( \sum_{m,n=0}^\infty
\th^{2k} (m+\half)^k (n+\half)^k |c_{mn}|^2 \biggr)^{1/2}
$$
soient finies. L'isomorphisme est donn\'e par la d\'ecomposition
$f = \sum_{m,n\in\N^N} c_{mn} f_{mn}$  dans la base des
$\{f_{mn}\}$.

Pour $N > 1$, $\A_\th$ est isomorphe au produit tensoriel projectif
de $N$ alg\`ebres matricielles de cette sorte.
\end{prop}

\begin{defn}
\label{df:Gst}
Soit $\G_{st}$ (pour $s,t \in \R$) l'espace de Hilbert des \'el\'ements
$f \in \SS'(\R^2)$ pour lesquels
\begin{equation}
\label{normesobolev}
\|f\|_{st}^2 := \sum_{m,n=0}^\infty
\th^{s+t} (m+\half)^s (n+\half)^t |c_{mn}|^2<\infty.
\end{equation}
Les espaces de Hilbert $\G_{st}$, pour $s,t\in\R^N$,
sont d\'efinis par produit tensoriel:
$$
\G_{st}:=\G_{s_1t_1} \oxyox \G_{s_Nt_N}.
$$
En d'autres termes,
les \'el\'ements $(2\pi)^{-N/2} \th^{-(N+s+t)/2}
(m+\half)^{-s/2} (n+\half)^{-t/2} f_{mn}$ (avec une \'evidente notation
multi-indicielle), pour $m,n \in \N^N$, constituent une base orthonorm\'ee
de~$\G_{st}$.
\end{defn}

Il est clair que pour $q \leq s$ et $r \leq t$ dans~$\R^N$, on a
$\SS \subset \G_{st} \subseteq \G_{qr} \subset \SS'$ avec des inclusions denses
et continues. De plus, $\SS = \bigcap_{s,t\in\R^N} \G_{st}$
topologiquement (i.e. la topologie projective co\"incide avec la topologie
usuelle de $\SS$) et
$\SS' = \bigcup_{s,t\in\R^N} \G_{st}$ topologiquement (i.e. la topologie
inductive co\"incide avec celle du dual fort de $\SS'$). En particulier,
le d\'eveloppement
$f=\sum_{m,n\in\N^N} c_{mn}f_{mn}$ pour $f \in \SS'$ converge dans la
topologie du dual fort.

Pour deux espaces $F$, $G$, tels que $f\mop g$ soit d\'efini pour
tout $f\in F$, $g\in G$, on d\'esignera par $F\mop G$ l'espace lin\'eairement
engendr\'e par l'ensemble $\{f\mop g:f\in F,\,g\in G\}$.

En utilisant la caract\'erisation matricielle du produit de Moyal,
il est possible de montrer que $\SS\mop\SS=\SS$. Autrement dit,
l'espace $\SS$ poss\`ede la propri\'et\'e de factorisation forte.

\begin{prop} {\rm \cite[p.~877]{Phobos}}
\label{pr:factorization}
L'alg\`ebre $\A_\th=(\SS,\mop)$ poss\`ede la propri\'et\'e de
factorisation  (non unique): pour tout $h \in \SS$ il existe $f,g \in \SS$ tel que
$h = f \mop g$.
\end{prop}
La preuve de ce r\'esultat est constructive, nous reproduisons ici
les arguments principaux.
\begin{proof}[Preuve]
Soit $h\in\SS$ et soit $c=\{c_{mn}\}$ la suite correspondante.
Posons
$$
d_m:=(\sup\{|c_{jk}|:j\in\N,k\leq m\})^{1/2}.
$$
Alors, la suite diagonale $d:=\{d_m\,\delta_{mn}\}$ et la suite
$b:=\{c_{mn}/d_m\}$ sont \`a d\'ecroissance rapide et satisfont
$bd=c$.
\end{proof}

\subsubsection{2.2.1.3 Alg\`ebres des multiplicateurs}

\begin{defn}
\label{df:Moyal-alg}
En utilisant la propri\'et\'e clef de trace (lemme \ref{trace}), le
produit de Moyal va pouvoir \^etre d\'efini par dualit\'e sur certains
sous-ensembles de $\SS'$.
En notant $\<T, g> \in \C$ l'\'evaluation d'une distribution
temp\'er\'ee $T \in \SS'$ sur une fonction test $g \in \SS$,
on d\'efinit en utilisant la continuit\'e du produit d\'eform\'e
sur $\SS$, les distributions $T \mop f$ et
$f \mop T$ pour $f \in \SS$, par $\<T \mop f, g> := \<T, f \mop g>$
et $\<f \mop T, g> := \<T, g \mop f>$. L'involution de $\SS$
est aussi \'etendue sur $\SS'$ par
$\<T^*,g> := \overline{\<T,g^*>}$.

Des propri\'et\'es r\'egularisantes du produit de
Moyal (voir paragraphe suivant), on tire que si $T \in \SS'$ et
$f \in \SS$, alors $T \mop f,\,f \mop T \in C^\infty(\R^{2N})$~\cite{Phobos}.

Les alg\`ebres des multiplicateurs (gauche et droit) de $\A_\th$
sont les sous-ensembles de $\SS'$ d\'efinis par
\begin{align*}
\M_L^\th
&:= \set{T \in \SS'(\R^{2N}) : T \mop h \in \SS(\R^{2N})
\text{ pour tout } h \in \SS(\R^{2N})},
\\
\M_R^\th
&:= \set{T \in \SS'(\R^{2N}) : h \mop T \in \SS(\R^{2N})
\text{ pour tout } h \in \SS(\R^{2N})}.
\end{align*}
L'alg\`ebre bilat\`ere des multiplicateurs consiste en leur intersection:
$$
\M^\th := \M_L^\th \cap \M_R^\th.
$$
\end{defn}

On peut alors d\'efinir les produits $T\mop S$ et $S\mop T$
pour $T\in\SS'$ et $S\in\M^\th$. De plus, on montre
(voir \cite{Deimos}) que
$\M^\th$ est une $*$-alg\`ebre unif\`ere, localement convexe,
semi-r\'eflexive, compl\`ete, nucl\'eaire et qu'elle est munie d'un produit hypocontinu
et d'une involution continue. De plus $\M^\th$ est la compactification
maximale de $\A_\th$ d\'efinie par dualit\'e (voir {\rm\cite[Sec.~1.3]{Polaris}}).
Cette alg\`ebre contient en plus des fonctions constantes, les ondes planes,
les fonctions \`a croissance polynomiale
ainsi que toutes les d\'eriv\'ees de la distribution de Dirac.
Il est aussi \`a noter que pour des valeurs diff\'erentes du param\`etre de
d\'eformation, $\M_\th$ et $\M_{\th'}$ sont isomorphes mais pas identique
en tant que sous ensemble de $\SS'$. En particulier l'exponentielle imaginaire
quadratique
$$
h_\beta(x):=\exp\big(i\beta\sum_{j=1}^Nx_jx_{j+N}\big),
$$
appartient \`a $\M_\th$ si et seulement si $|\beta|\ne 2/\th$
(voir \cite{Nereid}).

Rappelons aussi que lorsque $\th = 0$, le r\^ole de $\M^\th$
est jou\'e par
$\Oh_M$ (``$M$'' pour multiplicateur), l'ensemble des
fonctions ind\'efiniment diff\'erentiables sur $\R^{2N}$ ayant toutes
leurs d\'eriv\'ees \`a croissances polynomiales.

\smallskip

Il existe un autre moyen de d\'efinir le produit de Moyal d'une
paire de distributions appartenant aux espaces de type Sobolev
$\G_{st}$~\cite{Phobos}. Si $f = \sum_{m,n} c_{mn}f_{mn}\in\G_{st}$,
$g = \sum_{m,n} d_{mn}f_{mn}\in\G_{qr}$ et si $t + q \geq 0$, alors
pour $a_{mn} := \sum_k c_{mk} d_{kn}$, la s\'erie
$h := \sum_{m,n} a_{mn} f_{mn}$ converge dans $\G_{sr}$ et $f \mop g = h$.
Cette propri\'et\'e est obtenue en utilisant
\begin{equation}
\label{sobo}
\|f \mop g\|_{st} \leq \|f\|_{sq} \, \|g\|_{rt}
\quad\text{si } q + r \geq 0.
\end{equation}
Ces estimations, que nous r\'e\'etablirons dans le cas $s=t=0$, sont
une cons\'equence directe de l'in\'egalit\'e de Cauchy--Schwarz
pour le produit scalaire sous-jacent \`a la norme (\ref{normesobolev}).
En particulier,  $\G_{t,-t}$ est une alg\`ebre de Banach, pour tout $t \in \R^N$.

Cette premi\`ere unitalisation est clairement trop
``grosse'' pour la construction de triplets spectraux. En effet, une fois
repr\'esent\'ee sur $L^2(\R^{2N})$, $\M^\th$ n'est pas une alg\`ebre
d'op\'erateurs born\'es. En particulier $x^\mu\in\M^\th$ et
$$
L^\th_{x^\mu}=M_{x^\mu}-\frac{i\th}{2}\frac{\pa}{\pa(Sx)^\mu},
$$
n'est pas born\'e.\\
Nous introduisons maintenant une deuxi\`eme compactification,
qui par construction sera repr\'e\-sen\-t\'ee sur $L^2(\R^{2N})$ par des
op\'erateurs born\'es.

\begin{defn}
Soit $A_\th$ la $C^*$-alg\`ebre
$$
A_\th := \set{T \in \SS' : T \mop g \in L^2(\R^{2N})
\text{ pour tout } g \in L^2(\R^{2N})},
$$
munie de la norme op\'eratorielle
$\|L^\th(T)\|_{\mathrm{op}} :=
\sup\set{\|T \mop g\|_2/\|g\|_2 : 0 \neq g \in L^2(\R^{2N})}$.

Evidemment, $\A_\th = \SS \hookto A_\th$, mais $\A_\th$ n'est pas dense dans
$A_\th$. Nous noterons  $A^0_\th$ la fermeture (en norme) de $\A_\th$ dans $A_\th$.
\end{defn}

Les lemmes suivants montrent que $L^2(\R^{2N})$, tout comme
$L^1(\R^{2N})$ sont inclus dans $A_\th$.

\begin{lem} {\rm\cite{Phobos}}
\label{lm:norm-HS}
Si $f,g \in L^2(\R^{2N})$, alors $f \mop g \in L^2(\R^{2N})$ et
$\|L^\th_f\|_{\mathrm{op}} \leq (2\pi\th)^{-N/2} \|f\|_2$.
\end{lem}

\begin{proof}[Preuve]
En d\'eveloppant  $f,g \in L^2(\R^{2N})$ sur la base orthonorm\'ee
$\{\a_{nm}\} := (2\pi\th)^{-N/2} \{f_{nm}\}$, $f = \sum_{m,n} c_{mn} \a_{mn}$ et
$g = \sum_{m,n} d_{mn} \a_{mn}$, on a
\begin{align*}
\|f\mop g\|_2^2
&= (2\pi\th)^{-2N} \biggl\| \sum_{m,l}
\Bigl( \sum_n c_{mn}\,d_{nl} \Bigr) f_{ml} \biggr\|_2^2
= (2\pi\th)^{-N} \sum_{m,l} \Bigl|\sum_n c_{mn}\,d_{nl}\Bigr|^2
\\
&\leq (2\pi\th)^{-N} \sum_{m,j} |c_{mj}|^2 \sum_{k,l} |d_{kl}|^2
= (2\pi\th)^{-N} \|f\|_2^2 \, \|g\|_2^2,
\end{align*}
d'apr\`es l'in\'egalit\'e de Cauchy--Schwarz.
\end{proof}

\begin{lem}
Si $f\in L^1(\R^{2N})$, alors $f\in A_\th$ et
$$
\|L_f\|\leq(\pi\th)^{-N} \|f\|_1.
$$
\end{lem}

\begin{proof}[Preuve]
R\'e\'ecrivons le produit de Moyal comme
$$
f\mop g(x)=(\pi\th)^{-2N}\int d^{2N}y\, f(y)\,e^{-\frac{2i}{\th}xSy}\,
\widehat{g}\big(\frac{2}{\th}S(y-x)\big),
$$
o\`u $\widehat{g}$ d\'esigne la transform\'ee de Fourier de $g$. En remarquant
ensuite que l'op\'erateur $T_y$, d\'efini par
$$
(T_y\psi)(x):=(\pi\th)^{-N\,}e^{-\frac{2i}{\th}xSy}\,
\widehat{\psi}\big(\frac{2}{\th}S(y-x)\big),
$$
est unitaire, on obtient le r\'esultat car
$$
L_f\psi =(\pi\th)^{-N}\int d^{2N}\,f(y)\, (T_y\psi).
$$
\end{proof}

\begin{rem}
\label{wigner}
Le lemme \ref{lm:norm-HS} peut aussi s'obtenir en utilisant la
repr\'esentation irr\'educti\-ble de Schr\"odinger, \`a la place de la
repr\'esentation r\'eguli\`ere gauche (d\'eg\'en\'er\'ee avec multiplicit\'e
infinie). Soit $\sigma^\th(.)$ la repr\'esentation de Schr\"odinger
(voir \cite{Johnny,FollandPhase} par exemple), agissant sur $L^2(\R^N)$. On peut alors
construire un op\'erateur unitaire $W$, la transformation
de Wigner~\cite{FollandPhase,Deimos}, de $L^2(\R^{2N})$ sur
$L^2(\R^N) \ox L^2(\R^N)$, tel que
$$
W\,L^\th(f)\,W^{-1} = \sigma^\th(f) \ox {\bf 1}.
$$
Dans la repr\'esentation irr\'eductible, $\sigma^\th(f)$ pour $f\in L^2(\R^{2N})$
est un op\'erateur de Hilbert-Schmidt avec $\|\sigma^\th(f)\|_2
=(2\pi\th)^{-N/2} \|f\|_2$. On obtient alors directement le r\'esultat:
$$
\|L^\th(f)\|_{L^2(\R^{2N})}=\|\sigma^\th(f) \|_{L^2(\R^{N})}
\leq\|\sigma^\th(f) \|_2=(2\pi\th)^{-N/2} \|f\|_2.
$$
\end{rem}

\begin{prop} {\rm\cite{Kammerer,Deimos}}
\label{pr:algebra}
$(A_\th,\|.\|_{\mathrm{op}})$ est une $C^*$-alg\`ebre unif\`ere
d'op\'erateurs sur $L^2(\R^{2N})$, isomorphe \`a $\L(L^2(\R^N))$ et
incluant $L^2(\R^{2N})$. De plus, $\A_\th \hookto A_\th$ contin\^ument,
mais $\A_\th$ n'est pas dense dans~$A_\th$, c'est-\`a-dire
$A_\th^0 \subsetneq A_\th$.
\end{prop}

La preuve de ce r\'esultat est directement bas\'ee sur l'isomorphisme
de Wigner $W$. En effet, pour $f\in\SS$, $\sigma^\th(f)$ est un op\'erateur
compact (en particulier \`a trace), et donc
$$
A^0_\th=\set{W^{-1}(T \ox 1)W : T \text{ compact}},
$$
alors que
$$A_\th=\set{W^{-1}(T \ox 1)W : T \text{ born\'e}}.
$$

$\A_\th$ est donc une alg\`ebre de Fr\'echet, dont la topologie
est plus fine que celle donn\'ee par la norme op\'eratorielle. Nous allons voir
maintenant qu'elle est aussi stable par calcul fonctionnel holomorphe
dans sa $C^*$-compl\'etion $A^0_\th$.

\begin{prop}
\label{pr:pre-S}
$\A_\th$ est une pr\'e-$C^*$-alg\`ebre de Fr\'echet sans unit\'e.
\end{prop}

\begin{proof}[Preuve]
Nous adaptons ici les arguments du cas commutatif non compact
\cite[p.~135]{Polaris}. Pour montrer que $\A_\th$ est stable sous
calcul fonctionnel holomorphe, il est suffisant de v\'erifier que si
$f \in \A_\th$ avec $1 + f$ inversible dans $A_\th$
avec inverse $1 + g$, alors la quasi-inverse $g$ de $f$ appartient \`a
$\A_\th$. De
$$
(1+f)\mop(1+g)=1+f + g + f \mop g = 1,
$$
on obtient
$f \mop f + g \mop f + f \mop g \mop f = 0$. Il est alors suffisant de
montrer que $f \mop g \mop f \in \A_\th$, car la relation pr\'ec\'edente impliquera
$g \mop f \in \A_\th$ et donc
$g = -f - g \mop f \in \A_\th$.

Etant donn\'e que, $A_\th \subset \G_{-r,0}$ pour tout
$r > N$ \cite[p.~886]{Deimos}, et que $\SS = \bigcap_{s,t\in\R} \G_{s,t}$,
alors pour $s,t$ arbitraires et $p,q$ positifs
$$
f \mop g \mop f \in
\G_{s,p+r} \mop \G_{-r,0} \mop \G_{q,t} \subset \G_{s,t},
$$
en utilisant le fait que $\G_{s,q}\mop\G_{r,t}\subset\G_{s,t}$ si
$q+r\geq 0$ (cf. \'equation \ref{sobo}).
\end{proof}

\smallskip

Finalement, les deux compactifications diff\'erentes
$\M^\th$ et $A_\th$ de $\A_\th$, contiennent toutes deux un groupe d'unitaires \`a
$2N$ param\`etres, engendr\'e par les ondes planes:
\begin{equation}
\exp(ik\,\.) \mop \exp(il\,\.)
= e^{-\frac{i}{2}\th\,k\.Sl} \, \exp(i(k+l)\.).
\label{eq:Moyal-planewave}
\end{equation}
La pertinence de la pr\'esence de
ce groupe d'unitaires r\'eside dans le fait que l'alg\`ebre de ce groupe est isomorphe
\`a l'alg\`ebre du $2N$-tore non commutatif, qui se trouve \^etre plong\'e dans $\M^\th$
et $A_\th$.
Notons aussi que dans la repr\'esentation adjointe, ce groupe repr\'esente
le groupe des translations de~$\R^{2N}$:
$$
\bigl(\exp(ik\,\.)\mop f \mop \exp(-ik\,\.)\bigr)(x) = f(x + \th Sk),
$$
pour $f \in L^2(\R^{2N})$.

\subsubsection{2.2.1.4 Autres alg\`ebres}
\begin{defn}
\label{psido}
Un op\'erateur pseudodiff\'erentiel ($\PsiDO$) $A \in \PsiDO$ sur $\R^k$ est un
op\'erateur lin\'eaire pouvant s'\'ecrire comme
$$
A\,h(x) = (2\pi)^{-k} \iint_{\R^{2k}} \,d^k\xi \,d^ky\,
\sigma[A](x,\xi)\, h(y)\, e^{i\xi\.(x-y)} .
$$
Soit $\Psi^d := \set{A\in\Psi DO : \sigma[A] \in S^d}$ la classe des
$\Psi$DOs d'ordre $d$, avec
$$
S^d := \set{\sigma \in \Coo(\R^k \times \R^k) :
|\del_x^\a \del_\xi^\b \sigma(x,\xi)|
\leq C_{K\a\b}(1 + |\xi|^{2})^{(d-|\b|)/2} \text{ pour tout } x \in K},
$$
o\`u $K$ est n'importe quel sous-ensemble compact de $\R^k$, $\a,\b\in\N^k$, et
les $C_{K\a\b}$ sont des constantes positives. On notera aussi
$\Psi^{\infty} := \bigcup_{d\in\R} \Psi^d$ et
$\Psi^{-\infty} := \bigcap_{d\in\R}\Psi^d$. Un
$\Psi$DO $A$ est dit r\'egularisant si
$A \in \Psi^{-\infty}$, de mani\`ere \'equivalente~\cite{Hormander,Shubin}, si
$A$ s'\'etend en une application lin\'eaire continue de $\Coo(\R^k)'$, le dual de
l'espace des fonctions ind\'efiniment diff\'erentiables,
dans  $\Coo(\R^k)$.
\end{defn}

\begin{lem}
\label{lm:cojoreg}
Si $f \in \SS$, alors $L^\th_f$ est un $\PsiDO$ r\'egularisant.
\end{lem}

\begin{proof}[Preuve]
{}A partir de sa d\'efinition~\eqref{eq:moyal-prod-slick}, on obtient directement
que l'op\'erateur de multiplication twist\'ee \`a gauche par $f\in\SS$ est un
op\'erateur  pseudodiff\'erentiel
de symbole $f(x - \frac{\th}{2}S\xi)$. Clairement, $L^\th_f$ d\'efinit
une application continue $\Coo(\R^{2N})' \to\Coo(\R^{2N})$,
\'etant donn\'e que $\Coo(\R^{2N})'\hookto \SS'$, avec une
inclusion continue et que $\SS\mop\SS'\in \Coo(\R^{2N})$
\cite[th\'eor\`eme 1]{Phobos}.
On obtient aussi le r\'esultat par
$$
|\del_x^\a \del_\xi^\b f(x - \tfrac{\th}{2} S\xi)|
\leq C_{K\a\b} (1 + |\xi|^2)^{(d-|\b|)/2},
$$
pour tout $\a,\b \in \N^{2N}$, tout compact $K \subset \R^{2N}$
et tout $d \in \R$, car $f\in \SS$.
\end{proof}

\begin{rem}
A la diff\'erence du cas d'une vari\'et\'e compacte, les $\PsiDO$
r\'egularisants sur une vari\'et\'e non compacte
ne sont pas  n\'ecessairement des op\'erateurs compacts. En
effet, pour tout $n\in\N$, $L^\th(f_{nn})$ poss\`ede
la valeur propre $1$ avec multiplicit\'e
infinie et ne peut alors pas \^etre compact.
\end{rem}

Pour les produits de Moyal g\'en\'eriques $\Mop$, pour lesquels
l'antisym\'etrie est la seule contrainte sur la matrice de d\'eformation,
les op\'erateurs $L^\Th_f$, $R^\Th_f$ sont aussi des op\'erateurs
pseudo\-diff\'erentiels de symboles
\begin{equation}
\label{eq:symbole}
\sigma[L^\Th_f]=f(x-\thalf\Th\xi)\sepword{et}
\sigma[R^\Th_f]=f(x+\thalf\Th\xi).
\end{equation}
\begin{defn}
\label{df:distr-zoo}
Pour $m \in \N$, $f \in C^m(\R^k)$ et $\ga,l \in \R$, soit
$$
q_{\ga lm}(f) := \sup\set{(1 + |x|^2)^{(-l+\ga|\a|)/2}|\del^\a f(x)| :
x\in\R^k,\ |\a| \leq m};
$$
et soit $\underline{\V}^m_{\ga,l}$, respectivement $\V^m_{\ga,l}$,
l'espace des fonctions appartenant \`a $C^m(\R^k)$ pour lesquelles
$$
(1 + |x|^2)^{(-l+\ga|\a|)/2} \,\del^\a f(x)
$$
s'annule \`a l'infini pour tout $|\alpha|\leq m$, respectivement est fini
pour tout $x \in \R^k$, norm\'es par $q_{\ga lm}$. On d\'efinit alors
$$
\V_\ga := \bigcup_{l\in\R}\,\bigcap_{m\in\N} \V^m_{\ga,l},
\sepword{et plus g\'en\'eralement,}
\V_{\ga,l} := \bigcap_{m\in\N} \V^m_{\ga,l},
$$
en particulier $\V_\ga = \bigcup_{l\in\R} \V_{\ga,l}$. Soient aussi,
$\Oh_C := \V_0$ et $\Oh_r := \V_{0,r}$. On a
$$
\SS   = \bigcap_{m\in\N}\,\bigcap_{l\in\R} \underline{\V}^m_{0,l}.
$$
Finalement, en utilisant les notations de Laurent Schwartz~\cite{Schwartz},
on d\'esignera par $\B := \Oh_0$, l'espace des fonctions ind\'efiniment
diff\'erentiables ayant toutes leurs d\'eriv\'ees born\'ees.
\end{defn}

Nous aurons aussi besoin de
$\dot{\B} := \bigcap_{m\in\N} \underline{\V}^m_{0,0}$, l'espace
des fonctions lisses s'annulant \`a l'infini ainsi que toutes leurs d\'eriv\'ees
et l'espace $\D_{L^2}$, des \'el\'ements de
$L^2(\R^{2N})$ dont toutes les d\'eriv\'ees sont de carr\'e sommable
(intersection des espaces de Sobolev) \cite{OrtnerW,Schwartz}.
En utilisant le lemme de Sobolev,
on peut montrer que les \'el\'ements de  $\D_{L^2}$ sont en fait des
fonctions lisses et que de plus, $\D_{L^2}\subset\dot{\B}$
\cite{Schwartz}: si $f \in \D_{L^2}$, sa transform\'ee de
Fourier $\F(f)$ v\'erifie $(1+|\xi|^{2n}) \F(f) \in L^2(\R^{2N})$ pour tout
$n\in\N$ et par l'in\'egalit\'e de Cauchy--Schwarz $\F(f) \in L^1(\R^{2N})$,
donc $f$ tend vers z\'ero \`a l'infini.

Nous allons reproduire maintenant les arguments, bas\'es encore une fois
sur la d\'efinition du produit de Moyal en terme d'int\'egrales oscillantes,
qui montrent que l'espace $\Oh_C$ est stable sous produit de Moyal.
Le produit de Moyal poss\`ede ainsi, sous certaines conditions, le m\^eme comportement
que le produit ordinaire.
Cette proposition, extraite de \cite{Amalthea}, g\'en\'eralise le
lemme \ref{lem:B} dans le cas sp\'ecifique du plan de Moyal.

\begin{prop} {\rm \cite{Amalthea}}
\label{pr:hector}
L'espace $\Oh_C$ est une $*$-alg\`ebre associative pour le produit de Moyal.
Plus pr\'ecis\'ement, le produit de Moyal d\'efinit une application jointement
continue de $\Oh_r \x \Oh_s$ \`a valeur dans $\Oh_{r+s}$, pour tout $r,s \in \R$. De plus,
$\A_\th$ est un id\'eal bilat\`ere essentiel de $\Oh_C$.
\end{prop}

\begin{proof}[Preuve]
Soit $f \in \Oh_r$ et $g \in \Oh_s$. D'apr\`es la r\`egle de Leibniz
pour le produit de Moyal, on a
$\del^\a(f \mop g) =
\sum_{\b+\ga=\a} \binom{\a}{\b}\, \del^\b f \mop \del^\ga g$. Il est
alors suffisant de montrer qu'il existe des constantes $C_{rsm}$ telles que
\begin{equation}
(1 + |x|^2)^{-(r+s)/2} |(\del^\b f \mop \del^\ga g)(x)|
\leq C_{rsm} \,q_{0rm}(f) \,q_{0sm}(g)
\label{eq:hectoreq}
\end{equation}
pour tout $x \in \R^{2N}$ et avec $m \geq |\b| + |\ga|$ suffisamment grand. Pour
$k \in \N$ (que l'on d\'eterminera plus tard), on a
\begin{align*}
(\del^\b f\mop \del^\ga g)(x)
&= (\pi\th)^{-2N} \iint \,d^{2N}y \,d^{2N}z\,\frac{\del^\b f(x+y)}{(1+|y|^2)^k}\,
\frac{\del^\ga g(x+z)}{(1+|z|^2)^k}\,
(1+|y|^2)^k (1+|z|^2)^k e^{\frac{2i}{\th}y\.Sz}
\\
&= (\pi\th)^{-2N} \iint \,d^{2N}y \,d^{2N}z\,\frac{\del^\b f(x+y)}{(1+|y|^2)^k}\,
\frac{\del^\ga g(x+z)}{(1+|z|^2)^k}\,
P_k(\del_y,\del_z) \bigl[e^{\frac{2i}{\th}y\.Sz}\bigr]
\\
&= (\pi\th)^{-2N} \iint \,d^{2N}y\,d^{2N}z\,e^{\frac{2i}{\th}y\.Sz}\,P_k(-\del_y, -\del_z)
\biggl[ \frac{\del^\b f(x+y)}{(1+|y|^2)^k}\,
\frac{\del^\ga g(x+z)}{(1+|z|^2)^k}\biggr] ,
\end{align*}
o\`u $P_k$ est un polyn\^ome de degr\'e $2k$ dans ses deux variables $y$ et $z$.
En utilisant ensuite
$$
|\del^\a((1 + |x|^2)^{-k})| \leq c_{\a,k} (1 + |x|^2)^{-k}
$$
on obtient
\begin{align*}
&|\del^\b f \mop \del^\ga g|(x)
\leq \sum_{k',k''\leq 2k} C'_{k'k''}
\iint \,d^{2N}y\,d^{2N}z\,\left|\frac{\del^{\b+k'}f(x+y)}{(1+|y|^2)^k}
\frac{\del^{\ga+k''}g(x+z)}{(1+|z|^2)^k}\right|
\\
&\qquad \leq C''_{rsm} \,q_{0rm}(f) \,q_{0sm}(g)\,
\iint \,d^{2N}y\,d^{2N}z\,\frac{(1 + |x+y|^2)^{r/2}}{(1+|y|^2)^k}
\frac{(1 + |x+z|^2)^{s/2}}{(1+|z|^2)^k}
\\
&\qquad \leq C'''_{rsm} \,q_{0rm}(f) \,q_{0sm}(g)\,
(1+|x|^2)^{(r+s)/2} \int \,d^{2N}y\,(1+|y|^2)^{r/2-k}
\int \,d^{2N}z\,(1+|z|^2)^{s/2-k} ,
\end{align*}
avec $m \geq |\b| + |\ga| + 2k$. L'in\'egalit\'e de Cauchy
$1 + |x + y|^2 \leq 2(1 + |x|^2)(1 + |y|^2)$ a \'et\'e utilis\'ee pour extraire
la variable  $x$. Pour $k > N + \max\{r,s\}/2$ (et donc avec
$m \geq |\b| + |\ga| + 2N + \max\{r,s\}$), les
int\'egrales sont finies. La continuit\'e dans la topologie produit
s'obtient alors \`a partir des estimations~\eqref{eq:hectoreq}.

Que $\SS$ soit un id\'eal bilat\`ere de $\Oh_C$ est une cons\'equence de l'inclusion
$\Oh_C \subset \M^\th$. Rappelons qu'un id\'eal $\mathcal{I}$ d'une alg\`ebre
$A$ est dit essentiel si $\mathcal{I}\cap \mathcal{I}'\ne \{0\}$ pour tout
autre id\'eal $\mathcal{I}'\ne\{0\}$. Rappelons aussi que cette condition
est \'equivalente~\cite[proposition~1.8]{Polaris} \`a $a\,\mathcal{I}\ne 0$
pour tout $0\ne a\in A$.\\
Supposons alors qu'il existe un \'el\'ement $g\in\Oh_C$,
qui soit tel que $g\mop\SS=0$. L'ensemble $\{f_{mn}\}$ \'etant une
base de $\SS$,  on a alors $g\mop f_{mn}=0$ pour tout
$m,n\in\N^{2N}$. Ainsi, dans la d\'ecomposition $g=\sum_{m,n}c_{mn}f_{mn}$
(en tant qu'\'el\'ement de $\SS'$), tous les coefficients $c_{mn}$ doivent
\^etre nuls et donc $g=0$, qui contredit l'hypoth\`ese.
\end{proof}

Finalement, le r\'esultat suivant (bas\'e sur des estimations de type
Calder\'on--Vaillancourt) donne des conditions suffisantes pour
que des fonctions lisses appartiennent \`a $A^0_\th$ ou~$A_\th$.

\begin{thm} {\rm \cite{FollandPhase,Howe}}
\label{th:CVai}
L'inclusion $\V_{0,0}^{2N+1} \subset A_\th$ est satisfaite, en particulier
on a $\B \subset A_\th$. De plus
$\underline{\V}_{00}^{2N+1} \subset A^0_\th$, et donc
$\dot{\B} \subset A^0_\th$.
\end{thm}

Nous avons montr\'e dans le lemme \ref{lem:B} (et aussi dans la proposition
\ref{pr:hector}), que $\B$ est une $*$-alg\`ebre
par produit de Moyal, pour lequel $\A_\th$ est un id\'eal
bilat\`ere dense (car $\B\subset \M^\th$). De plus,
$\D_{L^2} \subset \dot{\B} \subset \M^\th$. Nous allons montrer
maintenant que $(\D_{L^2},\mop)$ est aussi une $*$-alg\`ebre.

\begin{lem}
\label{lm:DL2}
$(\D_{L^2},\mop)$ est une $*$-alg\`ebre avec produit et
involution continus. De plus, $(\D_{L^2},\mop)$ est un id\'eal
bilat\`ere dans~$(\B,\mop)$.
\end{lem}

\begin{proof}[Preuve]
Que les \'el\'ements de $\D_{L^2}$ soient stables sous produit de Moyal,
est une cons\'equence de la r\`egle de Leibniz et du lemme~\ref{lm:norm-HS}:
$$
\|\del^\a(f \mop g)\|_2 \leq (2\pi\th)^{N/2} \sum_{|\b|\leq|\a|}
\binom{\a}{\b}\, \|\del^\b f\|_2 \,\|\del^{\a-\b}g\|_2.
$$
La continuit\'e jointe est encore obtenue par ces estimations, en utilisant le fait
que $\D_{L^2}$ est un espace de Fr\'echet (voir \cite{Schwartz}).
La continuit\'e de l'involution $f \mapsto f^*$ est imm\'ediate.

La propri\'et\'e d'id\'eal bilat\`ere de $\D_{L^2}$ pour $\B$, vient
directement de la stabilit\'e de ces deux espaces sous d\'erivations
et de l'inclusion $\B \subset A_\th$, car alors
$$
\|\del^\a f \mop \del^\b g\|_2\leq\|L_{\del^\a f}\|\|g\|_2 < \infty
$$
pour tous $f \in \B$, $g \in \D_{L^2}$ et tous $\a,\b \in \N^{2N}$.
\end{proof}

\subsubsection{2.2.1.5 Le plongement unif\`ere pr\'ef\'er\'e}

Les alg\`ebres $\M^\th$, $A_\th$ sont trop grosses pour la construction
de triplets spectraux. En particulier, $L^\th(\M^\th)$ contient des op\'erateurs non
born\'es et $A_\th$, \'etant isomorphe \`a $\L(L^2(\R^N))$, a une cohomologie
de Hochschild inint\'eressante.
Un plongement dans une alg\`ebre \`a unit\'e
plus utile est donn\'ee par $\A_\th\hookrightarrow\Aun_\th := (\B,\mop)$.
Cette derni\`ere alg\`ebre poss\`ede de plus une caract\'erisation intrins\`eque: Le
commutant de $L^\th(\SS)$ (i.e. le commutant de $\A_\th$ dans
la repr\'esentation r\'eguli\`ere gauche) dans la sous-alg\`ebre des op\'erateurs
born\'es lisses (en norme) par rapport \`a l'action du groupe
d'Heisenberg, est exactement $R^\th(\B)$ \cite{MeloM}.

\begin{prop}
\label{pr:pre-Aun}
$\Aun_\th$ est une pr\'e-$C^*$-alg\`ebre de Fr\'echet \`a unit\'e.
\end{prop}

\begin{proof}[Preuve]
Nous avons d\'ej\`a obtenu que l'espace de Fr\'echet $\B$
est une \'etoile alg\`ebre sous produit de Moyal, produit
continu dans la topologie d\'efinie par la famille de semi-normes
$q_{00m}$, $m \in \N$. Aussi, les \'el\'ements de $\B$ sont exactement
les \'el\'ements de $A_\th$ lisses (en norme) par rapport \`a
l'action $\a$ (fortement continue sur $\Aun_\th$ par rapport aux semi-normes
$q_{00m}$) de $\R^l$ par translation \cite[th\'eor\`eme 7.1]{RieffelDefQ}.
Soit maintenant $G$ une fonction holomorphe sur un voisinage du spectre
de $L^\th_f$, $f\in\Aun_\th$. Alors $G(f)\in\Aun_\th$, car $\a(G(f))=G(\a(f))$
et donc $L^\th_{G(f)}$ est lisse en norme par rapport \`a l'action $\a$.
\end{proof}

La $C^*$-compl\'etion de $\Aun_\th$ contient $A_\th^0$ (celle de $\A_\th$),
mais il n'est pas certain qu'elle soit identique \`a $A_\th$. Aussi,
$\Aun_\th$ munie de la topologie que nous avons consid\'er\'ee ici,
n'est pas un espace s\'eparable. Il existe d'autres topologies sur $\Aun_\th$
(relativement naturelles), la rendant s\'eparable \cite[p.~203]{Schwartz}.

Finalement et en anticipant sur le paragraphe suivant, une des raisons
principales pour choisir $\Aun_\th$ comme compactification, provient du fait qu'elle
est maximale en tant que sous-alg\`ebre de $A_\th$, stable par commutation avec
l'op\'erateur de Dirac sur $\R^{2N}$: pour tout $f\in\Aun_\th$, $n\in\N$, on a
$$
\ad_{\dslash}^n(L^\th_f)\in L^\th(\B)\ox M_{2^N}(\C).
$$

En plus d'\^etre isomorphe \`a
l'espace des spineurs `lisses', $\D_{L^2}$ poss\`ede toutes les propri\'et\'es
n\'ecessaires \`a la construction d'un triplet spectral sans unit\'e.
\begin{cly}
\label{cr:pre-DL2}
$(\D_{L^2},\mop)$ est une pr\'e-$C^*$-alg\`ebre de Fr\'echet sans unit\'e, dont la
$C^*$-compl\'e-tion est $A^0_\th$.
\end{cly}

\begin{proof}[Preuve]
Les arguments sont tout \`a fait analogues \`a ceux
de la preuve de la proposition~\ref{pr:pre-S}. Tout d'abord, on a
$\SS \subset \D_{L^2} \subset A^0_\th$ avec des inclusions denses,
de telle sorte que $A^0_\th$ est aussi la $C^*$-compl\'etion de
$(\D_{L^2},\mop)$.
En effet, la deuxi\`eme inclusion s'obtient en remarquant que si
$f \in \D_{L^2}$, alors $W\,L^\th(f)\,W^{-1} = \sigma^\th(f) \ox 1$
o\`u $\sigma^\th(f)$ est Hilbert--Schmidt donc compact.

Deuxi\`emement, si $f \in \D_{L^2}$ a un quasi-inverse $g \in A^0_\th$, alors
la proposition pr\'ec\'edente montre que $g \in \Aun_\th$. En effet, \'etant donn\'e que
$(\D_{L^2},\mop)$ est un id\'eal dans~$\Aun_\th$ d'apr\`es le lemme~\ref{lm:DL2},
on en d\'eduit que $f \mop g \mop f \in \D_{L^2}$, qui est suffisant pour \'etablir que
$g \in \D_{L^2}$.
\end{proof}

Nous allons introduire maintenant un des outils essentiels pour la construction de
triplets spectraux associ\'es aux plans de Moyal. La terminologie ainsi que
les concepts que nous allons \'evoquer sont issus de l'article \cite{RennieLocal},
dans lequel Rennie a propos\'e d'\'equiper une alg\`ebre non commutative
sans unit\'e $\A$ (par exemple $\A_\th$) d'un ``id\'eal local'' $\A_c \subset
\A$. Cette notion constitue une g\'en\'eralisation non commutative
de $\Coo_c(M)$, l'espace des fonctions \`a support compact.
{\it Une alg\`ebre de Fr\'echet $\A$ est dite locale}, si elle poss\`ede un id\'eal
dense $\A_c$ avec unit\'es locales; {\it une alg\`ebre $\A_c$ poss\`ede des
unit\'es locales}, si pour chaque sous-ensemble fini  $\{a_1,\dots,a_k\}\subset\A_c$,
il existe $u \in \A_c$ tel que $ua_i = a_iu = a_i$ pour tout $i = 1,\dots,k$.

Le produit de Moyal n'est \'evidemment pas local dans le sens ordinaire,
le produit de deux fonctions (ou distributions) \`a supports disjoints
n'\'etant pas nul a priori. De plus, il n'y a pas d'id\'eaux connus
pour $\A_\th$ et $\D_{L^2}$ (autres que $\A_\th$), elles ne sont certainement pas locales
au sens de Rennie~\cite{RennieLocal}.

On peut cependant d\'efinir une notion plus faible de localit\'e,
qui se trouve \^etre particuli\`erement pertinente dans notre
situation:

\begin{defn}
\label{unitloc}
Une alg\`ebre de Fr\'echet  $\A$ est dite \textit{quasi-locale} s'il existe
une \textit{sous-$*$-alg\`ebre} dense $\A_c\subset\A$ poss\'edant des unit\'es locales.
Le candidat naturel pour le plan de Moyal est:
$$
\A_c := \bigcup_{K\in\N} \A_{c,K},  \sepword{o\`u}  \A_{c,K} :=
\biggl\{f\in\SS : f = \sum_{0\leq|m|,|n|\leq K} c_{mn} f_{mn}\biggr\}.
$$
C'est-\`a-dire, $\A_c$ est l'alg\`ebre des combinaisons lin\'eaires finies des
$\set{f_{mn} : m,n \in \N^N}$; elle est \'evidemment dense dans $\A_\th$ et dans
$\D_{L^2}$ et poss\`ede un syst\`eme
d'unit\'es locales donn\'e par $\{e_{_K}\}$, o\`u $e_{_K}:=\sum_{|n|\leq K}f_{nn}$.
Ainsi, $\A_\th$ comme $\D_{L^2}$ sont quasi-locales.
\end{defn}

\subsection{Les triplets associ\'es aux plans de Moyal}
\label{bour}

Nous allons donner maintenant les triplets spectraux sans unit\'e associ\'es
aux plans de Moyal. Nous verrons qu'ils
satisfont \`a toutes les conditions donn\'ees lors du paragraphe
\ref{axiomes}.

Soient $\bar{\A}_\th=(\D_{L^2},\mop)$, avec sous-alg\`ebre dense
$\A_\th = (\SS(\R^{2N}),\mop)$ et plongement unif\`ere pr\'ef\'er\'e
$\Aun_\th := (\B(\R^{2N}),\mop)$. L'espace de Hilbert
des sections de carr\'e int\'egrable du
fibr\'e (trivial) des spineurs est dans ce cas $\H := L^2(\R^{2N}) \ox \C^{2^N}$.
La repr\'esentation de $\bar{\A}_\th$ (aussi celles de
$\A_\th$ et de $\Aun_\th$) est donn\'ee par
$\pi^\th \: \A_\th \to \L(\H) : f \mapsto L^\th_f \ox 1_{2^N}$, o\`u
$L^\th_f$ agit sur l'espace de Hilbert r\'eduit
$\H_r := L^2(\R^{2N})$ par multiplication twist\'ee.

Les op\'erateurs $\pi^\th(f)$ sont born\'es, car ils
agissent diagonalement sur~$\H$ et
$$
\|L^\th_f\| \leq (2\pi\th)^{-N/2} \|f\|_2
$$
pour $f\in\bar{\A}_\th$
(lemme~\ref{lm:norm-HS}) et car $\Aun_\th\subset A_\th$ (th\'eor\`eme
\ref{th:CVai}).

Comme dans le cadre g\'en\'erique des d\'eformations isospectrales,
l'op\'erateur auto-adjoint $\D$ n'est pas d\'eform\'e: il sera l'op\'erateur
de Dirac (Euclidien) ordinaire $\dslash := -i\, \del_\mu\otimes \ga^\mu$,
o\`u les matrices (Hermitiennes) de Dirac $\ga^1,\dots,\ga^{2N}$, satisfaisant
$\{\ga^\mu, \ga^\nu\} = +2\,\delta^{\mu\nu}$, repr\'esentent (irr\'eductiblement)
l'alg\`ebre de Clifford $\Cl(\R^{2N})$ associ\'ee \`a $(\R^{2N},\eta)$, avec
$\eta$ la m\'etrique Euclidienne standard.

Pour la graduation de l'espace de Hilbert $\chi$, on prendra la chiralit\'e
associ\'ee \`a l'alg\`ebre de Clifford:
$$
\chi := \ga_{2N+1} := {\bf 1}_{\H_r} \ox (-i)^N \ga^1 \ga^2 \dots \ga^{2N}.
$$
Donc,
$\chi^2 = (-1)^N (\ga^1\dots\ga^{2N})^2 = (-1)^{2N} = {\bf 1}$ et
$\chi \ga^\mu = -\ga^\mu \chi$.

La structure r\'eelle $J$ est la conjugaison de charge pour les spineurs
sur $\R^{2N}$.
Il est suffisant de supposer que $J^2 = \pm 1$ (o\`u les signes sont
donn\'es dans \cite{ConnesReal,Polaris}) et que
$$
J (1_{\H_r} \ox \ga^\mu) J^{-1} = -{\bf 1}_{\H_r} \ox \ga^\mu,
$$
les conditions sur les autres signes seront alors automatiquement garanties.
En g\'en\'eral, dans une repr\'esentation donn\'ee, la conjugaison de charge
s'\'ecrit
\begin{equation}
J := CK,
\label{eq:charge-conjn}
\end{equation}
o\`u $C$ est une matrice unitaire $2^N \x 2^N$ et $K$ est
la conjugaison complexe. Une forme explicite pour $C$ dans une
repr\'esentation particuli\`ere (o\`u toutes les matrices $\ga^\mu$
sont soit purement imaginaires soit purement r\'eelles suivant que
$\mu$ est pair ou impair) peut \^etre trouv\'ee dans~\cite{ZJ}.

La propri\'et\'e fondamentale de la structure r\'eelle est l'\'echange
des multiplications twist\'ees \`a droite et \`a gauche:
\begin{equation}
J(L^\th(f^*) \ox 1_{2^N}) J^{-1} = R^\th(f) \ox 1_{2^N}.
\label{eq:right-Moyal}
\end{equation}
Cette relation s'obtient en utilisant le fait que la
conjugaison complexe est une involution pour le produit de Moyal (\'equation
\ref{pro1}) et que $L^\th_f$ agit diagonalement.

\medskip

Le lemme \ref{lem:proo} propri\'et\'e (\ref{pro3}) implique que $[\dslash,\pi^\th(f)] =
-i L^\th(\del_\mu f)\ox \gamma^\mu =: \pi^\th(\dslash(f))$ qui par
le th\'eor\`eme~\ref{th:CVai} est born\'e pour tout
$f \in \Aun_\th = \B(\R^{2N})$, exactement comme dans le cas commutatif.

\subsection{V\'erification des axiomes}

\subsubsection{2.2.3.1 Compacit\'e et dimension spectrale}
\label{sec:axiom-comp}

Dans ce paragraphe, nous allons d\'evelopper des outils et des techniques
qui g\'en\'eralisent ceux de la th\'eorie de la diffusion
en m\'ecanique quantique. La r\'ef\'erence principale utilis\'ee est
le livre de B. Simon~\cite[Chapitre~4]{SimonTrace}. Nous utiliserons
quelquefois la convention $\L^\infty(\H) := \K(\H)$, avec
$\|A\|_\infty := \|A\|_{\mathrm{op}}$.

Alternativement \`a \ref{psido}, on peut d\'efinir  l'op\'erateur
pseudodiff\'erentiel  $g(-i\nb)$ sur $\H_r$ pour $g\in L^\infty(\R^{2N})$, par
$$
g(-i\nb)\psi := \F^{-1}(g\,\F\psi),
$$
o\`u $\F$ est la transform\'ee de Fourier ordinaire. Ces deux d\'efinitions
co\"incident \'evidemment:
$$
g(-i\nb)\psi(x)
= (2\pi)^{-2N}\iint\,d^{2N}\xi\,d^{2N}y\, e^{i\xi\.(x-y)}\,g(\xi)\psi(y),
$$
mais permet d'obtenir directement l'in\'egalit\'e $\|g(-i\nb)\|_\infty \leq \|g\|_\infty$,
car alors
$$
\|g(-i\nb)\psi\|_2 = \|\F^{-1}M_g\F\psi\|_2\leq \|M_g\|\|\psi\|_2=
\|g\|_\infty\|\psi\|_2.
$$

\begin{thm}
\label{th:compacite}
Soient $f\in \A$ et $\la \notin \spec \dslash$. En d\'esignant par
$R_\dslash(\la)$ la r\'esolvante de l'op\'erateur de Dirac, alors
$\pi^\th(f)\, R_\dslash(\la)$ est un op\'erateur compact.
\end{thm}

En utilisant la premi\`ere \'equation r\'esolvante, $R_\dslash(\la) =
R_\dslash(\la') + (\la' - \la) R_\dslash(\la) R_\dslash(\la')$, on peut
toujours supposer que $\la = i\mu$ avec $\mu \in \R^*$. Ce th\'eor\`eme
va \^etre d\'emontr\'e par interpolation complexe,
en utilisant le r\'esultat g\'en\'eral sur l'invariance
de la norme de Hilbert--Schmidt pour les d\'eformations isospectrales
(th\'eor\`eme \ref{th:HS-norm}) et la borne pour la
norme op\'eratorielle des
op\'erateurs de multiplication twist\'ee (lemme \ref{lm:norm-HS}).

\begin{lem}
\label{lm:comp}
Lorsque $f \in \SS$ et $0 \neq \mu \in \R$, alors
$$
\pi^\th(f)  R_\dslash(i\mu)    \in \K(\H)
\iff
\pi^\th(f) |R_\dslash(i\mu)|^2 \in \K(\H).
$$
\end{lem}

\begin{proof}[Preuve]
La direction `$\Rightarrow$' est \'evidente car
$R_\dslash(-i\mu)$ est un op\'erateur normal born\'e; en supposant
que $\pi^\th(f)  R_\dslash(i\mu) $ est compact, on en d\'eduit que
$\pi^\th(f) |R_\dslash(i\mu)|^2 =\pi^\th(f)  R_\dslash(i\mu)R_\dslash(-i\mu)$ l'est aussi.
R\'eciproquement, si $\pi^\th(f)|R_\dslash(i\mu)|^2$ est compact, alors
$\pi^\th(f)|R_\dslash(i\mu)|^2 \pi^\th(f^*)$ est compact et
$\pi^\th(f)R_\dslash(i\mu)$ est lui aussi compact, sachant qu'un op\'erateur
$T$ est compact si et seulement si $TT^*$ l'est.
\end{proof}

L'int\'er\^et de ce lemme r\'eside dans la nature diagonale de l'action de
$\pi^\th(f)|R_\dslash(i\mu)|^2$ sur $\H = \H_r \ox \C^{2^N}$;
il sera alors possible de remplacer, sans aucune modification dans
nos conclusions  $\H$ par $\H_r$,
$\pi^\th(f)$ par $L^\th_f$ et de consid\'erer le Laplacien
$\Delta := -\sum_{\mu=1}^{2N} \del_\mu^2$ \`a la place du carr\'e
de l'op\'erateur de Dirac $\dslash^2$.

\begin{lem}
\label{lm:HSO}
Si $f,g\in L^2(\R^{2N})$, alors $L^\th_fg(-i\nb),\,M_fg(-i\nb)
\in \L^2(\H_r)$ avec
$$
\|L^\th_f\,g(-i\nb)\|_2=\|M_f\,g(-i\nb)\|_2=
(2\pi)^{-N}\|f\|_2\,\|g\|_2.
$$
\end{lem}
\begin{proof}[Preuve]
La premi\`ere \'egalit\'e a d\'ej\`a \'et\'e \'etablie lors du th\'eor\`eme \ref{th:HS-norm}.
Pour la seconde, il suffit de remarquer que
$K_{M_f\,g(-i\nb)}(x,y)=(2\pi)^{-N}f(x)\widehat{g}(x-y)$,
o\`u $\widehat{g}$ d\'esigne la transform\'ee de Fourier de $g$ et d'appliquer
le th\'eor\`eme 2.11~\cite{SimonTrace} sur la caract\'erisation des op\'erateurs de
Hilbert--Schmidt; i.e. un op\'erateur sur un espace de Hilbert s\'eparable $L^2(X,\mu)$
est de Hilbert-Schmidt si et seulement si c'est un op\'erateur \`a noyau dont le noyau
est de carr\'e sommable sur $(M\x M,\mu\x\mu)$.
Ainsi,
$$
\|M_f\,g(-i\nb)\|_2^2=(2\pi)^{-2N}\int d^{2N}x\,d^{2N}y\,|f|^2(x)\,|\widehat{g}|^2(x-y)=
(2\pi)^{-2N}\,\|f\|_2\,\|g\|_2.
$$
\end{proof}

\begin{lem}
\label{lm:schatten}
Si $f \in L^2(\R^{2N})$ et $g \in L^p(\R^{2N})$ avec $2 \leq p < \infty$,
alors $L^\th_f\,g(-i\nb) \in \L^p(\H_r)$ et
$$
\|L^\th_f\,g(-i\nb)\|_p
\leq (2\pi)^{-N(1/2+1/p)} \th^{-N(1/2-1/p)} \,\|f\|_2\,\|g\|_p.
$$
\end{lem}

\begin{proof}[Preuve]
Le cas $p = 2$ (avec l'\'egalit\'e) a \'et\'e trait\'e dans le lemme pr\'ec\'edent. Pour
$p = \infty$, on a $\|L^\th_f\,g(-i\nb)\|_\infty \leq
(2\pi\th)^{-N/2} \|f\|_2 \,\|g\|_\infty$, car
$\|L^\th_f\,g(-i\nb)\|_\infty \leq
\|L^\th_f\|_\infty \,\|g(-i\nb)\|_\infty$.

Le r\'esultat pour $2 < p < \infty$ sera obtenu par interpolation complexe
(voir \cite{SimonTrace,ConnesBook}). A cette fin, notons que l'on peut toujours
supposer $g \geq 0$: en d\'efinissant la fonction $a$ par
$|a| = 1$ et $g = a|g|$, nous avons
\begin{align*}
\|L^\th_f\,g(-i\nb)\|_2^2
&= \Tr(|L^\th_f\,g(-i\nb)|^2)
= \Tr(\bar{g}(-i\nb)\,L^\th_{f^*}\,L^\th_f\,g(-i\nb))
\\
&= \Tr(|g|(-i\nb)\,\bar{a}(-i\nb)\,L^\th_{f^*}
\,L^\th_f\,a(-i\nb)\,|g|(-i\nb))
\\
&= \Tr(\bar{a}(-i\nb)\,|g|(-i\nb)\,L^\th_{f^*}
\,L^\th_f\,|g|(-i\nb)\,a(-i\nb))
\\
&= \Tr(|L^\th_f\,|g|(-i\nb)|^2) = \|L^\th_f\,|g|(-i\nb)\|_2^2,
\end{align*}
et
\begin{align*}
\|L^\th_f\,g(-i\nb)\|_\infty
&= \|L^\th_f\,a(-i\nb)\,|g|(-i\nb)\|_\infty
=\|L^\th_f\,|g|(-i\nb)\,a(-i\nb)\|_\infty
\\
&\leq \|L^\th_f\,|g|(-i\nb)\|_\infty \,\|a(-i\nb)\|_\infty
= \|L^\th_f\,|g|(-i\nb)\|_\infty.
\end{align*}
Ensuite et pour une fonction positive, born\'ee et \`a support compact $g$,
on d\'efinit l'application
$$
F_p : z \mapsto  L^\th_f\, g^{zp}(-i\nb)
: S = \set{z\in\C : 0 \leq \Re z \leq \half} \to \L(\H_r).
$$
Pour tout $y\in \R$, on a
$F_p(iy) = L^\th_f \, g^{iyp}(-i\nb) \in \L^\infty(\H_r)$
car $g^{iyp}\in L^\infty(\R^{2N})$.
De plus, $\|F_p(iy)\|_\infty \leq(2\pi\th)^{-N/2} \|f\|_2$.

On a aussi, d'apr\`es le lemme~\ref{lm:HSO}, $F_p(\half+iy) \in \L^2(\H_r)$ et
$$
\|F_p(\half+iy)\|_2 = (2\pi)^{-N} \|f\|_2 \,\|g^{p/2}\|_2.
$$
On obtient alors
par interpolation complexe (voir par exemple \cite[Chap.~9]{ReedSII}
ou~\cite{SimonTrace}) que $F(z) \in \L^{1/\Re z}(\H_r)$, pour tout
$z$ dans la bande $S$. De plus,
$$
\|F_p(z)\|_{1/\Re z}
\leq \|F(0)\|_\infty^{1-2\Re z}\,\|F(\half)\|_2^{2\Re z}
= \|f\|_2 (2\pi\th)^{-\frac{N}{2}(1-2\Re z)}(2\pi)^{-2N\Re z}\,
\|g^{p/2}\|_2^{2\Re z}.
$$
En appliquant cette estimation pour $z = 1/p$, on obtient pour de telles fonctions $g$:
$$
\|L^\th_f\,g(-i\nb)\|_p =  \|F(1/p)\|_p
\leq(2\pi)^{-N(1/2+1/p)}\th^{-N(1/2-1/p)}\|f\|_2\,\|g\|_p .
$$
Finalement, le r\'esultat est obtenu en utilisant la densit\'e des
fonctions \`a support compact dans $L^p(\R^{2N})$.
\end{proof}

\begin{rem}
Dans le cas commutatif, si $f$ et $g$ sont born\'ees sur $\R^k$, alors
$$
\|f(x)\,g(-i\nb)\|_\infty \leq \|f\|_\infty\,\|g\|_\infty.
$$
On obtient ainsi, par interpolation complexe~\cite{BirmanKS,ReedSII,SimonTrace},
une estimation `sym\'e\-tri\-que' en $f$ et $g$:
$$
\|f(x)\,g(-i\nb)\|_p \leq (2\pi)^{-k/p} \,\|f\|_p\,\|g\|_p,
$$
pour $p \geq 2$. Pour $g(x)=(1+|x|^2)^{-1}$ et $f\in\SS$, on obtient
alors que $M_fR_\dslash(z)$ est compact.
\end{rem}

\begin{lem}
\label{lm:neo-schatten}
Si $f \in \SS$ et $0 \neq \mu \in \R$, alors
$\pi^\th(f)\,|R_\dslash(i\mu)|^2 \in \L^p(\H)$ pour tout $p > N$.
\end{lem}

\begin{proof}[Preuve] Nous avons:
$$
\pi^\th(f)\,|R_\dslash(i\mu)|^2
= (L^\th_f \ox 1_{2^N})\,(\dslash-i\mu)^{-1} (\dslash+i\mu)^{-1}
= L^\th_f \,(-\del^\nu\del_\nu +\mu^2)^{-1} \ox 1_{2^N}.
$$
Ainsi, cet op\'erateur agit diagonalement sur $\H_r \ox \C^{2^N}$ et le
lemme~\ref{lm:schatten} implique que
$$
\bigl\| L^\th_f \,(-\del^\nu\del_\nu +\mu^2)^{-1} \bigr\|_p
\leq (2\pi)^{-N(1/2+1/p)}\th^{-N(1/2-1/p)}\,\|f\|_2
\biggl(\int\frac{d^{2N}\xi}{(\xi^\nu\xi_\nu + \mu^2)^p}\biggr)^{1/p},
$$
qui est fini pour $p > N$.
\end{proof}

\begin{proof}[Preuve du th\'eor\`eme~\ref{th:compacite}]
D'apr\`es le  lemme~\ref{lm:comp}, il \'etait suffisant de montrer que\\
$\pi^\th(f)\,|R_\dslash(i\mu)|^2$ est compact pour
$0\ne\mu$ r\'eel.
\end{proof}

Nous allons \'etablir le r\'esultat principal de ce paragraphe:

\begin{thm}
\label{th:dim-spectrale}
La dimension spectrale du $2N$-plan de Moyal est~$2N$.
\end{thm}

Nous allons commencer par \'etablir des propri\'et\'es d'existence.
D'apr\`es le lemme~\ref{lm:schatten} et parce que
$[\dslash,\pi^\th(f)] = -iL^\th(\del_\mu f) \ox \ga^\mu$, on voit que
$\pi^\th(f)(\dslash^2 + \eps^2)^{-l}$ et
$[\dslash,\pi^\th(f)]\,(\dslash^2 + \eps^2)^{-l}$ appartiennent \`a $\L^p(\H)$
pour tout $p > N/l$ (nous supposerons toujours $\eps > 0$). Dans le prochain lemme,
nous allons montrer que $[|\dslash|,\pi^\th(f)]\, (\dslash^2 + \eps^2)^{-l}$
satisfait aux m\^emes propri\'et\'es de sommabilit\'e.

\begin{lem}
\label{lm:commutateur}
Lorsque $f \in \SS$ et $\half \leq l \leq N$, alors
$[|\dslash|,\pi^\th(f)]\,(\dslash^2 + \eps^2)^{-l} \in \L^p(\H)$
pour tout $p > N/l$.
\end{lem}

\begin{proof}[Preuve]
Nous allons utiliser l'identit\'e spectrale, valable pour n'importe quel op\'erateur
positif $A$,
$$
A = \frac{1}{\pi} \int_0^\infty \,\frac{d\mu}{\sqrt{\mu}}
\frac{A^2}{A^2 + \mu},
$$
ainsi que l'identit\'e (valable aussi pour tout $A$, $B$ avec
$\la \notin \spec A$):
\begin{equation}
[B, (A-\la)^{-1}] = (A - \la)^{-1} [A,B] (A - \la)^{-1}.
\label{eq:deriv-inv}
\end{equation}
Pour tout $\rho > 0$,
\begin{align}
\label{eq:spectid}
[|\dslash|, \pi^\th(f)]
&= [|\dslash|+\rho, \pi^\th(f)]
= \frac{1}{\pi} \int_0^\infty \frac{d\mu}{\sqrt{\mu}}\,\biggl[
\frac{(|\dslash|+\rho)^2}{(|\dslash|+\rho)^2+\mu},
\pi^\th(f) \biggr] \,
\nonumber\\
&= \frac{1}{\pi} \int_0^\infty\, \frac{d\mu}{\sqrt{\mu}}\,
\biggl(1 - \frac{(|\dslash|+\rho)^2}{(|\dslash|+\rho)^2+\mu}\biggr)
\bigl[(|\dslash|+\rho)^2, \pi^\th(f)\bigr]
\frac{1}{(|\dslash|+\rho)^2+\mu}
\nonumber\\
&= \frac{1}{\pi} \int_0^\infty\,d\mu\,
\frac{\sqrt{\mu}}{(|\dslash|+\rho)^2+\mu}
\bigl[(|\dslash|+\rho)^2,\pi^\th(f)\bigr]
\frac{1}{(|\dslash|+\rho)^2+\mu}
\nonumber\\
&= \frac{1}{\pi} \int_0^\infty\,d\mu
\frac{\sqrt{\mu}}{(|\dslash|+\rho)^2+\mu}
\biggl(-\pi^\th(\del^\mu \del_\mu f)
- 2i(L^\th(\del_\mu f) \ox \ga^\mu) \dslash
+ 2\rho \bigl[ |\dslash|, \pi^\th(f) \bigr] \biggr)
\nonumber\\
&\hspace{13em} \x
\frac{1}{(|\dslash|+\rho)^2+\mu}.
\end{align}
La derni\`ere \'egalit\'e implique que
\begin{align*}
\bigl\| [|\dslash|,\pi^\th(f)]\,(\dslash^2 + \eps^2)^{-l} \bigr\|_p
&\leq \frac{1}{\pi} \int_0^\infty d\mu
\biggl\| \frac{\sqrt{\mu} }{(|\dslash|+\rho)^2+\mu}
\Bigl( -\pi^\th(\del^\mu \del_\mu f)
- 2i(L^\th(\del_\mu f) \ox \ga^\mu) \dslash
\\
&\hspace{4em} + 2\rho \bigl[ |\dslash|, \pi^\th(f) \bigr] \Bigr)
\frac{1}{(|\dslash|+\rho)^2+\mu}\,(\dslash^2 + \eps^2)^{-l}\biggr\|_p.
\end{align*}

Ainsi, la preuve se r\'eduit \`a montrer que pour tout $f \in \SS$,
\begin{equation}
\label{eq:integrale}
\frac{1}{\pi} \int_0^\infty d\mu
\biggl\| \frac{\sqrt{\mu}}{(|\dslash|+\rho)^2+\mu} \,\pi^\th(f)\dslash
\,\frac{1}{(|\dslash|+\rho)^2+\mu} \,(\dslash^2 + \eps^2)^{-l}
\biggr\|_p  < \infty.
\end{equation}

Etant donn\'e que les normes de Schatten sont des
normes sym\'etriques
$$
\|ABC\|_p\leq\|A\|\|B\|_p\|C\|,
$$
et que seul l'espace
de Hilbert restreint est concern\'e, l'expression~\eqref{eq:integrale}
est major\'ee par
\begin{align*}
\frac{1}{\pi} \int_0^\infty\,d\mu\,\sqrt{\mu}\,
&\biggl\| \frac{1}{(|\dslash|+\rho)^2+\mu} \biggr\|^{3/2}
\biggl\| \frac{\dslash}{(\dslash^2 + \eps^2)^{1/2}} \biggr\|
\biggl\| \pi^\th(f) \frac{1}{(\dslash^2 + \eps^2)^{l-1/2}}
\frac{1}{((|\dslash|+\rho)^2+\mu)^{1/2}} \biggr\|_p
\\
&\leq \frac{1}{\pi} \int_0^\infty\, \frac{\sqrt{\mu} \,d\mu}{(\mu+\rho^2)^{3/2}}
\bigl\| \pi^\th(f)\, (\dslash^2 + \eps^2)^{-l+1/2}
((|\dslash|+\rho)^2 + \mu)^{-1/2} \bigr\|_p.
\end{align*}

En utilisant le lemme~\ref{lm:schatten}, on peut estimer la d\'ependance
en $\mu$ de la derni\`ere norme:
\begin{align*}
\bigl\| \pi^\th(f) &((|\dslash|+\rho)^2+\mu)^{-1/2}
(\dslash^2+\eps^2)^{-l+1/2} \bigr\|_p
\\
&\leq (2\pi)^{-N(1/2+1/p)}\th^{-N(1/2-1/p)} \|f\|_2\,
\bigl\| ((|\xi|+\rho)^2 +\mu)^{-1/2} (|\xi|^2 + \eps^2)^{-l+1/2}
\bigr\|_p
\\
&\leq C(p,\th) \bigl\| ((|\xi|+\rho)^2 +\mu)^{-1/2} \bigr\|_q \,
\bigl\| (|\xi|^2+\eps^2)^{-l+1/2} \bigr\|_r,
\end{align*}
avec $p^{-1} = q^{-1} + r^{-1}$ choisis de mani\`ere appropri\'ee. Ces int\'egrales
sont finies pour $q > 2N$ et $r > 2N/(2l-1)$ (pour $l = \half$, il faut prendre
$r = \infty$ et $q = p$). Pour de telles valeurs,
\begin{align*}
&\bigl\| \pi^\th(f) ((|\dslash|+\rho)^2+\mu)^{-1/2}
(\dslash^2+\eps^2)^{-l+1/2} \bigr\|_p
\\
&\leq C(p,\th,N;f) \|(|\xi|^2+\eps^2)^{-l+1/2}\|_r \, \Omega_{2N}^{1/q}
\biggl( \int_0^\infty dR\frac{R^{2N-1}}{((R+\rho)^2+\mu)^{q/2}}
\biggr)^{1/q}
\\
&= C(p,\th,N;f) \|(|\xi|^2+\eps^2)^{-l+1/2}\|_r \, \pi^{N/q} \,
\frac{\Ga^{1/q}(\tfrac{q}{2} - N)}{\Ga^{1/q}(\tfrac{q}{2})}\,
\mu^{-1/2 + N/q} \\
&=: C'(p,q,\th,N;f) \,\mu^{-1/2 + N/q}.
\end{align*}
Finalement, l'int\'egrale~\eqref{eq:integrale} est plus petite que
$$
C'(p,q,\th,N;f) \int_0^\infty d\mu
\frac{\mu^{N/q}}{(\mu+\rho^2)^{3/2}} ,
$$
donc finie pour $q > 2N$ et $p > N/l$. La preuve est alors compl\`ete.
\end{proof}

Le lemme suivant est une g\'en\'eralisation de l'in\'egalit\'e de Cwikel
\cite{Cwikel,SimonTrace,Weidl} pour les op\'erateurs de multiplication
twist\'ee. Cette in\'egalit\'e permet d'obtenir dans le cas commutatif
une estimation sur les valeurs singuli\`eres des op\'erateurs compacts du type
$M_fg(-i\nb)$. Ces estimations donnent alors des conditions
suffisantes, sur les fonctions $f$ et $g$,
pour que $M_fg(-i\nb)$ appartienne aux classes de Schatten faibles $\L^{(p,\infty)}(\H)$.

\begin{lem}
\label{lm:Cwikel}
Si  $f \in \SS$, alors
$\pi^\th(f) \,(|\dslash|+\eps)^{-1}\, \pi^\th(f^*) \in \L^{(2N,\infty)}(\H)$.
\end{lem}

\begin{proof}[Preuve]
Comme remarqu\'e pr\'ec\'edemment, il suffit de remplacer
$\dslash^2$ par $\Delta$, $\pi^\th(f)$ par $L^\th_f$
et $\H$ par $\H_r$. Soit
$g(-i\nb) := (\sqrt{\Delta} + \eps)^{-1}$. Etant donn\'e que la fonction $g$ est positive,
elle peut \^etre d\'ecompos\'ee comme $g = \sum_{n\in\Z} g_n$ o\`u
$$
g_n(x) := \begin{cases} g(x) &\text{si  $2^{n-1} < g(x) \leq 2^n$}, \\
0 &\text{sinon}. \end{cases}
$$

Pour $n \in \Z$, soient $A_n$ et $B_n$ les op\'erateurs d\'efinis par
$$
A_n := \sum_{k\leq n} L^\th_f \,g_k(-i\nb) \,L^\th_{f^*},  \quad
B_n := \sum_{k>n}     L^\th_f \,g_k(-i\nb) \,L^\th_{f^*}.
$$
La norme op\'eratorielle de $A_n$ peut \^etre estim\'ee par
\begin{align*}
\|A_n\|_\infty
&\leq \|L^\th_f\|^2\, \biggl\|\sum_{k\leq n} g_k(-i\nb)\biggr\|_\infty
\leq (2\pi\th)^{-N} \|f\|_2^2
\biggl\| \sum_{k\leq n} g_k \biggr\|_\infty\\
&\leq (2\pi\th)^{-N} \|f\|_2^2 \,2^n =: 2^n\,c_1(\th,N;f).
\end{align*}
La norme trace de $B_n$ peut, quant \`a elle, \^etre calcul\'ee en utilisant le lemme~\ref{lm:HSO}:
\begin{align*}
\|B_n\|_1
&= \biggl\|\Bigl(\smash[b]{\sum_{k>n}g_k(-i\nb)}\Bigr)^{1/2}
L^\th_{f^*} \biggr\|_2^2
= \biggl\| L^\th_f \Bigl(\smash[b]{\sum_{k>n}g_k(-i\nb)}\Bigr)^{1/2}
\biggr\|_2^2
= (2\pi)^{-2N} \|f\|_2^2 \,
\biggl\| \Bigl(\smash[b]{\sum_{k>n} g_k}\Bigr)^{1/2} \biggr\|_2^2
\\
&= (2\pi)^{-2N} \|f\|_2^2\, \biggl\| \sum_{k>n} g_k \biggr\|_1
= (2\pi)^{-2N} \|f\|_2^2 \,\sum_{k>n} \|g_k\|_1
\\
&\leq (2\pi)^{-2N} \|f\|_2^2 \,
\sum_{k>n} \|g_k\|_\infty \,\nu\{\supp(g_k)\},
\end{align*}
o\`u  $\nu$ est la mesure Lebesgue sur $\R^{2N}$. Par d\'efinition,
$\|g_k\|_\infty \leq 2^k$ et
\begin{align*}
\nu\{\supp(g_k)\}
&= \nu\set{x \in \R^{2N} : 2^{k-1} < g(x) \leq 2^k}
\leq \nu\set{x \in \R^{2N} : (|x|+\eps)^{-1} \geq 2^{k-1}}
\\
&\leq 2^{2N(1-k)} \,c_2.
\end{align*}
Ainsi,
\begin{align*}
\|B_n\|_1
&\leq (2\pi)^{-2N} \|f\|_2^2 \, 2^{2N} c_2\, \sum_{k>n} 2^{k(1-2N)}
\\
&< \pi^{-2N} \,c_2 \,\|f\|_2^2 \, 2^{n(1-2N)}
=: 2^{n(1-2N)} \,c_3(N;f).
\end{align*}
Dans la derni\`ere \'egalit\'e, nous avons somm\'e une s\'erie
g\'eom\'etrique, convergente puisque $N > \half$.

On peut alors estimer la $m$-i\`eme valeur singuli\`ere $\mu_m$ de $B_n$
(les valeurs singuli\`eres \'etant compt\'ees avec leur propre multiplicit\'e):
$\|B_n\|_1 = \sum_{k=0}^\infty \mu_k(B_n)$ et pour
$m = 1,2,3,\dots$, on a
$\|B_n\|_1 \geq \sum_{k=0}^{m-1} \mu_k(B_n) \geq m\,\mu_m(B_n)$. Ainsi,
$\mu_m(B_n) \leq  \|B_n\|_1\, m^{-1}\leq 2^{n(1-2N)} \,c_3 \,
m^{-1}$. Finalement, l'in\'egalit\'e de Fan
\cite[th\'eor\`eme~1.7]{SimonTrace} implique
\begin{align*}
\mu_m(L^\th_f \,g(-i\nb) \,L^\th_{f^*})
&= \mu_m(A_n + B_n) \leq \mu_1(A_n) + \mu_m(B_n)
\\
&\leq \|A_n\| +  \|B_n\|_1 \,m^{-1}
\leq 2^n\,c_1 + 2^{n(1-2N)} \,c_3 \,m^{-1}.
\end{align*}
Pour un $m$ donn\'e, on peut choisir $n \in \Z$ de telle sorte que $2^n \leq m^{-1/2N} < 2^{n+1}$.
Alors
$$
\mu_m(L^\th_f \,g(-i\nb) \,L^\th_{f^*})
\leq  c_1 \, m^{-1/2N} + c_3 \, m^{-(1-2N)/2N} m^{-1}
=: c_4(\th,N;f) \, m^{-1/2N}.
$$
Cette derni\`ere in\'egalit\'e implique directement le r\'esultat, car alors
$$
L^\th_f\,(\sqrt{\Delta}+\eps)^{-1}\,L^\th_{f^*} \in \L^{(2N,\infty)}(\H_r).
$$
\end{proof}

\begin{cly}
\label{cr:fragmentation}
Si  $f,g \in \SS$, alors
$\pi^\th(f)\, (|\dslash|+\eps)^{-1} \,\pi^\th(g) \in \L^{(2N,\infty)}(\H)$.
\end{cly}

\begin{proof}[Preuve]
Il suffit de consid\'erer des combinaisons lin\'eaires de
$\pi^\th(f \pm g^*)\,(|\dslash|+\eps)^{-1}\,\pi^\th(f^* \pm g)$ et de
$\pi^\th(f \pm ig^*)\,(|\dslash|+\eps)^{-1}\,\pi^\th(f^* \mp ig)$,
pour obtenir le r\'esultat.
\end{proof}

\begin{cly}
\label{cr:spec-dim-one}
Pour tout $h \in \SS$,
$\pi^\th(h)\,(|\dslash|+\eps)^{-1},
\,\,\pi^\th(h)\,(\dslash^2+\eps^2)^{-1/2} \in \L^{(2N,\infty)}(\H)$.
\end{cly}

\begin{proof}[Preuve]
D'apr\`es la propri\'et\'e de factorisation forte
(proposition \ref{pr:factorization}), pour tout $h\in\SS$, il existe
$f,g\in\SS$ tels que $h = f \mop g$ . Ainsi
$$
\pi^\th(h)\, (|\dslash|+\eps)^{-1}
= \pi^\th(f)\, (|\dslash|+\eps)^{-1}\,\pi^\th(g)
+ \pi^\th(f)\, [\pi^\th(g), (|\dslash|+\eps)^{-1}],
$$
et nous obtenons \`a partir de l'identit\'e \eqref{eq:deriv-inv}
$$
\pi^\th(h) \,(|\dslash|+\eps)^{-1}
= \pi^\th(f) \,(|\dslash|+\eps)^{-1} \,\pi^\th(g)
+ \pi^\th(f) \,(|\dslash|+\eps)^{-1} \,[|\dslash|, \pi^\th(g)]\,
(|\dslash|+\eps)^{-1}.
$$
Par des arguments similaires \`a ceux des lemmes
\ref{lm:schatten} et \ref{lm:commutateur}, le dernier terme appartient
\`a l'espace $\L^p(\H)$ pour $p > N$ et donc
\`a~$\L^{(2N,\infty)}(\H)$.

La deuxi\`eme assertion est obtenue en remarquant (par calcul fonctionnel)
que l'op\'erateur $(|\dslash|+\eps)(\dslash^2+\eps^2)^{-1/2}$ est born\'e.
\end{proof}
Le lemme suivant est la derni\`ere propri\'et\'e d'existence
dont nous avons besoin.

\begin{lem}
\label{lm:existence}
Si $f\in \SS$, alors $\pi^\th(f) (|\dslash|+\eps)^{-2N}$ et
$\pi^\th(f) (\dslash^2+\eps^2)^{-N}$ sont dans $\L^{(1,\infty)}(\H)$.
\end{lem}

\begin{proof}[Preuve]
Par calcul fonctionnel, il est suffisant de montrer le r\'esultat pour
$\pi^\th(f) (|\dslash|+\eps)^{-2N}$. Nous allons
factoriser $f \in \SS$ (proposition~\ref{pr:factorization}), en utilisant la
notation:
\begin{align*}
f &= f_1 \mop f_2 = f_1 \mop f_{21} \mop f_{22}
= f_1 \mop f_{21} \mop f_{221} \mop f_{222}
\\
&= \cdots = f_1 \mop f_{21} \mop f_{221} \mop\cdots\mop
f_{22\cdots 21} \mop f_{22\cdots 22}.
\end{align*}
Ainsi,
\begin{align}
\pi^\th(f) \,(|\dslash|+\eps)^{-2N}
&= \pi^\th(f_1) \,(|\dslash|+\eps)^{-1} \,\pi^\th(f_2)
\,(|\dslash|+\eps)^{-2N+1}
\nonumber \\
&\qquad + \pi^\th(f_1) \,(|\dslash|+\eps)^{-1}
\,[|\dslash|,\pi^\th(f_2)] \,(|\dslash|+\eps)^{-2N}.
\label{eq:modtrace}
\end{align}

D'apr\`es le lemme~\ref{lm:schatten},
$\pi^\th(f_1)(|\dslash|+\eps)^{-1} \in \L^p(\H)$ pour tout $p > 2N$;
et par le lemme~\ref{lm:commutateur}, le terme
$[|\dslash|,\pi^\th(f_2)] (|\dslash|+\eps)^{-2N}$ appartient \`a $\L^q(\H)$
pour tout $q > 1$. Alors, le dernier terme du membre de droite de l'\'equation
\eqref{eq:modtrace} appartient \`a $\L^1(\H)$. Introduisons la relation
d'\'equivalence $A \sim B$ pour $A,B \in \K(\H)$ lorsque $A - B$ est
\`a trace. On a alors
$$
\pi^\th(f) (|\dslash|+\eps)^{-2N}
\sim \pi^\th(f_1) (|\dslash|+\eps)^{-1} \pi^\th(f_2)
(|\dslash|+\eps)^{-2N+1},
$$
donc,
\begin{align*}
\pi^\th(f) &(|\dslash|+\eps)^{-2N}
\sim \pi^\th(f_1) (|\dslash|+\eps)^{-1} \pi^\th(f_2)
(|\dslash|+\eps)^{-2N+1}
\\
&= \pi^\th(f_1) (|\dslash|+\eps)^{-1} \pi^\th(f_{21})
(|\dslash|+\eps)^{-1} \pi^\th(f_{22}) (|\dslash|+\eps)^{-2N+2}
\\
&\qquad + \pi^\th(f_1) (|\dslash|+\eps)^{-1} \pi^\th(f_{21})
(|\dslash|+\eps)^{-1} \,[|\dslash|, \pi^\th(f_{22})]\,
(|\dslash|+\eps)^{-2N+1}
\\
&\sim \pi^\th(f_1) (|\dslash|+\eps)^{-1} \pi^\th(f_{21})
(|\dslash|+\eps)^{-1} \pi^\th(f_{22}) (|\dslash|+\eps)^{-2N+2}
\sim \cdots \\
&\sim \pi^\th(f_1) (|\dslash|+\eps)^{-1} \pi^\th(f_{21})
(|\dslash|+\eps)^{-1} \pi^\th(f_{221}) (|\dslash|+\eps)^{-1}
\dots \pi^\th(f_{22\cdots 22}) (|\dslash|+\eps)^{-1}.
\end{align*}

La deuxi\`eme relation est satisfaite car
$\pi^\th(f_1)(|\dslash|+\eps)^{-1}\pi^\th(f_{21})(|\dslash|+\eps)^{-1}
\in \L^p(\H)$ pour $p > N$ d'apr\`es le lemme~\ref{lm:schatten} et
$[|\dslash|,\pi^\th(f_{22})] (|\dslash|+\eps)^{-2N+1} \in \L^q(\H)$
pour $q > 2N/(2N - 1)$ d'apr\`es le lemme~\ref{lm:commutateur}.
Les autres \'equivalences sont obtenues \`a partir d'arguments similaires.
Le corollaire~\ref{cr:fragmentation}, l'in\'egalit\'e de H\"older pour
les classes de Schatten faibles (voir par exemple
\cite[proposition~7.16]{Polaris}) ainsi que l'inclusion
$\L^1(\H) \subset \L^{(1,\infty)}(\H)$ conduisent finalement au r\'esultat.
\end{proof}

\smallskip

Nous allons calculer maintenant des traces de Dixmier. En utilisant
la trace r\'egularis\'ee d'op\'erateurs pseudodiff\'erentiels
d\'efinie par
$$
\Tr_\La(A) := (2\pi)^{-2N} \iint_{|\xi|\leq\La} \,d^{2N}\xi \,d^{2N}x\,
\sigma[A](x,\xi),
$$
le r\'esultat que nous allons obtenir peut \^etre conjectur\'e car
$$
\lim_{\La\to\infty} \Tr_\La(\.)/\log(\La^{2N})
$$
est heuristiquement reli\'ee \`a la trace de Dixmier et car
nous avons
\begin{align*}
\lim_{\La\to\infty} &\frac{1}{2N\log\La}
\Tr_\La \bigl(\pi^\th(f)(\dslash^2 + \eps^2)^{-N}\bigr)
\\
&= \lim_{\La\to\infty}\,\frac{2^N}{2N(2\pi)^{2N}\log\La}
\iint_{|\xi|\leq\La}\,d^{2N}\xi\,d^{2N}x\, f(x - \tfrac{\th}{2}S\xi)\,
(|\xi|^2 + \eps^2)^{-N}\\
&= \frac{2^N\,\Omega_{2N}}{2N\,(2\pi)^{2N}} \int d^{2N}x\,f(x) .
\end{align*}

Cependant, pour \'etablir rigoureusement ce r\'esultat dans
le contexte du produit de Moyal, il est n\'ecessaire d'adopter une 
strat\'egie
plus subtile. Nous allons ainsi calculer la trace de Dixmier de l'op\'erateur
$\pi^\th(f)\,(\dslash^2 + \eps^2)^{-N}$, comme le r\'esidu de la trace
ordinaire d'une famille m\'eromorphe d'op\'erateurs.
Dans cette perspective, les r\'esultats r\'ecents de \cite{CareyPS},
qui \'etendent le th\'eor\`eme de trace de Connes
(voir~\cite{ConnesAction} et~\cite[Chap.~7]{Polaris}), seront
d\'eterminants.

Dans le langage de~\cite{HigsonRes}, notre premi\`ere t\^ache va \^etre
de v\'erifier que la \textit{dimension analytique} de $\A_\th$
est \'egale \`a $2N$, c'est-\`a-dire que pour
$f \in \A_\th$, l'op\'erateur $\pi^\th(f)\,(\dslash^2 + \eps^2)^{-z/2}$
est \`a trace lorsque $\Re z>2N$.

\begin{lem}
\label{lm:in-extremis}
Si $f \in \SS$, alors $L^\th_f\,(\dslash^2 + \eps^2)^{-z/2}$ est
\`a trace pour $\Re z>2N$, et
$$
\Tr[L^\th_f\,(\tri+\eps^2)^{-z/2}] = (2\pi)^{-2N} \iint d^{2N}\xi \,d^{2N}x
f(x)\, (|\xi|^2 + \eps^2)^{-z/2}.
$$
\end{lem}

\begin{proof}[Preuve]
Soit $S^{d,l}$ la classe des symboles  (d\'efinition \ref{SHUbin})
de Shubin (ou symboles GLS)~\cite{GrossmannLS, Shubin}
et $\Psi^{ld}(\R^{2n})$ la classe correspondante d'op\'erateurs pseudodiff\'erentiels.
Il est d\'emontr\'e dans \cite{DimassiS}, que si
$\sigma(x,\xi) \in S^{d,l}(\R^{2n})$ pour $d,l < -n$, alors l'op\'erateur
pseudodiff\'erentiel associ\'e \`a ce symbole est \`a trace. Qui plus est, on a
$$
\Tr A = (2\pi)^{-n} \iint \,d^nx\,d^n\xi\,\sigma(x,\xi) .
$$

En utilisant la formule du produit pour les symboles
d'op\'erateurs pseudodiff\'erentiels, on obtient pour
$p > N$,
\begin{align*}
\sigma \bigl[L^\th_f(\Delta + \eps^2)^{-p}\bigr](x,\xi)
&= \sum_{\a\in\N^N} \frac{(-i)^{|\a|}}{\a!} \;
\del_\xi^\a \sigma[L^\th_f](x,\xi) \,
\del_x^\a \sigma\bigl[(\Delta + \eps^2)^{-p}\bigr](x,\xi)
\\
&= \sigma[L^\th_f](x,\xi) \,
\sigma\bigl[(\Delta + \eps^2)^{-p}\bigr](x,\xi)
\\
&= f(x - \tfrac{\th}{2}S\xi)\,(|\xi|^2 + \eps^2)^{-p}.
\end{align*}
Ainsi, toujours avec $p > N$,
\begin{align*}
\Tr\bigl(L^\th_f(\Delta + \eps^2)^{-p}\bigr)
&= (2\pi)^{-2N}
\iint \,d^{2N}\xi \,d^{2N}x\,
f(x - \tfrac{\th}{2} S\xi)\,(|\xi|^2 + \eps^2)^{-p}
\\
&= (2\pi)^{-2N}
\iint\xi \,d^{2N}x\, f(x)\, (|\xi|^2 + \eps^2)^{-p} \,d^{2N}.
\tag*\qed
\end{align*}
\hideqed
\end{proof}

Nous continuons avec un r\'esultat technique, dans l'esprit de
\cite{RennieLocal}. Consid\'erons le syst\`eme d'unit\'es approch\'ees
$\{\ee_K\}_{K\in\N} \subset \A_c$ avec
$\ee_K := \sum_{0\leq|n|\leq K} f_{nn}$. Les \'el\'ements $\ee_K$
sont des projecteurs ordonn\'es:
$\ee_K \mop \ee_L = \ee_L \mop \ee_K = \ee_K$
pour $K \leq L$ et sont de plus des unit\'es locales pour~$\A_c$
(d\'efinition \ref{unitloc}).

\begin{lem}
\label{lm:traceclass}
Si $f \in \A_{c,K}$,
$$
\pi^\th(f) \,(\dslash^2+\eps^2)^{-N} - \pi^\th(f) \,
\bigl( \pi^\th(\ee_K)(\dslash^2+\eps^2)^{-1}\pi^\th(\ee_K) \bigr)^N
\in \L^1(\H).
$$
\end{lem}

\begin{proof}[Preuve]
Pour simplifier les notations, nous utiliserons $e := \ee_K$ et
$e_n := \ee_{K+n}$. Etant donn\'e que $\pi^\th(f)$ est born\'e et
$f=f\mop e$, nous pouvons supposer $f = e \in \A_{c,K}$.

De $e_n \mop e = e \mop e_n = e$, on d\'eduit
\begin{equation}
\pi^\th(e) (\dslash+\la)^{-1} \bigl( 1 - \pi^\th(e_n) \bigr)
= \pi^\th(e) \,(\dslash+\la)^{-1} \,[\dslash,\pi^\th(e_n)]
\,(\dslash+\la)^{-1}.
\label{eq:difference}
\end{equation}
De plus, $\pi^\th(e)\,[\dslash, \pi^\th(e_n)] =
[\dslash,\pi^\th(e\mop e_n)] - [\dslash,\pi^\th(e)]\,\pi^\th(e_n) = 0$,
car $\pi^\th(e_n)$, pour
$n \geq1$, est une unit\'e locale pour $[\dslash,\pi^\th(e)]$ (voir l'\'equation
\ref{eq:plusun}):
$$
[\dslash,\pi^\th(e)]\,\pi^\th(e_n) = [\dslash,\pi^\th(e)].
$$
On obtient alors
\begin{align*}
& A_n := \pi^\th(e) (\dslash+\la)^{-1} [\dslash,\pi^\th(e_n)]
(\dslash+\la)^{-1}
\nonumber\\
&= \pi^\th(e) (\dslash+\la)^{-1} [\dslash,\pi^\th(e_1)]
(\dslash+\la)^{-1} [\dslash,\pi^\th(e_n)] (\dslash+\la)^{-1}
\nonumber\\
&= \pi^\th(e) (\dslash+\la)^{-1} [\dslash,\pi^\th(e_1)] \pi^\th(e_2)
(\dslash+\la)^{-1} [\dslash,\pi^\th(e_n)] (\dslash+\la)^{-1}
= \cdots
\nonumber\\
&= \bigl( \pi^\th(e) (\dslash+\la)^{-1} \bigr)
\bigl( [\dslash,\pi^\th(e_1)] (\dslash+\la)^{-1} \bigr)
\bigl( [\dslash,\pi^\th(e_2)] (\dslash+\la)^{-1} \bigr)
\cdots \bigl( [\dslash,\pi^\th(e_n)] (\dslash+\la)^{-1} \bigr).
\end{align*}

Pour $n = 2N$, $A_{2N}$ appara\^it comme un produit de $2N+1$ termes,
chacun appartenant \`a $\L^{2N+1}(\H)$ d'apr\`es le lemme~\ref{lm:schatten}. Ainsi,
d'apr\`es l'in\'egalit\'e de H\"older, $A_{2N}$ est \`a trace et donc
$\pi^\th(e) (\dslash+\la)^{-1} (1 - \pi^\th(\ee_{2N})) \in \L^1(\H)$.
D'o\`u
\begin{align}
\pi^\th(e)\, &(\dslash^2+\eps^2)^{-1} \bigl(1-\pi^\th(\ee_{4N})\bigr)
\nonumber\\
&= \pi^\th(e) (\dslash-i\eps)^{-1}
\bigl(1-\pi^\th(\ee_{2N}) + \pi^\th(\ee_{2N})\bigr)
(\dslash+i\eps)^{-1} \bigl(1-\pi^\th(\ee_{4N})\bigr)
\nonumber\\
&= \pi^\th(e)(\dslash-i\eps)^{-1} \bigl(1-\pi^\th(\ee_{2N})\bigr)
(\dslash+i\eps)^{-1} \bigl(1-\pi^\th(\ee_{4N})\bigr)
\nonumber\\
&\qquad + \pi^\th(e) (\dslash-i\eps)^{-1} \pi^\th(\ee_{2N})
(\dslash+i\eps)^{-1} \bigl(1-\pi^\th(\ee_{4N})\bigr) \in \L^1(\H),
\label{eq:difference2}
\end{align}
c'est-\`a-dire que $\pi^\th(e) (\dslash^2+\eps^2)^{-1} \sim
\pi^\th(e) (\dslash^2+\eps^2)^{-1} \pi^\th(\ee_{4N})$, avec la
relation d'\'equivalence d\'efinie pr\'ec\'edemment.
En it\'erant ces \'equivalences, on obtient
\begin{align*}
\pi^\th(e) (\dslash^2+\eps^2)^{-N}
&\sim \pi^\th(e) (\dslash^2+\eps^2)^{-1} \pi^\th(\ee_{4N})
(\dslash^2+\eps^2)^{-N+1}
\\
&\sim \pi^\th(e) (\dslash^2+\eps^2)^{-1} \pi^\th(\ee_{4N})
(\dslash^2+\eps^2)^{-1} \pi^\th(\ee_{8N}) (\dslash^2+\eps^2)^{-N+2}
\sim \cdots
\\
&\sim \pi^\th(e) (\dslash^2+\eps^2)^{-1} \pi^\th(\ee_{4N})
(\dslash^2+\eps^2)^{-1} \pi^\th(\ee_{8N}) \cdots
(\dslash^2+\eps^2)^{-1} \pi^\th(\ee_{4N^2}).
\end{align*}
En utilisant ensuite l'identit\'e \eqref{eq:deriv-inv}, le dernier terme
devient
\begin{align*}
&\pi^\th(e) (\dslash+i\eps)^{-1} \pi^\th(e) (\dslash-i\eps)^{-1}
\pi^\th(\ee_{4N}) (\dslash^2+\eps^2)^{-1} \pi^\th(\ee_{8N}) \cdots
(\dslash^2+\eps^2)^{-1} \pi^\th(\ee_{4N^2})
\\
&\quad + \pi^\th(e) (\dslash+i\eps)^{-1} [\dslash,\pi^\th(e)]
(\dslash^2+\eps^2)^{-1} \pi^\th(\ee_{4N}) (\dslash^2+\eps^2)^{-1}
\pi^\th(\ee_{8N}) \cdots (\dslash^2+\eps^2)^{-1} \pi^\th(\ee_{4N^2}).
\end{align*}

Le second terme de la derni\`ere expression est \`a trace, car
compos\'e d'un produit de $N$ termes, chacun appartenant \`a
$\L^p(\H)$ pour $p > N$, ainsi que d'un terme dans $\L^q(\H)$ pour $q > 2N$,
d'apr\`es le lemme~\ref{lm:schatten}. En commutant
$\pi^\th(e)$ encore une fois, on obtient en utilisant la relation d'ordre
sur les unit\'es locales $\ee_K$
\begin{align*}
&\pi^\th(e) (\dslash+i\eps)^{-1} \pi^\th(e) (\dslash-i\eps)^{-1}
\pi^\th(\ee_{4N}) (\dslash^2+\eps^2)^{-1} \pi^\th(\ee_{8N}) \cdots
(\dslash^2+\eps^2)^{-1} \pi^\th(\ee_{4N^2})
\\
&= \pi^\th(e) (\dslash^2+\eps^2)^{-1} \pi^\th(e)
(\dslash^2+\eps^2)^{-1} \pi^\th(\ee_{8N}) \cdots
(\dslash^2+\eps^2)^{-1} \pi^\th(\ee_{4N^2})
\\
&\quad + \pi^\th(e) (\dslash^2+\eps^2)^{-1} [\dslash,\pi^\th(e)]
(\dslash-i\eps)^{-1} \pi^\th(\ee_{4N}) (\dslash^2+\eps^2)^{-1}
\pi^\th(\ee_{8N}) \cdots (\dslash^2+\eps^2)^{-1} \pi^\th(\ee_{4N^2}).
\end{align*}
Le dernier terme \'etant encore \`a trace, il en r\'esulte que
$$
\pi^\th(e) (\dslash^2+\eps^2)^{-N}
\sim \pi^\th(e) (\dslash^2+\eps^2)^{-1} \pi^\th(e)
(\dslash^2+\eps^2)^{-1} \pi^\th(\ee_{8N}) \cdots
(\dslash^2+\eps^2)^{-1} \pi^\th(\ee_{4N^2}).
$$
On obtient finalement le r\'esultat en appliquant cet algorithme
$(N-1)$ fois:
$$
\pi^\th(e) (\dslash^2+\eps^2)^{-N}
\sim \bigl( \pi^\th(e) (\dslash^2+\eps^2)^{-1} \pi^\th(e) \bigr)^N.
\eqno\qed
$$
\hideqed
\end{proof}

\begin{cly}
\label{cr:killproj}
$\Tr_\omega\bigl(\pi^\th(g)\,[\pi^\th(f),(\dslash^2+\eps^2)^{-N}]\bigr)
= 0$ pour tout $g \in \SS$ et tout projecteur $f \in \A_c$.
\end{cly}

\begin{proof}[Preuve]
C'est une cons\'equence du lemme pr\'ec\'edent, g\'en\'eralis\'e
\`a n'importe quel projecteur $f$ de l'alg\`ebre quasi-locale $\A_c$ et appliqu\'e
\`a $\pi^\th(g\mop f)\,(\dslash^2+\eps^2)^{-N}$ et \`a son adjoint.
\end{proof}

Nous disposons dor\'enavant de tous les outils n\'ecessaires pour
\'evaluer la trace de Dixmier.

\begin{prop}
\label{pr:calcul}
Soit $f \in \SS$, alors chaque trace de Dixmier $\Tr_\omega$ de
$\pi^\th(f)\,(\dslash^2 + \eps^2)^{-N}$ est ind\'ependante de~$\eps$ et de $\omega$,
et
$$
\Tr_\omega \bigl( \pi^\th(f)\,(\dslash^2 + \eps^2)^{-N} \bigr)
= \frac{2^N\,\Omega_{2N}}{2N\,(2\pi)^{2N}} \int \,d^{2N}x\,f(x)
= \frac{1}{N!\,(2\pi)^N} \int \,d^{2N}x\,f(x) ,
$$
o\`u $\Omega_{2N}$ d\'esigne l'hypersurface de la boule unit\'e de $\R^{2N}$.
\end{prop}

\begin{proof}[Preuve]
Nous allons commencer par d\'emontrer ce r\'esultat pour
$f \in \A_c$. Soit $e$ une unit\'e locale pour $f$ (qui en poss\`ede toujours par
hypoth\`ese),
i.e. $e \mop f = f \mop e = f$. D'apr\`es les lemmes ~\ref{lm:existence}
et~\ref{lm:traceclass} et parce que $\L^1(\H)$ est contenu dans le noyau de
la trace Dixmier, on obtient
$$
\Tr_\omega( \pi^\th(f) \,(\dslash^2+\eps^2)^{-N}) = \Tr_\omega \bigl(
\pi^\th(f) \,(\pi^\th(e) (\dslash^2+\eps^2)^{-1} \pi^\th(e))^N \bigr).
$$
Le lemme~\ref{lm:traceclass} appliqu\'e \`a $f = e$ implique alors que
l'op\'erateur positif $\bigl( \pi^\th(e) (\dslash^2+\eps^2)^{-1} \pi^\th(e) \bigr)^N$
appartient \`a $\L^{1+}(\H)$ puisqu'\'egal \`a
$\pi^\th(e) (\dslash^2+\eps^2)^{-N}$ plus un terme appartenant \`a $\L^1(\H)$.
Ainsi, \cite[th\'eor\`eme~5.6]{CareyPS} implique (\'etant donn\'e que la limite converge,
toutes les traces de Dixmier donneront le m\^eme r\'esultat):
\begin{align}
\Tr_\omega \bigl( \pi^\th(f) \,(\dslash^2+\eps^2)^{-N} \bigr)
&= \lim_{s\downarrow 1} (s-1) \Tr\bigl[ \pi^\th(f)\,
(\pi^\th(e) (\dslash^2+\eps^2)^{-1} \pi^\th(e))^{Ns} \bigr]
\nonumber \\
&= \lim_{s\downarrow 1} (s-1) \Tr\bigl( \pi^\th(f) \pi^\th(e)
(\dslash^2+\eps^2)^{-Ns} \pi^\th(e) + E_{Ns} \bigr),
\label{eq:decompo}
\end{align}
o\`u
$$
E_{Ns} := \pi^\th(f) \,
\bigl( \pi^\th(e) (\dslash^2+\eps^2)^{-1} \pi^\th(e) \bigr)^{Ns}
- \pi^\th(f)\pi^\th(e) (\dslash^2+\eps^2)^{-Ns} \pi^\th(e).
$$
Le lemme~\ref{lm:traceclass} montre encore que $E_N \in \L^1(\H)$.

Pour $s > 1$,  le premier terme de $\pi^\th(f) \,\bigl(\pi^\th(e)
(\dslash^2+\eps^2)^{-1} \pi^\th(e) \bigr)^{Ns}$ de $E_{Ns}$ est \`a trace.
En effet, d'apr\`es le lemme~\ref{lm:schatten} et puisque
$\pi^\th(e) (\dslash^2+\eps^2)^{-1} \in \L^p(\H)$ pour $p > N$, nous avons
$\pi^\th(e) (\dslash^2+\eps^2)^{-1} \pi^\th(e) \in \L^{Ns}(\H)$. Cet op\'erateur
\'etant positif, on conclut que
$$
\bigl( \pi^\th(e) (\dslash^2+\eps^2)^{-1} \pi^\th(e) \bigr)^{Ns}
\in \L^1(\H).
$$
Le second terme,
$\pi^\th(f) \pi^\th(e) (\dslash^2+\eps^2)^{-Ns} \pi^\th(e)$ appartient aussi \`a
$\L^1(\H)$, car
$$
\|\pi^\th(e) (\dslash^2+\eps^2)^{-Ns} \pi^\th(e)\|_1
= \|(\dslash^2+\eps^2)^{-Ns/2} \pi^\th(e)\|_2^2
= \|\pi^\th(e) (\dslash^2+\eps^2)^{-Ns/2}\|_2^2
$$
qui est fini d'apr\`es le lemme~\ref{lm:HSO}. Ainsi, $E_{Ns} \in \L^1(\H)$ pour tout
$s \geq 1$. De l'\'equation \eqref{eq:decompo}, il r\'esulte que
\begin{align*}
\Tr_\omega \bigl( \pi^\th(f) \,(\dslash^2+\eps^2)^{-N} \bigr)
&= \lim_{s\downarrow 1} (s-1) \Tr \bigl(
\pi^\th(f) \pi^\th(e) (\dslash^2+\eps^2)^{-Ns} \pi^\th(e) \bigr)
\\
&= \lim_{s\downarrow 1}
(s-1) \Tr \bigl(\pi^\th(f) (\dslash^2+\eps^2)^{-Ns} \bigr).
\end{align*}

En appliquant le lemme~\ref{lm:in-extremis}, nous obtenons
\begin{align*}
\Tr_\omega \bigl( \pi^\th(f) \,(\dslash^2+\eps^2)^{-N} \bigr)
&= \lim_{s\downarrow 1} (s-1) \Tr(1_{2^N})
\Tr\bigl(L^\th_f (\Delta +\eps^2)^{-Ns}\bigr)
\\
&= 2^N(2\pi)^{-2N} \lim_{s\downarrow 1} (s-1)
\iint \,d^{2N}\xi \,d^{2N}x\,f(x)\, (|\xi|^2+\eps^2)^{-Ns}
\\
&= \frac{1}{N!\,(2\pi)^N} \int \,d^{2N}x\,f(x) ,
\end{align*}
o\`u l'identit\'e
$$
\int \,d^{2N}\xi\,(|\xi|^2+\eps^2)^{-Ns}
= \pi^N \frac{\Ga(N(s-1))}{\Ga(Ns)\,\eps^{2N(s-1)}},
$$
ainsi que le comportement $\Ga(N\a) \sim 1/N\a $ pour $ \a \downarrow 0$,
ont \'et\'e utilis\'es. La proposition est alors d\'emontr\'ee pour $f \in \A_c$.

Soit finalement $f$ arbitraire dans $\SS$.
Rappelons aussi que $\{e_K\}_K$ forme un syst\`eme d'unit\'es approch\'ees
pour $\A_\th$. D'apr\`es la propri\'et\'e de factorisation forte
$f = g \mop h$ avec
$g,h \in \SS$, le corollaire~\ref{cr:killproj} implique
\begin{align*}
\bigl| \Tr_\omega &\bigl( (\pi^\th(f) - \pi^\th(\ee_K \mop f \mop \ee_K))
(\dslash^2+\eps^2)^{-N} \bigr) \bigr|
\\
&= \bigl| \Tr_\omega \bigl( (\pi^\th(f) - \pi^\th(\ee_K \mop f))\,
(\dslash^2+\eps^2)^{-N} \bigr) \bigr|
\\
&= \bigl| \Tr_\omega \bigl( (\pi^\th(g) - \pi^\th(\ee_K \mop g))\,
\pi^\th(h) (\dslash^2+\eps^2)^{-N} \bigr) \bigr|
\\
&\leq \| \pi^\th(g) -\pi^\th(\ee_K \mop g) \|_\infty \,
\Tr_\omega \bigl| \pi^\th(h)\,(\dslash^2+\eps^2)^{-N} \bigr|.
\end{align*}
Etant donn\'e que $\|\pi^\th(g) -\pi^\th(\ee_K \mop g)\|_\infty
\leq (2\pi\th)^{-N/2}\|g - \ee_K \mop g\|_2$ tend vers z\'ero lorsque
$K$ cro\^it, la preuve est compl\`ete car $\ee_K\mop f\mop \ee_K$
appartient \`a $\A_c$ et car
\begin{align}
\int \,d^{2N}x\,[\ee_K\mop f\mop \ee_K](x) \to \int d^{2N}x\,f(x)\, ,\,\,\,
K\uparrow\infty.
\tag*{\qed}
\end{align}
\hideqed
\end{proof}

\begin{rem}
A partir d'arguments similaires, on peut aussi montrer que pour tout
$f\in\SS$,
$$
\Tr_\omega \bigl( \pi^\th(f)\,(|\dslash| + \eps)^{-2N} \bigr)
= \Tr_\omega \bigl( \pi^\th(f)\,(\dslash^2 + \eps^2)^{-N} \bigr).
$$
\end{rem}

En conclusion: {\it les dimensions spectrale et analytique des plans de Moyal
co\"incident}. Le lemme~\ref{lm:existence}, la proposition~\ref{pr:calcul} ainsi
que la remarque pr\'ec\'edente terminent la preuve
du th\'eor\`eme~\ref{th:dim-spectrale}. \qed

\subsubsection{2.2.3.2 La condition de r\'egularit\'e}

\begin{thm}
\label{th:regular}
Pour $f \in \Aun_\th$, les op\'erateurs born\'es $\pi^\th(f)$ et
$[\dslash,\pi^\th(f)]$ appartiennent \`a l'intersection des domaines
des puissances de la d\'erivation
$\delta(T) := [|\dslash|,T]$.
\end{thm}

Dans le cas compact commutatif, la d\'emonstration de ce r\'esultat
\cite{ConnesMIndex,Polaris} est grandement simplifi\'ee, en remarquant que
$$
\bigcap_{n\in\N}\Dom(\delta^n)=\bigcap_{n,m\in\N}\big(\Dom(L^n)\cap
\Dom(R^n)\big),
$$
o\`u les op\'erateurs (sur $\L(\H)$) $L$ et $R$ sont d\'efinis par
\begin{equation}
\label{eq:lr-smooth}
L(\.) := |\dslash|^{-1}\,[\dslash^2,\.],  \quad
R(\.) := [\dslash^2,\.]\,|\dslash|^{-1}.
\end{equation}
Cette simplification vient du fait que les commutateurs avec $\dslash^2$
sont beaucoup plus faciles \`a
manipuler que ceux avec $|\dslash|$. Cependant, dans le cas non compact
(commutatif ou non), les op\'erateurs $L$ et $R$ ne sont pas bien d\'efinis,
car $|\dslash|^{-1}$ ne l'est pas. Une preuve analogue devrait exister,
en d\'efinissant \`a la place de $L$ et $R$ les transformations
$L_\la$ et $R_\la$, pour $\la$ r\'eelle, par
$$
L_\la(\.) := (|\dslash| + i\la)^{-1}\,[\dslash^2,\.],  \quad
R_\la(\.) := [\dslash^2,\.]\,(|\dslash| - i\la)^{-1}.
$$

Nous allons d\'emontrer cette condition par une approche plus directe,
dont la vertu est de
rester valable dans le cas commutatif compact ou non.

\begin{proof}[Preuve du th\'eor\`eme~\ref{th:regular}]
Comme pr\'ec\'edemment, puisque
$[\dslash,\pi^\th(f)] = -iL^\th(\del_\mu f) \ox \ga^\mu$, il est
suffisant de d\'emontrer le r\'esultat pour $\pi^\th(f)$.
Pour tout $n \in \N$ et $\rho > 0$,  en it\'erant
l'identit\'e spectrale \eqref{eq:spectid} $n$ fois, on obtient pour
$\delta^n(\pi^\th(f))$:
$$
\frac{1}{\pi^n} \int_0^\infty \!\!\cdots \int_0^\infty \,d\la_n\dots d\la_1\,
\prod_{i=1}^n \frac{\sqrt{\la_i}}{(|\dslash|+\rho)^2+\la_i} \,
(\ad(|\dslash|+\rho)^2)^n \,(\pi^\th(f))
\prod_{i=1}^n \frac{1}{(|\dslash|+\rho)^2+\la_i} ,
$$
avec une notation \'evidente pour les commutateurs
it\'er\'es  $n$ fois.

De $[\dslash^2,\pi^\th(f)] = \dslash^2(f) + 2\dslash(f)\,\dslash$,
avec la notation $\dslash(f) := -iL^\th(\del_\mu f) \ox \ga^\mu$,
on peut v\'erifier que le terme de plus haut degr\'e en $\dslash$ dans
le d\'eveloppement de $(\ad(|\dslash|+\rho)^2)^n \,(\pi^\th(f))$ est
$2^n\dslash^n(f)\,\dslash^n$. Pour le reste de la d\'emonstration,
nous ne consid\'ererons que les termes de plus haut degr\'e en $\dslash$.
Comme dans la preuve du lemme~\ref{lm:commutateur}, tous les commutateurs
$[|\dslash|, \pi^\th(f)]$ qui apparaissent \`a cause de la pr\'esence
artificielle de~$\rho$, seront trait\'es comme une somme de deux
op\'erateurs d'ordre un.
Ainsi,
\begin{align}
\frac{1}{\pi^n} & \int_0^\infty \!\!\cdots \int_0^\infty d\la_n\dots d\la_1\,
\prod_{i=1}^n \frac{\sqrt{\la_i}}{(|\dslash|+\rho)^2+\la_i}
2^n \dslash^n(f) \dslash^n
\prod_{j=1}^n \frac{1}{(|\dslash|+\rho)^2+\la_j}
\nonumber \\
&=\frac{1}{\pi^n} \int_0^\infty \!\!\cdots \int_0^\infty \,d\la_n\dots d\la_1\,
2^n \dslash^n(f) \,\dslash^n \prod_{i=1}^n
\frac{\sqrt{\la_i}}{((|\dslash|+\rho)^2+\la_i)^2}
              \label{eq:chorizo} \\
&\qquad + \frac{1}{\pi^n} \int_0^\infty \!\!\cdots \int_0^\infty\,d\la_n\dots d\la_1\,
\biggl[ \prod_{i=1}^n \frac{1}{(|\dslash|+\rho)^2+\la_i},
2^n \dslash^n(f) \biggr] \dslash^n \prod_{i=1}^n
\frac{\sqrt{\la_i}}{(|\dslash|+\rho)^2+\la_i} .
\nonumber
\end{align}

En utilisant $\int_0^\infty\,d\la\, t(\la+t^2)^{-2} \sqrt{\la} = \pi/2$,
le premier terme du membre de droite de l'\'equation \eqref{eq:chorizo}
est \'egal \`a
$$
2^n \dslash^n(f)  \frac{\dslash^n}{(|\dslash|+\rho)^n}
\biggl( \frac{1}{\pi} \int_0^\infty \,d\la\,
\frac{|\dslash|+\rho}{((|\dslash|+\rho)^2+\la)^2}
\,\sqrt{\la} \biggr)^n
=\dslash^n(f)  \frac{\dslash^n}{(|\dslash|+\rho)^n},
$$
qui est un op\'erateur born\'e.

Pour les autres termes, il faut remarquer que le commutateur
$\bigl[\prod_i ((|\dslash|+\rho)^2+\la_i)^{-1}, \dslash^n(f)\bigr]$
peut se r\'e\'ecrire comme
$$
- \prod_{i=1}^n ((|\dslash|+\rho)^2+\la_i)^{-1} \,
\biggl[ \prod_{j=1}^n ((|\dslash|+\rho)^2+\la_j), \dslash^n(f) \biggr]
\prod_{k=1}^n ((|\dslash|+\rho)^2+\la_k)^{-1},
$$
et le terme de plus haut degr\'e en $\dslash$ est, \`a une constante pr\`es
$$
\prod_{i=1}^n ((|\dslash|+\rho)^2+\la_i)^{-1} \,
\dslash^{n+1}(f)\,\dslash^{2n-1}
\prod_{k=1}^n ((|\dslash|+\rho)^2+\la_k)^{-1}.
$$
Ainsi la preuve se r\'eduit \`a montrer que la norme suivante est finie
\begin{align*}
&\biggl\| \int_0^\infty \!\!\cdots \int_0^\infty \,d\la_n\dots d\la_1\,
\prod_{i=1}^n \frac{\sqrt{\la_i}}{(|\dslash|+\rho)^2+\la_i}\,
\dslash^{n+1}(f)\,\dslash^{3n-1}
\biggl( \prod_{j=1}^n \frac{1}{(|\dslash|+\rho)^2+\la_j} \biggr)^2
 \biggr\|
\\
&\leq \|\dslash^{n+1}(f)\| \int_0^\infty \!\!\cdots \int_0^\infty
\prod_{i=1}^n \,d\la_i\,\biggl( \biggl\|
\frac{\dslash^{3-1/n}}{((|\dslash|+\rho)^2+\la_i)^{3/2-1/2n}} \biggr\|
\biggr\| \frac{\sqrt{\la_i} }{(|\dslash|+\rho)^2+\la_i} \biggr\|^{3/2+1/2n}
 \biggr)
\\
&\leq \|\dslash^{n+1}(f)\| \biggl( \int_0^\infty\,d\la\,
\frac{\sqrt{\la}}{(\rho^2 + \la)^{3/2+1/2n}}  \biggr)^n.
\end{align*}
Cette derni\`ere int\'egrale est finie pour tout $n\in\N$ ainsi que la derni\`ere norme,
car $\del^\a f \in \Aun_\th \subset A_\th$ pour $|\a| \leq n+1$;
la preuve est alors compl\`ete.
\end{proof}

\subsubsection{2.2.3.3 La condition de finitude}

\begin{lem}
Les vecteurs lisses pour l'op\'erateur de Dirac sont donn\'es par
$$
\H^\infty
:= \bigcap_{k\in\N} \Dom(\dslash^k) \simeq \D_{L^2} \ox \C^{2^N}.
$$
\end{lem}

\begin{proof}[Preuve]
Puisque $\D_{L^2}$ est le domaine commun des d\'eriv\'ees
partielles $\del_\mu$, o\`u $\mu = 1,\dots,2N$ et
$\dslash = -i\del_\mu \ox \ga^\mu$, la conclusion est imm\'ediate.
\end{proof}

D'apr\`es le lemme \ref{lm:DL2}, $\bar{\A}_\th := \D_{L^2}$ est un id\'eal dans
$\Aun_\th$. Alors, $\H^\infty$ est un $\bar{\A}_\th$-pullback d'un module libre \`a gauche
sur $\Aun_\th$.

De plus, il y a sur $\H^\infty$ une structure Hermitienne naturelle
\`a valeur dans $\bar{\A}_\th$, donn\'ee par
$$
\roundbraket{\xi}{\eta}' := \sum_{j=1}^{2^N} \xi_j \mop \eta_j^*,
\sepword{pour tout} \xi,\eta \in \H^\infty.
$$
Puisque $\D_{L^2} \subset \M^\th$, le couplage Hermitien
$\roundbraket{\pi^\th(a)\xi}{\eta}' := a \mop \roundbraket{\xi}{\eta}'$
est \`a valeur dans $\A_\th$ si $a \in \A_\th$.
La proposition~\ref{pr:calcul} et le lemme \ref{lem:proo} impliquent alors
\begin{align*}
\Tr_\omega\bigl( \pi^\th(\roundbraket{\xi}{\eta}')\,
(\dslash^2 + \eps^2)^{-N} \bigr)
&= \frac{2^N\,\Omega_{2N}}{2N\,(2\pi)^{2N}} \sum_{j=1}^{2^N}
\int d^{2N}x\,(\xi_j \mop \eta_j^*)(x) \
\\
&= \frac{1}{N!\,(2\pi)^N} \sum_{j=1}^{2^N}
\int \,d^{2N}x\,\eta_j^*(x)\,\xi_j(x) .
\end{align*}
Ainsi,
$\roundbraket{\xi}{\eta} := N!\,(2\pi)^N \roundbraket{\xi}{\eta}'$
est une structure Hermitienne satisfaisant \`a l'\'equation~\eqref{eq:abs-cont}. Son
unicit\'e peut \^etre obtenue en reproduisant la preuve du cas
commutatif~\cite[p.~501]{Polaris}.

\begin{rem}
De la d\'efinition des espaces $\Oh_r$, il est clair que
$\Oh_r \subset \D_{L^2}$ si et seulement si $r < -N$. Soit alors
$\NN := \bigcup_{r<-N} \Oh_r \subset \D_{L^2}$. L'utilisation de la
proposition~\ref{pr:hector}, indique que l'espace $\NN$ est une $*$-alg\`ebre pour le
produit~$\mop$ et qu'il est un id\'eal dans
$\Aun_\th = \B$. Cet espace de fonctions a d\'ej\`a \'et\'e utilis\'e en
physique, en particulier dans~\cite{LangmannM}.
\end{rem}

\subsubsection{2.2.3.4 Les autres axiomes}

\begin{description}

\item[$\quad \bullet$]
Les propri\'et\'es de commutation/anticommutation
avec la structure r\'eelle  $J$, sont \'evidem\-ment satisfaites car ni
$J$, ni $\chi$, ni $\dslash$ n'ont \'et\'e d\'eform\'es. L'axiome de
r\'ealit\'e est alors satisfait.

\item[$\quad \bullet$]
La condition de premier ordre est une cons\'equence de
l'\'equation \eqref{eq:right-Moyal},
$$
[[\dslash, \pi^\th(f)], J \pi^\th(g) J^{-1}]
= [\pi^\th(\dslash(f)), J \pi^\th(g) J^{-1}]
= -i\,[L^\th(\del_\mu f) \ox \ga^\mu, R^\th(g^*) \ox 1] = 0,
$$
car par associativit\'e du produit twist\'e, les op\'erateurs de
multiplication twist\'ee \`a droite et \`a gauche commutent.

\item[$\quad \bullet$]
La condition d'orientation n\'ecessite un $2N$-cycle de Hochschild
sur $\Aun_\th = \B$. En utilisant le plongement du tore non commutatif
$\Coo(\T_\Theta^k)$ dans $\B$ (vu comme sous-alg\`ebre des fonctions
p\'eriodiques avec une p\'eriode fix\'ee), on obtient le cycle d\'esir\'e
\`a partir des g\'en\'erateurs $u_j(x):=e^{-ix_j}$, $j = 1,\dots,2N$ (qui satisfont
\'evidemment aux relations canoniques du tore NC). On obtient alors
$$
\frac{(-i)^N}{(2N)!} \sum_\sigma (-1)^\sigma
(u_{\sigma(1)} \mop u_{\sigma(2)} \mop\cdots\mop
u_{\sigma(2N)})^{\star_{\Theta}-1}
[\dslash,u_{\sigma(1)}]\cdots[\dslash,u_{\sigma(2N)}] = \chi,
$$
o\`u la repr\'esentation $\pi^\th$ est sous-entendue.

\end{description}

\begin{thm}
\label{th:main}
Les plans de Moyal $(\SS,\D_{L^2},\B,\H,\dslash,J,\chi)$ sont
des triplets spectraux r\'eels, sans unit\'e, connexes de dimension
spectrale~$2N$.
\qed
\end{thm}

\subsection{Remarques sur les contraintes impos\'ees}

Il est normal de se questionner sur la pertinence des choix que nous
avons fait, qui \'evidemment ne peuvent \^etre absolument uniques. Cependant,
le contenu analytique  des donn\'ees $(\A_\th, \bar{\A}_\th,\Aun_\th, \H, $ $\D, J, \chi)$
est s\'ev\`erement contraint par les axiomes.

Tout d'abord et comme remarqu\'e pr\'ec\'edemment, le choix de l'alg\`ebre
$\bar{\A}_\th$ est presque exclusivement donn\'e par la condition de finitude.
En d'autres termes, c'est l'op\'erateur $\D$ qui nous nous dit quelle
doit \^etre l'alg\`ebre sans unit\'e \`a consid\'erer.

Pour le choix du plongement $\A_\th\hookrightarrow\Aun_\th$,
le fait que $\Aun_\th$ soit repr\'esent\'ee
par des op\'erateurs born\'es donne une contrainte `sup\'erieure'.
La condition d'orientabilit\'e
donne quant \`a elle un plongement `minimal'. En effet, $\Aun_\th$ doit
permettre de construire `le bon' cycle de Hochschild. Dans le
cas des plans de Moyal, nous avons vu que l'appartenance
des ondes planes au plongement choisi est une contrainte d\'ecisive.

Concernant l'alg\`ebre interm\'ediaire $\A_\th$, les contraintes impos\'ees
sont de nature purement analytique ($a(\D-\lambda)^{-k}\in\L^{(1,\infty)}$
pour tout $a\in\A_\th$). Au vu des techniques utilis\'ees dans le cas des plans de Moyal,
l'alg\`ebre $\A_\th$ doit d'une part \^etre inclue dans l'intersection des fonctions
sommable et de carr\'e sommable et d'autre part doit satisfaire \`a la
propri\'et\'e de factorisation forte. Pour l'heure, seule  $(\SS,\mop)$
est connue pour satisfaire \`a toutes ces conditions.

Notons finalement que les contraintes impos\'ees par la condition
de r\'egularit\'e sont en fait relativement faibles.

\section{Application aux d\'eformations isospectrales}

Nous allons regarder maintenant les points clefs de la construction
de triplets spectraux sans unit\'e associ\'es aux d\'eformations isospectrales
non compactes.

Rappelons que la construction de nouveaux
triplets spectraux (\`a unit\'e) \cite{CL}, fut la motivation premi\`ere pour introduire les
d\'eformations isospectrales p\'eriodiques.
Ainsi, \`a une d\'eformation isospectrale p\'eriodique et compacte,
on peut associer canoniquement un triplet spectral satisfaisant
\`a toutes les conditions requises \cite{ConnesReal}.
En effet, d'une part la construction de Connes--Dubois-Violette
\cite{CDV} donne directement acc\`es \`a une `pr\'e-$C^*$-alg\`ebre
lisse' $\Coo(M_\Th)$ et d'autre part, $\Coo(M_\Th)$ ayant une unit\'e
et le reste du triplet n'\'etant pas d\'eform\'e, tous les autres
axiomes sont directement satisfaits.

Nous n'allons donner que des r\'esultat partiels dans les cas g\'en\'eriques
non compacts (p\'eriodi\-ques et non p\'eriodiques). Il y a deux raisons
principales \`a cela. Tout d'abord, comparativement aux plans
de Moyal, la `technologie alg\'ebrique' n'est pas suffisamment
\'elabor\'ee; ce travail est en cours. Deuxi\`emenent, il r\'esulte de
l'\'etude pr\'ec\'edente, que la condition d\'eterminante pour
l'\'elaboration de triplets spectraux sans unit\'e est celle
de la dimension spectrale.

Nous allons alors nous borner \`a d\'emontrer qu'elle
est satisfaite aussi dans les cas courbes. Pour ce faire, l'outil
principal sera une estimation du comportement
du noyau de la chaleur sur sa diagonale. Nous verrons aussi au chapitre
\ref{NCQFT}, que son comportement hors de la diagonale est
primordial pour l'\'etude des th\'eories quantiques des champs sur
d\'eformations isospectrales.

\subsection{Noyau de la chaleur et classes de Schatten}
\label{section26}

Dans ce paragraphe, nous allons \'etablir des estimations pour les
normes de Schatten des op\'era\-teurs
$M_f(1 + \Delta)^{-k}$ et $L_f(1 + \Delta)^{-k}$ agissant sur
l'espace de Hilbert r\'eduit $\H_r := L^2(M,\mu_g)$. Ici, $\Delta$
d\'esigne encore le Laplacien $(d + d^*)^2$ restreint aux  0-formes.
De telles estimations seront obtenues en utilisant le comportement
du noyau de la chaleur sur sa diagonale, la pr\'esentation en terme de transform\'ee
de Laplace des op\'erateurs $(1+\Delta)^{-k}$,  ainsi que
la proposition~\ref{pr:kernel} et le th\'eor\`eme~\ref{th:HS-norm}.

Nous avons besoin de faire de plus amples hypoth\`eses sur le `comportement
de la g\'eom\'etrie \`a l'infini'. Dor\'enavant, $K_t(p,p')$ d\'esignera le noyau
de la chaleur, i.e. le noyau du semi-groupe de la chaleur $e^{-t\Delta}$.
Rappelons qu'en toute g\'en\'eralit\'e, $K_t(p,p')$ est une fonction
ind\'efiniment diff\'erentiable, strictement positive et sym\'etrique
sur $M \x M \x (0, \infty)$ \cite[th\'eor\`eme~5.2.1]{Davies}.

\vspace{0.7cm}
Dans le reste de ce chapitre, nous allons supposer que
$M$ est une vari\'et\'e Riemannienne connexe, compl\`ete,
de dimension $n \geq 2$ et de plus, qu'une des deux
hypoth\`eses suivantes soit satisfaite.

Soit $B(p,r) := \set{p'\in M : d_g(p,p') < r}$ la boule g\'eod\'esique
centr\'ee en~$p$ et de rayon $r$. La constante isop\'erim\'etrique $\I(M)$
de la vari\'et\'e,  est donn\'ee \cite[p.~96]{Chavel} par
$$
\I(M) := \inf_\Omega \frac{A(\del\Omega)^n}{V(\Omega)^{n-1}},
$$
o\`u $\Omega$ varie dans l'ensemble des sous-vari\'et\'es  de $M$ ayant
une fermeture compacte et un bord lisse.
Ici, $V(\Omega)$ et $A(\del\Omega)$ sont respectivement
le volume et l'aire Riemannien de $\Omega$ et de $\del\Omega$.

Les hyphoth\`eses suppl\'ementaires sur le comportement
de la g\'eom\'etrie \`a l'infini sont:
{\it
\begin{enumerate}
\item[(I)]
$M$ n'est pas compacte et il existe $a > 0$ tel que
$\sup_{p\in M} \I(B(p,a))^{-1} < \infty$. (Cette propri\'et\'e
est en particulier v\'erifi\'ee lorsque $M$ a une constante
isop\'erim\'etrique strictement positive: $\I(M) > 0$.)

\item[(II)]
$M$ a une courbure de Ricci born\'ee inf\'erieurement, c'est-\`a-dire,
$\Ricci(p,p) \geq (n - 1)\,\chi$, pour tout $p \in M$ et pour une certaine
constante~$\chi$. \\
D'apr\`es \cite[th\'eor\`eme 3.9]{Chavel2}, ceci entra\^ine que
$\sup_{p\in M} V(B(p,a))^{-1} < \infty$, pour un certain $a > 0$.
\end{enumerate}
}

Dans\cite[th\'eor\`eme~8, p.~198]{Chavel}, il est d\'emontr\'e que si $M$
satisfait \`a l'hypoth\`ese~(I), alors le noyau de la chaleur est born\'e inf\'erieurement:
pour tout $p \in M$ et $r > 0$ pour lequel $\overline{B(p,r)}$ appartient
\`a l'image de l'application exponentielle $\exp_p$, on a
\begin{equation}
\label{eq:k1}
K_t(p,p) \leq C_1(n)\, (t^{-n/2} + r^{-(n+2)}\,t) \, \I(B(p,r))^{-1}.
\end{equation}

Des estimations diff\'erentes sont donn\'ees dans  \cite[lemme 15]{DaviesBound}
(voir aussi \cite{Davies}) lorsque la condition (II) est satisfaite:
en se donnant $\eps > 0$, il existe une constante $c_\eps$ telle que
pour tout $t > 0$ et $p\in M$, on ait
\begin{equation*}
0 \leq K_t(p,p) \leq c_\eps(n)\,V(B(p,t^{1/2}))^{-1} \,e^{(\eps-E)t},
\end{equation*}
o\`u $E := \inf\spec(\Delta) \geq 0$. Etant donn\'e que
\cite[proposition~4.1]{CheegerGromovTaylor},
$$
V(B(p,r)) \geq c\,r^n \,V(B(p,1))  \sepword{pour} 0 < r < 1,
$$
on obtient alors
\begin{equation}
\label{eq:k2}
K_t(p,p) \leq
\begin{cases}
C_2(\eps)\, t^{-n/2}\, V(B(p,1))^{-1}\, e^{(\eps-E)t}, & t \leq 1, \\
C_3(\eps)\, V(B(p,1))^{-1}\, e^{(\eps-E)t},            & t > 1.
\end{cases}
\end{equation}

Que les suprema soient finis dans les hyphoth\`eses (I) ou (II) ne
joue aucun r\^ole dans les estimations \eqref{eq:k1} et \eqref{eq:k2},
mais c'est une condition suffisante pour obtenir le lemme \ref{lm:gro}.

\begin{lem}
Supposons que la courbure de Ricci de $(M,g)$ soit born\'ee inf\'erieurement.
Alors le semi-groupe de la chaleur est une contraction
pr\'eservant la positivit\'e sur $L^p(M,\mu_g)$, pour
$1 \leq p \leq \infty$.
\end{lem}
\begin{proof}[Preuve]
La propri\'et\'e de pr\'eservation de la positivit\'e ($\phi\geq0$ implique
$e^{-t\tri}\phi\geq0$) est une propri\'et\'e
g\'en\'erale des Laplacien de Dirichlet (voir \cite[th\'eor\`eme 5.2.1]{Davies} ou
\cite{DaviesS}); on peut alors appliquer le crit\`ere de Beurling--Deny
\cite[th\'eor\`eme~XIII.51]{ReedSIV}: un semi-groupe pr\'eservant la positivit\'e
est une contraction sur $L^\infty(M,\mu_g)$ pour tout $t > 0$ si et seulement si
il est une contraction sur chaque $L^p(M,\mu_g)$ pour tout $t > 0$.
Alors, pour $\phi \in L^\infty(M,\mu_g)$, on a
$$
\|e^{-t\Delta}\phi\|_\infty
= \sup_{p\in M} \int_M K_t(p,p') \,\phi(p') \,\mu_g(p')
\leq \sup_{p\in M} \|\phi\|_\infty \int_M K_t(p,p') \,\mu_g(p')
= \|\phi\|_\infty.
$$
Nous avons utilis\'e ici la positivit\'e du noyau de la chaleur $K_t$
et de la forme volume Riemannienne $\mu_g$, ainsi que la loi de
conservation de probabilit\'e $\int_M K_t(p,p') \,\mu_g(p') = 1$ qui est
satisfaite pour toute vari\'et\'e compl\`ete avec courbure de Ricci born\'ee
inf\'erieurement (voir \cite[th\'eor\`eme~5,
p.~191]{Chavel} ou \cite[th\'eor\`eme~5.2.6, p.~153]{Davies}).
\end{proof}

\begin{lem}
\label{lm:gro}
Supposons que  $M$ satisfasse (I) ou (II); alors $(1+\Delta)^{-k}$ est un
op\'erateur born\'e de $L^2(M,\mu_g)$ vers $L^\infty(M,\mu_g)$ pour tout $k > n/4$.
\end{lem}

\begin{proof}[Preuve]
Soit $\phi \in L^2(M,\mu_g)$. En utilisant l'in\'egalit\'e de Cauchy--Schwarz,
la positivit\'e ainsi que la sym\'etrie de $K_{(1+\Delta)^{-k}}$, la positivit\'e de
$\mu_g$, la r\`egle de produit pour les noyaux d'op\'erateurs et la pr\'esentation
en terme de transform\'ee de Laplace
$$
(1+\Delta)^{-2k} =
\Gamma(2k)^{-1} \int_0^\infty t^{2k-1}\,e^{-t\,(1+\Delta)} \,dt,
$$
nous obtenons
\begin{align*}
\|(1+\Delta)^{-k}\phi\|_\infty^2
&=
\sup_{p\in M} \biggl| \int_M\,\mu_g(p')\, K_{(1+\Delta)^{-k}}(p,p') \,\phi(p')
 \biggr|^2
\\
&\leq \|\phi\|_2^2 \, \sup_{p\in M} \int_M \,\mu_g(p')\,
|K_{(1+\Delta)^{-k}}(p,p')|^2
\\
&= \|\phi\|_2^2 \, \sup_{p\in M} \int_M \,\mu_g(p')\,
K_{(1+\Delta)^{-k}}(p,p')\, K_{(1+\Delta)^{-k}}(p',p)
\\
&= \|\phi\|_2^2 \, \sup_{p\in M} K_{(1+\Delta)^{-2k}}(p,p)
\\
&= \|\phi\|_2^2 \, \sup_{p\in M} \frac{1}{\Gamma(2k)}
\int_0^\infty \,dt\, t^{2k-1} \,e^{-t} \,K_t(p,p) .
\end{align*}

Par hypoth\`ese, les estimations \eqref{eq:k1} et \eqref{eq:k2} donnent
\begin{equation*}
K_t(p,p) \leq c_1\,(t^{-n/2} + c_2\,t)\, \max(e^{(\eps_0-E)t}, 1)
\end{equation*}
pour des constantes $c_1$, $c_2$, ind\'ependantes de $p$ et pour un
$\eps = \eps_0 < 1$ fix\'e.
Alors, l'int\'egrale sur $t$ est finie lorsque $k > n/4$, donc
$\|(1+\Delta)^{-k}\phi\|_\infty \leq c(k)\, \|\phi\|_2$.
\end{proof}

Nous arrivons au r\'esultat principal de ce paragraphe.

\begin{prop}
\label{pr:HiSc}
Si $M$ satisfait (I) ou (II), alors pour tout  $f\in L^2(M,\mu_g)$
et tout $k > n/4$, l'op\'erateur
$M_f\,(1+\Delta)^{-k}$ est Hilbert--Schmidt et satisfait
$$
\|M_f\,(1+\Delta)^{-k}\|_2 \leq C_k(n)\,\|f\|_2.
$$
\end{prop}

\begin{proof}[Preuve]
Que $M_f\,(1+\Delta)^{-k}$ soit un op\'erateur de Hilbert--Schmidt
est une cons\'equence du principe de factorisation de
Grothendieck \cite{Gro}: se donnant deux op\'erateurs born\'es
$$
A \in \L(L^2(X,\mu), L^\infty(X,\mu))\sepword{ et}
B\in \L(L^\infty(X,\mu), L^2(X,\mu)),
$$
leur produit $BA$ est
un op\'erateur de Hilbert--Schmidt sur $L^2(X,\mu)$.

Etant donn\'e que pour $f \in L^2(M,\mu_g)$, $M_f$ est born\'e
de $L^\infty(M,\mu_g)$ vers $L^2(M,\mu_g)$, le
lemme~\ref{lm:gro} montre que $M_f\,(1+\Delta)^{-k}$ est
Hilbert--Schmidt pour $k > n/4$. Pour l'estimation de la norme
de Hilbert--Schmidt nous utilisons encore  \eqref{eq:k1}, \eqref{eq:k2}, ainsi
que la repr\'esentation en termes de transform\'ee de Laplace:
\begin{align*}
\|M_f\,(1+\Delta)^{-k}\|_2^2
&= \int_{M\x M} \,\mu_g(p) \,\mu_g(p')\,
|f(p)|^2 |K_{(1+\Delta)^{-k}}(p,p')|^2
\\
&= \int_M \,\mu_g(p)\,|f(p)|^2 \, K_{(1+\Delta)^{-2k}}(p,p)
\\
&= \frac{1}{\Gamma(2k)}\int_M \,\mu_g(p)\,|f(p)|^2
\int_0^\infty\,dt\, t^{2k-1} \,e^{-t} \, K_t(p,p)
\leq C_k(n)^2 \,\|f\|_2^2,
\end{align*}
o\`u nous avons utilis\'e les m\^emes estimations sur l'int\'egrale en $t$ que
dans le lemme~\ref{lm:gro}, la sym\'etrie et la r\`egle du produit
pour le noyau $K_{(1+\Delta)^{-k}}$.
\end{proof}

\begin{rem}
Au vu des techniques employ\'ees, on peut g\'en\'eraliser
ce r\'esultat au moins pour les op\'erateurs de la forme
$M_f\,h(\sqrt{\Delta})$ o\`u $h$ est la transform\'ee de Laplace
d'une fonction qui se comporte comme $t^{k-1}$ lorsque $t \downarrow 0$,
pour $k > n/4$ et qui d\'ecro\^it suffisamment vite \`a l'infini.
\end{rem}

Nous allons, toujours par interpolation complexe,
obtenir des propri\'et\'es d'existence analogues \`a celles obtenues
lors du paragraphe pr\'ec\'edent, pour les autres classes de Schatten.

\begin{thm}
\label{th:inter}
Si $M$ satisfait (I) ou (II), lorsque $f \in L^p(M,\mu_g)$ pour $2 \leq p < \infty$ et $k > n/2p$,
on a $M_f\,(1+\Delta)^{-k} \in \L^p(\H_r)$.
\end{thm}

\begin{proof}[Preuve]
Le cas $p = 2$ a \'et\'e obtenu dans la proposition~\ref{pr:HiSc}.
Pour $p =\infty$, on a
$$
\|M_f\,(1+\Delta)^{-k}\| \leq \|M_f\| \,\|(1+\Delta)^{-k}\|
\leq \|f\|_\infty \,\sup_{x\in\R^+} (1+x)^{-k}
= \|f\|_\infty.
$$
Nous allons encore \'etablir le r\'esultat pour $2 < p < \infty$ par interpolation
complexe. Premi\`erement, on peut toujours supposer $f$
positive, car
$$
\|M_f\| = \|M_{|f|}\|,  \qquad
\|M_f\,(1+\Delta)^{-k}\|_2 = \|M_{|f|}\,(1+\Delta)^{-k}\|_2.
$$
Alors, pour $f \geq 0$ appartenant \`a $L^p(M,\mu_g)$, on
d\'efinit l'application
$$
F_p : z \mapsto M_f^{pz} \,(1+\Delta)^{-kpz},
$$
pour tout $z$ dans la bande
$S := \set{z \in \C : 0 \leq \Re z \leq \half}$. Pour tout $y \in \R$,
l'op\'erateur $F_p(iy) = M_{f^{ipy}}\,(1+\Delta)^{ikpy}$ est born\'ee, avec
$\|F_p(iy)\| \leq 1$; et pour $z = \half + iy$,
la proposition~\ref{pr:HiSc} montre que
$$
\|F_p(\half + iy)\|_2 = \|M_{f^{p/2}}\,(1+\Delta)^{-kp/2}\|_2
\leq C_{kp/2}(n)\, \|f^{p/2}\|_2 = C_{kp/2}(n)\, \|f\|_p^{p/2},
$$
qui est fini car $k > n/2p$. Par interpolation complexe
\cite{SimonTrace} on obtient que $F_p(z) \in \L^{1/\Re z}(\H_r)$ pour tout
$z \in S$, et
\begin{align*}
\|F_p(z)\|_{1/\Re z}
&\leq \|F_p(0)\|_\infty^{1-2\Re z}\, \|F_p(\half)\|_2^{2\Re z}
\leq \|M_{f^{p/2}}(1+\Delta)^{-kp/2}\|_2^{2\Re z}
\\
&\leq C_{kp/2}(n)^{2\Re z}\, \|f^{p/2}\|_2^{2\Re z}
= C_{kp/2}(n)^{2\Re z}\, \|f\|_p^{p\Re z}.
\end{align*}
Ainsi, pour $z = 1/p$
$$
\|F_p(1/p)\|_p = \|M_f\,(1+\Delta)^{-k}\|_p
\leq C_{kp/2}(n)^{2/p}\, \|f\|_p,
$$
qui conclut la preuve.
\end{proof}

Nous allons ensuite obtenir l'analogue de ce r\'esultat
pour les op\'erateurs d\'eform\'es, en utilisant la propri\'et\'e
fondamentale de l'invariance de la norme de Hilbert-Schmidt pour les
d\'eformations isospectrales (lemme \ref{th:HS-norm}), ainsi que la
borne sur la norme de $L_f$ (proposition \ref{pr:kernel}).
Nous ne formulerons cette proposition que pour les op\'erateurs
de multiplication \`a gauche, mais  \'evidemment, elle sera vraie
aussi dans le cas de la multiplication twist\'ee \`a droite.

\begin{prop}
\label{pr:interpolation}
Soient $2 \leq p < \infty$
et $f \in \Coo_c(M)$ o\`u $M$ satisfait \`a (I) ou (II).
Alors, dans le cas d'une action $\a$ isom\'etrique,  lisse
et propre du groupe Ab\'elien $\R^l$,
$L_f\,(1+\Delta)^{-k} \in \L^p(\H_r)$ pour tout $k > n/2p$.
\end{prop}

\begin{proof}[Preuve]
La preuve \'etant essentiellement la m\^eme que celle du th\'eor\`eme
pr\'ec\'edent, nous allons seulement l'esquisser.
Les th\'eor\`emes~\ref{th:HS-norm} et~\ref{pr:HiSc} impliquent que pour
$k > n/4$,
$$
\|L_f\,(1+\Delta)^{-k}\|_2 = \|M_f\,(1+\Delta)^{-k}\|_2
\leq C_k(n)\, \|f\|_2.
$$
De plus, par (\ref{eq:borne}),
$$
\|L_f\,(1+\Delta)^{-k}\| \leq \|L_f\| \leq \widetilde C_r(l) \,
\sup_{p\in M} \int_{\R^l} |(1+\Delta_y)^r \a_{\half\Th y}f(p)| \,d^ly
=: \omega(f;r,l,n),
$$
qui est fini lorsque $r > l/2$. En d\'efinissant
$G_p(z) := L_f\,(1+\Delta)^{-kpz}$ pour $z \in S$ et $k > n/2p$, on
conclut que pour tout $y \in \R$,
$$
\|G_p(iy)\| = \|L_f\,(1+\Delta)^{-ikpy}\| \leq \omega(f;r,l,n),
$$
et
$$
\|G_p(\half + iy)\|_2 = \|L_f\,(1+\Delta)^{-kpy/2}\|_2
\leq C_{kp/2}(n)\, \|f\|_2.
$$
Une fois encore, on obtient le r\'esultat par interpolation complexe:
\begin{align}
\|L_f\,(1+\Delta)^{-k}\|_p
&= \|G_p(p^{-1})\|_p
\leq \|G_p(0)\|_\infty^{1-2/p}\, \|G_p(2^{-1})\|_2^{2/p}
\notag \\
&\leq \omega(f;r,l,n)^{1-2/p}\, C_{kp/2}(n)^{2/p}\, \|f\|_2^{2/p}.
\tag*{\qed}
\end{align}
\hideqed
\end{proof}

\begin{rem}
Les conclusions de la proposition pr\'ec\'edente peuvent s'\'etendre au cas $p=1$
dans de nombreuses situations. En particulier, pour les plans de Moyal en utilisant la propri\'et\'e
de factorisation forte (Proposition 2.2.6), et dans le cas
g\'en\'eral lorsque $f=g\mop h$ pour $g,h\in\Coo_c(M)$.
En effet, pour $k>n/2$, avec $n\geq 2$, en \'ecrivant $k=1+\delta$ on a
$$
L_f(1+\tri)^{-k}=L_g(1+\tri)^{-1}L_h(1+\tri)^{-\delta}+L_g(1+\tri)^{-1}[\tri,L_h](1+\tri)^{-1-\delta}.
$$
On obtient le r\'esultat en utilisant la proposition pr\'ec\'edente, les in\'egalit\'es de H\"older
et en remarquant que le commutateur $[\tri,L_h]$ est \'egal \`a une somme d'op\'erateurs
de multiplication twist\'e fois un op\'erateur diff\'erentiel d'ordre un, qui multipli\'e avec le
facteur suppl\'ementaire $(1+\tri)^{-1}$ donne un op\'erateur born\'e.
\end{rem}

Nous allons montrer maintenant que la proposition pr\'ec\'edente
s'\'etend naturellement au fibr\'e spinoriel.

\begin{cly}
\label{cl:interpolation}
Soit $2 \leq p < \infty$ et $f \in \Coo_c(M)$. Sous les hypoth\`eses
de la proposition pr\'ec\'edente,  $L_f\,(1 + \Dslash^2)^{-k}$ et
$L_f\,(1 + |\Dslash|)^{-2k}$ appartiennent \`a $\L^p(\H)$ pour tout $k > n/2p$.
\end{cly}

\begin{proof}
L'op\'erateur $(1 + \Dslash^2)^{k}(1 + |\Dslash|)^{-2k}$ \'etant
born\'e, il est suffisant de consid\'erer le cas $L_f\,(1 + \Dslash^2)^{-k}$.
Pour cet op\'erateur, en utilisant la formule de Lichnerowicz 
\begin{equation}
\Dslash^2 = \Delta + \tfrac{1}{4} R,
\label{eq:Lich-formula}
\end{equation}
o\`u $R$ est le scalaire de courbure, on obtient le r\'esultat par comparaison
$$
(1 + \Dslash^2)^{-1}
= (1 + \Delta)^{-1} (1 - \tfrac{1}{4} R(1 + \Dslash^2)^{-1}).
$$
\end{proof}

Avant de clore ce paragraphe, nous allons montrer que les commutateurs
suivant poss\`edent les m\^emes propri\'et\'es de sommabilit\'e que celles de
$L_f(1 + \Dslash^2)^{-k}$.

\begin{lem}
\label{lem:plenty}
Si $f \in \Coo_c(M)$ et $2 \leq p < \infty$, alors les op\'erateurs
\begin{align*}
& [\Dslash, L_f]\,(1 + \Dslash^2)^{-k},
&& [|\Dslash|, L_f]\,(1 + \Dslash^2)^{-k},
&& [(1 + \Dslash^2)^{\half}, L_f]\,(1 + \Dslash^2)^{-k},
\\
& [\Dslash, L_f]\,(1 + |\Dslash|)^{-2k},
&& [|\Dslash|, L_f]\,(1 + |\Dslash|)^{-2k},
&& [(1 + \Dslash^2)^{\half}, L_f]\,(1 + |\Dslash|)^{-2k}
\end{align*}
appartiennent tous \`a $\L^p(\H)$ lorsque $k > n/2p$.
\end{lem}

\begin{proof}
Comme remarqu\'e pr\'ec\'edemment,
il est suffisant d'\'etablir le lemme dans le cas
$(1 + \Dslash^2)^{-k}$.

Pour $[\Dslash,L_f](1 + \Dslash^2)^{-k}$, c'est une cons\'equence
directe de la propri\'et\'e d'isom\'etrie de l'action: \'etant donn\'e
que $\Dslash$ commute avec (le rel\`evement de) l'action, on obtient
$$
[\Dslash, L_f] = L_{[\Dslash, M_f]} = L_{\Dslash f},
$$
Ainsi, la preuve de la proposition \ref{pr:interpolation} s'applique avec
$\Dslash f$ \`a la place de $f$ car (avec un petit abus de notation) 
$\Dslash f \in \Coo_c(M)$.

Pour $[|\Dslash|,L_f]$, la preuve se d\'eduit des pr\'ec\'edentes, en
utilisant l'identit\'e spectrale pour un op\'erateur positif $A$:
\begin{equation}
A = \frac{1}{\pi} \int_0^\infty \frac{A^2}{A^2 + \la}
\frac{d\la}{\sqrt{\la}}\ .
\label{eq:spectral-sqrt}
\end{equation}
Ainsi, pour n'importe quel r\'eel strictement positif $\rho$,
\begin{align*}
& [|\Dslash|,L_f] = [|\Dslash|+\rho, L_f]
= \frac{1}{\pi} \int_0^\infty \frac{1}{(|\Dslash|+\rho)^2 + \la}
\bigl[ (|\Dslash|+\rho)^2, L_f \bigr]
\frac{1}{(|\Dslash|+\rho)^2 + \la} \,\sqrt{\la}\,d\la
\\
&= \frac{1}{\pi} \int_0^\infty \frac{1}{(|\Dslash|+\rho)^2 + \la}
\big(\Dslash[\Dslash, L_f] + [\Dslash, L_f]\Dslash
 + 2\rho |\Dslash| L_f - 2\rho L_f|\Dslash|\big)
\frac{1}{(|\Dslash|+\rho)^2 + \la} \,\sqrt{\la}\,d\la.
\end{align*}
Consid\'erons les diff\'erents termes: vu que
$[\Dslash,L_f] = L_{\Dslash f}$, ils sont tous du m\^eme ordre en
$\Dslash$; cela nous permet de ne traiter 
en d\'etail que le premier terme, pour les
autres la preuve est identique.

En commutant $[\Dslash,L_f]$ avec le facteur
$((|\Dslash|+\rho)^2 + \la)^{-1}$ \`a sa gauche, le premier terme de la
derni\`ere \'equation devient:
\begin{align*}
&\frac{1}{\pi} \int_0^\infty
\frac{|\Dslash|+\rho}{((|\Dslash|+\rho)^2 + \la)^2}
\,\sqrt{\la}\,d\la\
\frac{\Dslash}{|\Dslash|+\rho} \,[\Dslash,L_f]
\\
&\qquad + \frac{1}{\pi} \int_0^\infty
\frac{1}{((|\Dslash|+\rho)^2 + \la)^2}
\Dslash \bigl[ (|\Dslash^2| + \rho)^2, [\Dslash,L_f] \bigr]
\frac{1}{(|\Dslash|+\rho)^2 + \la} \,\sqrt{\la}\,d\la
\\
&= \frac{1}{2} \frac{\Dslash}{|\Dslash|+\rho} \,[\Dslash,L_f]
 + \frac{1}{\pi} \int_0^\infty \frac{1}{((|\Dslash|+\rho)^2 + \la)^2}
\Dslash \Bigl( \Dslash[\Dslash,[\Dslash,L_f]]
+ [\Dslash,[\Dslash,L_f]]\Dslash
\\
&\hspace{12em} + 2\rho|\Dslash|\,[\Dslash,L_f]
- 2\rho[\Dslash,L_f]\,|\Dslash| \Bigr)
\frac{1}{(|\Dslash|+\rho)^2 + \la} \,\sqrt{\la}\,d\la.
\end{align*}
Vu que $\Dslash (|\Dslash|+\rho)^{-1}$ est born\'e,
le corollaire~\ref{cl:interpolation} montre que
$$
\frac{\Dslash}{|\Dslash|+\rho}\,[\Dslash,L_f]\,(1 + \Dslash^2)^{-k}
\in \L^p(\H)  \sepword{lorsque} k > n/p.
$$
Pour les quatre autres termes, par exemple pour le premier, on obtient
(et similairement pour les trois autres):
\begin{align*}
\biggl\| \frac{1}{\pi} \int_0^\infty
&\frac{\Dslash^2}{((|\Dslash|+\rho)^2 + \la)^2}\,
[\Dslash, [\Dslash, L_f]] (1 + |\Dslash|)^{-k}
\frac{1}{(|\Dslash|+\rho)^2 + \la} \,\sqrt{\la}\,d\la \biggr\|_p
\\
&\leq \bigl\| [\Dslash, [\Dslash, L_f]](1 + |\Dslash|)^{-k} \bigr\|_p
\frac{1}{\pi} \int_0^\infty
\biggl\| \frac{\Dslash^2}{(|\Dslash|+\rho)^2 + \la} \biggr\|
\biggl\|\frac{1}{(|\Dslash|+\rho)^2 + \la}\biggr\|^2
\,\sqrt{\la}\,d\la
\\
&\leq \bigl\| [\Dslash, [\Dslash, L_f]](1 + |\Dslash|)^{-k} \bigr\|_p
\frac{1}{\pi} \int_0^\infty \frac{\sqrt{\la}}{(\rho^2 + \la)^2} \,d\la
\\
&= \frac{1}{2\rho}\,
\bigl\| L_{\Dslash^2 f}\, (1 + |\Dslash|)^{-k} \bigr\|_p
\end{align*}
qui est fini d'apr\`es le m\^eme corollaire.

Pour $[(1 + \Dslash^2)^{1/2},L_f]$, la preuve est essentiellement la
m\^eme, en utilisant la repr\'esentation spectrale
\eqref{eq:spectral-sqrt} 
appliqu\'ee \`a l'op\'erateur positif $(1 + \Dslash^2)^{1/2}$.
\end{proof}

\subsection{Classes de Schatten faibles}

Nous allons maintenant d\'emontrer que les
d\'eformations isospectrales des vari\'et\'es
spinorielles non compactes ont pour dimension spectrale 
celle classique, i.e. celle de la vari\'et\'e.

La proposition suivante utilise l'estimation (2.3.1) ou (2.3.2) 
pour donner une version am\'elior\'ee de l'in\'egalit\'e de Cwikel
(lemme 2.2.32).

\begin{prop}
\label{pr:weak-schatten}
Soit $f\in\Coo_c(M)$. Alors 
$$
L_f\, (1 + \Delta)^{-1/2} \,L_{\bar f} \in \L^{n,\infty}(\H_r).
$$
\end{prop}
\begin{proof}
Choisissons un nombre $m$ tel que $0 < m < 1$. On d\'efinit les
op\'erateurs positifs
\begin{align*}
A_k
&:= L_f \int_0^{m^{2k}} t^{-1/2}\, e^{-t(1+\Delta)} \,dt \ L_{\bar f},
\\
B_k
&:= L_f \int_{m^{2k}}^1 t^{-1/2}\, e^{-t(1+\Delta)} \,dt \ L_{\bar f},
\\
C &:= L_f \int_1^\infty t^{-1/2}\, e^{-t(1+\Delta)} \,dt \ L_{\bar f},
\end{align*}
pour tout $k \in \N$ (la valeur de~$k$ la plus commode sera choisie
plus tard). Leur somme est \'evidemment
$A_k + B_k + C = \Ga(\half)\, L_f (1 + \Delta)^{-1/2} L_{\bar f}$ pour
tout $k \in \N$.

Montrons tout d'abord que $C$ appartient \`a toutes les classes de
Schatten $\L^p(\H_r)$ pour
$p\geq 1$. En effet, en utilisant le th\'eor\`eme~\ref{th:HS-norm}
ainsi que (2.3.1) ou (2.3.2), on obtient
\begin{align*}
\|C\|_1
&= \biggl\| L_f\biggl( \int_1^\infty t^{-1/2} e^{-t(1 + \Delta)} \,dt
\biggr)^{1/2} \biggr\|_2^2
= \biggl\| M_f\biggl( \int_1^\infty t^{-1/2} e^{-t(1 + \Delta)} \,dt
\biggr)^{1/2} \biggr\|_2^2
\\
&= \Tr\biggl(
M_{|f|^2} \int_1^\infty t^{-1/2} e^{-t(1 + \Delta)} \,dt \biggr)
\\
&= \int_M |f(p)|^2 \int_1^\infty t^{-1/2} e^{-t}\, K_t(p,p) 
\,\mu_g(p) \,dt
\\
&\leq c \int_M |f(p)|^2 \,\mu_g = c\, \|f\|_2^2.
\end{align*}
Ainsi $C \in \L^n(\H_r) \subset \L^{n,\infty}(\H_r)$.

De plus, on obtient l'estimation suivante pour la norme
op\'eratorielle de $A_k$:
$$
\|A_k\|
\leq \|L_f\|^2 \int_0^{m^{2k}} t^{-1/2} \|e^{-t(1 + \Delta)}\| \,dt
\leq \|L_f\|^2 \int_0^{m^{2k}} t^{-1/2} \,dt = 2\,\|L_f\|^2\,m^k.
$$
D'apr\`es le th\'eor\`eme~\ref{th:HS-norm}, 
on peut aussi estimer $B_k$ en norme de trace:
\begin{align*}
\|B_k\|_1
&= \int_{m^{2k}}^1 t^{-1/2} e^{-t}
\int_M |f(p)|^2 \,K_t(p,p) \,\mu_g(p) \,dt
\\
&\leq c \,\int_M |f(p)|^2 \,\mu_g(p)
\int_{m^{2k}}^1 t^{-(n+1)/2} \,dt
\\
&= c \, \|f\|_2^2 \, \frac{2}{n-1}\, (m^{-k(n-1)} - 1)
\\
&\leq c' \, \|f\|_2^2 \, m^{-k(n-1)},
\end{align*}
car $m < 1$.

Par l'in\'egalit\'e de Fan ---voir \cite{SimonTrace}---,
on peut estimer la $j$-i\`eme valeur singuli\`ere de\\
$D := A_k + B_k$:
\begin{align*}
\mu_j(D) &= \mu_j(A_k + B_k) \leq \mu_1(A_k) + \mu_j(B_k)
\\
&\leq \|A_k\| + j^{-1}\|B_k\|_1
\\
&\leq 2\,\|L_f\|^2\,m^k + c'\,\|f\|_2^2 \,\, j^{-1}\,m^{k(1-n)} .
\end{align*}
Finalement, se donnant $j$ et $m < 1$, on peut choisir $k \in \N$ qui
soit tel que 
$m^k \leq j^{-1/n} < m^{k-1}$. Ainsi, 
$j^{-1}\,m^{-k(n-1)} < m^{(k-1)n}\, m^{-k(n-1)} = m^{-n}\,m^{k}$ et donc
$$
\mu_j(D) \leq c(f,n,m) \, j^{-1/n},
$$
qui termine la preuve car
$L_f\,(1 + \Delta)^{-1/2}\,L_{\bar f} = \Ga(\half)^{-1} (C + D)$.
\end{proof}

Ce r\'esultat a un corollaire imm\'ediat.

\begin{cly}
\label{cl:weak-schatten}
Soit $f,h \in \Coo_c(M)$. Alors  
$L_f\,(1 + \Delta)^{-1/2}\,L_h \in \L^{n,\infty}(\H_r)$.
\end{cly}

\begin{proof}
On obtient le r\'esultat par polarisation, i.e. en additionnant\\
$L_{(f\pm\bar h)} \,(1 + \Delta)^{-1/2}\, L_{(\bar f\pm h)}$ et 
$L_{(f\pm i\bar h)} \,(1 + \Delta)^{-1/2}\, L_{(\bar f\mp ih)}$.
\end{proof}

Une fois encore, ce r\'esultat se rel\`eve \`a l'espace de Hilbert des
spineurs de carr\'e int\'egrable.

\begin{cly}
\label{cl:weak-schatten-bis}
Soit $f,h \in \Coo_c(M)$. Alors
$L_f \,(1 + \Dslash^2)^{-1/2}\, L_h$ et
$L_f \,(1 + |\Dslash|)^{-1}\, L_h$ sont dans $\L^{n,\infty}(\H)$.
\end{cly}

\begin{proof}
D\'ecomposons le deuxi\`eme op\'erateur comme
\begin{align*}
L_f \,(1 + |\Dslash|)^{-1}\, L_h
&= L_f(1 + \Dslash^2)^{-1/2}L_h\frac{(1 + \Dslash^2)^{1/2}}
{1 + |\Dslash|} + L_f(1 + \Dslash^2)^{-1/2}
\biggl[\frac{(1 + \Dslash^2)^{1/2}}{1 + |\Dslash|}, L_h\biggr]
\\
&= L_f (1 + \Dslash^2)^{-1/2} L_h
\frac{(1 + \Dslash^2)^{1/2}}{1 + |\Dslash|}
- L_f (1 + |\Dslash|)^{-1} [|\Dslash|,L_h] (1 + |\Dslash|)^{-1}
\\
&\qquad\qquad
+ L_f(1 + \Dslash^2)^{-1/2}\big[(1 + \Dslash^2)^{1/2},L_h\big]
(1 + |\Dslash|)^{-1}.
\end{align*}
Vu que $L_f(1 + |\Dslash|)^{-1}$, $[|\Dslash|,L_h](1 + |\Dslash|)^{-1}$,
$L_f(1 + \Dslash^2)^{-1/2}$ et
$[(1 + \Dslash^2)^{1/2},L_h](1 + |\Dslash|)^{-1}$ appartiennent tous \`a
$\L^{2n}(\H)$ d'apr\`es le lemme~\ref{lem:plenty}, et comme
$(1 + \Dslash^2)^{1/2}(1 + |\Dslash|)^{-1}$ est born\'e,
il est suffisant de d\'emonter le corollaire pour
$L_f \,(1 + \Dslash^2)^{-1/2}\, L_h$.

Par l'identit\'e \eqref{eq:spectral-sqrt} et la formule de
Lichnerowicz une fois encore, on obtient
\begin{align*}
L_f\, (1 + \Dslash^2)^{-1/2} \,L_h
&= L_f\, \frac{1}{\pi} \int_0^\infty
\frac{(1 + \Dslash^2)^{-1}}{(1 + \Dslash^2)^{-1} + \la}
\,\frac{d\la}{\sqrt{\la}} \, L_h
\\
&= L_f\, \frac{1}{\pi} \int_0^\infty
\frac{(1 + \Delta)^{-1} (1 - \frac{1}{4}R (1 + \Dslash^2)^{-1})}
{(1 + \Delta)^{-1} (1 - \frac{1}{4}R (1 + \Dslash^2)^{-1}) + \la}
\,\frac{d\la}{\sqrt{\la}} \,L_h
\\
&= L_f\, \frac{1}{\pi} \int_0^\infty \biggl(
\frac{(1 + \Delta)^{-1}}{(1 + \Delta)^{-1} + \la}
+ \frac{1}{4} \frac{(1 + \Delta)^{-2}}{(1 + \Delta)^{-1} + \la} R
\frac{(1 + \Dslash^2)^{-1}}{(1 + \Dslash^2)^{-1} + \la}
\\
&\hspace{6em} - \frac{1}{4} (1 + \Delta)^{-1} R
\frac{(1 + \Dslash^2)^{-1}}{(1 + \Dslash^2)^{-1} + \la} \biggl)
\frac{d\la}{\sqrt{\la}} \,L_h
\\
&= L_f\, (1 + \Delta)^{-1/2} \,L_h
 + \frac{1}{4\pi} L_f \int_0^\infty \biggl(
\frac{(1 + \Delta)^{-2}}{(1 + \Delta)^{-1} + \la} R
\frac{(1 + \Dslash^2)^{-1}}{(1 + \Dslash^2)^{-1} + \la}
\\
&\hspace{6em} - (1 + \Delta)^{-1} R
\frac{(1 + \Dslash^2)^{-1}}{(1 + \Dslash^2)^{-1} + \la} \biggr)
\frac{d\la}{\sqrt{\la}} \,L_h.
\end{align*}
Le premier terme appartient \`a $\L^{n,\infty}(\H)$ d'apr\`es 
le corollaire~\ref{cl:weak-schatten}, les deux autres sont dans $\L^n(\H)$
car
\begin{align*}
\biggl\| L_f & \int_0^\infty
\frac{(1 + \Delta)^{-2}}{(1 + \Delta)^{-1} + \la} R
\frac{(1 + \Dslash^2)^{-1}}{(1 + \Dslash^2)^{-1} + \la}
\frac{d\la}{\sqrt{\la}} \,L_h \biggr\|_n
\\
&\leq \|L_f(1 + \Delta)^{-2}\|_n \|R(1 + \Dslash^2)^{-1}\| \|L_h\|
\int_0^\infty \biggl\| \frac{1}{(1 + \Delta)^{-1} + \la} \biggr\|
\biggl\| \frac{1}{(1 + \Dslash^2)^{-1} + \la} \biggr\|
\frac{d\la}{\sqrt{\la}}
\\
&\leq 4\, \|L_f(1 + \Delta)^{-2}\|_n \|L_h\|
\int_0^\infty \frac{1}{(1 + \la)^2} \frac{d\la}{\sqrt{\la}},
\end{align*}
qui est fini d'apr\`es la proposition~\ref{pr:interpolation}. 
Aussi, la proposition~\ref{pr:interpolation} et le 
corollaire~\ref{cl:interpolation}, impliquent que
\begin{align*}
\biggl\| L_f & \int_0^\infty (1 + \Delta)^{-1} R
\frac{(1 + \Dslash^2)^{-1}}{(1 + \Dslash^2)^{-1} + \la}
\frac{d\la}{\sqrt{\la}} L_h \biggr\|_n
\\
&\leq \|L_f\,(1 + \Delta)^{-1}\|_n \,\|R(1 + \Dslash^2)^{-1}\,L_h\|
\int_0^\infty \frac{1}{1 + \la} \,\frac{d\la}{\sqrt{\la}}
\end{align*}
est fini. Finalement, comme $\L^n(\H) \subset \L^{n,\infty}(\H)$, 
la preuve est compl\`ete.
\end{proof}

\subsection{Calcul de la trace de Dixmier: le cas p\'eriodique}
\label{sec:Trw-periodic}

\subsubsection{Une discussion heuristique}

Dans ce paragraphe, nous allons voir que la trace de Dixmier donne
lieux \`a un invariant de la d\'eformation, tenant exactement
le m\^eme r\^ole que la trace ordinaire.
Avant de d\'emontrer cette propri\'et\'e, c'est-\`a-dire
$$
\Trw(L_f\,(1 + \Dslash^2)^{-n/2}) = \Trw(M_f\,(1 + \Dslash^2)^{-n/2}),
\sepword{pour tout} f \in \Coo_c(M),
$$
(ou au niveau scalaire, i.e. lorsque $L_f$ agit sur $\H_r$, avec
$(1 + \Delta)^{-n/2}$ \`a la place de $(1 + \Dslash^2)^{-n/2}$),
nous allons donner un argument heuristique pour montrer combien ce
r\'esultat est naturel.

Pour cette fin, nous allons utiliser la d\'efinition de $L_f$
pour $f \in \Coo_c(M)$, en terme d'int\'egrale \`a valeur op\'erateur
(\'equation 1.2.7).
En utilisant cette pr\'esentation, la propri\'et\'e de
trace de la trace de Dixmier ainsi que la commutativit\'e de
l'op\'erateur de Dirac (ou du Laplacien) avec les unitaires $V_z$,
le r\'esultat serait imm\'ediat si nous pouvions \'echanger la trace de
Dixmier avec les int\'egrales de Lebesgue:
\begin{align*}
\Trw((L_f\,(1 + \Dslash^2)^{-n/2})
&= (2\pi)^{-l}\Trw \biggl( \int_{\R^{2l}} e^{-iy z}\,
 V_{\half\Th y}\, M_f \,V_{-\half\Th y-z}
\,d^ly \,d^lz\,(1 + \Dslash^2)^{-n/2} \biggr)
\\
&= (2\pi)^{-l} \int_{\R^{2l}} e^{-iy z} \Trw(
V_{\half\Th y}  M_f \,(1 + \Dslash^2)^{-n/2}\, V_{-\half\Th y-z})
\,d^ly \,d^lz
\\
&= (2\pi)^{-l} \int_{\R^{2l}} e^{-iy z} \Trw(
M_f \,(1 + \Dslash^2)^{-n/2}\, V_{-z}) \,d^ly \,d^lz
\\
&= \Trw \biggl( M_f \,(1 + \Dslash^2)^{-n/2} \int_{\R^l}
\delta_0(z)\, V_{-z} \,d^lz \biggr)
\\
&= \Trw\big( M_f\,(1 + \Dslash^2)^{-n/2}\big).
\end{align*}

Cependant, l'\'echange de la trace de Dixmier avec l'int\'egrale est
loin d'\^etre rigoureux: ces int\'egrales sont
oscillantes et la trace de Dixmier n'ob\'eit par en g\'en\'eral
\`a la convergence domin\'ee.

Pour les d\'eformations non p\'eriodiques, nous allons d\'emontrer ce
r\'esultat en passant par un calcul de r\'esidu de fonction z\'eta.
Nous allons commencer par l'\'etablir dans le cas plus simple des 
d\'eformations p\'eriodiques.

\subsubsection{Le cas p\'eriodique non compacte}

La d\'ecomposition en sous-espaces spectraux de $f \in \Coo_c(M)$
donne un acc\`es direct \`a la Dixmier-tra\c cabilit\'e des
op\'erateurs $L_f\,(1 + \Delta)^{-n/2}$
agissant sur $\H_r = L^2(M,\mu_g)$ et $L_f\,(1 + \Dslash^2)^{-n/2}$
agissant sur $\H = L^2(M,S)$, ainsi qu'\`a la valeur de leur trace de
Dixmier.

\begin{prop}
\label{pr:periodic}
Soit $\a$ une action effective lisse et isom\'etrique de $\T^l$ sur $M$,
avec $l \geq 2$, et soit $f \in \Coo_c(M)$. Alors, l'op\'erateur
$L_f\,(1 + \Delta)^{-n/2}$ est Dixmier-traable sur $\H_r$, et la
valeur de sa trace de Dixmier est ind\'ependante de $\omega$:
$$
\Trw(L_f\,(1 + \Delta)^{-n/2})
= C'(n)\,\delta_{0,r}\,\int_Mf_r \, \mu_g
= C'(n)\,\int_M f_0 \, \mu_g,
$$
o\`u $C'(n) := \Omega_n/n\,(2\pi)^n$, $\Omega_n$ est le volume de la
sph\`ere unit\'e de $\R^n$, et $f = \sum_r f_r$ est la d\'ecomposition
de $f$ en composantes homog\`enes.
\end{prop}

\begin{proof}
Chaque $f_r$ satisfait \`a $\a_z(f_r) = e^{-iz r}\,f_r$ pour tout
$z \in \T^l$. Vu que $[M_{f_r}, V_z] = M_{f_r} (1 - e^{-iz r}) V_z$,
on en d\'eduit que $[M_{f_r}, V_{-\half\Th r}] = 0$ par
antisym\'etrie de la matrice de d\'eformation.

D'apr\`es \cite[Prop.~15]{RennieSum}, $M_f\,(1 + \Delta)^{-n/2}$ 
appartient \`a $\L^{1,\infty}(\H_r)$ et de plus
$$
\|M_f\,(1 + \Delta)^{-n/2}\|_{1,\infty} \leq C_1(n)\, \|f\|_\infty.
$$
Cette estimation est obtenue par une partition (finie) de l'unit\'e
sur le compact $\supp f$ et en appliquant le th\'eor\`eme de Weyl. 
On obtient alors,
\begin{align*}
\|L_f\,(1 + \Delta)^{-n/2}\|_{1,\infty}
&\leq \sum_{r\in\Z^l}
\|M_{f_r} V_{-\half\Th r}\,(1 + \Delta)^{-n/2}\|_{1,\infty}
\\
&\leq \sum_{r\in\Z^l} \|M_{f_r}\,(1 + \Delta)^{-n/2}\|_{1,\infty}
\\
&\leq C_1(n) \sum_{r\in\Z^l} \|f_r\|_\infty,
\end{align*}
car chaque $f_r$ est \`a support (compact) contenu dans
$\T^l \cdot (\supp f)$. Ces estimations donnent la 
Dixmier-tra\c cabilit\'e, \'etant donn\'e que la d\'ecomposition en
sous-espaces spectraux est convergente dans la norme $\|\cdot\|_\infty$.

Pour le calcul de la trace Dixmier, il suffit de remarquer que pour tout
 $z \in \T^l$,
\begin{align*}
\Trw(L_{f_r}\,(1 + \Delta)^{-n/2})
&=\Trw(V_z\,M_{f_r}\,V_{-\half\Th r}\,(1+\Delta)^{-n/2}\,V_{-z})
\\
&= \Trw(M_{\a_z(f_r)}\,V_{-\half\Th r}\,(1 + \Delta)^{-n/2})
\\
&= e^{-iz r} \Trw(M_{f_r}\,V_{-\half\Th r}\,(1+\Delta)^{-n/2}),
\end{align*}
qui est diff\'erent de z\'ero si et seulement si $r = 0$.
Ainsi,
$$
\Trw\big(L_{f_r}\,(1 + \Delta)^{-n/2}\big)
= \Trw\big(M_{f_0}\,(1 + \Delta)^{-n/2}\big)\,\delta_{0,r}
= C'(n)\,\delta_{0,r}\,\int_M f_0 \, \mu_g.
$$
La derni\`ere \'egalit\'e est obtenue (c.f. \cite{RennieSum})
par calcul du r\'esidu de Wodzicki de l'op\'erateur\\
$M_f(1 + \Delta)^{-n/2}$.
\end{proof}

\begin{cly}
\label{cl:periodic}
Sous les hypoth\`eses de la proposition pr\'ec\'edente, l'op\'erateur
$L_f\,(1 + \Dslash^2)^{-n/2}$ est Dixmier-tra\c cable sur $\H$ pour
$f \in \Coo_c(M)$; de plus, la valeur de sa trace de Dixmier est
ind\'ependante de $\omega$:
$$
\Trw(L_f\,(1 + \Dslash^2)^{-n/2}) =
C(n)\,\delta_{0,r}\,\int_M f_r \,\mu_g  = C(n)\,\int_M f_0 \,\mu_g,
$$
o\`u $C(n) := 2^{\piso{n/2}} \Omega_n/n\,(2\pi)^n$, avec
$2^{\piso{n/2}}$ le rang du fibr\'e spinoriel.
\end{cly}

\begin{proof}
En utilisant la formule de Lichnerowicz $\Dslash^2 = \Delta +
\tfrac{1}{4} R$, la Dixmier-tra\c cabilit\'e est obtenue par comparaison:
\begin{equation}
\label{lich}
(1 + \Dslash^2)^{-1} = (1 + \Delta)^{-1}
\bigl(1 - \tfrac{1}{4} R\,(1 + \Dslash^2)^{-1} \bigr).
\end{equation}
Pour le calcul de la trace de Dixmier, on peut appliquer les arguments
de la proposition pr\'ec\'edente. On obtient le r\'esultat en
remarquant que, modulo le facteur $2^{\piso{n/2}}$, les symboles
principaux de $(1 + \Dslash^2)^{-n/2}$
et de $(1 + \Delta)^{-n/2}$ sont identiques (cf. \'equation
\eqref{lich}).
Ainsi les op\'erateurs $M_{f_r}(1 + \Dslash^2)^{-n/2}$ et
$M_{f_r}(1 + \Delta)^{-n/2}$ ont le m\^eme r\'esidu de Wodzicki, \`a
un facteur constant pr\`es.
\end{proof}

\subsection{Calcul de la trace de Dixmier: le cas non p\'eriodique}
\label{sec:Trw-aperiodic}

Dans ce paragraphe, nous allons \'etablir le th\'eor\`eme suivant:

\begin{thm}
\label{th:Dix-tr}
Soit $M$ une vari\'et\'e Riemannienne \`a spin, non compacte,
sans bord, connexe, compl\`ete, satisfaisant \`a (I) ou (II),  avec
un scalaire de courbure born\'e et munie d'une action lisse propre et 
isom\'etrique de $\R^l$. Si $f \in \Coo_c(M)$, alors
$L_f\,(1 + |\Dslash|)^{-n}$ appartient \`a $\L^{1,\infty}(\H)$ 
et la valeur de sa trace Dixmier est donn\'ee par
$$
\Trw \bigl( L_f\, (1 + |\Dslash|)^{-n} \bigr)
= \Trw \bigl( M_f\, (1 + |\Dslash|)^{-n} \bigr)
= C(n) \int_M f(p) \,\mu_g(p),
$$
o\`u $C(n) = 2^{\piso{n/2}}\,\Omega_n/n\,(2\pi)^n$.
\end{thm}

Dans le cas non p\'eriodique, la vari\'et\'e $M$ est n\'ecessairement
de la forme $V \x \R^l$, o\`u le groupe $\R^l$ agit par translation
sur le second facteur.
En effet, une action propre du groupe additif $\R^l$ 
est automatiquement libre, car le seul sous-groupe compact de $\R^l$ 
est le sous-groupe trivial $\{0\}$. Ainsi la projection sur l'espace
des orbites $\pi: M \to M/\R^l$ d\'efinit une projection de 
$\R^l$-fibr\'e principal \cite[Thm.~1.11.4]{DuistermaatK}. 
Remarquons que l'action soit propre, a \'et\'e cruciallement
utilis\'e dans la proposition~\ref{pr:kernel} pour montrer que
les op\'erateurs de multiplication twist\'ee sont born\'es.
Or un $\R^l$-fibr\'e principal poss\`ede une section (lisse) globale,
il est donc automatiquement trivial (voir \cite[\S 16.14.5]{DieudonneIII}).

Ainsi $M = V \x \R^l$, o\`u $V$ est une vari\'et\'e lisse, non
n\'ecessairement compacte de dimension $k = n - l$, qui est munie
d'une structure Riemannienne, induite de celle de $M$, et $\pi: M \to
V$ est juste la projection sur le premier facteur.
Si $\{\phi_j\}_{j\in J}$ est une partition de l'unit\'e  localement
finie sur $V$, o\`u chaque $\phi_j$ est lisse et \`a support compact, on peut
d\'efinir une partition de l'unit\'e $\a$-invariante $\{\psi_j\}$ sur
$M$, en posant $\psi_j := \phi_j \circ \pi$. Pour tout $f \in
\Coo_c(M)$, la somme
$f = \sum_j f \psi_j$ est finie car $\supp f$ est compact; \'etant
donn\'e que $\psi_j$ est $\a$-invariant, on obtient directement
$$
L_f = \sum_j L_{f\psi_j} = \sum_j L_f\,M_{\psi_j}.
$$
Pour manipuler des op\'erateurs du type $L_f\,h(\Dslash)$, 
on peut sans perte de g\'en\'eralit\'e se restreindre a une seule
carte de~$V$.

Notons $\hat x := (x^1,\dots,x^k)$ et
$\bar x := (x^{k+1},\dots,x^n)$ respectivement les
coordonn\'ees transverses et longitudinales sur~$M$.
Il est aussi imm\'ediat de remarquer que l'op\'erateur $L_f$
est pseudodiff\'erentiel, de symbole
\begin{equation}
\sigma[L_f](\hat x,\bar x;\hat\xi,\bar\xi)
= f(\hat x,\bar x - \half\Th\bar\xi).
\end{equation}
En effet, pour tout $\psi \in \H$, la d\'efinition~\ref{df:left-multn}
montre que
\begin{align*}
L_f\psi(\hat x,\bar x)
&= (f \Mop \psi)(\hat x,\bar x)
= (2\pi)^{-l} \int_{\R^{2l}} e^{-i\bar\xi \bar y}
\a_{\half\Th\bar\xi}(f)(\hat x,\bar x)\,
V_{-\bar y}\psi(\hat x,\bar x) \,d^l\bar\xi \,d^l\bar y
\\
&= (2\pi)^{-l} \int_{\R^{2l}} e^{-i\bar\xi \bar y}
f(\hat x,\bar x - \half\Th\bar\xi)\,
\psi(\hat x,\bar x + \bar y) \,d^l\bar\xi \,d^l\bar y
\\
&= (2\pi)^{-n} \int_{\R^{2n}}
e^{-i\bar\xi(\bar y-\bar x)} e^{-i\hat\xi(\hat y-\hat x)} \,
f(\hat x,\bar x - \half\Th\bar\xi)\, \psi(\hat y,\bar y)
\,d^l\bar\xi \,d^l\bar y \,d^k\hat\xi \,d^k\hat y.
\end{align*}

\begin{prop}
\label{pr:matrix-basis}
Sous les hyphoth\`ese du th\'eor\`eme~\ref{th:Dix-tr}, si $f \in \Coo_c(M)$
alors $L_f(1 + |\Dslash|)^{-n}$ appartient \`a $\L^{1,\infty}(\H)$.
\end{prop}

\begin{proof}
Pour $\hat x$ fix\'e, la fonction $\bar x \mapsto f(\hat x,\bar x)$
est dans $\Coo_c(\R^l)$, et donc peut \^etre d\'ecompos\'ee dans la
base de Wigner $\{f_{mn}\}$, avec $m,n \in \N^{l/2}$:
$$
f(\hat x,\bar x) = \sum_{m,n} c_{mn}(\hat x) \,f_{mn}(\bar x),
$$
o\`u les coefficients matriciels $c_{mn}$ sont des \'el\'ements de 
$\Coo_c(V)$.

Se donnant deux fonctions de la sorte,
$f(\hat x,\bar x) = \sum c_{mn}(\hat x) \,f_{mn}(\bar x)$,\\
$h(\hat x,\bar x) = \sum d_{mn}(\hat x) \,f_{mn}(\bar x)$,
leur produit twist\'e devient un produit matriciel dans les 
variables~$\bar x$:
\begin{equation}
(f \Mop h)(\hat x,\bar x)
= \sum_{m,n,k} c_{mk}(\hat x)\,d_{kn}(\hat x) \,f_{mn}(\bar x).
\label{eq:Moyal-matricial}
\end{equation}
Ainsi, l'op\'erateur $L_f$ peut \^etre vu comme appartenant \`a
l'alg\`ebre $M_\infty(\Coo_c(V))$ avec des \'el\'ements de matrices (\`a
valeurs dans $\Coo_c(V)$) \`a d\'ecroissance rapide.

Ainsi, on peut \'etendre la propri\'et\'e de factorisation forte
(proposition 2.2.6) \`a ce contexte: pour tout
$f \in \Coo_c(M)$, il existe $h, k \in \Coo(M)$ qui sont des fonctions
Schwartz dans les variables $\bar x$, \`a support compact sur $V$ et tel que
\begin{equation}
f(\hat x,\bar x) = (h \Mop k)(\hat x,\bar x).
\label{eq:strong-factorn}
\end{equation}
Par factorisations r\'ep\'et\'ees (permettant d'\'ecrire $f$
comme un produit de $n$ fonctions de la sorte) et par commutateurs
it\'er\'es (exactement comme dans le Corollaire 2.2.34 et le lemme
2.2.35), on peut faire appara\^itre 
$L_f(1 + |\Dslash|)^{-n}$ comme un produit de $n$ termes
de la forme $L_h (1 + |\Dslash|)^{-1} L_k$, chacun d'entre eux
appartenant \`a $\L^{n,\infty}(\H)$ 
d'apr\`es le corollaire~\ref{cl:weak-schatten-bis}, plus un extra terme
dans $\L^1(\H)$. Finalement, l'in\'egalit\'e de H\"older pour les
classes de Schatten faibles, implique que
$L_f(1 + |\Dslash|)^{-n} \in \L^{1,\infty}(\H)$.
\end{proof}

Nous allons introduire une famille d'unit\'es locales,
en g\'en\'eralisant la construction du paragraphe 2.2.

\begin{defn}
La vari\'et\'e  $V$ pouvant \^etre vue comme une union de
compacts $C_i$, chacun \'etant contenu dans l'int\'erieur de $C_{i+1}$,
on d\'efinit les fonctions $\chi_i$, par: $\chi_i := 1$ 
sur~$C_i$ et $\chi_i := 0$ ailleurs. 
Pour tout $K \in \N$, soit la fonction $e_K$ d\'efinie par
$$
e_K(\hat x,\bar x)
:= \sum_{|n|\leq K} \chi_K(\hat x) \,f_{nn}(\bar x),
$$
o\`u $|n| = n_1 +\cdots+ n_{l/2}$. Ainsi $e_K$ est r\'e\'el,
$e_K \Mop e_K = e_K$ d'apr\`es \eqref{eq:Moyal-matricial} et
$L_{e_K}$ est un projecteur orthogonal sur 
$\H$. Soit ensuite, $f_K := e_K \Mop f \Mop e_K$, ou plus explicitement
\begin{equation}
f_K(\hat x,\bar x)
:= \sum_{|m|,|n|\leq K} \chi_K(\hat x)\,c_{mn}(\hat x) \,f_{mn}(\bar x).
\label{eq:f-trunc}
\end{equation}
Par construction, $e_K \Mop f_K = f_K \Mop e_K = f_K$.
\end{defn}

L'op\'erateur $L_{e_K} (1 + |\Dslash|)^{-n} L_{e_K}$ est Dixmier-tra\c cable:
dans la proposition~\ref{pr:weak-schatten} ainsi que dans la suite, 
on peut remplacer $f$ par $e_K$ m\^eme si il n'appartient pas \`a
$\Coo_c(M)$, \'etant donn\'e qu'il reste \`a d\'ecroissance rapide.
La propri\'et\'e de trace de la trace de Dixmier implique alors que
$$
\Trw\bigl( L_{f_K} (1 + |\Dslash|)^{-n} \bigr) =
\Trw\bigl( L_{f_K} L_{e_K} (1 + |\Dslash|)^{-n} L_{e_K} \bigr).
$$
Comme $L_{f_K}$ est born\'e, le th\'eor\`eme~5.6 de \cite{CareyPS} montre
que si la limite
$$
\lim_{s\downarrow 1} (s - 1)
\Tr\bigl(L_{f_K}(L_{e_K} (1 + |\Dslash|)^{-n} L_{e_K})^s \bigr),
$$
existe, alors elle co\"incidera avec la valeur de toute trace de
Dixmier de l'op\'erateur $L_{f_K} (1 + |\Dslash|)^{-n}$.

\begin{lem}
\label{lm:zeta-control}
La norme trace
\begin{equation}
\bigl\| L_{f_K} (L_{e_K} (1 + |\Dslash|)^{-n} L_{e_K})^s
- L_{f_K} (1 + |\Dslash|)^{-ns} \bigl\|_1
\label{eq:zeta-control}
\end{equation}
est une fonction born\'ee de~$s$, pour $1 \leq s \leq 2$.
\end{lem}

\begin{proof}
Posons $s =: 1 + \eps$, avec $0 < \eps \leq 1$. Nous allons utiliser
la repr\'esentation spectrale suivante, g\'en\'eralisant
\eqref{eq:spectral-sqrt}, pour une puissance fractionnaire d'un
op\'erateur positif $A$:
$$
A^\eps = \frac{\sin\pi\eps}{\pi}
\int_0^\infty A \, (1 + \la A)^{-1} \, \la^{-\eps} \,d\la.
$$
Etant donn\'e que $L_{e_K}$ est un projecteur orthogonal et que
$L_{f_K} L_{e_K} = L_{f_K}$, on a
\begin{align*}
L_{f_K} (L_{e_K} (1 + |\Dslash|)^{-n} L_{e_K})^s
&= L_{f_K} L_{e_K} (1 + |\Dslash|)^{-n} L_{e_K} \,
(L_{e_K} (1 + |\Dslash|)^{-n} L_{e_K}\big)^\eps
\\
&= L_{f_K} (1 + |\Dslash|)^{-n}
(L_{e_K} (1 + |\Dslash|)^{-n} L_{e_K})^\eps.
\end{align*}
Ainsi,
\begin{align}
L_{f_K} & (L_{e_K} (1 + |\Dslash|)^{-n} L_{e_K})^s
- L_{f_K} (1 + |\Dslash|)^{-ns}
\label{eq:power} \\
&=
L_{f_K} (1 + |\Dslash|)^{-n}\, \frac{\sin\pi\eps}{\pi}
\int_0^\infty \biggl(
\frac{L_{e_K} (1 + |\Dslash|)^{-n} L_{e_K}}
     {1 + \la L_{e_K} (1 + |\Dslash|)^{-n} L_{e_K}}
- \frac{(1 + |\Dslash|)^{-n}}{1 + \la(1 + |\Dslash|)^{-n}} \biggr)
\,\la^{-\eps} \,d\la.
\nonumber
\end{align}
La premi\`ere fraction entre parenth\`ese peut \^etre r\'e\'ecrite comme
$$
\bigl( (1 + |\Dslash|)^n + \la T_K \bigr)^{-1} \,T_K,
$$
o\`u
$$
T_K := (1 + |\Dslash|)^n L_{e_K} (1 + |\Dslash|)^{-n} L_{e_K}.
$$
Comme $L_{e_K}$ est un projecteur, on obtient
\begin{align}
T_K &= L_{e_K}^2
+ [(1 + |\Dslash|)^n, L_{e_K}]\, (1 + |\Dslash|)^{-n} L_{e_K}
\nonumber \\
&= L_{e_K} + \sum_{0\leq k<r\leq n}  \binom{n}{r}
|\Dslash|^k \,[|\Dslash|, L_{e_K}] \,|\Dslash|^{r-k-1}
(1 + |\Dslash|)^{-n} \,L_{e_K}
\nonumber \\
&=: L_{e_K} + \sum_{0\leq k<r\leq n} A_{rk}.
\label{eq:teux}
\end{align}
La propri\'et\'e cruciale pour montrer que la diff\'erence
\eqref{eq:power} est uniform\'ement (en $\eps$) \`a trace, est que, \`a
la seule exception du premier terme de~\eqref{eq:teux} qui est
seulement born\'e, tous les autres $A_{rk}$ sont compacts. 
Plus pr\'ecis\'ement, en utilisant la
proposition~\ref{pr:interpolation} (ainsi que quelques commutateurs),
on en d\'eduit que chaque $A_{rk} \in \L^p(\H)$ pour tout $p > n$.

En suivant une id\'ee de Rennie \cite[Thm.~12]{RennieSum}, 
on peut r\'eduire la diff\'erence des fractions dans l'\'equation
\eqref{eq:power} de la mani\`ere suivante:
\begin{align*}
& \frac{L_{e_K} (1 + |\Dslash|)^{-n} L_{e_K}}
     {1 + \la L_{e_K} (1 + |\Dslash|)^{-n} L_{e_K}}
- \frac{(1 + |\Dslash|)^{-n}}{1 + \la(1 + |\Dslash|)^{-n}}
\\
&= ((1 + |\Dslash|)^n + \la T_K)^{-1} \,T_K
- ((1 + |\Dslash|)^n + \la)^{-1}
\\
&= \bigl( ((1 + |\Dslash|)^n + \la T_K)^{-1}
   - ((1 + |\Dslash|)^n + \la)^{-1} \bigr) \,T_K
+ \bigl( (1 + |\Dslash|)^n + \la \bigr)^{-1} \,(T_K - 1)
\\
&= ((1 + |\Dslash|)^n + \la)^{-1} (\la - \la T_K)
((1 + |\Dslash|)^n + \la T_K)^{-1} \,T_K
+ \bigl( (1 + |\Dslash|)^n + \la \bigr)^{-1} \,(T_K - 1)
\\
&= ((1 + |\Dslash|)^n + \la)^{-1}  \,(T_K - 1)\,
\bigl(1 - ((1 + |\Dslash|)^n + \la T_K)^{-1} \,\la T_K \bigr)
\\
&= ((1 + |\Dslash|)^n + \la)^{-1}  \,(T_K - 1)\,
\bigl(1 + \la L_{e_K} (1 + |\Dslash|)^{-n} L_{e_K} \bigr)^{-1} \,.
\end{align*}
Ainsi, nous obtenons
\begin{align*}
L_{f_K} & (L_{e_K} (1 + |\Dslash|)^{-n} L_{e_K})^s
- L_{f_K} (1 + |\Dslash|)^{-ns}
\\
&= L_{f_K} (1 + |\Dslash|)^{-n} \frac{\sin\pi\eps}{\pi}
\int_0^\infty \frac{1}{(1 + |\Dslash|)^n + \la} (T_K - 1)
\frac{1}{1 + \la L_{e_K} (1 + |\Dslash|)^{-n} L_{e_K}}
\,\la^{-\eps} \,d\la
\\
&= L_{f_K} (1 + |\Dslash|)^{-n} \frac{\sin\pi\eps}{\pi}
\int_0^\infty \frac{1}{(1 + |\Dslash|)^n + \la} \,L_{e_K} (T_K - 1)
\frac{1}{1 + \la L_{e_K} (1 + |\Dslash|)^{-n} L_{e_K}}
\,\la^{-\eps} \,d\la
\\
&\qquad + L_{f_K} \frac{\sin\pi\eps}{\pi} \int_0^\infty
\biggl[ L_{e_K},
\frac{(1 + |\Dslash|)^{-n}}{(1 + |\Dslash|)^n + \la} \biggr]
(T_K - 1) \frac{1}{1 + \la L_{e_K} (1 + |\Dslash|)^{-n} L_{e_K}}
\,\la^{-\eps} \,d\la.
\end{align*}

Nous allons montrer maintenant que le deuxi\`eme terme du membre de
droite est uniform\'ement born\'e en norme de trace. En \'ecrivant
\begin{align*}
\bigl[L_{e_K},
& (1 + |\Dslash|)^{-n} ((1 + |\Dslash|)^n + \la)^{-1} \bigr]
\\
&= [L_{e_K}, (1 + |\Dslash|)^{-n}]\, ((1 + |\Dslash|)^n + \la)^{-1}
+ (1 + |\Dslash|)^{-n} \,[L_{e_K}, ((1 + |\Dslash|)^n + \la)^{-1}],
\end{align*}
pour le premier de ces termes, on obtient l'estimation suivante:
\begin{align*}
& \biggl\| L_{f_K} [L_{e_K}, (1 + |\Dslash|)^{-n}]\,
\frac{\sin\pi\eps}{\pi} \int_0^\infty \!
\frac{1}{(1+|\Dslash|)^n + \la} L_{e_K} (T_K - 1)
\frac{\la^{-\eps}}{1 + \la L_{e_K} (1 + |\Dslash|)^{-n} L_{e_K}}
\,d\la \biggr\|_1
\\
&\leq \bigl\| L_{f_K} [L_{e_K}, (1 + |\Dslash|)^{-n}] \bigr\|_1
\\
&\qquad \x \frac{\sin\pi\eps}{\pi}
\int_0^\infty \|((1+|\Dslash|)^n + \la)^{-1}\|\, \|L_{e_K}(T_K-1)\|
\bigl\| (1 + \la L_{e_K} (1+|\Dslash|)^{-n} L_{e_K})^{-1} \bigr\|
\,\la^{-\eps} \,d\la
\\
&\leq \|L_{e_K} (T_K - 1)\|\,
\bigl\| L_{f_K} [L_{e_K}, (1 + |\Dslash|)^{-n}] \bigr\|_1
\frac{\sin\pi\eps}{\pi} \int_0^\infty \frac{\la^{-\eps}}{1+\la} \,d\la
\\
&= \|L_{e_K} (T_K - 1)\|\,
\bigl\| L_{f_K} [L_{e_K}, (1 + |\Dslash|)^{-n}] \bigr\|_1 =: C_1.
\end{align*}
La constante $C_1$ est finie (et ind\'ependante de $\eps$) car
\begin{equation}
L_{f_K} \,[L_{e_K}, (1 + |\Dslash|)^{-n}]
= L_{f_K} \sum_{0\leq k<r\leq n} \binom{n}{r}
\frac{|\Dslash|^k}{(1 + |\Dslash|)^n} \,[|\Dslash|, L_{e_K}]\,
\frac{|\Dslash|^{r-k-1}}{(1 + |\Dslash|)^n},
\label{eq:comm-expan}
\end{equation}
et car chaque terme appartient \`a $\L^1(\H)$, 
d'apr\`es la proposition~\ref{pr:interpolation} et
l'in\'egalit\'e de H\"older. D'une fa\c con analogue, on peut montrer
que la norme de trace de
$$
L_{f_K} (1 + |\Dslash|)^{-n} \,\frac{\sin\pi\eps}{\pi} \int_0^\infty
\! [L_{e_K}, ((1 + |\Dslash|)^n + \la)^{-1}]\, L_{e_K} (T_K - 1)
\frac{\la^{-\eps}}{1 + \la L_{e_K} (1 + |\Dslash|)^{-n} L_{e_K}}
\,d\la
$$
est born\'ee par une constante $C_2 := \|L_{f_K} (1 + |\Dslash|)^{-n}\|
\, \|[L_{e_K}, (1 + |\Dslash|)^{-n}]\|_1$, ind\'ependante de~$\eps$.

En utilisant le d\'eveloppement \eqref{eq:teux} de $T_K$,
nous obtenons finalement
\begin{align*}
& \bigl\| L_{f_K} (L_{e_K} (1 + |\Dslash|)^{-n} L_{e_K})^s
- L_{f_K} (1 + |\Dslash|)^{-ns} \bigr\|_1
\\
&\leq \sum_{0\leq k<r\leq n} \biggl\| L_{f_K} (1 + |\Dslash|)^{-n}
\frac{\sin\pi\eps}{\pi} \int_0^\infty \frac{1}{(1+|\Dslash|)^n + \la}
L_{e_K} A_{rk}
\frac{\la^{-\eps}}{1 + \la L_{e_K} (1 + |\Dslash|)^{-n} L_{e_K}}
\,d\la \biggr\|_1
\\
&\hspace{6em} + C_1 + C_2
\\
&\leq \sum_{0\leq k<r\leq n}
\bigl\| L_{f_K} (1 + |\Dslash|)^{-n} \bigr\|_{p/(p-1)}
\, \|L_{e_K} A_{rk}\|_p \frac{\sin\pi\eps}{\pi} \int_0^\infty
\frac{\la^{-\eps}}{1 + \la} \,d\la \, + C_1 + C_2
\\
&= \sum_{0\leq k<r\leq n}
\bigl\| L_{f_K} (1 + |\Dslash|)^{-n} \bigr\|_{p/(p-1)}
\, \|L_{e_K} A_{rk}\|_p + C_1 + C_2,
\end{align*}
qui est fini pour $p > n$.
\end{proof}

\begin{proof}[Preuve du th\'eor\`eme~\ref{th:Dix-tr}]
Pour $1 < s \leq 2$, l'op\'erateur $L_{f_K}(1 + |\Dslash|)^{-ns}$
apparaissant dans~\eqref{eq:zeta-control} est \`a trace, \'etant
\'egale au produit de
$L_{f_K}(1 + |\Dslash|)^{-n} \in \L^{1,\infty}(\H)$ par\\
$L_{e_K} (1 + |\Dslash|)^{-n(s-1)} \in \L^p(\H)$ pour
$p > 1/(s - 1)$, plus un commutateur \`a trace. 
La diff\'erence des traces
$$
\Tr\bigl( L_{f_K} (L_{e_K} (1 + |\Dslash|)^{-n} L_{e_K})^s \bigr)
- \Tr\bigl( L_{f_K} (1 + |\Dslash|)^{-ns} \bigr)
$$
est alors une fonction born\'ee de $s$, pour $1 \leq s \leq 2$. Ainsi,
\begin{equation}
\lim_{s\downarrow 1} (s - 1)
\Tr\bigl( L_{f_K} (L_{e_K} (1 + |\Dslash|)^{-n} L_{e_K})^s \bigr)
= \lim_{s\downarrow 1} (s - 1)
\Tr\bigl( L_{f_K} (1 + |\Dslash|)^{-ns} \bigr).
\label{eq:zeta-res}
\end{equation}

{}L'expression du noyau de
$L_{f_K}(1 + |\Dslash|)^{-ns}$ et le lemme~\ref{lm:invkernel} donnent 
\begin{align*}
\Tr\bigl( L_{f_K} (1 + |\Dslash|)^{-ns} \bigr)
&= \int_M K_{L_{f_K}(1+|\Dslash|)^{-ns}}(p,p) \,\mu_g(p)
\\
&= (2\pi)^{-l} \int_M \int_{\R^{2l}} e^{-iy z} f((-\half\Th y)\.p)\,
K_{(1+|\Dslash|)^{-ns}}(z\.p,p) \,d^ly \,d^lz \,\mu_g(p)
\\
&= (2\pi)^{-l} \int_M \int_{\R^{2l}} e^{-iy z} f(p')\,
K_{(1+|\Dslash|)^{-ns}}(z\.p',p') \,d^ly \,d^lz \,\mu_g(p')
\\
&= \int_M f_K(p') \,K_{(1+|\Dslash|)^{-ns}}(p',p') \,\mu_g(p')
\\
&= \Tr\bigl( M_{f_K} (1 + |\Dslash|)^{-ns} \bigr).
\end{align*}
Le calcul du terme de droite de \eqref{eq:zeta-res} est alors direct:
\begin{equation}
\lim_{s\downarrow 1} (s - 1)\,
\Tr\bigl( M_{f_K}\, (1 + |\Dslash|)^{-ns} \bigr)
= \Trw \bigl( M_{f_K}\, (1 + |\Dslash|)^{-n} \bigr)
= C(n) \int_M f_K(p) \,\mu_g(p).
\label{eq:res-value}
\end{equation}
La derni\`ere \'egalit\'e est une propri\'et\'e connue de la trace de
Dixmier dans le cas commutatif non compact \cite{RennieSum}. 
La constante de proportionnalit\'e  
$C(n) = 2^{\piso{n/2}}\,\Omega_n/n\,(2\pi)^n$
est la m\^eme que dans le corollaire~\ref{cl:periodic}.

Il reste a s'affranchir de la troncature induite par les projecteurs
$L_{e_K}$. Notons tout d'abord que
$$
\Trw\bigl( (L_f - L_{f_K})\, (1 + |\Dslash|)^{-n} \bigr)
= \Trw\bigl( (1 - L_{e_K}) L_f\, (1 + |\Dslash|)^{-n} \bigr),
$$
car $L_f[L_{e_K}, (1 + |\Dslash|)^{-n}]$ est \`a trace 
(voir \eqref{eq:comm-expan}) et car 
$L_{e_K}$ est un idempotent. 
En utilisant une fois de plus la propri\'et\'e de
factorisation,
$f = h \Mop k$, on obtient
\begin{equation}
\bigl|\Trw\bigl((L_f - L_{f_K})\, (1+|\Dslash|)^{-n}\bigr)\bigr|
\leq \|L_h - L_{e_K \Mop h}\| \,
\bigl| \Trw\bigl( L_k\, (1 + |\Dslash|)^{-n} \bigr) \bigr|.
\label{eq:eval-trunc}
\end{equation}
Le terme de droite tend vers z\'ero lorsque $K \to \infty$, 
en vertu de l'estimation \eqref{eq:borne} pour la norme de
l'op\'erateur de multiplication twist\'e. En r\'e\'ecrivant 
\eqref{eq:res-value} comme
$$
\Trw \bigl( L_{f_K}\, (1 + |\Dslash|)^{-n} \bigr)
= C(n) \int_M f_K(p) \,\mu_g(p),
$$
on obtient que le terme de gauche converge 
vers $\Trw(L_f\,(1 + |\Dslash|)^{-n})$
lorsque $K \to \infty$. Pour le terme de droite, le fait que les
coefficients $c_{mn}(\hat x)$ dans \eqref{eq:f-trunc} sont \`a
d\'ecroissance rapide assure que $f_K \to f$ dans $L^1(M,\mu_g)$. 
En prenant la limite $K \to \infty$ pour les deux membres de
\eqref{eq:eval-trunc} on obtient le r\'esultat:
$$
\Trw \bigl( L_f\, (1 + |\Dslash|)^{-n} \bigr)
= C(n) \int_M f(p) \,\mu_g(p).
\eqno \qed
$$
\hideqed
\end{proof}

\chapter{Fonctionnelles d'actions}
\label{action}

L'importante intersection entre la g\'eom\'etrie non commutative et la
physique des interactions fondamentales \cite{CIKS, AliAlain, AliAlain2,
ConnesAction, ConnesGrav, ConnesL1, ConnesL2, IochumS, Sirius} a plusieurs
origines.

D'une part, il y a les motivations conceptuelles inh\'erentes \`a la notion
d'espace quantique, qui m\^eme en l'absence d'avanc\'ee majeure,
fournit un cadre prometteur pour la r\'esolution du probl\`eme de la
quantification de la gravitation et/ou de la compr\'ehension de l'origine
des divergences qui apparaissent en th\'eorie quantique des champs ordinaire.
En particulier, il n'est pas encore clair que les divergences ultraviolettes dont sont
affubl\'ees ces th\'eories, viennent de la non prise en compte de la nature
quantique de l'espace-temps, ou si elles sont une caract\'eristique fondamentale,
et donc sont \`a consid\'erer avec plus d'\'egard \cite{CMarco1, CMarco2, CMarco3}.

D'autre part, au niveau classique cette fois, la g\'eom\'etrie non commutative
a permis des avanc\'ees conceptuelles importantes: interpr\'etation g\'eom\'etrique
du m\'ecanisme de Higgs et unification des quatre interactions fondamentales.
Ces avanc\'ees sont cons\'equentes \`a la constructions de `fonctionnelles
d'actions pour champs de jauge non commutatif',
g\'en\'eralisant celles de Yang--Mills et d'Einstein--Hilbert
dans un cadre g\'eom\'etrique plus \'etendu.\\
Dans ce chapitre, nous allons voir deux constructions
d'actions non commutatives, au travers des exemples
de d\'eformations isospectrales que sont les plans de Moyal:
l'action de Connes--Lott et l'action spectrale.\medskip

L'action de Connes--Lott, introduite dans \cite{ConnesL1,
ConnesL2}, a permis de d\'ecrire le mod\`ele standard de la physique
des particules (MS) dans un cadre
non commutatif. Le triplet spectral d\'ecrivant ce mod\`ele est le produit
d'un triplet spectral commutatif $(\Coo(M),L^2(M,S),\Dslash)$ avec un
triplet spectral fini (de dimension spectrale nulle),
o\`u l'alg\`ebre est la somme directe
$\A_F=\mathbb{H}\oplus\C\oplus M_3(\C)$ ($\mathbb{H}$ est le corps
des quaternions), l'op\'erateur de Dirac est
la matrice de masse fermionique et l'espace de repr\'esentation est
le corps des complexes \`a
la puissance du nombre de particules du mod\`ele (en comptant tous les
degr\'es de libert\'es: couleur, isospin....). Ces triplets portent le nom
\'eloquent de triplets spectraux presque commutatifs. Le groupe de jauge du mod\`ele
standard est alors obtenu comme le groupe des unitaires de l'alg\`ebre
$\A_F$.\\
Une fonctionnelle d'action est ensuite construite \`a partir de `l'int\'egrale
non commutative'
$$
\A\ni a\mapsto \Tr_\omega\Big(\pi(a)(1+\D^2)^{-n/2}\Big),
$$
\'evalu\'ee sur le carr\'e de la courbure $F$ d'un `champ de jauge
non commutatif' repr\'esent\'e $A$:
$$
\tpi\big(\Omega^1\A\big)\ni A=
\sum_{j\in J} \tpi(a_j\delta b_j):=
\sum_{j\in J} \pi(a_i)[\D,\pi(b_i)],
$$
et la courbure, par analogie avec sa d\'efinition ordinaire, est
donn\'ee par
$$
F=[\D,A]+A^2.
$$
Pour que cette action soit analytiquement bien
d\'efinie dans le cas sans unit\'e, i.e. pour que
$$
\Tr_\omega\Big(\pi(.)(1+\D^2)^{-n/2}\Big)\in\A^{*},
$$
il faut ne consid\'erer que des champs de jauge prenant
valeur dans l'alg\`ebre sans unit\'e. Cette restriction peut
para\^itre excessive, car ne permettant pas aux
potentiels vecteurs d'\^etre de ``pure jauge'', i.e. d'avoir la forme
$$
A_{pj}=\pi(u)^*[\D,\pi(u)],
$$
pour $u$ un unitaire de l'alg\`ebre unif\`ere $\Aun$.
Nous verrons comment, avec l'action spectrale, on peut
contourner cet obstacle.\\
Concernant le mod\`ele standard,  cette nouvelle
description donne enfin une interpr\'etation g\'eom\'e\-tri\-que du champ de Higgs.
Ce dernier fera alors int\'egralement partie de la connection de
``Yang--Mills'' g\'en\'eralis\'ee. Plus heuristiquement, il sera interpr\'et\'e comme une
connection, au sens math\'ematique aussi bien qu'au sens litt\'eral,
entre le monde des fermions droits et celui des fermions gauches.\\
Il y a aussi une avanc\'ee au niveau ph\'enom\'enologique: les param\`etres du potentiel
de Higgs ne sont plus libres, on obtient des contraintes sur sa masse \cite{CIKS}.

\medskip

Alors que l'action de Connes--Lott n'est finalement qu'une
g\'en\'eralisation abstraite de celle de Yang--Mills,
l'action spectrale est fond\'ee sur des concepts diff\'erents.
Le point de d\'epart de cette construction fut la recherche d'une fonctionnelle d'action
ne d\'ependant que du spectre de l'op\'erateur de Dirac
g\'en\'eralis\'e $\D_A$, covariant par rapport \`a l'action du groupe
des unitaires de l'alg\`ebre. Dans le cas d'un triplet
spectral unital, l'action spectrale est alors d\'efinie comme
le nombre de valeurs propres de l'op\'erateur $\D_A=\D+A+\eps JAJ^{-1}$ inf\'erieures
en module \`a une constante $\Lambda$:
$$
S_\Lambda(\D,A)=\#\{\lambda_k:\,|\lambda_k|\leq\Lambda\}.
$$
Lorsque l'on consid\`ere le triplet spectral du MS,
l'action spectrale unifie au niveau classique le mod\`ele standard complet
(Yang--Mills--Higgs) avec la gravitation d'Einstein--Weyl.
Cette approche donne elle aussi une interpr\'etation
g\'eom\'etrique du m\'ecanisme de Higgs, ainsi que des contraintes sur la masse
de cette particule scalaire \cite{CIKS}.

\section{Action de Connes--Lott}

La construction (et le calcul) de l'action
de Connes--Lott pour un triplet spectral sans unit\'e, sera donn\'ee
au travers de l'exemple des plans de Moyal non d\'eg\'en\'er\'es.
Le r\'esultat que nous allons obtenir est tout \`a fait naturel, dans le sens
o\`u l'action obtenue est celle de Yang--Mills pour laquelle tous les
produits point \`a point ont \'et\'e remplac\'es par des produits de
Moyal. Le point clef du calcul d'une telle action est la d\'etermination
d'un id\'eal, appel\'e Junk, de l'alg\`ebre diff\'erentielle universelle
associ\'ee \`a l'alg\`ebre non commutative d\'ecrivant le mod\`ele.
Nous verrons que cet id\'eal poss\`ede la m\^eme structure
que dans tous les exemples connus: g\'eom\'etries presque commutatives,
tores non commutatifs. Cependant, les \'el\'ements caract\'eristiques de cet id\'eal
sont propres \`a chaque exemple.
L'outil de base va une fois encore \^etre la base de Wigner de l'oscillateur
harmonique.

\subsection{L'alg\`ebre diff\'erentielle universelle}

Nous allons nous borner \`a d\'ecrire une th\'eorie de jauge $U(1)$
sur plan de Moyal. Le cas g\'en\'eral $U(n)$ peut \^etre obtenu
en tensorisant l'alg\`ebre $\A_\th=(\SS(\R^{2N}),\mop)$ par l'alg\`ebre
finie $M_n(\C)$. C'est -\`a-dire que nous utiliserons ici $\A_F=\C$.\\
Aussi, pour des raisons de coh\'erence analytique, nous ne consid\'ererons
ni son extension $\bar{\A}_\th=(\D_{L^2},\mop)$ ni son plongement unif\`ere
$\Aun_\th=(\B,\mop)$ (voir paragraphe \ref{bour}), pour lesquelles l'application
$$
\Tr_\omega\Big(\pi(.)(1+\D^2)^{-n/2}\Big),
$$
n'est pas d\'efinie.\\
Les r\'ef\'erences concernant la construction utilis\'ee dans ce paragraphe sont
\cite{ConnesL1,ConnesL2,ConnesBook,Sirius}. Soit
$$
\Omega^\8 \A_\th := \bigoplus_{p\in\N} \Omega^p\A_\th,
$$
l'alg\`ebre
diff\'erentielle gradu\'ee universelle sur $\A_\th$. Par d\'efinition,
$\Omega^p \A_\th$, $p\geq 1$,  est d\'efinie par symboles et relations
$$
\Omega^p \A_\th :=
\set{f_0\,\delta f_1 \cdots \delta f_p : f_i \in \A_\th},
$$
et la seule contrainte sur
le symbole $\delta$ est de satisfaire \`a la r\`egle de Leibniz
$$
\delta(f_1\mop f_2) = (\delta f_1)\,f_2 + f_1\,\delta f_2,
$$
ainsi $\delta$ peut \^etre \'etendue sur tout $\Omega^\8 \A_\th$.
Etant donn\'e que $\A_\th$ n'a pas d'unit\'e,
on d\'efinit \cite[III.1.$\a$]{ConnesBook} $\Omega^0\A_\th := \A_\th\oplus \C$,
qui est la compactification maximale de $\A_\th$ et on pose
$\delta(0\oplus 1) := 0$. Pour obtenir une $*$-alg\`ebre gradu\'ee, on pose
aussi $(\delta f)^* := \delta{f^*}$.

La repr\'esentation $\pi^\th$ de $\A_\th$ dans $\L(\H)$ s'\'etend
naturellement sur tout $\Omega^\8\A_\th$, par
$$
\tpi^\th : \Omega^p\A_\th \to \L(\H) :
f_0\,\delta f_1 \cdots \delta f_p \mapsto
i^p\,\pi^\th(f_0)\,[\dslash,\pi^\th(f_1)]\cdots[\dslash,\pi^\th(f_p)].
$$

On a \'evidemment:
\begin{lem}
Lorsque $f_i \in \A_\th$, alors
$$
\tpi^\th(f_0\,\delta f_1 \cdots \delta f_p) =
L^\th(f_0 \mop \del_{\mu_1}f_1 \mop\cdots\mop \del_{\mu_p}f_p)
\ox \ga^{\mu_1} \cdots \ga^{\mu_p}.
$$
\end{lem}

\begin{proof}[Preuve]
C'est une simple cons\'equence de
$[\dslash, L^\th_f \ox 1_{2^N}] = -i L^\th(\del_\mu f) \ox \ga^\mu$
et de $L^\th_f\, L^\th_g = L^\th(f \mop g)$.
\end{proof}

Pour palier au fait que la repr\'esentation $\tpi^\th$ n'est pas a priori
fid\`ele (bien que $\pi^\th$ le soit), on introduit l'id\'eal bilat\`ere
gradu\'e de $\Omega^\8\A_\th$,
$$
\Junk := \bigoplus_{p\in \N} J^p :=
\bigoplus_{p\in\N} J_0^p + \delta J_0^{p-1},
$$
o\`u $J_0^p$ est le noyau de $\tpi^\th$ relev\'e \`a $\Omega^p\A_\th$,
$$
J_0^p := \set{\omega\in \Omega^p\A_\th :\tpi^\th(\omega) = 0},
$$
et finalement, on d\'efinit l'alg\`ebre quotient
$$
\Omega_\dslash\,\A_\th := \tpi^\th(\Omega^\8\A_\th)/\tpi^\th(\Junk)=
\tpi^\th(\Omega^\8\A_\th)/\bigoplus_{p=1}^\infty\tpi^\th(\delta J_o^p).
$$
Le 2-junk, seule composante non triviale
n\'ecessaire \`a la construction de l'action de Connes--Lott,
est dans notre cas particuli\`erement simple car isomorphe \`a
$\pi^\th(\A_\th)$.  

\medskip

En utilisant la base de Wigner $\{f_{mn}\}$, nous allons
dans le lemme suivant, exhiber des \'el\'ements particuliers de $\tpi^\th(J^2)$,
suffisants pour le caract\'eriser enti\`erement.

\begin{lem}
\label{lm:basic-junk}
Pour $m,n,k,l \in \N^N$, soit
$\omega_{mnkl} := f_{mk}\,\delta f_{kn} - f_{ml}\,\delta f_{ln} \in
\Omega^1\A_\th$ (sans sommation sur~$k$ et~$l$). Alors
$$
\tpi^\th(\omega_{mnkl}) = 0  \sepword{et}
\tpi^\th(\delta\omega_{mnkl})
= \tfrac{2}{\th}(|k| - |l|)\, L^\th(f_{mn}) \ox 1_{2^N}.
$$
\end{lem}
La premi\`ere composante du Junk est par d\'efinition triviale
$$
\tpi^\th(J^1)=\tpi^\th(J_0^1)+\tpi^\th(\delta J_0^0)
=\tpi^\th(\delta J_0^0)=0,
$$
car la repr\'esentation $\pi^\th$ est fid\`ele:
$J_0^0=\ker\pi=0$.
Les \'el\'ements de $\tpi^\th(J^2)$ sont les \'el\'ements
$\delta\omega\in\Omega^2\A_\th$ qui sont tels que
$\tpi^\th(\omega)=0$, alors que $\tpi^\th(\delta\omega)\ne0$.
Les \'el\'ements $\omega_{mnkl}$ d\'efinis lors de l'\'enonc\'e
du lemme pr\'ec\'edent appartiennent alors \`a $\tpi^\th(J^2)$.
\begin{proof}[Preuve du lemme \ref{lm:basic-junk}]
En utilisant les fonctions de cr\'eation et d'annihilation~\eqref{eq:crea-annl},
on peut r\'e\'ecrire l'op\'erateur de Dirac en termes de Moyal-commutateurs;
en adoptant la convention
$\del_{a_j} = \del/\del a_j$ et
$\del_{a_j^*} = \del/\del a_j^*$, pour $j = 1,\dots,N$,  on obtient
$$
\dslash = -\frac{i}{\sqrt{2}} \sum_j \ga^j(\del_{a_j} + \del_{a_j^*})
+ i\ga^{j+N} (\del_{a_j}-\del_{a_j^*})
= -i \sum_j (\ga^{a_j}\,\del_{a_j} + \ga^{a_j^*}\,\del_{a_j^*}),
$$
o\`u $\ga^{a_j} := \frac{1}{\sqrt{2}}(\ga^j + i\ga^{j+N})$ et
$\ga^{a_j^*} := \frac{1}{\sqrt{2}}(\ga^j - i\ga^{j+N})$.

La propri\'et\'e (\ref{pro4}), appliqu\'ee \`a $a_j$ et $a_j^*$
donne
$$
\del_{a_j} = -\frac{1}{\th} \ad_\mop a_j^*
:= -\frac{1}{\th} [a_j^*,\.\,]_\mop,  \qquad
\del_{a_j^*} = \frac{1}{\th} \ad_\mop a_j
:= \frac{1}{\th} [a_j,\.\,]_\mop
$$
et donc
$$
\dslash = - \frac{i}{\th} \sum_j
(\ga^{a_j^*} \ad_\mop a_j - \ga^{a_j} \ad_\mop a_j^*).
$$

Soit $u_j := (0,0,\dots,1,\dots,0)$ le $j$-i\`eme vecteur de base standard
de~$\R^N$. A partir de la d\'efinition \eqref{eq:basis} des \'el\'ements
de la base de Wigner $\{f_{mn}\}$, on
obtient directement
\begin{align}
\label{riw}
a_j^* \mop f_{mn} &= \sqrt{\th(m_j+1)}\, f_{m+u_j,n},
& f_{mn} \mop  a_j^* &= \sqrt{\th n_j}\, f_{m,n-u_j},
\\
\label{riv}
a_j \mop f_{mn} &= \sqrt{\th m_j}\, f_{m-u_j,n},
& f_{mn} \mop a_j &= \sqrt{\th(n_j+1)}\, f_{m,n+u_j}.
\end{align}
Ainsi,
\begin{align}
[\dslash,L^\th(f_{mn})]
&= -\frac{i}{\th} \sum_j \ga^{a_j} \bigl(\sqrt{\th n_j}\, L^\th(f_{m,n-u_j})
- \sqrt{\th(m_j+1)}\, L^\th(f_{m+u_j,n}) \bigr)
\nonumber \\
&\hspace{4em} + \ga^{a_j^*} \bigl( \sqrt{\th m_j}\, L^\th(f_{m-u_j,n})
- \sqrt{\th(n_j+1)}\, L^\th(f_{m,n+u_j}) \bigr).
\label{eq:plusun}
\end{align}

Nous pouvons dor\'enavant calculer $\tpi^\th(\omega_{mnkl})$ et
$\tpi^\th(\delta\omega_{mnkl})$. Premi\`erement,
\begin{align*}
\tpi^\th(\omega_{mnkl})
&= \tpi^\th(f_{mk}\,\delta f_{kn} - f_{ml}\,\delta f_{ln})
= L^\th(f_{mk} \mop \del_\mu f_{kn} - f_{ml} \mop \del_\mu f_{ln})
\ox \ga^\mu
\\
&= \frac{1}{\th} \sum_j \bigl(
\sqrt{\th n_j}\, L^\th(f_{mk} \mop f_{k,n-u_j})
- \sqrt{\th(k_j+1)}\, L^\th(f_{mk} \mop f_{k+u_j,n})
\\
&\hspace{4em} - \sqrt{\th n_j}\, L^\th(f_{ml} \mop f_{l,n-u_j})
+ \sqrt{\th(l_j+1)}\,L^\th(f_{ml} \mop f_{l+u_j,n})\bigr) \ox \ga^{a_j}
\\
&\qquad + \bigl(
\sqrt{\th k_j}\, L^\th(f_{mk} \mop f_{k-u_j,n})
- \sqrt{\th(n_j+1)}\, L^\th(f_{mk} \mop f_{k,n+u_j})
\\
&\hspace{4em} - \sqrt{\th l_j}\, L^\th(f_{ml} \mop f_{l-u_j,n})
+ \sqrt{\th(n_j+1)}\, L^\th(f_{ml} \mop f_{l,n+u_j}) \bigr)
\ox \ga^{ a_j^*}
\\
&= 0,
\end{align*}
en utilisant la propri\'et\'e d'unit\'es matricielles des $f_{mn}$ pour le
produit de Moyal. Deuxi\`emement,
$$
\tpi^\th(\delta\omega_{mnkl})
= \tpi^\th(\delta f_{mk}\,\delta f_{kn} - \delta f_{ml}\,\delta f_{ln})
= L^\th(\del_\mu f_{mk} \mop \del_\nu f_{kn} -
\del_\mu f_{ml} \mop \del_\nu f_{ln}) \ox \ga^\mu \ga^\nu,
$$
qui est \'egal \`a
\begin{align*}
&\frac{1}{\th^2} \biggl\{ \sum_j \Bigl( \bigl(
\sqrt{\th k_j}\, L^\th(f_{m,k-u_j})
- \sqrt{\th(m_j+1)}\, L^\th(f_{m+u_j,k}) \bigr) \ox \ga^{a_j}
\\
&\hspace{5em} + \bigl( \sqrt{\th m_j}\, L^\th(f_{m-u_j,k})
- \sqrt{\th(k_j+1)}\, L^\th(f_{m,k+u_j}) \bigr) \ox \ga^{ a_j^*}
\Bigr)
\\
&\hspace{3em} \sum_p \Bigl( \bigl(
\sqrt{\th n_p}\, L^\th(f_{k,n-u_p})
- \sqrt{\th(k_p+1)}\, L^\th(f_{k+u_p,n}) \bigr) \ox \ga^{a_p}
\\
&\hspace{5em} + \bigl(
\sqrt{\th k_p}\, L^\th(f_{k-u_p,n})
- \sqrt{\th(n_p+1)}\, L^\th(f_{k,n+u_p}) \bigr) \ox \ga^{a_p^*}
\Bigr)
\\
&\qquad - \sum_j \Bigl( \bigl(
\sqrt{\th l_j}\, L^\th(f_{m,l-u_j})
- \sqrt{\th(m_j+1)}\, L^\th(f_{m+u_j,l}) \bigr) \ox \ga^{a_j}
\\
&\hspace{5em} + \bigl(
\sqrt{\th m_j}\, L^\th(f_{m-u_j,l})
- \sqrt{\th(l_j+1)}\, L^\th(f_{m,l+u_j}) \bigr) \ox \ga^{a_j^*}
\Bigr)
\\
&\hspace{3em} \sum_p \Bigl( \bigl(
\sqrt{\th n_p}\, L^\th(f_{l,n-u_p})
- \sqrt{\th(l_p+1)}\, L^\th(f_{l+u_p,n}) \bigr) \ox \ga^{a_p}
\\
&\hspace{5em} + \bigl(
\sqrt{\th l_p}\, L^\th(f_{l-u_p,n})
-\sqrt{\th(n_p+1)}\, L^\th(f_{l,n+u_p}) \bigr) \ox \ga^{a_p^*}
\Bigr) \biggr\}.
\end{align*}

En utilisant encore la propri\'et\'e d'unit\'es matricielles,
l'expression pr\'ec\'edente se simplifie
\begin{align}
\tpi^\th(\delta\omega_{mnkl})
&= \frac{1}{\th} \sum_j \bigl(
k_j L^\th(f_{mn}) \ox \ga^{a_j} \ga^{ a_j^*}
+ (k_j+1) L^\th(f_{mn}) \ox \ga^{a_j^*} \ga^{a_j}
\nonumber \\
&\hspace{3em} - l_j L^\th(f_{mn}) \ox \ga^{a_j} \ga^{ a_j^*}
- (l_j+1) L^\th(f_{mn}) \ox \ga^{ a_j^*} \ga^{a_j} \bigr)
\nonumber \\
&= \frac{1}{\th}\, L^\th(f_{mn}) \ox  \sum_j
(k_j - l_j)\,(\ga^{a_j} \ga^{ a_j^*} + \ga^{ a_j^*} \ga^{a_j})
\nonumber \\
&= \frac{2}{\th} \sum_j (k_j - l_j)\, L^\th(f_{mn}) \ox 1_{2^N}.
\tag*{\qed}
\end{align}
\hideqed
\end{proof}

\begin{prop}
Il y a une identification naturelle entre
$\tpi^\th(J^2)$ et $\pi^\th(\A_\th) = L^\th(\A_\th) \ox 1_{2^N}$.
\end{prop}

\begin{proof}[Preuve]
Tout $\omega \in \tpi^\th(J^2) \subset \tpi^\th(\Omega^2\A_\th)$ peut \^etre
\'ecrit comme $\omega = \sum_{j\in I}
L^\th(\del_\mu f_j) L^\th(\del_\nu g_j) \ox \ga^\mu\ga^\nu$ o\`u $I$
est un ensemble fini, et satisfait
$\sum_{j\in I} L^\th(f_j \mop \del_\mu g_j) \ox \ga^\mu = 0$. D'apr\`es la r\`egle
de Leibniz,
\begin{align*}
\omega &= \sum_{j\in I}
L^\th(\del_\mu(f_j \mop \del_\nu g_j) - f_j \mop \del_\mu\del_\nu g_j)
\ox \ga^\mu \ga^\nu
= -\sum_{j\in I} L^\th(f_j \mop \del_\mu \del_\nu g_j)
\ox \ga^\mu \ga^\nu\\
& = -\sum_{j\in I} L^\th(f_j \mop \del_\mu \del_\nu g_j)
\ox \eta^{\mu\nu}\,1_{2^N}.
\end{align*}

Ainsi
$\tpi^\th(J^2) \subset \pi^\th(\A_\th) = L^\th(\A_\th) \ox 1_{2^N}$.

Soit maintenant
$\omega_{mnkl} := f_{mk}\,\delta f_{kn} - f_{ml}\,\delta f_{ln}$, le
lemme pr\'ec\'edent montre que $\tpi^\th(\omega_{mnkl}) = 0$ et
$\tpi^\th(\delta\omega_{mnkl}) = \frac2{\th} \sum_{j=1}^N (k_j - l_j)
\,L^\th(f_{mn}) \ox 1_{2^N}$, qui est non nul si $|l| \neq |k|$. Ainsi,
$L^\th(f_{mn}) \ox 1_{2^N}$ appartient \`a $\tpi^\th(J^2)$ pour tout
$m,n \in \N^N$. Puisque $\{f_{mn}\}$ est une base de $\A_\th$,
on en d\'eduit que $\pi^\th(\A_\th) \simeq
L^\th(\A_\th) \ox 1_{2^N} \subset \tpi^\th(J^2)$, qui conclut la preuve.
\end{proof}

Il est ais\'e de g\'en\'eraliser ce r\'esultat pour caract\'eriser toutes les
composantes du Junk.

\begin{cly}
Pour $p \geq 2$, $\tpi^\th(J^p)$ est lin\'eairement engendr\'e par les \'el\'ements de
$\tpi^\th(\Omega^p\A_\th)$ ayant la forme
$L^\th_f \ox \ga^{\mu_1} \dots \ga^{\mu_k}$, avec $k \leq p - 2$ et de m\^eme
parit\'e que~$p$.
\end{cly}
En particulier, en utilisant l'antisym\'etrie des matrices de Clifford, on
obtient que $\tpi^\th(J^p)=0$ pour $p>2N$.
\begin{rem}
Pour les triplets spectraux commutatifs $(\Coo(M),L^2(M,S),\Dslash)$
et pour ceux des tores non commutatifs
$(\Coo(\T^l_\Th),L^2(\T^l)\ox \C^{2^\lfloor l/2\rfloor},\dslash)$
(o\`u  $\lfloor l/2\rfloor$ d\'esigne la partie enti\`ere de $l/2$),
le 2-Junk est aussi isomorphe \`a l'alg\`ebre de d\'epart. En fait, l'inclusion
$\widetilde{\pi}(J^2)\subset \pi(\A)$, pour $\A=\Coo(M)$ ou $\A=\Coo(\T^l_\Th)$
est automatiquement satisfaite car elle n'utilise pas la commutativit\'e/non
commutativit\'e de l'alg\`ebre mais seulement la r\`egle de Leibniz. Pour obtenir
l'inclusion inverse, les \'el\'ements caract\'eristiques du 2-Junk sont, dans les cas
commutatifs
$$
(\delta f)f-f(\delta f), \sepword{pour} f\in\Coo(M),
$$
et pour les tores non commutatifs
$$
u_j(\delta u_j^{-1})+(\delta u_j)u_j^{-1},\,\,j=1,\cdots,l,
$$
o\`u les $u_j$ sont les unitaires engendrant le tore non commutatif,
i.e. satisfaisant \`a la relation $u_iu_j=e^{-i\Th_{ij}}\,u_ju_i$.
\end{rem}

\subsection{Fonctionnelle d'action}

Ayant identifi\'e les \'el\'ements de $\tpi^\th(J^2)$, on va
pouvoir caract\'eriser les \'el\'ements du quotient $\Omega_{\dslash}^2\A_\th$.
Pour cependant ne pas avoir \`a manipuler des classes d'\'equivalence de
`formes universelles', il est commode d'introduire un produit scalaire,
qui permettra de choisir sans ambigu\"it\'e un unique repr\'esentant dans chaque
classe d'\'equivalence.

Soit $\Ht_p$ l'espace de Hilbert obtenu en compl\'etant
$\tpi^\th(\Omega^p \A_\th)$ par rapport \`a la norme
induite par le produit scalaire
$$
\braket{\tpi^\th(\omega)}{\tpi^\th(\omega')}_p
:= \Tr_\omega \bigl( \tpi^\th(\omega)^* \,\tpi^\th(\omega')\,
(\dslash^2 + \eps^2)^{-N} \bigr),
$$
pour $\omega,\omega' \in \Omega^p \A_\th$. Cette forme sesquilin\'eaire
d\'efinit une pr\'e-action $I(\eta)$ lorsque $p = 2$ et
$\omega' = \omega = \delta\eta + \eta^2$:
\begin{equation}
\label{eq:pre-action}
I(\eta) := \Tr_\omega \bigl( \tpi^\th(\omega)^* \,\tpi^\th(\omega)\,
(\dslash^2 + \eps^2)^{-N} \bigr).
\end{equation}
Soit $P$ le projecteur orthogonal sur $\Ht_p$ dont l'image est le compl\'ement
orthogonal de $\tpi^\th(\delta J_0^{p-1})$ et soit
$\H_p := P\Ht_p$. Alors $P$ prolonge sur $\H_p$, l'application quotient de
$\tpi^\th(\Omega^p \A_\th)$ vers $\Omega_\dslash^p\,\A_\th$, qui est
identifi\'e \`a un sous-espace dense de~$\H_p$. La possible ambigu\"it\'e dans
\eqref{eq:pre-action} due au fait que $\tpi^\th$ n'est pas fid\`ele, dispara\^it
en d\'efinissant une fonctionnelle d'action (action de Yang--Mills
non commutative) par:
\begin{equation}
YM(\a)
:= \frac{N!\,(2\pi)^N}{8g^2}\, \braket{P\tpi^\th(F)}{P\tpi^\th(F)}_2
\label{eq:Yang-Mills},
\end{equation}
o\`u $\Omega^1_\dslash\,\A_\th \ni \a = \tpi^\th(\eta)$ et
$F = \delta\eta + \eta^2$ est la courbure de la 1-forme~$\eta$ et
$g$ est une constante de couplage. Il est d\'emontr\'e dans \cite{ConnesL1,Sirius},
que $YM(\a)$ co\"incide avec l'infimum des pr\'e-actions sur l'ensemble
des 1-formes $\eta \in \Omega^1\A_\th$ ayant la m\^eme image dans
$\Omega^1_\dslash\,\A_\th$:
$$
YM(\a)
= \frac{N!\,(2\pi)^N}{8g^2} \inf\set{I(\eta) : \tpi^\th(\eta) = \a}.
$$
Ce r\'esultat justifie la notation $YM(\a)$, car cette fonctionnelle
positive en $\eta$, ne d\'epend que de sa classe d'\'equivalence dans
$\Omega^1_\dslash\A_\th$, c'est-\`a-dire~$\a$.

\begin{thm}
Soit $\eta = -\eta^* \in \Omega^1\A_\th$. Alors
l'action de Yang--Mills non commutative
$YM(\a)$ de la connection universelle $\delta + \eta$, avec
$\a = \tpi^\th(\eta)$, est \'egale \`a
\begin{equation}
\label{NCYM}
YM(\a) = -\frac{1}{4g^2} \int \,d^{2N}x\,F^{\mu\nu} \mop F_{\mu\nu}(x)
= -\frac{1}{4g^2} \int \,d^{2N}x\,F^{\mu\nu}(x) \,F_{\mu\nu}(x) ,
\end{equation}
o\`u $F_{\mu\nu} :=
\half(\del_\mu A_\nu - \del_\nu A_\mu + [A_\mu,A_\nu]_\mop)$ et
$A_\mu$ est d\'efini par $\a = L^\th(A_\mu) \ox \ga^\mu$.
\end{thm}

\begin{proof}[Preuve]
Si $\eta = \sum_{j\in I} f_j\,\delta g_j$ avec $f_j,g_j \in \SS$
pour un ensemble fini $I$, alors
$\a = \sum_{j\in I} L^\th_{f_j} L^\th_{\del_\mu g_j} \ox \ga^\mu$
$=\sum_{j\in I}L^\th(f_j \mop \del_\mu g_j) \ox \ga^\mu$ et donc
$A_\mu := \sum_{j\in I} f_j \mop \del_\mu g_j$. Ainsi,
\begin{align*}
\tpi^\th(\delta\eta + \eta^2)
&= \tpi^\th(\delta f_j\,\delta g_j
             + (f_j\,\delta g_j) (f_k\,\delta g_k))
= \tpi^\th(\delta f_j\,\delta g_j
             + f_j\,\delta(g_j \mop f_k) \delta g_k
             - (f_j \mop g_j)\,\delta f_k\,\delta g_k)
\\
&= L^\th(\del_\mu f_j \mop \del_\nu g_j
+ f_j \mop \del_\mu(g_j \mop f_k ) \mop \del_\nu g_k
- f_j \mop g_j \mop \del_\mu f_k \mop \del_\nu g_k) \ox \ga^\mu\ga^\nu
\\
&= L^\th(\del_\mu f_j \mop \del_\nu g_k
+ f_j \mop \del_\mu g_j \mop f_k \mop \del_\nu g_k) \ox \ga^\mu\ga^\nu
\\
&= L^\th(\del_\mu(f_j \mop \del_\nu g_j)
+ f_j \mop \del_\mu g_j \mop f_k \mop \del_\nu g_k) \ox \ga^\mu\ga^\nu
- L^\th(f_j \mop \del_\mu \del_\nu g_j) \ox \eta^{\mu\nu}\,1_{2^N}
\\
&= L^\th(\del_\mu A_\nu + A_\mu \mop A_\nu) \ox \half[\ga^\mu,\ga^\nu]
+ \eta^{\mu\nu} L^\th(\del_\mu A_\nu + A_\mu\mop A_\nu) \ox 1_{2^N}
\\
&\hspace{4em}
- \eta^{\mu\nu} L^\th(f \mop \del_\mu \del_\nu g) \ox 1_{2^N}.
\end{align*}
Les deux derniers termes \'etant dans $\tpi^\th(J^2)$, on obtient
\begin{align*}
P(\tpi^\th(F))
&= P\bigl( L^\th(\del_\mu A_\nu + A_\mu \mop A_\nu)
\ox \half[\ga^\mu,\ga^\nu] \bigr)
\\
&= P\bigl( L^\th(\half
(\del_\mu A_\nu - \del_\nu A_\mu + [A_\mu,A_\nu]_\mop)
\ox \ga^\mu \ga^\nu \bigr)
\\
&= P(L^\th(F_{\mu\nu}) \ox \ga^\mu \ga^\nu)
            = L^\th(F_{\mu\nu}) \ox \ga^\mu \ga^\nu.
\end{align*}
La derni\`ere \'egalit\'e provient du fait que tout
$\omega = \omega_{\mu\nu} \ox \ga^\mu \ga^\nu
\in \tpi^\th(\Omega^2\A_\th)$ peut \^etre uniquement d\'ecompos\'e en
$$
\omega = \omega_{\mu\nu} \ox \half(\ga^\mu \ga^\nu - \ga^\nu \ga^\mu)
+ \omega_{\mu\nu} \ox \half(\ga^\mu \ga^\nu + \ga^\nu \ga^\mu)
$$
dans $\tpi^\th(\Omega^2\A_\th)_a \oplus \tpi^\th(\Omega^2\A_\th)_s
= \tpi^\th(\Omega^2\A_\th)_a \oplus \tpi^\th(J^2)$, somme directe des parties
antisym\'etrique et sym\'etrique.
Puisque $A_\mu = -A_\mu^*$, on obtient $F^*_{\mu\nu} = -F_{\mu\nu}$
et
$P(\tpi^\th(F))^* = L^\th(F_{\mu\nu}) \ox \ga^\mu \ga^\nu$.

Finalement, en remarquant que
$$\Tr_\omega \bigl( L^\th(F_{\mu\nu} \mop F_{\rho\sigma})\,
(-\Delta+\eps^2)^{-N} \ox \ga^\mu \ga^\nu \ga^\rho \ga^\sigma \bigr)
= \Tr_\omega \bigl( L^\th(F_{\mu\nu} \mop F_{\rho\sigma})\,
(-\Delta+\eps^2)^{-N} \bigr)\, \Tr(\ga^\mu\ga^\nu\ga^\rho\ga^\sigma),
$$
que $\Tr(\ga^\mu\ga^\nu\ga^\rho\ga^\sigma) =
2^N(\eta^{\mu\nu}\eta^{\rho\sigma} - \eta^{\mu\rho} \eta^{\nu\sigma} +
\eta^{\mu\sigma}\eta^{\nu\rho})$ et $F_{\mu\nu} = -F_{\nu\mu}$,
on obtient \`a partir de la proposition~\ref{pr:calcul} (avec $-i\nb$ \`a la place de
$\dslash$)
$$
YM(\a) = - \frac{2\,N!\,(4\pi)^N}{8g^2}
\Tr_\omega\Bigl( L^\th(F_{\mu\nu} \mop F^{\mu\nu})\,
(-\Delta + \eps^2)^{-N} \Bigr)
= -\frac{1}{4g^2} \int\,d^{2N}x\, (F_{\mu\nu} \mop F^{\mu\nu})(x) .
$$
D'apr\`es la propi\'et\'e de trace (lemme~\ref{trace}), on peut remplacer
le produit de Moyal par celui ordinaire, pour obtenir (\ref{NCYM}).
\end{proof}

\begin{rem}
Dans nos conventions $F^*_{\mu\nu} = -F_{\mu\nu}$ et
on peut v\'erifier la positivit\'e de l'action ainsi d\'efinie:
$$
YM(\a) = \frac{1}{4g^2}
\int \,d^{2N}x\,\sum_{\mu,\nu=1}^{2N} |F_{\mu\nu}(x)|^2 .
$$
\end{rem}

\section{Action spectrale}

Apr\`es avoir revu les motivations de Chamseddine et Connes
pour d\'efinir, dans le cas d'un triplet spectral avec unit\'e, une
fonctionnelle d'action \`a partir du spectre de l'op\'erateur de
Dirac g\'en\'eralis\'e, nous proposerons une d\'efinition
de l'action spectrale dans le cas sans unit\'e.
Nous nous focaliserons ensuite sur les aspects techniques,
permettant de calculer une telle action dans le cas des d\'eformations
isospectrales. Faute de temps, nous ne pr\'esenterons ici
que les r\'esultats obtenus dans le cas des plans de Moyal.
Modulo une complexit\'e d'ordre calculatoire grandissante,
la strat\'egie que nous allons suivre sera cependant directement applicable
aux cas courbes, p\'eriodiques ou non.

\subsection{Motivations et g\'en\'eralit\'es}

Pour un triplet spectral \`a unit\'e $(\A,\H,\D)$, Chamseddine et
Connes \cite{AliAlain, AliAlain2} ont propos\'e
une `action physique' ne d\'ependant que du spectre de
l'op\'erateur $\D$; {\it le principe d'action spectrale}
\begin{equation}
S_\Lambda(\D,A):=\Tr \left(\chi_{_{[0,1]}}(\D_A^2/\Lambda^2)\right),
\end{equation}
i.e. le nombre de valeurs propres inf\'erieures \`a une \'echelle
de masse $\Lambda$.
Ici, $\D_A$ est l'op\'erateur de ``Dirac'' covariant
$\D_A:=\D +A+\epsilon JAJ^{-1}$, o\`u $A$ est une 1-forme
diff\'erentielle universelle repr\'esent\'ee
$A\in\tilde{\pi}(\Omega^1\A)$.  $\tilde{\pi}$
d\'esigne toujours la repr\'esentation relev\'ee \`a l'alg\`ebre
diff\'erentielle universelle $\Omega^\bullet\A$:
$$
\tilde{\pi}(a_0\delta a_1\cdots\delta a_p)
:=\pi(a_0)[\D, \pi(a_1)]\cdots[\D, \pi(a_p)], \,\,\,a_i\in\A, \;i=1,\cdots,p.
$$
$J$ est la structure r\'eelle du triplet (la conjugaison de charge
pour les spineurs dans les cas commutatifs comme pour les
d\'eformations isospectrales) et $\epsilon\in\{+1,-1\}$ en fonction
de la dimension. Initialement \cite{AliAlain, AliAlain2},
dans le cas presque commutatif du mod\`ele standard, le calcul de cette action
utilisait \`a la place de la fonction caract\'eristique $\chi_{[0,1]}$,
n'importe quelle fonction lisse $\phi$ l'approximant.
Pour pouvoir calculer cette action asymptotiquement
par des techniques de transform\'ee de
Laplace, de plus amples conditions sur $\phi$ doivent \^etre impos\'ees \cite{Odysseus}.
Nous reverrons ces conditions lors de son calcul pour les plans de Moyal.
Dans le cas \`a unit\'e, $\D$ ainsi que sa perturbation born\'ee $\D_A$
sont \`a r\'esolvante compacte, donc
$\phi(\D_A^2/\Lambda^2)$ est \`a trace d\`es que $\phi$ est
suffisamment d\'ecroissante \`a l'infini; $r^{n-1}\phi(r^2)\in L^1(\R^+)$
est une condition suffisante dans le cas d'un
triplet spectral de dimension $n$.

Concernant l'op\'erateur de ``Dirac'' covariant $\D_A$,
le point de d\'epart r\'eside dans
l'analogie entre le groupe d'invariance
d'une th\'eorie de jauge coupl\'ee avec la gravitation
sur une vari\'et\'e Riemannienne $M$,  $G=U\rtimes \Diff(M)$
et le groupe d'automorphismes $\Aut(\A)=\Int(\A)\rtimes\Out(\A)$
d'une alg\`ebre $\A$, au travers des suites exactes de groupe
$$
1\rightarrow U\rightarrow G\rightarrow \Diff(M)\rightarrow 1,
$$
$$
1\rightarrow\Int(\A)\rightarrow\Aut(\A)\rightarrow\Out(\A)\rightarrow 1.
$$
Les deux constructions co\"incident en particulier lorsque
$\A=\Coo(M, M_n(\C))$, $n>1$:
$\Out(\A)=\Diff(M)$, $\Int(\A)=
\Coo(M,SU_n/\Z_n)$.
Pour une fonctionnelle d'action d\'efinie
sur un triplet spectral, il est alors naturel d'imposer que le groupe d'invariance soit
le groupe d'automorphismes de l'alg\`ebre.

Pour obtenir une th\'eorie de jauge avec champs de mati\`eres
lorsque l'alg\`ebre $\A$ est presque commutative, c'est-\`a-dire
lorsque $\A=\Coo(M)\otimes A_F$
(o\`u $A_F$ est une alg\`ebre finie, en l'occurrence
$\mathbb{H} \oplus\C \oplus M_3(\C)$ pour le mod\`ele standard de la
physique des particules \cite{AliAlain,AliAlain2,CIKS}), il faut repr\'esenter
les automorphismes int\'erieurs sur l'espace de Hilbert fermionique $\H$. En particulier,
il faut relever $\Int(\A)$ sur le groupe des unitaires $\mathcal{U}(\H)$
des op\'erateurs born\'es sur $\H$:
$$
\mathcal{U}(\A)\ni u\mapsto \sigma(u)=\pi(u)J\pi(u)J^{-1}\in\mathcal{U}(\H).
$$
Pour les tores non commutatifs, les plans de Moyal et certaines
g\'eom\'etries presque commutatives, cette repr\'esentation est la
repr\'esentation adjointe: $\pi(u)J\pi(u)J^{-1}\psi=
u\Mop \psi\Mop u^{*}$, $\psi\in\H$. Sous cette transformation,
l'op\'erateur $\D$ se transforme comme
\begin{equation}
\label{gaugetransform}
\D\rightarrow\sigma(u)\D\sigma(u)^{-1}=\D+\pi(u)[\D,\pi(u^*)]
+\epsilon J\pi(u)[\D,\pi(u^*)] J^{-1},
\end{equation}
o\`u le signe $\epsilon$ vient de la relation de commutation
$\D J=\epsilon J\D$,
(voir \cite{ConnesReal, Polaris} pour une table des signes).
Ainsi, $\D_A\rightarrow \D_{A'}$ avec
$A'=\pi(u)A\pi(u^*)+ \pi(u)[\D,\pi(u^*)]$ se transforme de mani\`ere
covariante.

Pour les g\'eom\'etries presque commutatives $\Coo(M)\otimes A_F$,
en particulier pour le mod\`ele standard, avec
$$
\D=\Dslash\otimes 1_{\H_F}+1\ox D_F,\,\,\,
\Dslash=-i{e^\mu}_a \gamma^a(\pa_\mu+ \omega_\mu),
$$
o\`u $\omega$ est la connection de spin et $D_F$ une matrice Hermitienne
(la matrice de masse fermionique),
$S_\Lambda(\D,A)$ est asymptotiquement calculable en utilisant le
d\'eveloppement du noyau de la chaleur. Notons d\`es \`a pr\'esent que $\Dslash_A^2$
peut \^etre vu comme un Laplacien g\'en\'eralis\'e:
$\Dslash_A^2=-\left(g^{\mu\nu}(\pa_\mu +B_\mu)(\pa_\nu +B_\nu)
+E\right)\equiv P$ o\`u $g^{\mu\nu}$ est le tenseur m\'etrique, $B_\mu$
est une connection contenant une partie spin et une partie Yang--Mills et $E$
est un endomorphisme du fibr\'e spinoriel.
On peut montrer formellement \cite{AliAlain,AliAlain2}, en d\'eveloppant
$\phi$ en s\'erie de Taylor, que $S_\Lambda(\D,A)$ est li\'ee aux coefficients
de Seeley--DeWitt $a_k(P,p)$ de la trace du semi-groupe de la chaleur
\begin{equation}
\Tr \left(e^{-tP}\right)\sim_{t\rightarrow 0}
(4\pi)^{-n/2}\sum_{l\in\N}t^{(l-n)/2}\int_M \mu_g(p)\,a_l(P,p),
\end{equation}
par la relation
entre la fonction Z\'eta et la trace de l'op\'erateur de la chaleur \cite{Gilkey}:
\begin{equation}
\label{zeta}
\zeta_P(s):=\Tr(P^{-s})={\textstyle{\frac{1}{\Gamma(s)}}}\int_0^\infty dt\, t^{s-1}
\Tr(e^{-tP}).
\end{equation}
Sur une vari\'et\'e sans bord, les coefficients de Seeley--De Witt s'annulent
$a_l(P,.)=0$, lorsque $l$ est impair.
On obtient alors dans le cas quadri-dimensionnel
\begin{equation}
S_\Lambda(\Dslash\otimes 1_{A_F},A)= (4\pi)^{-2}\sum_{l=0}^{2}\Lambda^{4-2l}
\; \phi_{2l}\;\int_M\mu_g(p)\, a_{2l}(P,p) +O(\Lambda^{-2}),
\end{equation}
o\`u les `moments' $\phi_i$ sont d\'efinis par
\begin{equation}
\label{moments}
\phi_0=\int_0^\infty dt\,t\,\phi(t),\;\;\phi_2=\int_0^\infty dt\,\phi(t),\;\;
\phi_{2(2l+2)}=(-1)^l\phi^{(l)}(0), \;l\geq 0.
\end{equation}
Des d\'erivations moins formelles de cette relation, n\'ecessitant
quelques hypoth\`eses suppl\'ementai\-res sur $\phi$, ont \'et\'e obtenues dans
\cite{Odysseus, Nestetal}. \\
Lorsque $M$ est de dimension quatre et
$A_F=\mathbb{H} \oplus\C \oplus M_3(\C)$, {\it l'action spectrale
unifie la gravitation d'Einstein plus Weyl et le mod\`ele standard
avec son secteur de Higgs et le m\'ecanisme de brisure spontan\'ee
de sym\'etrie} (voir \cite{AliAlain,AliAlain2,CIKS}). Comme nous le
verrons dans le cas des plans de Moyal, il n'y a pas de
restriction sur les dimensions, mais l'expression des coefficients (\ref{moments})
sera sensiblement diff\'erente.
\begin{rem}
La relation (\ref{zeta}) lie aussi la trace de Dixmier avec le d\'eveloppement
du noyau de la chaleur et donc l'action de Connes--Lott avec l'action spectrale.
\end{rem}

L'op\'erateur $\D$ n'est plus
\`a r\'esolvante compacte dans le cas non compact, et
il va falloir introduire une r\'egularisation suppl\'ementaire
pour d\'efinir l'action spectrale. Cette r\'egularisation sera donn\'ee
par un \'el\'ement positif de l'alg\`ebre $\A$ et par analogie
avec le cas commutatif non compact, pourra \^etre qualifi\'ee de
r\'egularisation spatiale.

\begin{defn}
Pour un triplet spectral sans unit\'e $(\A,\Aun,\H,\D)$ de dimension
spectrale $n$, l'{\it action spectrale} est donn\'ee par
\begin{equation}
S_\Lambda(\D,A,\rho):=\Tr_\H\left(\pi(\rho)\;\phi(\D_A^2/\Lambda^2)\right).
\end{equation}
Comme dans le cas unital, $\D_A=\D+A+\epsilon J AJ^{-1}$
($\epsilon\in\{+1,-1\}$ ) et $A\in \Omega_{\D}^1(\Aun)$ est dor\'enavant
une 1-forme auto-adjointe de l'alg\`ebre \`a unit\'e $\Aun$ repr\'esent\'ee
$$
A^*=A=\sum_{i\in I}\pi(b_0^i) \; [\D,\pi(b_1^i)],
$$
o\`u $I$ est un ensemble fini,
$b_0^i,b_1^i\in\Aun$, $0\leq\rho\in\A$ et de plus,
$0\leq\phi$ et $\Lambda$ sont comme dans le cas \`a unit\'e.
\end{defn}

\begin{rem}
i) Cette d\'efinition donne une importance suppl\'ementaire au
choix du plongement unif\`ere de l'alg\`ebre.
La 1-forme $A$ \'etant maintenant construite \`a partir de $\Aun$, toutes
les consid\'erations de sym\'etrie discut\'ees dans le cas compact,
restent alors valables. Ce point est crucial car l'existence d'unitaires dans
l'alg\`ebre est primordiale pour impl\'ementer l'invariance de jauge:
$S_\Lambda(\D,A,\rho)$ est par construction invariante
sous l'action des automorphismes int\'erieurs
relev\'es sur l'espace de Hilbert
$\pi(u)J\pi(u)J^{-1}$. Dor\'enavant la r\'egularisation $\rho$ se transforme aussi
$$
\begin{cases}
A & \rightarrow \; uAu^*+u[\D,u^*], \\
\rho & \rightarrow \;u\rho u^* .
\end{cases}
$$
ii) Les positivit\'es de $\rho$ et de $\phi$ sont suffisantes pour
donner lieu \`a une action d\'efinie positive.\newline
iii) D'autres r\'egularisations existent. En l'occurrence,
$\Tr\Big(\phi \big(\D_A^2\,\pi(\rho)^{-1} \big)\Big)$ avec $\rho$ un
\'el\'ement inversible et positif de l'alg\`ebre, est aussi bien d\'efinie.
Cependant, pour les plans de Moyal, ce type de
r\'egularisation ne donne pas un acc\`es direct \`a un d\'eveloppement asymptotique.
\end{rem}

Dans le cas d'une g\'eom\'etrie commutative (ou presque commutative)
non compacte et sans bord
$\A=\Coo_c(M)$, il est
ais\'e de montrer que $\rho \;e^{-t\,P}$,
$t\in\R^*_+$, o\`u $P$ est un Laplacien g\'en\'eralis\'e et
$\rho\in\Coo_c(M)$, est \`a trace.
Pour cela, on peut utiliser les
m\^emes techniques d'op\'erateur pseudodiff\'erentiel que celles que
nous utiliserons lors du paragraphe suivant.
La formule (\ref{zeta}) a dans ce cas un analogue
\begin{equation}
\label{zeta2}
\zeta_{\rho,P}(s):=\Tr(\rho \;P^{-s})=
{\textstyle{\frac{1}{\Gamma(s)}}}\int_0^\infty dt\, t^{s-1}
\Tr(\rho \;e^{-t\,P}).
\end{equation}
Pour $P=(\Dslash + A +\epsilon JAJ^{-1})^2$, on obtient
\begin{equation*}
S_\Lambda(\Dslash\otimes 1_{A_F},A)= (4\pi)^{-n/2}\sum_{l=0}^m
\Lambda^{n-2l}\; \phi_{2l}\;\int_M \mu_g(p)\,a_{2l}(P,p)\;\rho(p)
+\;O(\Lambda^{n-2(m +1)}),
\end{equation*}
o\`u $\{a_{2l}\}$ d\'esigne toujours les coefficients de Seeley--DeWitt, qui
dans le cas d'une vari\'et\'e non compacte ne sont plus que 
localement int\'egrables.
Cependant, $a_{2l}(P,.)\rho(.)$ est globalement int\'egrable car $\rho\in\Coo_c(M)$.
Les coefficients $\{\phi_{2l}\}$ ont la forme (\ref{moments})
dans le cas quadri-dimensionnel; leurs expressions en dimension
quelconque sera \'etabli lorsque l'on traitera le cas des plans de Moyal.

\subsection{Cas des plans de Moyal}
\subsubsection{3.2.2.1 D\'eveloppement du noyau de la
chaleur pour Laplaciens non commutatifs}

Dans ce paragraphe, nous allons obtenir un d\'eveloppement
similaire \`a celui du noyau de la chaleur, pour la trace d'un
semi-groupe r\'egularis\'e, g\'en\'er\'e par un Laplacien
non commutatif. Cette formule asymptotique
permettra de calculer (une forme pr\'eliminaire de) l'action spectrale
associ\'ee aux triplets spectraux des plans de Moyal g\'en\'eriques,
i.e. l'hypoth\`ese d'inversibilit\'e de la matrice de d\'eformation
sera relax\'ee.

Nous commen\c cons par \'etablir un r\'esultat g\'en\'erique
concernant la tra\c cabilit\'e de semi-groupes r\'egularis\'es.
Pour ce faire, nous allons utiliser certains faits de la th\'eorie des
semi-groupes \`a un param\`etre;
$e^{-t\,A}$ o\`u $A$ est un op\'erateur non born\'e (auto-adjoint) et positif,
sur un espace de Hilbert s\'eparable $\H$.

Sous la seule hyphoth\`ese de la positivit\'e du g\'en\'erateur $A$,
on montre que $e^{-z\,A}$ est une contraction
holomorphe pour $\Re(z)>0$ \cite[exemple 1.25, p.493]{Kato} \cite{Zagreb},
c'est-\`a-dire que le semi-groupe peut \^etre d\'efini par calcul fonctionnel holomorphe:
\begin{equation}
\label{Hcal}
e^{-t\,A}=\frac{1}{2i\pi}\int_{\Gamma}e^{-tz}\;\mathcal{R}_A(z) \; dz,
\end{equation}
o\`u $\Gamma$ est un contour (\'eventuellement infini) orient\'e entourant
le spectre de $A$ et qu'il est contractant:
$$
\| e^{-z\,A} \| \leq1.
$$

\begin{lem}
\label{zagreblike}
Soit $B$ un op\'erateur born\'e et $A$ un g\'en\'erateur positif,
densement d\'efini d'un semi-groupe holomorphe qui soit tel que
$B\mathcal{R}_A(z)^l\in \L^1(\H)$ pour un $l\in\N^*$ et un
$z\notin \spec(A)$.
Alors, pour tout $t>0$, $Be^{-t\,A}$ est \`a trace.
\end{lem}
\begin{proof}[Preuve]
Pour un $z_0\notin \spec(A)$, la propri\'et\'e de semi-groupe,
ensemble avec la premi\`ere \'equation r\'esolvante
et la pr\'esentation (\ref{Hcal}), impliquent
$$
B\;e^{-t\,A}=B\; (e^{-\frac{t}{l}\,A})^l=B\;\mathcal{R}_A(z_0)^l\;\left(\frac{1}{2i\pi}
\int_\Gamma dz\,e^{-\frac{t}{l}z}(1+(z-z_0)\mathcal{R}_A(z)) \right)^l,
$$
qui conclut la preuve car $\| \mathcal{R}_A(z) \| \leq \frac{M}{|z|}$
pour tout $z$ avec $\Re(z)>0$ \cite[proposition 1.13]{Zagreb} et donc
\begin{align}
\int_\Gamma |dz|\,e^{-\frac{t}{l}\,\Re(z)}\Big(1+|z-z_0|\;
\|\mathcal{R}_A(z)\|\Big)<\infty.
\tag*{\qed}
\end{align}
\hideqed
\end{proof}

\bigskip

Pour les produits de Moyal g\'en\'eriques,
$\left(\SS(\R^n),\Mop\right)$, $\left(\D_{L^2}(\R^n),\Mop\right)$ et
$\left(\Oh_0(\R^n),\Mop\right)$, munis de leurs
topologies usuelles, forment des alg\`ebres de Fr\'echet. En effet,
tout produit de Moyal g\'en\'erique se d\'ecompose en un
produit de Moyal symplectique et un produit point \`a point;
en d\'ecomposant la matrice $\Th$ en une somme directe d'une
matrice symplectique de dimension $2k$ et  une matrice \`a entr\'ee
nulle de dimension $n-2k$ \cite[proposition 2.7 et corollaire 2.8]{RieffelDefQ}.
Nous avons d\'ej\`a montr\'e que chacun des espaces $\SS(\R^n)$,
$\D_{L^2}(\R^n)$ et $\Oh_0(\R^n)$ sont stables sous produit de Moyal symplectique.
Il suffit alors de v\'erifier qu'ils le sont aussi sous produit point \`a point.
Pour $\SS(\R^n)$ et $\Oh_0(\R^n)$ ce fait est \'evident, alors que pour
$\D_{L^2}(\R^n)$, c'est une cons\'equence de l'inclusion $\D_{L^2}(\R^n)
\subset\Oh_0(\R^n)$ \cite{Schwartz}.
Ainsi $\Big(\mathcal{C},L^2(\R^n)\otimes\C^{2^m},\dslash\Big)$
d\'efinit aussi un triplet spectral
sans unit\'e, si $\mathcal{C}:=\Big(\left(\SS(\R^n),\Mop\right),
\left(\D_{L^2}(\R^n),\Mop\right),
\left(\Oh_0(\R^n),\Mop\right)\Big)$. Dans ce cas, le calcul de l'action de Connes--Lott
(\ref{NCYM}) n'est pas direct,
car fond\'e sur la base de Wigner qui n'a
plus d'analogue dans les cas d\'eg\'en\'er\'es. Nous allons voir que l'action
spectrale peut \^etre calcul\'ee sans difficult\'es suppl\'ementaires dans
tous les cas de figure (produits de Moyal symplectique ou d\'eg\'en\'er\'e).

Soit $\triangle^\Th$ un {\it Laplacien g\'en\'eralis\'e non commutatif} associ\'e
au produit de Moyal
\begin{align}
\label{genelap}
\triangle^\Th &:=-\Big(\eta^{\mu\nu}(\pa_\mu+L^\Th(\omega_\mu))
(\pa_\nu+L^\Th(\omega_\nu))+L^\Th(E)\Big)\otimes1_{2^m}, \\
\triangle^\Th &=:\triangle_r^\Th \otimes 1_{2^m}, \nonumber
\end{align}
agissant sur l'espace de Hilbert $\H=L^2(\R^n)\otimes\C^{2^m}=:\H_r\otimes\C^{2^m}$, o\`u
$\omega_\mu^*=-\omega_\mu$ et $E$ sont dans $\Oh_0(\R^n)$.
Ici $2^m$ est le rang du fibr\'e des spineurs, i.e. $m$ d\'esigne la partie enti\`ere
de $n/2$.

Pour $f\in\A_\Th:=\left(\SS(\R^n),\Mop\right)$, soit
$L^\Th(f)e^{-t\,\triangle^\Th}$  {\it le semi-groupe r\'egularis\'e} par $L^\Th_f$
associ\'e au Laplacien g\'en\'eralis\'e $\triangle^\Th$.

Le th\'eor\`eme suivant est le r\'esultat principal de ce paragraphe:
\begin{thm}
\label{HKNCP}
Pour $\triangle^\th$ d\'efini dans l'\'equation (\ref{genelap})
et avec $f\in\SS(\R^n)$,
$\Tr(\pi^\Th(f)\,e^{-t\,\triangle^\Th})$ poss\`ede un d\'eveloppement
asymptotique
\begin{equation}
\label{asympexp}
\Tr\left(\pi^\Th(f)\,e^{-t\,\triangle^\Th}\right)\sim_{t\rightarrow 0}\;
{\textstyle{2^m (\;\frac{1}{4\pi t})^{n/2}}}\sum_{l\in\N}
\;t^l\;\int_{\R^n}d^nx\,f(x)\;\tilde{a}_{2l}(x) ,
\end{equation}
avec
\begin{align*}
\tilde{a}_0(x)=&1,\\
\tilde{a}_2(x)=&E(x),\\
\tilde{a}_4(x)=&\tfrac{1}{2}\; E\Mop E(x)+\tfrac{1}{6}\;
\eta^{\mu\nu} E_{;\mu\nu}(x)+\tfrac{1}{12}\;
\Omega^{\mu\nu}\Mop \Omega_{\mu\nu}(x),\\
\tilde{a}_6(x)=&\tfrac{1}{6}\; E\Mop E\Mop E(x)
+\tfrac{1}{12}\;\eta^{\mu\nu}E_{;\mu}\Mop E_{;\nu}(x)
+\tfrac{1}{6}\;\eta^{\mu\nu} E\Mop E_{;\mu\nu}(x)\\
&+\tfrac{1}{60}\;\eta^{\mu\nu}\eta^{\rho\sigma}
 E_{;\mu\nu\rho\sigma}(x)
+\tfrac{1}{12} \;E\Mop \Omega^{\mu\nu}\Mop \Omega_{\mu\nu}(x)
+\tfrac{1}{45}\;\eta^{\rho\sigma}{ \Omega^{\mu\nu}}_{;\rho}
\Mop \Omega_{\mu\nu ;\sigma}(x)\\
&+\tfrac{1}{180}\;{\eta^{\rho\sigma} \Omega^{\mu\nu}}_{;\nu}
\Mop\Omega_{\mu\rho;\sigma}(x)
+\tfrac{1}{30}\;\eta^{\rho\sigma} \Omega^{\mu\nu}
\Mop \Omega_{\mu\nu;\rho\sigma}
-\tfrac{1}{30}\; \Omega^{\mu\nu}
\Mop \Omega_{\nu\rho}\Mop {\Omega^\rho}_\mu(x),
\end{align*}
o\`u $g_{;\mu}:=\pa_{\mu}g+ [\omega_{\mu},g]_{\Mop}$ et
$\Omega_{\mu\nu}:=\pa_{\mu}\omega_{\nu}-\pa_{\nu}\omega_{\mu}
+[\omega_{\mu},\omega_{\nu}]_{\Mop}$
est la courbure de la connection $\omega$.
\end{thm}

Nous allons commencer par montrer que $L^\Th(f)e^{-t\;\triangle_r^\Th}$
est \`a trace pour tout $t\in\R^*_+$, pour ensuite montrer que sa trace
poss\`ede un d\'eveloppement en puissance de $t$, pour $t\to0$:
\begin{equation}
\Tr\left( L^\Th(f)e^{-t\;\triangle_r^\Th}\right)
\sim_{t\rightarrow 0}\;{(\textstyle{\frac{1}{4\pi
t})^{n/2}}}\sum_{l\in\N}
\;t^l\;\int_{\R^n}d^nx\;f(x)\tilde{a}_{2l}(x),\nonumber
\end{equation}
o\`u les `invariants locaux' $\tilde{a}_l$ sont construits \`a partir
de la connection universelle repr\'esent\'ee $\omega_\mu$,
de l'endomorphisme (non local) $E$ et de leurs d\'eriv\'ees covariantes
dans la repr\'esentation adjointe:
$$
\nabla^\Th_\mu:=\pa_\mu+L^\Th(\omega_\mu)-R^\Th(\omega_\mu).
$$

Nous allons \'etablir la tra\c cabilit\'e de $L^\Th(f)e^{-t\triangle_r^\Th}$
par deux approches. La premi\`ere utilise la th\'eorie des semi-groupes,
alors que la seconde est bas\'ee sur des techniques d'op\'erateurs
pseudodiff\'erentiels. Toutes deux seront applicables aux cas
commutatifs non compacts, mais seule la premi\`ere pourra
s'\'etendre aux cas  des d\'eformations isospectrales non compactes.

\begin{thm}
\label{traceclass}
Soient $f\in\SS(\R^n)$,  $\omega_\mu ,E\in\Oh_0(\R^n)$ avec
$\omega_\mu^*=-\omega_\mu$ et $E=-h^*\Mop h$ pour un certain
$h\in\Oh_0(\R^n)$. Alors, pour tout $t>0$ le semi-groupe
de g\'en\'erateur $\tri^\Th$, r\'egularis\'e par $L^\Th_f$,
est \`a trace.
\end{thm}

\begin{proof}[Premi\`ere preuve du th\'eor\`eme \ref{traceclass}]
De $(L^\Th(h))^*=L^\Th(h^*)$, on en d\'eduit que  $L^\Th(E)$
est positif et donc que
$\triangle^\Th$ l'est aussi. D'apr\`es le lemme \ref{zagreblike},
il est suffisant de montrer que $L^\Th(f)R_{\triangle_r^\Th}(z)^l$ est
\`a trace pour $l>\frac{n}{2}$.

Remarquons que $\triangle^\Th$ peut s'\'ecrire comme un op\'erateur de Dirac
covariant au carr\'e:
\begin{align*}
\triangle^\Th &=\dslash^2_\omega -B,\cr
\dslash_\omega &:=-i\big(\pa_\mu+L^\Th(\omega_\mu)\big) \otimes \gamma^\mu,
\end{align*}
o\`u $B:=L^\Th(E)\otimes1_{2^m}-L^\Th(\pa_\mu(\omega_\nu)-\omega_\mu\Mop\omega_\nu)
\otimes(\eta^{\mu\nu}1_{2^m}-\gamma^\nu\gamma^\mu)$ est born\'e.

Supposons tout d'abord que $l=1$, $z=-1$, $B=0$. En utilisant les
notations
$$
\pi^\Th(\omega):=L^\Th(\omega_\mu)\otimes\gamma^\mu
\sepword{et}\pi^\Th(\dslash(f)):=L^\Th(\pa_\mu f)\otimes\gamma^\mu
$$
et en remarquant que tout $f\in\SS(\R^n)$ se factorise comme $f=f_1\Mop f_2$,
avec $f_1,f_2\in\SS(\R^n)$ \cite[proposition 2.7]{Himalia}, on obtient
\begin{align*}
\pi^\Th(f)\mathcal{R}_{\triangle^\Th}(-1)&=-\pi^\Th(f)\frac{1}{\dslash-i}
\left(1-\pi^\Th(\omega)\frac{1}{\dslash_\omega-i}\right)\frac{1}{\dslash_\omega+i}\\
&=-\pi^\Th(f_1)\frac{1}{\dslash-i}\;\pi^\Th(f_2)\left(1-\pi^\Th(\omega)\frac{1}{\dslash_\omega-i}\right)
\frac{1}{\dslash_\omega+i} \\
&\quad-\pi^\Th(f_1)\frac{1}{\dslash-i}\;\pi^\Th(\dslash(f_2))\frac{1}{\dslash-i}
\left(1-\pi^\Th(\omega)\frac{1}{\dslash_\omega-i}\right)
\frac{1}{\dslash_\omega+i} \\
&=-\pi^\Th(f_1)\frac{1}{\dslash-i}\;\pi^\Th(f_2)\frac{1}{\dslash+i}
\left(1-\pi^\Th(\omega)\frac{1}{\dslash_\omega+i}\right)\\
&\quad+\pi^\Th(f_1)\frac{1}{\dslash-i}\;\pi^\Th(f_2)\Mop\omega)\frac{1}{\dslash-i}
\left(1-\pi^\Th(\omega)\frac{1}{\dslash_\omega-i}\right)\frac{1}{\dslash_\omega+i}\\
&\quad-\pi^\Th(f_1)\frac{1}{\dslash-i}\;\pi^\Th(\dslash(f_2)\Mop\omega)\frac{1}{\dslash-i}
\left(1-\pi^\Th(\omega)\frac{1}{\dslash_\omega-i}\right)\frac{1}{\dslash_\omega+i}\;.
\end{align*}
D'apr\`es la proposition \ref{pr:interpolation},
$\pi^\Th(g)\mathcal{R}_\dslash(i) \; \pi^\Th(h)\mathcal{R}_\dslash(i) \in \L^p(\H)$
lorsque
$g,h\in\SS(\R^n)$ et $p>\frac{n}{2}$. Les op\'erateurs $\pi^\Th(\omega)$ et
$\mathcal{R}_{\dslash_\omega}(z)$ \'etant born\'es,
on obtient alors que $\pi^\Th(f)\mathcal{R}_{\triangle^\Th}(-1)\in \L^p(\H)$
pour le m\^eme $p$.
Pour $l\geq 1$, on obtient de semblables conclusions
en r\'ep\'etant $l$ fois cet algorithme:
$(A+C)^{-1}=A^{-1}(1-CA^{-1}(\cdots(1-C(A+C)^{-1})\cdots))$.

Le cas avec $B$ non nul est obtenu par la m\^eme astuce:
$$
\pi^\Th(f)\,\mathcal{R}_{\triangle^\Th}(-1)=\pi^\Th(f)
\,\mathcal{R}_{\dslash_\omega^2}(-1)\left(
1+B\,\mathcal{R}_{\triangle^\Th}(-1)\right).
$$
Pour $z$ quelconque dans l'ensemble r\'esolvant de $\triangle^\Th$,
on obtient le r\'esultat en utilisant la premi\`ere \'equation r\'esolvante.
\end{proof}

La deuxi\`eme preuve, bas\'ee sur un calcul fonctionnel
des op\'erateurs pseudodiff\'erentiels \cite{DimassiS},
utilise la d\'efinition des classes de symboles de Shubin
\cite{Shubin} ou classes GLS \cite{GrossmannLS}.
Ce type de calcul pseudodiff\'erentiel se trouve \^etre
particuli\`erement pertinent pour l'analyse sous-jacente
aux d\'eformations de type Moyal.
\begin{defn}
\label{SHUbin}
Soit $S^{\rho,\lambda}$ la classe des symboles de Shubin
\begin{align*}
S^{\rho,\lambda}&:=\left\{\sigma\in\CC(\R^{2n}):\forall \alpha,\beta\in\N^n,
\exists \, C_{\alpha\beta} \in \R^+ \right.\\
&\hspace{3cm}\left.\left|\pa_x^\alpha\pa_\xi^\beta\sigma(\xi,x)\right|\leq C_{\alpha,\beta}
(1+|x|^2)^{(\rho-|\alpha|)/2}(1+|\xi|^2)^{(\lambda-|\beta|)/2}\right\},
\end{align*}
et soit $\Psi^{\rho,\lambda}:=\left\{A\in\Psi DO:\sigma[A]\in S^{\rho,\lambda}\right\}$
la classe des $\Psi$DOs associ\'es.
\end{defn}
En r\'ealit\'e, $S^{\rho,\lambda}$
s'inscrit dans les classes de symboles d'H\"ormander (voir \cite[Chapitre XVIII]{Hormander}) $S(m,g)$
de fonction d'ordre
$$
m(\xi,x)=(1+|x|^2)^{\rho/2}\;(1+|\xi|^2)^{\lambda/2}
$$
et de m\'etrique
$$
g_{\xi,x}=(1+|\xi|^2)^{-1}|d\xi|^2+(1+|x|^2)^{-1}|dx|^2.
$$

\begin{proof}[Deuxi\`eme preuve du th\'eor\`eme \ref{traceclass}]
Premi\`erement, l'\'equation (\ref{eq:symbole}) ainsi que la formule du produit pour
les  $\Psi$DOs permet de calculer le symbole de $\triangle_r^\th$:
$$
\sigma[\triangle_r^\th](\xi,x)=\eta^{\mu\nu}\Big(\xi_\mu\xi_\nu+2i\omega_\mu(x-\thalf\Th\xi)\xi_\nu
-\pa_\mu\omega_\nu(x-\thalf\Th\xi) -\omega_\nu\Mop\omega_\nu(x-\thalf\Th\xi)\Big)
-E(x-\thalf\Th\xi).
$$
En utilisant ensuite le fait que
$\omega_\mu, E\in\Oh_0(\R^n)$, on voit que le `Laplacien'  $\triangle_r^\th$
appartient \`a $\Psi^{0,2}$, car $\Psi^{p,q}\subset\Psi^{s,t}$,
lorsque $p\leq s$, $q\leq t$.
Soit $\{f_N\}_{N \in \N}$ une famille de fonctions lisses \`a supports compacts, d\'efinies par
$f_N(x):=\chi_{_N}(x)\;e^{-x},$
o\`u $0\leq \chi_{_N} \leq 1$, $\chi_{_N} \in\Coo_c(\R)$ avec $\chi_{_N}(x)=0$ lorsque
$x\in]-\infty,-\epsilon]\cup[N,+\infty[$ pour un
$\epsilon >0$ fix\'e et $\chi_{_N}(x)=1$ pour $x\in[0,N-\epsilon]$. D'apr\`es
\cite[th\'eor\`eme 8.7]{DimassiS},
$f_N(t \triangle_r^\th)\in\Psi^{0,-\infty}$
et d'apr\`es le lemme \ref {lm:cojoreg}  $L^\Th(f)\in\Psi^{-\infty,0}$
pour tout $f\in\SS(\R^n)$. Ainsi, d'apr\`es \cite[lemme 18.4.3]{Hormander} on obtient
$L^\Th(f)f_N(t \triangle_r^\Th)\in\Psi^{-\infty,-\infty}$
et donc son symbole appartient \`a $\SS(\R^{2n})$.
Donc,
\begin{align*}
C&:=\sum_{|\alpha|+|\beta|\leq 2n+1}\|\pa_x^\alpha\pa_\xi^\beta\sigma
\left[L^\Th(f)f_N(t\triangle_r^\Th)\right]\|_1\\
&\leq\sum_{|\alpha|+|\beta|\leq 2n+1}C_{\alpha,\beta}\int d^nx \;d^n\xi\,
(1+|x|^2)^{(-l-|\alpha|)/2}(1+|\xi|^2)^{(-k-|\beta|)/2},
\end{align*}
pour des constantes $C_{\alpha,\beta}<\infty$ et pour tout $l,k\in\N$, donc $C<\infty$.
Finalement, le th\'eor\`eme 9.4 de \cite{DimassiS} montre que
$L^\Th(f)f_N(t \triangle_r^\Th)$ est \`a trace pour tout $N\in\N$.
En regardant les estimations de la preuve du th\'eor\`eme 8.7 de \cite{DimassiS}, on peut trouver
des constantes $C_{\alpha,\beta}$ ind\'ependantes de $N$
(\'etant donn\'e que $e^{-x}$ est \`a d\'ecroissance rapide
lorsque $x\rightarrow+\infty$, le support \`a droite
de $f_N$ ne joue aucun r\^ole), on obtient alors que $L^\Th(f)f_N(t\triangle_r^\Th)$
est uniform\'ement (par rapport \`a $N$) \`a trace.

Pour terminer la preuve, il reste \`a montrer que $\slim L^\Th(f)f_N(t\triangle_r^\Th)
=L^\Th(f) e^{-t\triangle_r^\Th}$, car la proposition 2 de \cite{DeSi} garantira que $L^\Th(f)
e^{-t\triangle_r^\Th}$ est \`a trace pour tout $t>0$.

Soit $\phi\in\H$ et soit $E_\lambda$ la famille spectrale de $\triangle_r^\Th$, alors
\begin{align*}
\|(\chi_{_N}(\triangle_r^\Th)-1)\phi\|_2^2&=\langle\phi|(\chi_{_N}(\triangle_r^\Th)-1)^2\phi\rangle
=\int_{\spec(\triangle_r^\Th)}d\langle\phi|E_\lambda\phi\rangle\,(\chi_{_N}(\lambda)-1)^2 \\
&\leq\int_{\spec(\triangle_r^\th)}d\langle\phi|E_\lambda\phi\rangle
=\langle\phi|\phi\rangle.
\end{align*}
On obtient alors par convergence domin\'ee, avec $\phi=e^{-t\,\triangle_r^\Th}\psi$
\begin{align*}
\lim_{N\rightarrow\infty}\|(\chi_{_N}(\triangle_r^\Th)-1)\;e^{-t\;\triangle_r^\Th}\psi\|_2^2
&=\lim_{N\rightarrow\infty}\int_{\spec(\triangle_r^\Th)}
d\langle e^{-t\triangle_r^\Th}\psi, \;E_\lambda e^{-t\,\triangle_r^\Th}\psi\rangle
(\chi_{_N}(\lambda)-1)^2 \\
&=\int_{\spec(\triangle_r^\th)}
d\langle e^{-t\triangle_r^\Th}\psi,\;E_\lambda e^{-t\,\triangle_r^\Th}\psi\rangle
\Big(\lim_{N\rightarrow\infty}(\chi_{_N}(\lambda)-1)^2 \Big)\\
&=0.
\end{align*}
Dans la derni\`ere \'egalit\'e, nous avons utilis\'e la positivit\'e
de la mesure spectrale, cons\'equent \`a $\spec(\triangle_r^\Th)\subset\R^+$.
\end{proof}

Nous allons maintenant obtenir un d\'eveloppement asymptotique
de la trace du semi-groupe r\'egularis\'e, en suivant une approche de
Vassilevich \cite{Vass}.
\begin{proof}[Preuve du th\'eor\`eme~\ref{HKNCP}]
Soit
\begin{align*}
X&:=2L^\Th(\omega_\mu)\pa^\mu+L^\Th(\pa_\mu\omega^\mu)+
L^\Th(\omega_\mu\Mop\omega^\mu)+L^\Th(E)\cr
Y&:=-\pa_\mu\pa^\mu,
\end{align*}
de telle sorte que $\triangle_r^\Th=Y-X$. La formule de Baker-Campbell-Hausdorff (BCH)
\begin{equation*}
e^T\;e^S=e^{T\,+\,S\,+\frac{1}{2} \,[T,\,S]\,+\,\frac{1}{12}\,[T,\,[T,\,S]]\,
+\,\frac{1}{12}\,[S,\,[S,\,T]]\,-\,\frac{1}{48}\,[T,\,[S,\,[T,\,S]]]\,+\,\cdots},
\end{equation*}
permet de r\'ecrire le semi-groupe comme
\begin{equation*}
e^{-t\,\triangle_r^\Th}=e^{t\,X \, + \, \frac{1}{2} t^2\,[X,\,Y] \, + \, \frac{1}{12}t^3\,[X,\,[X,\,Y]] \,
- \, \frac{1}{6}t^3\,[Y,\,[X,\,Y]] \, - \, \frac{1}{48}t^4[X,\,[Y,\,[X,\,Y]]] \,
+ \, \frac{1}{48}t^4\,[Y,\,[Y,\,[X,\,Y]]] \, + \, \cdots}\; \;e^{-t \, Y}.
\end{equation*}

Pour obtenir un d\'eveloppement en puissance de $t$ lorsque $t\to0$,
la strat\'egie consiste \`a d\'evelopper la premi\`ere exponentielle,
calculer les commutateurs, r\'eorganiser la suite des
termes ainsi obtenus, pour finalement calculer les symboles de
ces op\'erateurs pseudodiff\'erentiels.
La trace sera prise en int\'egrant les symboles sur le fibr\'e cotangent.\\
En r\'ealit\'e, la r\'eorganisation en puissances homog\`enes est sensiblement
plus \'elabor\'ee qu'un simple d\'eveloppement d'exponentielle.
Tous les op\'erateurs provenant de ce d\'eveloppement sont du type
$L^\Th(g) \; \pa^\alpha e^{-t\tri},\;\alpha\in\N^n$, avec $g\in\SS(\R^n)$.
Certains d'entre eux donneront des contributions nulles car
$\int d^n\xi\,\xi_{1}^{\alpha_1}\cdots \xi_{n}^{\alpha_n}\;e^{-t|\xi|^2} =\Pi_i^n \;
\frac{1}{2}(1+(-1)^{{\alpha_i}})\;\Gamma(\frac{n+1}{2})\;t^{-(\alpha_i+1)/2}$ est nul
lorsque au moins un des
$\alpha_i$ est impair. Lorsqu'ils sont tous pairs, $\vert\alpha\vert=\sum_i^n\alpha_i=2l$
est aussi pair et on a
$$
\int_{\R^n} d^n\xi\,\xi_{\mu_1}\cdots \xi_{\mu_{2l}} \;e^{-t|\xi|^2}= {\textstyle{
\left(\frac{\pi}{t}\right)^{n/2}}} \; (2t)^{-l} \; \sum_{\sigma\in S_{2l}}
{\textstyle{\frac{1}{2^l l!} }}\; \delta_{\sigma(\mu_1)\sigma(\mu_2)}\cdots
\delta_{\sigma(\mu_{2l-1})\sigma(\mu_{2l})},
$$
o\`u $\sigma$ varie dans le groupe des permutations de $2l$ \'el\'ements
$S_{2l}$. Ainsi, dans la r\'eorganisation
en puissances homog\`enes, il faut prendre en compte que
$t^l \; L^\Th(g) \; \pa^\alpha$ est un terme d'ordre effectif
$t^{l-\frac{|\alpha|}{2}}$ (ind\'ependamment du facteur global
${\textstyle{\left(\frac{\pi}{t}\right)^{n/2}}}$).
De plus, pour avoir un d\'eveloppement jusqu'\`a l'ordre $k$,
il faut utiliser la formule BCH \`a l'ordre $2k-1$.
L'ordre de la formule BCH est d\'efini comme le nombre
de commutateurs entrant en jeu dans la premi\`ere exponentielle.
Le terme venant de la formule
BCH \`a l'ordre $k$, ayant un nombre maximal de d\'eriv\'ees  est
$$
[t\,\pa^2,\;[t\,\pa^2,\;\cdots,\;[t\,\pa^2, \; t \,L^\Th(g)\pa ]\cdots]\;]\propto t^{k+1} \; L^\Th(g)
\;\pa^{k+1},
$$
et il correspond alors \`a un terme d'ordre effectif $t^{\frac{k+1}{2}}$.

Montrons en d\'etail comment cet algorithme fonctionne \`a l'ordre un.
Nous avons besoin de la formule BCH aussi \`a l'ordre un:
$e^{-t \, \triangle_r^\Th}=e^{t\,X\,-\,t\,Y}=e^{t\,X\,+\,\frac{1}{2}
[tX,\,tY]\,+\,\cdots}\;e^{-t\,Y}$. Puisque
\begin{align*}
[tX,tY]&=t^2\;\left[\pa_\nu\pa^\nu,2L^\Th(\omega_\mu)\pa^\mu+
L^\Th(\pa_\mu\omega^\mu) +L^\Th(\omega_\mu\Mop\omega^\mu)
+L^\Th(E)\right]\\
&=t^2\Big(2L^\Th(\pa_\nu\pa^\nu\omega_\mu)\pa^\mu
+4L^\Th(\pa_\nu\omega_\mu)\pa^\mu\pa^\nu
+L^\Th(\pa_\nu\pa^\nu\pa_\mu\omega^\mu)
+2L^\Th(\pa_\nu\pa_\mu\omega^\mu)\pa^\nu\\
&\quad +L^\Th(\pa_\nu\pa^\nu(\omega_\mu\Mop\omega^\mu))
+2L^\Th(\pa_\nu(\omega_\mu\Mop\omega^\mu))\pa^\nu
+L^\Th(\pa_\nu\pa^\nu E)+2L^\Th(\pa_\nu E)\pa^\nu\Big)\\
&=4t^2\;L^\Th(\pa_\nu\omega_\mu)\pa^\mu\pa^\nu+O(t^2),
\end{align*}
on a
\begin{align*}
L^\Th(f)\,e^{-t \,\triangle_r^\Th}&=L^\Th(f)e^{t\left(2L^\Th(\omega_\mu)\pa^\mu
+L^\Th(\pa_\mu\omega^\mu)+
L^\Th(\omega_\mu\Mop\omega^\mu)+L^\Th(E)\right)
+2t^2L^\Th(\pa_\nu\omega_\mu)\pa^\mu\pa^\nu+\cdots}\;
e^{t\pa_\mu\pa^\mu}\\
&=L^\Th(f)\Big(1+t\big(2L^\Th(\omega_\mu)\pa^\mu+L^\Th(\pa_\mu\omega^\mu)+
L^\Th(\omega_\mu\Mop\omega^\mu)+L^\Th(E) \big)\\
& \quad +2t^2\big( L^\Th(\pa_\nu\omega_\mu)\pa^\mu\pa^\nu
+ L^\Th(\omega_\mu\Mop\omega_\nu)\pa^\mu\pa^\nu\big)
+O(t^2)\Big) \;  e^{t\,\pa_\mu\pa^\mu}.
\end{align*}
o\`u le dernier terme en facteur de $t^2$ vient du d\'eveloppement
de $e^{t\,X}$ \`a l'ordre deux. Ainsi,
\begin{align*}
\sigma\left[L^\Th(f)\,e^{-t \,\triangle_r^\Th}\right](\xi,x)
&=\Big(f(x-\thalf\Th\xi)
+t\big(2f\Mop\omega_\mu(x-\thalf\Th\xi)(-i\xi)^\mu
+\;f\Mop\pa_\mu\omega^\mu(x-\thalf\Th\xi)\\
&\quad+f\Mop\omega_\mu\Mop\omega^\mu(x-\thalf\Th\xi)
+f\Mop E(x-\thalf\Th\xi) \big)\\
&\quad +\;2t^2\big( f\Mop\pa_\nu\omega_\mu(x-\thalf\Th\xi)(-i\xi)^\mu(-i\xi)^\nu\\
&\quad+ \;f\Mop\omega_\mu\Mop\omega_\nu(x-\thalf\Th\xi)(-i\xi)^\mu(-i\xi)^\nu\big)
+O(t^2)\Big) \;e^{-t\,\xi_\mu\xi^\mu}.
\end{align*}
Il ne reste finalement qu'\`a int\'egrer
$\sigma\left[L^\Th(f)\,e^{-t \,\triangle_r^\Th}\right](\xi,x)$.
On obtient, en effectuant la translation $x\rightarrow x+\thalf\Th\xi$,
\begin{align*}
&\Tr \left(L^\Th(f)\,e^{-t\,\triangle_r^\Th}\right)\cr
&=(2\pi)^{-n}\;{\displaystyle \iint}d^nx \; d^n\xi\,
\Big(f(x)+t\big(2f\Mop\omega_\mu(x)(-i\xi)^\mu +f\Mop\pa_\mu\omega^\mu(x)
+f\Mop\omega_\mu\Mop\omega^\mu(x)\cr
&\hspace{1cm}+f\Mop E(x) \big)
+\;2t^2\big( f\Mop\pa_\nu\omega_\mu(x)(-i\xi)^\mu(-i\xi)^\nu
+ f\Mop\omega_\mu\Mop\omega_\nu(x)(-i\xi)^\mu(-i\xi)^\nu\big)
\Big) \; e^{-t\,\xi_\mu\xi^\mu}  \cr
&\hspace{1cm}+\;O(t^{-\frac{n}{2}+2}) \cr
&=(4\pi t)^{-\frac{n}{2}}\;{\displaystyle \int}d^nx\,f(x)\;
\Big(1+t\big(\pa_\mu\omega^\mu(x)+
\omega_\mu\Mop\omega^\mu(x)+E(x) -\pa_\mu\omega^\mu(x)
-\omega_\mu\Mop\omega^\mu(x)\big)\Big) \cr
&\hspace{1cm}+\;O(t^{{-\frac{n}{2}}+2}) \cr
&=(4\pi t)^{-\frac{n}{2}}\;{\displaystyle \int}d^nx\,f(x)\;
\big(1+tE(x)\big)+\;O(t^{{-\frac{n}{2}}+2}).
\end{align*}
Les autres coefficients sont obtenus par un calcul similaire,
c'est-\`a-dire que l'on obtient g\'en\'erique\-ment
$$
L^\Th(f)\,e^{-t\,\triangle_r^\Th}\sim_{t\rightarrow 0}
L^\Th(f)\Big(\sum_{l\in\N}\;t^l\;\sum_{\alpha\in\N^n,|\alpha|\leq l}
L^\Th(g_{\alpha,l})\;t^{|\alpha|/2}\;\pa^\alpha\Big) \; e^{t\,\pa_\mu\pa^\mu},
$$
o\`u $g_{\alpha,l}\in\SS$ et la s\'erie en puissance
de $t$ a \'et\'e corrig\'ee par l'ordre des d\'eriv\'ees, en accord avec
la discussion pr\'ec\'edente.\\
Ici $\sim$  signifie que nous avons un d\'eveloppement asymptotique par rapport
\`a la topologie de la norme trace:
$$
\|L^\Th(f)e^{-t\triangle_r^\Th}-L^\Th(f)\Big(\sum_{l\leq
N}\;t^l\;\sum_{\alpha\in\N^n,|\alpha|\leq l}
L^\Th(g_{\alpha,l})\;t^{|\alpha|/2}\;\pa^\alpha\Big) \; e^{t\,\pa_\mu\pa^\mu}\|_1 =O(t^{N+1}),
$$
la convergence, pour $t\to0$, est alors garantie par le th\'eor\`eme \ref{traceclass};
ce qui conclut la preuve du th\'eor\`eme \ref{HKNCP}.
\end{proof}
\begin{rem}
Ces calculs syst\'ematiques montrent aussi que les autres coefficients
$\tilde{a}_{2l},\;l>3$ ont la m\^eme forme canonique:
le produit de Moyal remplace le produit point \`a point dans toutes
les expressions.
\end{rem}
Nous allons utiliser ce d\'eveloppement pour calculer
l'action spectrale, mais il servira aussi
dans le chapitre \ref{NCQFT}, \`a
calculer des corrections quantiques pour des th\'eories de champs
sur d\'eformations isospectrales.

\subsubsection{3.2.2.2 Calcul de l'action spectrale}

En utilisant le r\'esultat pr\'ec\'edent, nous allons calculer une
forme pr\'eliminaire de l'action spectrale, pr\'eliminaire dans le sens o\`u seule
la repr\'esentation r\'eguli\`ere sera utilis\'ee.
Les raisons de l'obstruction li\'ee \`a la repr\'esentation adjointe seront explicit\'ees \`a la
fin de ce paragraphe.

La relation (\ref{zeta2}) est en fait relativement g\'en\'erale.
Pour tout op\'erateur born\'e $S$ et tout op\'erateur inversible
$T$ qui sont tels que $S\;T^{-s}$ soit \`a trace, on a
\begin{equation}
\label{zeta3}
\zeta_{S,T}(s):=\Tr(S\;T^{-s})={\textstyle \frac{1}{\Gamma(s)}}\int_0^\infty dt\,
t^{s-1}\Tr\left(S \;e^{-t\,T}\right).
\end{equation}
Avec cette relation et le r\'esultat du paragraphe pr\'ec\'edent,
on peut obtenir une expression asymptotique (pour $\Lambda\to\infty$)
de l'action spectrale des plans de Moyal.\\
Pour obtenir l'expression des coefficients $\phi_{2k}$
dans n'importe quelle dimension, nous allons utiliser
des techniques de transform\'ee de Laplace,
inspir\'ees par l'approche de Nest--Vogt--Werner
dans l'article \cite{Nestetal}.
Aussi, \cite{Widder} est la r\'ef\'erence principale sur la transform\'ee
de Laplace qui sera suivie ici.
Nous supposerons que $\phi$ poss\`ede la propri\'et\'e suivante
\begin{equation}
\label{condition}
\Coo(\R^+)\ni\phi=\int_0^\infty \,ds\,e^{-sz}\hat{\psi}(s) ,\,\,\,\,
\hat{\psi} \in \SS(\R^+):=\set{g\in\SS: g(x)=0, x \leq 0}
\end{equation}
Toute fonction poss\'edant cette propri\'et\'e admet une extension analytique
sur le demi plan complexe $\Re(z)\geq 0$.
Par cons\'equent, toute fonction $m$ fois diff\'erentiable $\psi$ qui
soit telle que $\psi^{(m)}=\phi$ est la transform\'ee de
Laplace d'une fonction $\hat{\psi}$ qui, par diff\'erentiation, satisfait \`a
\begin{equation*}
 \phi(z)=\psi^{(m)}(z)=(-1)^m\int_0^\infty ds\,e^{-sz}\; s^m\;\hat{\psi}(s),\;\Re z>0.
\end{equation*}
Avec $\triangle^\Th$ d\'efini dans l'\'equation (\ref{genelap}),
en utilisant $\phi(\triangle_r^\Th)=(-1)^m\int_0^\infty ds\,
e^{-s\,\triangle_r^\Th}\,s^m\,\hat{\psi}(s) $ ainsi que la positivit\'e de $\rho=g^*\Mop g$,
$g\in\SS(\R^n)$, on obtient
$$
\Tr\left(L^\Th(\rho)\;\phi\left(\triangle_r^\Th/\Lambda^2\right)\right)=(-1)^m
\Tr\left(L^\Th(g)\int_0^\infty dt\,e^{-t\,\triangle_r^\Th/\Lambda^2}t^m
\hat{\psi}(t) L^\Th(g^*)\right).
$$
Soit $\{\Phi_p\}_{p\in\N}$ une base orthonorm\'ee de $\H_r$ et soit
$0\leq B_t :=L^\Th(g)\,e^{-t\,\triangle_r^\Th/\Lambda^2}\,L^\Th(g^*)$, alors
\begin{align*}
\left|\Tr\left(L^\Th(\rho)\;\phi\left(\triangle_r^\Th/\Lambda^2\right)\right)\right|&=
\big|\lim_{N\rightarrow \infty} \int_0^\infty dt\;
\sum_{p\leq N}\langle\Phi_p, \;B_t \;\Phi_p\rangle \;t^m\hat{\psi}(t)\big|\\
&\leq\int_0^\infty dt\,\|B_t\|_{_1}\;t^m|\hat{\psi}(t)|
=\int_0^\infty dt \,\|B_t\|_{_1}\;
t^m|\hat{\psi}(t)|.
\end{align*}
Nous allons estimer $\|B_t\|_{_1}$. Pour  $t>\epsilon$, avec $\epsilon$ fix\'e
et arbitrairement petit, on a
$$
\|B_t\|_{_1}=\|L^\Th(g)e^{-t\,\triangle_r^\Th/2\Lambda^2}\|_{_2}^2\leq
\|L^\Th(g)e^{-\epsilon \,\triangle_r^\Th/2\Lambda^2}\|_{_2}^2 \;
\|e^{-(t-\epsilon)\,\triangle_r^\Th/2\Lambda^2}\|.
$$
Vu que $(t-\epsilon)\triangle_r^\Th$ est positif, on obtient
$\|e^{-(t-\epsilon)\,\triangle_r^\Th/2\Lambda^2}\|\leq1$.
Ainsi, pour tout $t>\epsilon$, $\|B_t\|_{_1}\leq C$ uniform\'ement en $t$.
Pour $t\leq\epsilon$, le calcul asymptotique du paragraphe pr\'ec\'edent
montre que $\|B_t\|_{_1}=O(t^{-n/2})$. Ainsi
$\Tr\left( \int_0^\infty dt B_t \;t^m\hat{\psi}(t) \right) <\infty$ et  par convergence
domin\'ee on obtient
\begin{align*}
&\hspace{-1cm}\Tr\left(L^\Th(\rho)\;\phi(\triangle_r^\Th/\Lambda^2)\right) \\
&=(-1)^m\int_0^\infty dt
\Tr\left(L^\Th(\rho) e^{-t\,\triangle_r^\Th/\Lambda^2}\right)t^m\hat{\psi}(t)\\
&=(-1)^m(4\pi)^{-n/2}\int_0^\infty  dt\sum_{l=0}^{m}\Lambda^{n-2l}\;t^{m+l-n/2}
\hat{\psi}(t) \int_{\R^n}d^nx \,\rho\Mop\;\tilde{a}_{2l}(x) +\;O(\Lambda^{n-2(m+1)})\\
&=(4\pi)^{-n/2}\;\sum_{l=0}^{m}\Lambda^{n-2l}\,\phi_{2l}
\int_{\R^n}d^nx\,\rho\Mop\;\tilde{a}_{2l}(x)
\;+\;O(\Lambda^{n-2(m+1)}),\\
\end{align*}
o\`u $\phi_{2l}$ est d\'efini par
\begin{equation}
\label{moments2}
\phi_{2l}:=(-1)^m\int_0^\infty dt\, t^{m+l-n/2}\,\hat{\psi}(t) .
\end{equation}
Lorsque $n=2m$, $\phi_{2l}$ peut \^etre mis sous la forme
plus famili\`ere (\ref{moments}):
\begin{equation}
\label{moments3}
\phi_{2l}=
\begin{cases}
{\textstyle{\frac{1}{\Gamma(m-l)}}}\;\int_0^\infty  dt\,\phi(t) \; t^{m-1-l},\;& {\rm pour}\;
l=0,\cdots,m-1,\cr
(-1)^{l}\;\phi^{(l-m)}(0),& {\rm pour}\; l=m,\cdots,n.
\end{cases}
\end{equation}
Lorsque $n$ est impair, la forme des coefficients
$\phi_{2l}$ n'est pas aussi explicite car faisant
intervenir des d\'eriv\'ees fractionnaires de $\phi$.
Il est alors pr\'ef\'erable de s'en tenir \`a la d\'efinition (\ref{moments2}).

En r\'esum\'e:

\begin{thm}
Soient $\rho\in\SS(\R^n)$, $A=-iL^\Th(A_\mu)\otimes \gamma^\mu$,
$A_\mu^*=-A_\mu\in\Oh_0(\R^n)$,
$\phi\in\mathcal{C}^\infty(\R^+)$
une fonction positive satisfaisant \`a la condition (\ref{condition}) et
$\dslash_A=\dslash +A$. Alors $L^\Th(\rho)\,\phi(\dslash_A^2/\Lambda^2)$
est \`a trace. De plus, pour $\Lambda\to\infty$:
$$
S_\Lambda(\dslash,A,\rho)=2^m\;
 (4\pi)^{-n/2}\;\sum_{l=0}^{m}\Lambda^{n-2l}
\; \phi_{2l}\;\int_{\R^n} d^nx\,\rho(x)\;\tilde{a}_{2l}(x)
\;+\;O(\Lambda^{n-2(m+1)}),
$$
o\`u les `moments' $\phi_{2l}$ sont d\'efinis dans (\ref{moments2}) ou dans
(\ref{moments3}) suivant la parit\'e de la dimension et les coefficients
$\tilde{a}_{2l}(x)$  sont donn\'es dans le th\'eor\`eme \ref{HKNCP},
avec le remplacement suivant dans l'\'equation (\ref{genelap}):
$$
\begin{cases}
L^\Th(\omega_\mu) & \rightarrow \;L^\Th(A_\mu),\\
L^\Th(E) \otimes 1_{2^m}& \rightarrow \; \left(L^\Th(\pa_\mu A_\nu)
+L^\Th(A_\mu\Mop A_\nu)\right)
\otimes\frac{1}{2}(\gamma^\mu\gamma^\nu-\gamma^\nu\gamma^\mu ).
\end{cases}
$$
De plus, tous les termes lin\'eaires en $E$ dans
$\tilde{a}_{2l}$ sont nuls.
\end{thm}

\begin{rem}
Lorsque l'op\'erateur de Dirac est sym\'etris\'e, $\D_A=\D+A+\epsilon JAJ^{-1}$,
il faut remplacer $L^\Th(A_\mu)$ par $L^\Th(A_\mu) - R^\Th(A_\mu)$,
vu que $\epsilon J\; (L^\Th(A_\mu)\otimes \ga^\mu)\;J^{-1}
=R^\Th(A_\mu^*) \otimes\ga^\mu$. Le comportement
en $t$ des termes du type $\Tr \big(L^\Th(f)R^\Th(g)(-i\pa)^\alpha e^{t\pa_\mu
\pa^\mu }\big)$ doit alors \^etre \'etabli. Plus pr\'ecis\'ement,
$$
\sigma[L^\Th(f)R^\Th(g)](\xi,x)=(2\pi)^{-n}\int \,d^ny\,d^n\eta\,
e^{i(\eta-\xi).(x-y)}\,f(x-\tfrac{1}{2}\Th\eta)\,
g(y+\tfrac{1}{2}\Th\xi),
$$
et on a
\begin{align*}
\Tr \big(L^\Th(f)R^\Th(g)(-i\pa)^\alpha e^{t\pa_\mu
\pa^\mu }\big)&=\int \, d^nx \, d^n\xi\,
\sigma[L^\Th(f)R^\Th(g)(-i\pa)^\alpha e^{t\pa_\mu
\pa^\mu }](\xi,x) \\
&=\int \,d^nx\,d^n\xi\,
f(x-\tfrac{1}{2}\Th \xi)\,g(x+\tfrac{1}{2}\Th \xi)\,\xi^\alpha \,e^{-t|\xi|^2},
\end{align*}
l'invariance par translation $x \rightarrow
x+\tfrac{1}{2}\Th\xi$ crucialement utilis\'ee dans la preuve du th\'eor\`eme
\ref{HKNCP} n'est maintenant plus valable.
Nous verrons dans le chapitre suivant (paragraphe \ref{PNP}),
que ce point est intimement li\'e au ph\'enom\`ene de m\'elange UV/IR.
\end{rem}

\subsubsection{3.2.2.3 Application au cas quadri-dimensionnel}

Dans le cas quadri-dimensionnel, en ne gardant que les termes en puissance positive
de $\Lambda$, on obtient
$$
S_\Lambda(\dslash,A,\rho)={\textstyle{\frac{1}{4\pi^2}}}\left(\Lambda^4\phi_0\;
\int_{\R^4} d^4x\,\rho(x) +{\textstyle{\frac{\phi(0)}{6}}}
\;\int_{\R^4} d^4x\,\rho(x) \;  F^{\mu\nu}
\Mop F_{\mu\nu}(x) \right) +O(\Lambda^{-2}),
$$
o\`u $F^{\mu\nu}:=\pa^\mu A^\nu-\pa^\nu A^\mu
+[A^\mu,A^\nu]_{\Mop}$.

En prenant, pour la r\'egularisation, la fonction caract\'eristique
$\chi_{_V}$ d'un sous-ensemble born\'e $V\subset \R^4$, la propri\'et\'e de trace du produit
de Moyal donne
\begin{equation}
\label{actionspect}
S_\Lambda(\dslash,A,\chi_{_V})={\textstyle{\frac{1}{4\pi^2}}}\left(\Lambda^4\,\phi_0\;
\int_{V}d^4x+{\textstyle{\frac{\phi(0)}{6}}}
\;\int_{V} d^4x\,F^{\mu\nu}
\Mop F_{\mu\nu}(x)  \right ) + O(\Lambda^{-2}).
\end{equation}
Modulo un terme cosmologique, c'est l'action de Yang--Mills
non commutative spatialement localis\'ee. Cette expression est sensiblement
diff\'erente du r\'esultat de l'action de Connes--Lott pour $\Th$ symplectique,
\`a cause de la brisure de la propri\'et\'e de trace due \`a la pr\'esence de
la r\'egularisation $\rho$. En effet, dans le cas de l'action de Connes--Lott,
le produit de Moyal entre les deux tenseurs de courbure peut \^etre omis:
$$
YM(\alpha)={\textstyle -\frac{1}{4g^2}}\int d^4x\,F^{\mu\nu} F_{\mu\nu}(x).
$$

\chapter{M\'elange UV/IR et noyau de la chaleur}
\label{NCQFT}

Dans ce chapitre, nous allons \'etudier certains aspects de la
th\'eorie quantique des champs sur d\'eformation isospectrale.
Nous nous restreindrons \`a l'\'etude d'une th\'eorie
scalaire en dimension quatre, mais les techniques que nous allons d\'evelopper
peuvent \^etre ais\'ement g\'en\'eralis\'ees en dimensions sup\'erieures et
pour les th\'eories de jauge. Le point essentiel sur lequel nous
allons nous focaliser
est le ph\'enom\`ene de m\'elange des divergences infrarouges et
ultraviolettes (m\'elange UV/IR). Nous donnerons une
interpr\'etation nouvelle de ce ph\'enom\`ene, en terme de noyau de la chaleur
sur et hors de sa diagonale. La vertu principale de l'approche que nous
allons consid\'erer (repr\'esentation en position), en plus de reproduire les
r\'esultats connus des paradigmes plats (plans de Moyal et tores
non commutatifs), est d'\^etre directement applicable aux espaces courbes
d\'eform\'es.
Nous ferons alors une \'etude syst\'ematique de ce ph\'enom\`ene pour
tous les types de d\'eformation isospectrale.

Une des motivations premi\`eres pour l'\'etude des th\'eories
des champs sur espace non commutatif, proche d'\^etre obsol\`ete
aujourd'hui, \'etait la construction de th\'eories sans divergence
ultraviolette. En effet, la notion de points (et donc celle
de petites distances) n'existant plus sur un espace non commutatif,
il semble raisonnable de s'attendre, lorsque l'on substitue un espace non commutatif
\`a un espace ordinaire,  \`a ce que les th\'eories
quantiques ne soient pas affect\'ees par des divergences ultraviolettes.\\
Ce ne fut pas le cas pour les premiers exemples \'etudi\'es,
les plans de Moyal ainsi que les tores non commutatifs \cite{CR, Filk, KW, MR,
MRS, Atlas}, o\`u en plus des divergences UV usuelles, des singularit\'es
d'un type nouveau apparurent, m\'elangeant courtes et grandes distances.

Pour ces d\'eformations plates,
la principale nouveaut\'e en rapport avec la renormalisation fut la coexistence de deux types
de diagrammes de Feynman, respectivement appel\'es planaires et non planaires (cf. paragraphe
4.1.1.2). Cette terminologie puise son origine dans
le fait que les diagrammes en question sont des rubans,
pouvant \^etre repr\'esent\'es sur des surfaces de Riemann.
Ces deux secteurs de la th\'eorie ont un comportement drastiquement diff\'erent.
Alors que les diagrammes planaires reproduisent \`a l'identique
les divergences des th\'eories commutatives, les diagrammes non planaires
sont caract\'eris\'es par des vertex qui d\'ependent des moments externes
au travers d'une phase. Ils sont g\'en\'eriquement finis
mais divergents pour des valeurs
exceptionnelles des moments. Par exemple, pour
une th\'eorie scalaire en dimension quatre avec
un terme d'interaction en $\vf^{\Mop 4}$, les diagrammes non planaires pr\'esents
dans la fonction \`a deux points divergent lorsque l'impulsion externe
tend vers z\'ero. C'est le fameux m\'elange infrarouge ultraviolet, qui rend
probl\'ematique la renormalisation de la th\'eorie, ind\'ependamment du
sch\'ema de r\'egularisation choisi.

En utilisant la repr\'esentation de Schwinger de l'action effective
dans l'approximation \`a une boucle,
nous allons montrer qu'en toute g\'en\'eralit\'e, deux secteurs aux comportements
fondamentalement diff\'erents, que nous appellerons planaire
et non planaire par analogie avec les cas plats, coexistent pour les th\'eories scalaires
sur toutes les d\'eformations isospectrales. On donnera une
interpr\'etation purement alg\'ebrique de la pr\'esence de ces deux secteurs
en termes d'op\'erateurs de multiplications twist\'ees \`a droite et \`a gauche.
La diff\'erence de leurs comportements, en particulier le ph\'enom\`ene de
m\'elange UV/IR, sera comprise en termes de noyau de la chaleur sur
et hors de sa diagonale.\\
Une vertu suppl\'ementaire de cette approche `sur espace de configuration',
est de mettre en lumi\`ere de nouvelles et/ou plus fines
manifestations du ph\'enom\`eme UV/IR, li\'ees \`a la g\'eom\'etrie ainsi qu'aux
propri\'et\'es arithm\'etiques des param\`etres de d\'eformation.\\
{\it Dans le cas non p\'eriodique, seules
les d\'eformations de rang sup\'erieur ou \'egal \`a quatre donneront lieu
\`a un secteur non planaire sans divergence.\\
Par cons\'equent, lorsque que le rang
de la matrice de d\'eformation est \'egal \`a deux, la th\'eorie ne sera
pas renormalisable, d\'ej\`a dans son approximation \`a une boucle.}\\
Ce r\'esultat semble \^etre contradictoire
avec ceux obtenus \`a partir des r\`egles de Feynman modifi\'ees dans
l'espace des moments, mais
l'apparente contradiction est simplement due au fait que la
singularit\'e IR (du m\'elange UV/IR) n'est pas
localement int\'egrable dans ces circonstances;
la fonction de Green associ\'ee ne d\'efinit alors pas une distribution.\\
Nous verrons dans le cas p\'eriodique, pour que la th\'eorie soit renormalisable,
qu'il est n\'ecessaire que {\it les entr\'ees de la matrice de d\'eformation
soient de nature arithm\'etique bien particuli\`ere}. Nous verrons
aussi que {\it la possible existence de points fixes pour l'action du groupe
peut \^etre responsable d'un nouveau type de divergences}.
Cette nouvelle manifestation du m\'elange UV/IR appara\^it d\'ej\`a pour
les arch\'etypes des d\'eformations isospectrales p\'eriodiques courbes que sont
les sph\`eres de Connes--Landi et leurs espaces ambiants.

\section{Th\'eorie $\vf^{\Mop 4}$ sur d\'eformations isospectrales}
\subsection{Action effective \`a une boucle}

Par souci de simplicit\'e mais aussi pour son int\'er\^et physique,
nous nous restreignons au cas quadri-dimensionnel;
$n=\dim(M)=4$. Il est n\'eanmoins clair que nos techniques s'appliquent
aux dimensions sup\'erieures sans modification essentielle.
Nous consid\'erons ici une fonctionnelle d'action classique pour un champ
scalaire r\'eel $\vf$
\begin{equation}
\label{eq:actioncl}
S[\vf] := \int_M\,\mu_g\,\Bigl[\thalf(\nabla^\mu\vf)\Mop
(\nabla_\mu\vf) +\thalf m^2\vf\Mop\vf+\frac{\lambda}{4!}\vf^{\Mop
4}\Bigr].
\end{equation}
Nous pouvons ajouter un terme de couplage avec la gravitation du type
$\xi R(\vf\Mop\vf)$, o\`u $R$ est la courbure scalaire et $\xi$ une
constante de couplage, sans aucun changement dans nos conclusions.
De plus, vu que le scalaire de courbure est invariant sous
l'action $\a$, ce terme n'est pas affect\'e par la d\'eformation
car
$R\Mop f=R. f$ pour tout $ f\in\Coo_c(M)$ et donc
en utilisant la propri\'et\'e de trace
$$
\int_M\mu_g\,R.(\vf\Mop\vf)=
\int_M\mu_g\,R\Mop\vf\Mop\vf=
\int_M\mu_g\,(R\Mop\vf).\vf=
\int_M\mu_g\,R\,.\vf\,.\vf.
$$
Pour les m\^emes raisons, le terme cin\'etique de l'action n'est pas non plus
affect\'e par la d\'eforma\-tion et $S[\vf]$ peut \^etre r\'e\'ecrite comme
\begin{equation}
S[\vf] = \int_M\,\mu_g\,\Bigl[\thalf\vf\tri\vf\
+ \thalf m^2\vf\,\vf+\frac{\lambda}{4!}(\vf\Mop\vf)\,(\vf\Mop\vf)\Bigr].
\end{equation}
(Rappelons que dans nos conventions le Laplacien est positif
$\tri=-\nabla^\mu\nabla_\mu$.)
\begin{rem}
Nous ne travaillons ici que dans le contexte de la th\'eorie
quantique des champs Euclidienne. L'extension au r\'egime
pseudo-Riemannien, par prolongement analytique
des quantit\'es Euclidiennes, n\'ecessitant de plus amples hypoth\`eses
sur la quadri-vari\'et\'e $M$, ne sera pas consid\'er\'ee ici.
Nous verrons cependant que le r\'egime Euclidien est
suffisant pour l'\'etude du ph\'enom\`ene de m\'elange UV/IR.
\end{rem}

\subsubsection{4.1.1.1 Calcul du potentiel effectif}

Notre objectif est de calculer la partie divergente de l'action
effective~$\Ga_{1l}[\vf]$ associ\'ee \`a l'action classique~$S[\vf]$,
dans son approximation \`a une boucle.
Dans la m\'ethode du champ de `background' (voir \cite{ZJ}
par exemple), la contribution \`a une boucle de l'action effective,
r\'esultant de l'int\'egration fonctionnelle
de l'exponentielle de la partie quadratique en $\vf_q$ de $S[\vf_q+\vf_c]$,
est donn\'ee par le d\'eterminant du potentiel effectif.
Ici $\vf_q$ est la partie quantique (sur laquel\-le l'int\'egration fonctionnelle
op\`ere) d'un champ quelconque d\'evelopp\'e au voisinage du
champ de `background' $\vf_c$ (partie classique). Pour que cette m\'ethode
soit valable, il faut en principe que le champ $\vf_c$ soit une solution
classique `des \'equations du mouvement'. L'existence de solutions pour de `telles
\'equations d'\'evolution' non lin\'eaires, est garantie par des th\'eor\`emes g\'en\'eraux
\cite{ReedSII}. Nous n'utiliserons pas explicitement ce fait car la
m\'ethode du champ de `background' nous servira essentiellement \`a extraire les parties
divergentes des fonctions de Green \`a $n$ points dans l'approximation
\`a une boucle.
Dans la suite, nous supprimerons l'indice $c$ et il sera toujours entendu
que  $\vf$ est le champ de `background'.

Ainsi, $\Ga_{1l}[\vf]$ est formellement donn\'e par
$\thalf \ln(\det H) $, o\`u $H$ est le potentiel effectif. Dans notre cas
(ainsi que dans le cas commutatif $\Th=0$), nous verrons que
$H=\tri+m^2+B$, o\`u $B$ est un op\'erateur positif et born\'e;
lorsque que la vari\'et\'e $M$ n'est pas compacte, le spectre essentiel
de $H$ n'est pas vide (typiquement l'intervalle $[m^2,+\infty[$).
Afin de manipuler des op\'erateurs ayant un spectre purement ponctuel
(discret et de multiplicit\'e finie) et ind\'ependamment de tout
sch\'ema de r\'egularisation, nous devons red\'efinir (formellement
pour l'instant) l'action effective \`a une boucle comme
$$
\Ga_{1l}[\vf] := \thalf\ln\det\big(HH_0^{-1}\big),
$$
o\`u $H_0^{-1}:=(\tri+m^2)^{-1}$ est le propagateur libre.
Nous ne sommes alors pas ``si loin'' d'avoir un d\'eterminant
bien d\'efini, car
$$
HH_0^{-1}=(H_0+B)H_0^{-1}=1+BH_0^{-1},
$$
et nous verrons que  $BH_0^{-1}$ appartient \`a toutes les classes de Schatten pour
$p>2$. Les cons\'equences physiques du remplacement de $H$ par $HH_0^{-1}$ sont
b\'enignes, cette normalisation correspondant \`a soustraire les amplitudes
vide/vide (diagrammes sans patte externe).\\
Nous d\'efinirons le logarithme du d\'eterminant, par la
repr\'esentation en `temps propre' de Schwin\-ger
\begin{equation}
\label{eq:actioneff}
\Ga_{1l}[\vf] = \frac{1}{2}\ln\big(\det(HH_0^{-1})\big)
:=-\frac{1}{2}\int_0^\infty\frac{dt}{t}\;\Tr\left(e^{-tH}-e^{-tH_0}\right).
\end{equation}
Avant de donner une signification pr\'ecise \`a cette expression,
c'est-\`a-dire avant de choisir un sch\'ema de r\'egularisation, nous allons
calculer le potentiel effectif~$H$.
Rappelons que le potentiel effectif (voir par exemple
\cite{ZJ}) est l'op\'erateur dont le noyau distributionnel
est donn\'e par la d\'eriv\'ee fonctionnelle seconde de l'action
classique
$$
K_H(p,p') := \frac{\delta^2S[\vf]}{\delta\vf(p)\delta\vf(p')},\;
\hspace{1cm} K_{H_0}(p,p') := \left.\frac{\delta^2S[\vf]}
{\delta\vf(p)\delta\vf(p')}\right|_{\lambda=0},
$$
o\`u les d\'eriv\'ees fonctionnelles sont d\'efinies au sens faible
$$
\left\langle\frac{\delta S[\vf]}{\delta\vf},\psi\right\rangle:=
\left.\frac{d S[\vf+t\psi]}{dt}\right|_{t=0},
$$
et o\`u le couplage est donn\'e par l'int\'egrale avec la forme
volume Riemannienne $\left\langle f,h\right\rangle=\int_M\,\mu_g \,f\,h$.

En utilisant la propri\'et\'e de trace on obtient alors
\begin{equation*}
\left.\frac{d S[\vf+t\psi]}{dt}\right|_{t=0}=
\left\langle \tri\vf+m^2\vf+\frac{\lambda}{3!}\vf^{\Mop 3},\psi\right\rangle.
\end{equation*}
D'o\`u
$$
\tilde{S}_p[\vf]:=\frac{\delta S[\vf]}{\delta\vf(p)}
=\tri\vf(p)+m^2\vf(p)+\frac{\lambda}{3!}\vf^{\Mop 3}(p).
$$
La d\'eriv\'ee fonctionnelle seconde donne quant \`a elle
\begin{align*}
\left\langle\frac{\delta^2S[\vf]}{\delta\vf(p)\delta\vf} ,\psi\right\rangle
&:=\left.\frac{d \tilde{S}_p[\vf+t\psi]}{dt}\right|_{t=0}\\
&=\left\langle \big(\tri+m^2+\frac{\lambda}{3!}(L_{\vf\Mop\vf}+R_{\vf\Mop\vf}
+R_{\vf}L_{\vf})\big)\delta_p^g,\psi\right\rangle,
\end{align*}
o\`u la distribution $\delta_p^g$ a \'et\'e  d\'efinie lors du paragraphe \ref{parahil}.
Finalement, en utilisant la relation
$$
K_A(p,p')=(A^*\delta^g_p)(p'),
$$
valable pour tout op\'erateur fermable sur $L^2(M,\mu_g)$,
on obtient pour le potentiel effectif
$$
H = \tri+m^2+\frac{\lambda}{3!}(L_{\vf\Mop\vf}+R_{\vf\Mop\vf}
+R_{\vf}L_{\vf}).
$$

Puisque $\vf$ est une fonction r\'eelle, les
op\'erateurs $L_\vf$ et $R_\vf$ sont auto-adjoints; on peut alors
v\'erifier la stricte positivit\'e de~$H$:
$$
L_{\vf\Mop\vf}+R_{\vf\Mop\vf}+L_\vf R_\vf=
\thalf(L_\vf+R_\vf)^*(L_\vf+R_\vf)+\thalf L_\vf^*L_\vf+\thalf R_\vf^*R_\vf.
$$

\smallskip
Nous arrivons au point crucial: \\
{\it l'existence du ph\'enom\`ene
de m\'elange des divergences ultraviolettes et infrarouges, vient de la
pr\'esence simultan\'ee des op\'erateurs de multiplication twist\'ee
\`a droite et \`a gauche dans le potentiel effectif.}\\
Plus sp\'ecifiquement, le produit des op\'erateurs
de multiplication twist\'ee \`a gauche et \`a droite $L_fR_h$ jouit
de propri\'et\'es r\'egularisantes importantes. La cons\'equence primordiale,
qui sera employ\'ee au paragraphe \ref{PNP}, est que la trace de
$L_f\,R_h\,e^{-t(\tri+m^2)}$ est r\'eguli\`ere lorsque
$t$ tend vers z\'ero, contrairement \`a
$\Tr(L_f\,e^{-t(\tri+m^2)})$,
$\Tr(R_f\,e^{-t(\tri+m^2)})$, $\Tr(M_f\,e^{-t(\tri+m^2)})$,
qui en $n$ dimensions se comportent comme~$t^{-n/2}$ lorsque $t\to 0$.
Nous verrons  qu'en fait, ces trois derni\`eres traces sont identiques.

\begin{rem}
Pour une th\'eorie $\vf^{\Mop 3}$ sur une vari\'et\'e de dimension six,
le potentiel effectif a la forme suivante
$$
H = \tri+m^2+\frac{\lambda}{2!}(L_{\vf}+R_{\vf}).
$$
M\^eme en l'absence du terme ``mixte'' $R_{\vf}L_{\vf}$, la th\'eorie
poss\`ede un secteur non planaire, mais les diagrammes non planaires
ne sont pr\'esents qu'au niveau de la fonction \`a deux points; le `tadpole'
n'est pas affect\'e par le m\'elange UV/IR.
\end{rem}

\subsubsection{4.1.1.2 Graphe planaire/non planaire versus repr\'esentation
r\'eguli\`ere gauche/droite}
\label{para}

La notion de diagramme de Feynman planaire et non planaire provient
des th\'eories matricielles \cite{Rivasseau, GrK}. Consid\'erons par
exemple un champ scalaire r\'eel  \`a valeurs matricielles sur une
vari\'et\'e Riemannienne $M$, i.e. $\phi^*=\phi\in M_N\big(\Coo_c(M)\big)$,
et une action classique
$$
S[\phi]:=\int_M \mu_g \Tr\big(\thalf \phi(\tri+m^2)\phi\big)+\frac{\lambda}{k!}
\Tr\big(\phi^k\big),
$$
o\`u le Laplacien agit sur chaque composante  du champ $\phi_{ij}$, $i,j=1,\cdots,N$.
Le propagateur de cette th\'eorie, i.e. le noyau de l'inverse de l'op\'erateur $(\tri+m^2)$
agissant sur l'espace vectoriel $M_N\big(\Coo_c(M)\big)$, porte des indices matriciels.
Il sera graphiquement repr\'esent\'e par une double ligne (un ruban), chacune portant
un indice matriciel \`a chaque extr\'emit\'es. Le vertex de cette th\'eorie, le noyau du terme d'interaction,
i.e. le noyau de la forme $k$-lin\'eaire $\int_M \mu_g \Tr\big(\phi_1\cdots\phi_k\big)$,
sera lui aussi repr\'esent\'e par $k$ doubles lignes, une ligne pour chaque indice matriciel,
se joignant \`a l'une de leurs extr\'emit\'es.
Les diagrammes de Feynman sont ensuite obtenus par contraction des vertex avec les
propagateurs. L'id\'ee est qu'un tel diagramme est dit planaire lorsque qu'il peut
\^etre trac\'e sur un plan (\'evidemment en l'absence de croisement de ruban), et
non planaire sinon. Plus pr\'ecis\'ement \cite[paragraphe II.5]{Rivasseau}, ces
diagrammes engendrent des surfaces de Riemann, en identifiant les pattes externes
avec `un point \`a l'infini'. Les diagrammes planaires seront ceux pour lesquels la
surface g\'en\'er\'ee est de genre $g=0$ (une 2-sph\`ere), et non planaire lorsque $g>0$.

\medskip

En utilisant la caract\'erisation matricielle du produit de Moyal symplectique
(cf. proposition \ref{rien}), on peut appliquer directement ces consid\'erations.
On pourra alors repr\'esenter les amplitudes de
transitions de ces th\'eories par des diagrammes en ruban \cite{GW, GW1, GW2, Filk, CR}.

\medskip

Le concept de graphes planaires et non planaires, ou plus
simplement de contribution planaires et non planaires, se g\'en\'eralise
\`a n'importe quelle th\'eorie quantique des champs non commutative,
en particulier aux th\'eories scalaires sur d\'eformations isospectrales. Par
th\'eorie quantique non commutative (scalaire), nous entendons une
th\'eorie o\`u les champs
classiques sont vus comme les \'el\'ements d'une alg\`ebre non commutative,
avec une action classique construite \`a partir d'une trace sur l'alg\`ebre.
Le cas vectoriel, pour lequel les champs classiques sont les \'el\'ements d'un module projectif de type fini
est sensiblement plus compliqu\'e.\\
En effet, la caract\'eristique essentielle utilis\'ee dans le cas matriciel est
l'invariance par permutation cyclique des arguments du vertex (dans une base ou une
repr\'esentation donn\'ee), cons\'equente \`a la cyclicit\'e de la trace. On repr\'esentera
alors les propagateurs et les vertex par des rubans, par analogie avec le cas matriciel.
Apr\`es contraction des vertex avec les propagateurs, on obtient deux types de
graphes:  planaires lorsque seuls les rubans des vertex qui sont adjacents ont \'et\'e
contract\'es et non planaires sinon.

\medskip

Nous allons \'evoquer maintenant le lien entre les secteurs planaire et non
planaire et les op\'erateurs de repr\'esentation r\'eguli\`ere gauche et droite.\\
Consid\'erons pour cela une alg\`ebre non commutative $\A$ munie
d'une trace finie $\tau$, ainsi qu'un `op\'erateur
cin\'etique' $M\in\L(\A)$ (non n\'ecessairement born\'e) jouant le r\^ole du
Laplacien. Pour tout $a\in\A$, on d\'efinit une action classique par
$$
S[a]:=\frac{1}{2}\tau\big(aMa\big)+\frac{\lambda}{k!}\tau\big(a^k\big).
$$
Tout d'abord, il est facile de montrer que
la structure du potentiel effectif, c'est-\`a-dire la pr\'esence simultan\'ee
d'op\'erateurs de multiplication \`a  gauche et \`a droite, donc l'existence de
deux secteurs distincts dans la th\'eorie, est tout \`a fait g\'en\'erale:
le potentiel effectif contiendra toujours des sommes et produits d'op\'erateurs
de repr\'esentation r\'eguli\`ere  gauche $L$ et droite $R$:
$$
H_a=M+\frac{\lambda}{(k-1)!}\sum_{j=0}^{k-2} L(a^j)\,R(a^{k-2-j}).
$$

Dans la m\'ethode du champ de `background', les contributions
\`a une boucle sont donn\'ees par l'int\'egrale fonctionnelle sur la partie quantique $a_q$,
d'un champ quelconque $\A\ni a=a_q+a_c$ d\'evelopp\'e autour d'un champ classique
$a_c$, de l'exponentielle (r\'eelle dans le r\'egime Euclidien) de la partie quadratique en $a_q$
de $S[a_q+a_c]$. Elles sont donn\'ees par le logarithme du d\'eterminant du potentiel effectif.
Un `d\'eveloppement en puissance de $a_c$' de $\thalf\ln\det^{reg}\big(H_{a_c}H_0^{-1}\big)$,
o\`u $\det^{reg}$ est un d\'eterminant r\'egularis\'e correspondant au sch\'ema de
r\'egularisation choisi et $H_0:=M$, fournit la cor\-respondance avec les parties divergentes des diagrammes
de Feynman (r\'egularis\'es) \`a une boucle en pr\'esence d'un champs de `background' $a_c$.\\
En effet, les diagrammes de Feynman \`a une boucle en pr\'esence d'un champs de `background'
sont obtenus par contraction des vertex et des propagateurs donn\'es par l'action
$S_{quad}[a_q]$, partie quadratique en $a_q$ de $S[a_q+a_c]$:
$$
S_{quad}[a_q]=\frac{1}{2}\tau\big(a_qMa_q\big)+\frac{\lambda}{2(k-1)!}
\sum_{j=0}^{k-2} \tau\big(a_q\,a_c^j\,a_q\,a_c^{k-2-j}\big).
$$
Dans ce cas, le propagateur est encore donn\'e par $M^{-1}$, les vertex par des graphes consistant en
$k-2$ rubans externes (attach\'es en leurs extr\'emit\'es \`a $a_q$) altern\'es avec deux rubans
internes (avec des indices libres \`a leurs extr\'emit\'es).\\
La similitude entre $H_{a_c}$ et $S_{quad}[a_q]$ donne ainsi la correspondance entre les
diagrammes planaire/ non planaire et les op\'erateurs de repr\'esentation r\'eguli\`ere
gauche/droite.

\bigskip

Nous allons terminer par une petite digression,
dans le but d'illustrer le caract\`ere r\'egularisant des
op\'erateurs $L_fR_h$. Nous allons observer
ce ph\'enom\`ene dans les cas `limites' des plans de Moyal non d\'eg\'en\'er\'es
($n=2N,\Th$ inversible), pour lesquels il est particuli\`erement
\'eloquent car les op\'erateurs $L_fR_h$ sont \`a trace
lorsque $f,h\in\SS(\R^{2N})$. Ce fait, connu des experts,
n'a \'et\'e que rarement rapport\'e dans la litt\'erature (\`a ma connaissance, la seule
publication le mentionnant est~\cite{Braunss}). Plusieurs arguments peuvent \^etre
invoqu\'es pour d\'emontrer cette propri\'et\'e.
Par exemple, on peut consid\'erer la d\'ecomposition
polaire $f_i=u_i\mop |f_i|_{alg}$ de $f_i\in\SS(\R^{2N})$, $i=1,2$ (ici $|.|_{alg}$
d\'esigne le module au sens de la $C^*$-alg\`ebre unif\`ere $A_\th$ et non pas au
sens des fonctions) et utiliser  le fait que $|f_i|_{alg}=g_i^*\mop g_i$ pour certains
$g_i\in\SS$. Alors, en utilisant la commutation de $L$ avec $R$, on obtient
\begin{align*}
\|L_{f_1}\,R_{f_2}\|_1&\leq \|L_{u_1}\| \,\|R_{u_2}\|\,
\|L_{g_1^*}\,R_{g_2^*}\,R_{g_2}\,L_{g_1}\|_1\\
&\leq \|L_{u_1}\| \,\|R_{u_2}\| \,\|R_{g_2}\,L_{g_1}\|_2^2,
\end{align*}
pour finalement v\'erifier que le noyau de $R_fL_g$, $f,h\in\SS(\R^{2N})$,
est de carr\'e sommable.\\
Il existe une mani\`ere de proc\'eder encore plus \'eloquente, utilisant
la base de Wigner de l'oscilla\-teur harmonique, qui montre \`a quel point
ce ph\'enom\`ene est proche de celui des
op\'erateurs de Toeplitz. En d\'eveloppant $f,h\in\SS(\R^{2N})$ dans cette base,
$$
f=\sum_{m,n}c_{mn}f_{mn},\hspace{2cm} h=\sum_{m,n}d_{mn}f_{mn},
$$
et en utilisant leurs propri\'et\'es d'unit\'es matricielles
$f_{mn}\Mop f_{kl}=\delta_{nk}f_{ml}$ et de base orthonorm\'ee
$(2\pi\th)^{-N}\langle f_{mn},f_{kl}\rangle=\delta_{m,k}\delta_{n,l}$,
on obtient
\begin{align*}
\Tr\big(L_fR_h\big)&=(2\pi\th)^{-N}\sum_{m,n,k,l,s,t}c_{kl}\,d_{st}\,
\langle f_{mn},f_{kl}\Mop f_{mn}\Mop f_{st}\rangle\\
&=\sum_{m,n}c_{mm}\,d_{nn}\\
&=(2\pi\th)^{-N}\int d^{2N}x\,f(x) \int d^{2N}y\,h(y) < \infty.
\end{align*}
On peut alors factoriser $HH_0^{-1}$ et extraire une partie finie dans
l'action effective:
\begin{align}
\label{bel}
H_0H^{-1}&=
\Big(1-\frac{\lambda}{3!}(L_{\vf\mop\vf}+R_{\vf\mop\vf})
\frac{1}{\tri+m^2 +\frac{\lambda}{3!}(L_{\vf\mop\vf}+R_{\vf\mop\vf})}\Big)\nonumber\\
&\hspace{2cm}\times\Big(1-\frac{\lambda}{3!}L_\vf R_\vf
\frac{1}{\tri+m^2
+\frac{\lambda}{3!}(L_{\vf\mop\vf}+R_{\vf\mop\vf}+L_\vf R_\vf)}\Big).
\end{align}
Or,
$$
1-\frac{\lambda}{3!}L_\vf R_\vf
\frac{1}{\tri+M^2
+\frac{\lambda}{3!}(L_{\vf\mop\vf}+R_{\vf\mop\vf}+L_\vf R_\vf)}
\in 1+\L^1(\H),
$$
son d\'eterminant est donc parfaitement bien d\'efini. Seul le d\'eterminant
de la premi\`ere parenth\`ese de l'\'equation (\ref{bel}) a besoin
d'\^etre r\'egularis\'e. Nous verrons que d\'eterminant de la deuxi\`eme
parenth\`ese de (\ref{bel}) contient l'int\'egralit\'e de la contribution
non planaire \`a la fonction \`a deux points. Pour la fonction
\`a quatre points, la partie non planaire finie r\'eside dans les deux
termes.

\subsubsection{4.1.1.3 Sch\'emas de r\'egularisation}

Retournons au calcul de~$\Ga_{1l}[\vf]$ dans le cas g\'en\'eral.
L'int\'egrale sur $t$ dans  (\ref{eq:actioneff}) est divergente \`a cause du
comportement du noyau de la chaleur
sur sa diagonale pour $t$ tendant vers z\'ero. On d\'efinit l'action effective \`a une boucle
r\'egularis\'ee par
\begin{equation}
\label{eq:actioneffreg}
\Ga^\ep_{1l}[\vf]:=-\frac{1}{2}\int_\ep^\infty\frac{dt}{t}\;
\Tr\left(e^{-tH}-e^{-tH_0}\right).
\end{equation}
D'autres sch\'emas de r\'egularisation existent. Par exemple
la r\'egularisation $\zeta$, le pendant en espace courbe
de la r\'egularisation dimensionnelle, obtenue en rempla\c cant dans
l'expression de l'action effective $1/t$ par
$\mu^{2\sigma}/t^{1-\sigma}$, o\`u $\mu$ est une \'echelle de masse
(n\'ecessaire pour conserver la bonne dimension) et  $\sigma$ est
un nombre complexe:
\begin{equation}
\label{eq:p}
\Ga^{\sigma,\mu}_{1l}[\vf]:=-\frac{1}{2}\int_0^\infty\frac{dt}{t}
\left(t\mu^2\right)^\sigma\;\Tr\left(e^{-tH}-e^{-tH_0}\right).
\end{equation}
En dimension $n$ (paire), l'int\'egrale \eqref{eq:p} est bien d\'efinie pour
$\Re(\sigma)>n/2-1$. On peut montrer qu'elle poss\`ede un prolongement analytique
en $\sigma$ sur $\C\setminus\{n/2-1,n/2-2,\cdots\}$.\\
Nous n'utiliserons cependant
cette r\'egularisation que lorsque l'on voudra comparer nos r\'esultats avec ceux
obtenus dans le cas du plan de Moyal, par r\`egles de Feynman
modifi\'ees (dans l'espace des impulsions) et r\'egularisation dimensionnelle.
Pour les autres applications, en particulier pour l'\'etude du m\'elange
UV/IR, la r\'egularisation (\ref{eq:actioneffreg}) sera suffisante.
Notons que $\ep$ peut \^etre pens\'e comme un cut-off en impulsion,
i.e. $\ep=\Lambda^{-2}$.

Pour montrer que les expressions~(\ref{eq:actioneffreg}) et~(\ref{eq:p})
sont maintenant bien d\'efinies, il est n\'ecessaire que la diff\'erence
de semi-groupe $e^{-tH}-e^{-tH_0}$ soit
\`a trace pour tout $t>0$. Notons aussi que la convergence de
l'int\'egrale en $t\to\infty$ est assur\'ee par le facteur global $e^{-tm^2}$.
Aussi, lorsque le spectre du Laplacien est
born\'e inf\'erieurement par une constante strictement positive,
on pourra construire des th\'eories
sans masse et sans divergence infrarouge. C'est en particulier
le cas pour le plan hyperbolique twist\'e $\HH^n_\Th$, vu
que le spectre ($L^2$) de $\tri$ sur $\HH^n$ est la demi-droite $[n^2/4,\infty[$.
\begin{lem}
La diff\'erence de semi-groupe $e^{-tH}-e^{-tH_0}$ est \`a trace pour tout
$t>0$.
\end{lem}
\begin{proof}[Preuve]
En utilisant la positivit\'e de $H$ ainsi que celle de $H_0$, la propri\'et\'e de
semi-groupe et le calcul fonctionnel holomorphe avec un chemin
$\ga$ entourant les spectres $\spec(H)\subset\R^+$ et
$\spec(H_0)\subset\R^+$, on a
$$
e^{-tH}-e^{-tH_0}=\frac{1}{(2i\pi)^2}
\int_{\ga\times\ga}dz_1\,dz_2\,e^{-t(z_1+z_2)/2}\,
\left(R_H(z_1)R_H(z_2)-R_{H_0}(z_1)R_{H_0}(z_2)\right),
$$
o\`u $R_A(z)=(z-A)^{-1}$ est la r\'esolvante de $A$. Or $H=H_0+B$ o\`u $B$
est born\'e. Puisque $R_H(z)=R_{H_0}(z)(1+BR_H(z))$, on obtient
\begin{align*}
R_H(z_1)R_H(z_2)-R_{H_0}(z_1)R_{H_0}(z_2)&=
R_{H_0}(z_1)R_{H_0}(z_2)BR_H(z_2)
+R_{H_0}(z_1)BR_H(z_1)R_{H_0}(z_2)\\
&\quad+R_{H_0}(z_1)BR_H(z_1)R_{H_0}(z_2)BR_H(z_2).
\end{align*}
La premi\`ere \'equation r\'esolvante et le fait que
$L_f(z-\tri)^{-k},R_f(z-\tri)^{-k}\in\L^p(\H)$, pour
$p>2/k,f\in\Coo_c(M)$ (proposition \ref{pr:interpolation}),
utilis\'ee avec l'in\'egalit\'e de H\"older pour les classes de
Schatten, permettent de montrer que
$$
\int_{\ga\times\ga}dz_1\,dz_2\,e^{-t(z_1+z_2)/2}
R_{H_0}(z_1)R_{H_0}(z_2)BR_H(z_2)
$$
ainsi que les autres termes, sont des int\'egrales au
sens de Bochner pour la norme trace. Ainsi,
$e^{-tH}-e^{-tH_0}$ est \`a trace comme \'enonc\'e.
\end{proof}

\subsection{D\'eveloppement en puissance du champ}

Pour d\'ecrire les divergences, nous allons \'etudier le
comportement de $\Ga^\ep_{1l}[\vf]$ lorsque $\ep\to 0$. Nous allons ensuite
montrer
qu'il existe pour toutes les d\'eformations isospectrales deux types de contributions
pour les fonctions de Green: les contributions planaires donnant lieu \`a des
singularit\'es UV ordinaires et celles non planaires exhibant le ph\'enom\`ene
de m\'elange UV/IR.\\
Travaillant en espace courbe, il n'est plus possible
de d\'efinir des diagrammes de Feynman dans l'espace des moments.
Nous allons cependant continuer \`a parler de contributions
planaires et non planaires (cf. paragraphe 4.1.1.2), en utilisant le clivage entre les op\'erateurs
de multiplication twist\'ee \`a droite et \`a gauche, clivage
qui co\"incide avec celui entre les diagrammes planaires et non
planaires dans les cas plats connus. Ce point deviendra plus limpide dans
les prochains paragraphes.

Int\'eress\'es par le
comportement en $\ep$ de $\Ga^\ep_{1l}[\vf]$
(nous ne consid\'erons que la partie potentiellement divergente de
l'action effective), nous avons besoin d'un d\'eveloppement en puissance
de $t$ pour $\Tr\left(e^{-tH}-e^{-tH_0}\right)$ lorsque $t\to 0$.
Ce d\'eveloppement est obtenu par des  techniques
d\'ej\`a utilis\'ees dans le calcul de l'action spectrale.
Parce que la dimension est quatre, la formule de
Baker--Campbell--Hausdorff (BCH) \`a l'ordre deux
$$
e^{-tH}=e^{-tB+\frac{t^2}{2}[\tri ,B]
-\frac{t^3}{6}[\tri ,[\tri ,B]]-\frac{t^3}{12}[B,[\tri ,B]]+\cdots}\;
e^{-tH_0},
$$
est suffisante pour capturer la structure des divergences. Il faut ensuite
d\'evelopper la premi\`ere exponentielle et ne garder que les termes qui,
sous la trace, donneront des termes d'ordre inf\'erieur ou \'egal \`a z\'ero
en $t$. Seuls quelques termes seront important, car en
dimension $n$
\begin{align}
\Tr(L_f\tri^k e^{-t\tri})&\simeq t^{-n/2-k},\,t\rightarrow 0,\nonumber\\
\Tr(R_f\tri^k e^{-t\tri})
&\simeq t^{-n/2-k},\,t\rightarrow 0.
\label{eq:dhk}
\end{align}
En effet, pour le cas ``gauche'' (le cas droit \'etant similaire), \'etant donn\'e
que $L_f(1+\tri)^{-k}\in\L^p(\H)$ pour tout $p>n/2k$ d'apr\`es la
proposition \ref{pr:interpolation}, on d\'eduit  que pour tout $\ep>0$
\begin{align}
\label{robe}
\|L_f\tri^ke^{-t\tri}\|_1&\leq\|L_f(1+\tri)^{-n/2-\ep}\|_1\,\,
\|\frac{\tri^k}{(1+\tri)^k}\|\|(1+\tri)^{n/2+k+\ep}e^{-t\tri}\|\nonumber\\
&\leq C(\ep)t^{-(n/2+k+\ep)}
\end{align}
o\`u la derni\`ere estimation a \'et\'e obtenue par calcul fonctionnel.
Dans le d\'eveloppement en puissance du champ, il faut alors corriger
la puissance en $t$ par l'ordre de l'op\'erateur diff\'erentiel et donc
$$
e^{-tH}=\left(1-tB+\frac{t^2}{2}[\tri ,B]
-\frac{t^3}{6}[\tri ,[\tri ,B]]+\frac{t^2}{2}B^2\right)
e^{-tH_0}\;+O(t);
$$
qui signifie que nous avons le d\'eveloppement suivant pour $t\to 0$:
\begin{equation}
\label{jupe}
\Tr\left(e^{-tH}-e^{-tH_0}\right)=
\Tr\left((-tB+\frac{t^2}{2}[\tri,B]
-\frac{t^3}{6}[\tri,[\tri,B]]+\frac{t^2}{2}B^2)
e^{-tH_0}\right)\;+O(t).
\end{equation}

Nous allons commencer par montrer que les commutateurs pr\'esents dans l'expression
\eqref{jupe} fournissent des contributions nulles \`a l'action effective.\\
Consid\'erons pour cela les termes
$\Tr\left([\tri,C]e^{-tH_0}\right)$, avec $C=B$ ou $C=[\tri,B]$.\\
En effet, \`a la condition que chacun des termes $\tri Ce^{-tH_0}$ et $C\tri e^{-tH_0}$ soit
\`a trace, on obtient le r\'esultat en utilisant la cyclicit\'e de la trace
ainsi que la commutation du Laplacien avec le semi-groupe de la chaleur:
$$
\Tr\left(\tri\,C\,e^{-tH_0}-C\,\tri\,e^{-tH_0}\right)=
\Tr\left(\tri\,C\,e^{-tH_0}-\tri\,C\,e^{-tH_0}\right)=0.
$$
La tra\c cabilit\'e des op\'erateurs $C\tri e^{-tH_0}$ s'obtient facilement
par calcul fonctionnel et en invoquant les m\^emes arguments que ceux
utilis\'es pour \'etablir l'estimation \eqref{robe}. Pour celle des termes
$\tri Ce^{-tH_0}$, nous allons utiliser la relation tautologique
$$
\tri \,C\,e^{-tH_0}=C\,\tri\, e^{-tH_0}+[\tri,C]\,e^{-tH_0}.
$$
Ainsi, il ne reste qu'\`a montrer que les op\'erateurs $[\tri,C]\,e^{-tH_0}$
sont eux aussi \`a trace.\\
Il est suffisant pour cela de calculer les commutateurs
$[\tri,L_{\vf\Mop\vf}]$, $[\tri,R_{\vf\Mop\vf}]$, $[\tri,L_\vf
R_\vf]$, etc.. La m\'ethode la plus simple est d'utiliser les formules (\ref{Lfint}) et
(\ref{Rfint}). Par $[V_z,\tri]=0$ pour tout $z\in\R^l$ (signature concr\`ete de
la propri\'et\'e d'isom\'etrie de l'action) et en choisissant un syst\`eme de coordonn\'ees
locales $\{x^\mu\}$, nous obtenons
\begin{align*}
[\tri,L_{\vf\Mop\vf}]&=
(2\pi)^{-l}\int_{\R^{2l}}\,d^ly\,d^lz\;e^{-i<y,z>}\;
V_{\thalf \Th y}\,[\tri,M_{\vf\Mop \vf}]\,V_{-\thalf\Th y-z}\\
&=
(2\pi)^{-l}\int_{\R^{2l}}\,d^ly\,d^lz\;e^{-i<y,z>}\;
V_{\thalf \Th y}\,\left(M_{\tri(\vf\Mop \vf)}
-2M_{\nabla^\mu(\vf\Mop\vf)}\nabla_\mu\right)\,V_{-\thalf\Th y-z}\\
&=L_{\tri(\vf\Mop \vf)}-2L_{\nabla^\mu(\vf\Mop\vf)}\nabla_\mu,
\end{align*}
et similairement,
\begin{align*}
[\tri,R_{\vf\Mop\vf}]=&\,\,R_{\tri(\vf\Mop \vf)}
-2R_{\nabla^\mu(\vf\Mop\vf)}\nabla_\mu,\\
[\tri,R_\vf L_\vf]=&\,\,R_\vf[\tri,L_\vf]+[\tri,R_\vf]L_\vf\\
=&\,\,R_\vf L_{\tri\vf}
+R_{\tri\vf}L_\vf-2R_{\nabla^\mu\vf}L_{\nabla_\mu\vf}
-2(R_\vf L_{\nabla^\mu\vf}+R_{\nabla^\mu\vf}L_\vf)\nabla_\mu.
\end{align*}
Le syst\`eme de coordonn\'ees utilis\'e doit \^etre compatible avec
la d\'eformation, c'est-\`a-dire qu'il doit \^etre d\'efini sur un voisinage
$U\subset M$ invariant par $\a$. On peut obtenir un tel syst\`eme,
en choisissant n'importe quel recouvrement d'ouverts
$\{U_I\}_{i\in I}$ de $M$, pour d\'efinir $\{\widetilde{U}_I\}_{i\in I}$ en
laissant $\R^l$ agir dessus: $\widetilde{U}_i:=\R^l.U_i$.

Les derniers commutateurs dont on a besoin sont:
\begin{align*}
[\tri,[\tri,L_{\vf\Mop\vf}]]=&\,\,L_{\tri\tri(\vf\Mop\vf)}
-4L_{\nabla^\mu\tri(\vf\Mop\vf)}\nabla_\mu
+4L_{\nabla^\mu\nabla^\nu(\vf\Mop\vf)}\nabla_\mu\nabla_\nu,\\
[\tri,[\tri,R_{\vf\Mop\vf}]]=&\,\,R_{\tri\tri(\vf\Mop\vf)}
-4R_{\nabla^\mu\tri(\vf\Mop\vf)}\nabla_\mu
+4R_{\nabla^\mu\nabla^\nu(\vf\Mop\vf)}\nabla_\mu\nabla_\nu,\\
[\tri,[\tri,R_\vf L_\vf]]=&\,\,R_\vf L_{\tri\tri\vf}+R_{\tri\tri\vf}L_\vf
+2R_{\tri\vf}L_{\tri\vf}\\
&-4R_{\nabla^\mu\vf}L_{\nabla_\mu\tri\vf}
-4R_{\nabla^\mu\tri\vf}L_{\nabla_\mu\vf}+4R_{\nabla^\mu\nabla^\nu\vf}
L_{\nabla_\mu\nabla_\nu\vf}\\
&-4(R_\vf L_{\nabla^\mu\tri\vf}+
R_{\nabla^\mu\tri\vf}L_\vf+R_{\tri\vf}L_{\nabla^\mu\vf}\\
&+R_{\nabla^\mu\vf}L_{\tri\vf}-2R_{\nabla^\nu\vf}L_{\nabla_\nu\nabla^\mu\vf}
-2R_{\nabla^\nu\nabla^\mu\vf}L_{\nabla_\nu\vf})\nabla_\mu\\
&+4(R_\vf L_{\nabla^\mu\nabla^\nu\vf}+R_{\nabla^\mu\nabla^\nu\vf}L_\vf
+2R_{\nabla^\mu\vf}L_{\nabla^\nu\vf})\nabla_\mu\nabla_\nu.
\end{align*}
Par calcul fonctionnel, on obtient alors que chaque terme des d\'eveloppements
pr\'ec\'edents de $[\tri,C]e^{-tH_0}$,
avec $C=B$ ou $C=[\tri,B]$ sont \`a trace, et donc $\Tr\big([\tri,C]e^{-tH_0}\big)=0$.

Finalement, on obtient
\begin{align*}
\Tr\Big(e^{-tH}-e^{-tH_0}\Big)=
&-t\frac{\lambda}{3!}\Tr\Big(\big(L_{\vf\Mop\vf}
+R_{\vf\Mop\vf}+R_\vf L_\vf\big)e^{-t(\tri+m^2)}\Big)\\
&+\frac{t^2}{2}\frac{\lambda^2}{(3!)^2}\Tr\Big(\big(L_{\vf^{\Mop 4}}
+R_{\vf^{\Mop 4}}+3R_{\vf\Mop\vf}L_{\vf\Mop\vf}\\
&\hspace{3cm}+2R_\vf L_{\vf^{\Mop 3}}
+2R_{\vf^{\Mop 3}}L_\vf\big)e^{-t(\tri+m^2)}\Big)+O(t).
\end{align*}

\subsection{Contributions planaires et non planaires}
\label{PNP}

Nous allons s\'eparer l'expression pr\'ec\'edente en deux parties.
Dans la premi\`ere, nous ne garderons que les termes du type
$L_fe^{-t\tri}$ et $R_fe^{-t\tri}$. Ces derniers constituent ``la partie
planaire'', car donnant des contributions du m\^eme
type que celles des th\'eories commutatives, comme on peut s'en
persuader en regardant l'\'equation ~(\ref{eq:deq}) ci-dessous.
Le deuxi\`eme type de contribution, correspondant \`a la partie
non planaire, consiste en les termes de la forme $L_f R_he^{-t\tri}$.

La contribution planaire \`a (la partie divergente de) l'action
effective \`a une boucle est alors donn\'ee par
\begin{align*}
\Ga_{1l,P}^\ep[\vf]:=\frac{1}{2}\int_\ep^\infty dt\,e^{-tm^2}\Big\lbrace
&\frac{\lambda}{3!}\Tr\Big(\big(L_{\vf\Mop\vf}
+R_{\vf\Mop\vf}\big)e^{-t\tri}\Big)\\
&-\frac{t}{2}\frac{\lambda^2}{(3!)^2}
\Tr\Big(\big(L_{\vf^{\Mop 4}}
+R_{\vf^{\Mop 4}}\big)e^{-t\tri}\Big)\Big\rbrace+O(\ep^0).
\end{align*}
Pour calculer ces traces, montrons que la trace est un d\'equantificateur
pour le produit d\'eform\'e:
\begin{equation}
\label{eq:deq}
\Tr\big(L_f\,e^{-t\tri}\big)=\Tr\big(R_f\,e^{-t\tri}\big)=\Tr\big(M_f\,e^{-t\tri}\big).
\end{equation}
Nous r\'ep\'etons les arguments pour l'obtention de
l'\'equation (\ref{eq:deq}), dans le cas de $L_fe^{-t\tri}$.
Pour $R_fe^{-t\tri}$ les arguments sont similaires. A partir de l'expression
du noyau distributionnel de l'op\'erateur
$L_f\,e^{-t\tri}$
$$
K_{L_f\,e^{-t\tri}}(p,p')=(2\pi)^{-l}\int_{\R^{2l}}\,d^ly\,d^lz\,e^{-i<y,z>}\,
f(-\thalf\Th y.p)\,K_t(z.p,p'),
$$
on a
\begin{align*}
\Tr\big(L_f\,e^{-t\tri}\big)&=\int_M\,\mu_g(p)\,K_{L_f\,e^{-t\tri}}(p,p)\\
&=(2\pi)^{-l}\int_M\,\mu_g(p)\int_{\R^{2l}}\,d^ly\,d^lz\,e^{-i<y,z>}\,
f(-\thalf\Th y.p)\,K_t(z.p,p).
\end{align*}
En utilisant l'invariance de la forme volume sous l'isom\'etrie
$p\rightarrow \thalf\Th y.p$ ainsi que $[e^{-t\tri},V_z]=0$,
traduit en terme de l'invariance de son noyau
\begin{equation}
\label{invprop}
K_t(z.p,z.p')=K_t(p,p'),
\end{equation}
on obtient
$$
\Tr\big(L_f\,e^{-t\tri}\big)=\int_M\,\mu_g(p)\,
f(p)\,K_t(p,p)=\Tr\big(M_f\,e^{-t\tri}\big).
$$
La partie planaire de l'action effective \`a une boucle est alors
\begin{equation}
\Ga_{1l,P}^\ep[\vf]=\int_\ep^\infty dt\,e^{-tm^2}\Big\lbrace
\frac{\lambda}{3!}\Tr\Big(M_{\vf\Mop\vf}e^{-t\tri}\Big)
-\frac{t}{2}\frac{\lambda^2}{(3!)^2}\Tr\Big(M_{\vf^{\Mop 4}}
e^{-t\tri}\Big)\Big\rbrace + O(\ep^0).
\end{equation}
D'apr\`es la relation
$$
K_{M_fe^{-t\tri}}(x,x)=f(x)K_t(x,x),
$$
et le d\'eveloppement du noyau de la chaleur sur sa diagonale
\`a l'ordre un
$$
K_t(x,x)=(4\pi t)^{-2}\bigl(1-\frac{t}{6}R(x)\bigr)+O(t^0),
$$
on obtient modulo des termes d'ordre $O(\ep^0)$:
\begin{equation}
\label{eq:P}
\Ga_{1l,P}^\ep[\vf]=\int_\ep^\infty dt\,\frac{e^{-tm^2}}{(4\pi t)^2}
\int_M\,\mu_g\,\Big(\frac{\lambda}{3!}\vf\Mop\vf
-t\Big(\frac{1}{6}\frac{\lambda}{3!}(\vf\Mop\vf)R
+\frac{1}{2}\frac{\lambda^2}{(3!)^2}\vf^{\Mop 4}\Big)\Big) +O(\ep^0).
\end{equation}
La partie planaire reproduit alors les divergences ordinaires en
$\frac{1}{\ep}$ et $|\ln\ep|$ correspondant,
dans le langage des diagrammes de Feynman,
aux fonctions \`a deux
et quatre points.

\smallskip

La contribution non planaire est quant \`a elle
\begin{align}
\hspace{-0.3cm}\Ga_{1l,NP}^\ep[\vf]:=&\frac{1}{2}\int_\ep^\infty dt\,e^{-tm^2}\Big\lbrace
\frac{\lambda}{3!}\Tr\Big(R_\vf L_\vf\,e^{-t\tri}\Big)\nonumber\\
&\quad-\frac{t}{2}\frac{\lambda^2}{(3!)^2}
\Tr\Big(\big(3R_{\vf\Mop\vf}L_{\vf\Mop\vf}
+2R_\vf L_{\vf^{\Mop 3}}+2R_{\vf^{\Mop 3}}L_\vf\big)e^{-t\tri}\Big)\Big\rbrace
+O(\ep^0).
\end{align}
En utilisant l'expression des noyaux des op\'erateurs $L_f$ et $R_f$
$$
K_{L_f}(p,p')=(2\pi)^{-l}\int_{\R^{2l}}\,
d^ly\,d^lz\,e^{-i<y,z>}f(-\thalf\Th y.p)\,\delta^g_{z.p}(p'),
$$
$$
K_{R_f}(p,p')=(2\pi)^{-l}\int_{\R^{2l}}\,
d^ly\,d^lz\,e^{-i<y,z>}f(z.p)\,\delta^g_{-\thalf\Th y.p}(p').
$$
on obtient le noyau de $L_fR_h e^{-t\tri}$ en terme de celui du semi-groupe de
la chaleur $K_t$:
$$
K_{L_fR_h e^{-t\tri}}(p,p')=(2\pi)^{-l}
\int_{\R^{2l}}\,d^ly\,d^lz\,e^{-i<y,z>}\,f((-\thalf\Th y-z).p)\,
h(z.p)\,K_t(-\thalf\Th y.p,p').
$$
Apr\`es quelques changements de variables,
la trace de $L_fR_h e^{-t\tri}$ devient
\begin{equation}
\label{eq:RL}
\Tr\big(L_fR_h e^{-t\tri}\big)=(2\pi)^{-l}
\int_M\mu_g(p)\,\int_{\R^{2l}}\,d^ly\,d^lz\,e^{-i<y,z>}\,
f(p)\,h(z.p)\,K_t(-\Th y.p,p).
\end{equation}
Puisque $K_t$ est
sym\'etrique et invariant sous $\a$, l'isom\'etrie $p\mapsto-z.p$
entra\^ine
\begin{equation}
\Tr\big(L_fR_h e^{-t\tri}\big)=\Tr\big(R_fL_h e^{-t\tri}\big).
\end{equation}
En utilisant cette derni\`ere relation, on obtient finalement pour $\Ga_{1l,NP}^\ep[\vf]$
\begin{align}
\label{bell}
\hspace{-0.4cm}\Ga_{1l,NP}^\ep[\vf]&=
\frac{1}{2(2\pi)^{l}}\int_\ep^\infty dt\,e^{-tm^2}
\int_M\mu_g(p)\,\int_{\R^{2l}}\,d^ly\,d^lz\,e^{-i<y,z>}
\Big\lbrace\frac{\lambda}{3!}\vf(p)\vf(z.p)\nonumber\\
&\quad-\frac{t}{2}\frac{\lambda^2}{(3!)^2}\Big(3\vf\Mop\vf(p)\vf\Mop\vf(z.p)
+4\vf(p)\vf^{\Mop 3}(z.p)
\Big)\Big\rbrace K_t(-\Th y.p,p)+O(\ep^0).
\end{align}
Nous verrons que le meilleur comportement du secteur non planaire
ainsi que le m\'elange UV/IR viennent de la pr\'esence du noyau de la
chaleur hors diagonale dans l'expression pr\'ec\'edente.

\section{D\'eformations non p\'eriodiques}
\subsection{NCQFT sur plan de Moyal dans l'espace de configuration}
\subsubsection{4.2.1.1 Cas non d\'eg\'en\'er\'e}

Lorsque $M=\R^4$ avec la m\'etrique plate, $l=4$ et lorsque $\R^4$ agit sur lui-m\^eme
par translation, la d\'eformation
isospectrale correspondante co\"incide avec le plan
de Moyal quadri-dimensionnel non d\'eg\'en\'er\'e $\R^4_\Th$.
Dans ce cas, le noyau de la chaleur est exactement donn\'e par
$$
K_t(x,y)=(4\pi t)^{-2}e^{-\frac{|x-y|^2}{4t}}.
$$
$\Ga^\ep_{1l,P}(\vf)$ et $\Ga^\ep_{1l,NP}(\vf)$ sont alors
calculables explicitement.\\
Pour la partie planaire, nous obtenons \`a partir de~(\ref{eq:P})
$$\Ga_{1l,P}^\ep[\vf]=\int_\ep^\infty\,dt\frac{e^{-tm^2}}{(4\pi t)^2}\,
\int_{\R^4}\,d^4x\,\Big(\frac{\lambda}{3!}\vf^2(x)
-\frac{t}{2}\frac{\lambda^2}{(3!)^2}(\vf\Mop\vf)^2(x)\Big)+O(\ep^0),
$$
qui donne des divergences ordinaires en $\ep^{-1}$ et $|\ln\ep|$ pour,
respectivement, les fonctions \`a deux et quatre points.

La partie non-planaire est alors donn\'ee par
\begin{align*}
\Ga_{1l,NP}^\ep[\vf]&=(2\pi)^{-4}\int_\ep^\infty dt\,
\frac{e^{-tm^2}}{(4\pi t)^2}\,
\int_{\R^{12}}\,d^4x\,d^4y\,d^4z\,e^{-i<y,z>}\,
e^{-\frac{|\Th y|^2}{4t}}
\Big(\frac{1}{2}\frac{\lambda}{3!}\vf(x)\vf(x+z)\\
&\quad-\frac{\lambda^2}{(3!)^2}\frac{t}{4}
\big(3\vf\Mop\vf(x)\vf\Mop\vf(x+z)
+4\vf(x)\vf^{\Mop 3}(x+z)\big)\Big)
+O(\ep^0).
\end{align*}
L'int\'egration  Gaussienne sur $y$ peut \^etre effectu\'ee
\begin{align*}
\Ga_{1l,NP}^\ep[\vf]&=(2\pi\th)^{-4}\int_\ep^\infty dt\,e^{-tm^2}\,
\int_{\R^8}\,d^4x\,d^4z\,e^{-t|\Th^{-1}(z-x)|^2}\\
&\quad\times\Big(
\frac{1}{2}\frac{\lambda}{3!}\vf(x)\vf(z)
-\frac{\lambda^2}{(3!)^2}\frac{t}{4}\big(3\vf\Mop\vf(x)\vf\Mop\vf(z)
+4\vf(x)\vf^{\Mop 3}(z)\big)\Big)+O(\ep^0),
\end{align*}
o\`u $\th:=(\det\Th)^{1/4}$. Finalement, l'int\'egration sur $t$ aboutit \`a
\begin{align*}
\Ga_{1l,NP}^\ep[\vf]&=(2\pi\th)^{-4}
\int_{\R^8}\,d^4x\,d^4z\,
\frac{e^{-\ep(m^2+|\Th^{-1}(z-x)|^2)}}{m^2+|\Th^{-1}(z-x)|^2}\\
&\quad\times\Big(
\frac{\lambda}{2.3!}\vf(x)\vf(z)
-\frac{\lambda^2}{(3!)^2}\frac{3\vf\Mop\vf(x)\vf\Mop\vf(z)
+4\vf(x)\vf^{\Mop 3}(z)}{4(m^2+|\Th^{-1}(z-x)|^2)}\Big)+O(\ep^0).
\end{align*}
Cette expression est r\'eguli\`ere lorsque $\ep$ tend vers z\'ero. Nous verrons
que malgr\'e ce qui est commun\'ement admis, ce n'est pas le cas lorsque $l=2$.

Rappelons que l'action effective 1PI (une particule irr\'eductible)
est reli\'ee aux fonctions de Green 1PI
par
$$
\Ga_{1l}^\ep[\vf]=\sum_{n=0}^\infty\frac{1}{n!}\langle G_n^\ep;\vf,\cdots,\vf\rangle,
$$
o\`u le couplage multilin\'eaire est donn\'e par
$$
\langle G_n^\ep;\vf_1,\cdots,\vf_n\rangle:=
\int_{M\x\cdots\x M}\mu_g(p_1)\cdots \mu_g(p_n)\,
G_n^\ep(p_1,\cdots,p_n)\,\vf_1(p_1)\cdots\vf_n(p_n).
$$
On peut alors extraire des formules pr\'ec\'edentes, par d\'erivation fonctionnelle,
les fonctions de Green
non planaires \`a deux et quatre points dans l'espace de configuration dans la limite
$\ep\rightarrow 0$:
\begin{align*}
&G_{1l,NP}^2(x,y)=(\pi\th)^{-4}\frac{\lambda}{96}
\frac{1}{m^2+|\Th^{-1}(x-y)|^2},\\
&G_{1l,NP}^4(x,y,z,u)=-(\pi\th)^{-8}
\frac{\lambda^2}{24}\Big(\frac{3}{2}\delta(x-y+z-u)
\,\int d^4v\,\frac{e^{2i<v,\Th^{-1}(u-z)>}}{(m^2+|\Th^{-1}(z-v-x)|^2)^2}\\
&\hspace{6cm}+\frac{e^{2i<x-y,\Th^{-1}(z-y)>}}
{(m^2+|\Th^{-1}(x-y+z-u)|^2)^2}\Big).
\end{align*}

{\it Le ph\'enom\`ene de m\'elange UV/IR dans l'espace de positions se manifeste
dans le comportement \`a l'infini des fonctions de corr\'elations:}\\
Par transform\'ee de Fourier, on observe que la lente d\'ecroissance
\`a l'infini de ces fonctions \'equivaut \`a une singularit\'e
infrarouge
$$
\widehat{G^2}_{1l,NP}(\xi,\eta) \propto \frac{m}{|\Th\xi|}
K_1(m|\Th\xi|)\delta(\xi+\eta),
$$
o\`u  $K_n(z)$ d\'esigne la $n$-i\`eme fonction de Bessel modifi\'ee.
On retrouve le m\'elange UV/IR dans sa forme habituelle (voir
par exemple \cite{FRR}) car
$$
\frac{m}{|\Th\xi|}K_1(m|\Th\xi|)\sim(|\Th\xi|)^{-2},\;|\xi|\to 0.
$$
Ce dernier r\'esultat co\"incide avec celui usuellement obtenu par calcul de diagrammes
de Feynman dans l'espace des moments.\\
Pour la fonction \`a quatre points, les deux approches ne co\"incident pas tout \`a
fait:  le d\'evelop\-pe\-ment que nous avons utilis\'e ici n'est
qu'\'equivalent au d\'eveloppement en diagrammes de Feynman, dans le sens o\`u ils
diff\`erent d'une quantit\'e finie. On a cependant co\"incidence pour la fonction
\`a deux points.

Les comportements des amplitudes en position et en impulsion
pour $\th\downarrow0$ pr\'esentent aussi d'int\'eressantes diff\'erences. Supposons
pour simplifier que $\Th$ ait la forme canonique:
$$
\Th =\th\, S,
$$
o\`u la matrice $S$ a \'et\'e d\'efinie dans l'\'equation \eqref{matS}.
En d\'eveloppant la fonction \`a deux points en puissance de~$\th$,
on trouve
$$
\frac{1}{\th^4 m^2 + \th^2|x|^2} = \frac{1}{\th^2|x|^2} \biggl(1 -
\frac{\th^2m^2}{|x|^2} + \frac{\th^4m^4}{|x|^4} - \cdots \biggr).
$$
On remarque que la d\'ependance
logarithmique en $\th$ du m\'elange UV/IR dans l'espace
des moments (en addition \`a celle quadratique),
trouv\'ee dans~\cite{FRR}, est apparemment absente dans l'espace
de configuration.
Aussi, \`a la seule exception du premier terme, la s\'erie pr\'ec\'edente
est compos\'ee de fonctions qui ne sont pas des distributions
temp\'er\'ees; elles n'ont alors pas de transform\'ees de Fourier.
En d'autres termes, {\it la `limite commutative' ne commute pas avec la
transform\'ee de Fourier.}

Le probl\`eme est en fait plus subtil. On peut se demander \`a quel
type de divergences le d\'eveloppement pr\'ec\'edent est associ\'e.
La r\'eponse est que le premier terme diverge dans l'infra\-rouge (en
position), le deuxi\`eme dans l'infrarouge et l'ultraviolet et tous les autres
dans l'ultra\-violet. Il est peut \^etre surprenant que l'on puisse retrouver le
r\'esultat exact \`a partir de cette s\'erie mal d\'efinie, en invoquant
pr\'ecis\'ement une correction aux divergences UV indiqu\'ees.
On peut en effet ``renormaliser'' (dans le sens d'Epstein et Glaser)
les fonctions $1/|x|^{2k+4}$, avec pour r\'esultat des distributions
temp\'er\'ees $[1/|x|^{2k+4}]_R$, qui d\'ependent d'une \'echelle
de masse $\mu$. Leurs transform\'ees de Fourier, $\widehat{[1/|x|^{2k+4}]_R}$,
ont \'et\'e calcul\'ees dans~\cite{Carme,NR}:
$$
\widehat{[1/|x|^{2k+4}]_R}(\xi) =
\frac{(-)^{k+1}|\xi|^{2k}}{4^{k+1}k!(k+1)!}
\biggl[2\ln\frac{|\xi|}{2\mu} - \Psi(k+1) - \Psi(k+2)\biggr].
$$
Le param\`etre de masse naturel dans notre contexte \'etant $1/\th m$,
c'est en l'identifiant avec le param\`etre libre $\mu$, que l'on peut
retrouver exactement le r\'esultat en re-sommant la s\'erie des
transform\'ees de Fourier des distributions $[1/|x|^{2k+4}]_R$:
$$
\frac{1}{\theta^2|\xi|^2}+
\frac{m^2}{2} \sum_{n=0}^\infty \frac{\th^{2n}m^{2n}|\xi|^{2n}}{4^n\,n!(n+1)!}
\biggl(\ln\frac{\th m|\xi|}{2} - \Psi(n+1) - \Psi(n+2)\biggr) =
\frac{m}{\th|\xi|}K_1(\th m|\xi|).
$$

\subsubsection{4.2.1.2 Cas d\'eg\'en\'er\'es}

L'effet du rang de la matrice de d\'eformation
sur la renormalisabilit\'e de la th\'eorie,
devient lui aussi plus clair dans l'espace de configuration.
Pour un plan de Moyal de dimension $n$ avec une matrice
de d\'eformation de rang $l$ et une th\'eorie $\vf^{\Mop 4}$
\`a une boucle, la fonction \`a deux points dans l'espace des
moments se comporte
comme $|\Th\xi|^{-n+2}$, lorsque $\xi\to 0$.
Cependant, \'etant donn\'e que $\Th\xi\in\im(\Th)=\R^l$, la singularit\'e
infrarouge n'est pas localement int\'egrable si $l\leq n-2$. Il s'ensuit
que la fonction de Green \`a deux points ne d\'efinit pas une distribution
temp\'er\'ee et n'a  pas de transform\'ee de Fourier. \\
{\it Pour le `tadpole', en dimension quatre, la contribution
non planaire en position reste alors divergente lorsque $l=2$.}\\
La situation est plus cl\'emente pour la fonction \`a quatre points, car
la singularit\'e IR est int\'egrable d\`es que $l\ne 0$.
Cet aspect du m\'elange va aussi se manifester
dans le cas g\'en\'eral des d\'eformations isospectrales non p\'eriodiques
o\`u nous verrons que la partie non planaire de l'action effective reste
divergente pour $l=2$.

Nous allons terminer ce paragraphe consacr\'e aux d\'eformations
isospectrales non p\'eriodiques plates, en illustrant cette discussion
par le calcul de la fonction \`a deux points
en r\'egularisation $\zeta$. Cela permettra de comparer
nos r\'esultat avec ceux pr\'esents dans la litt\'erature.
Pour le restant du paragraphe, nous allons consid\'erer une matrice de
d\'eformation du type
\begin{equation*}
    \Theta=\begin{pmatrix} 0 & 0 \\
                           0 & \theta S_2
           \end{pmatrix}, ~~~{\rm avec}~~~
     S_2 = \begin{pmatrix} 0 & 1 \\
                         -1 & 0
          \end{pmatrix}~.
\end{equation*}
Il sera alors commode d'introduire la notation
$\R^4\ni x=(\tilde{x},\bar{x})$, avec $\tilde{x}=(x^0,x^1)$ et
$\bar{x}=(x^2,x^3)$. De m\^eme, on \'ecrira $p=(\tilde p,\bar p)$
dans l'espace des moments.

Consid\'erons la r\'egularisation $\zeta$:
\begin{equation*}
    \Gamma_{1l}^{\sigma}[\vf] = -\frac{\mu^{2\sigma}}{2}\, \int_0^\infty
      dt\,t^{\sigma-1} ~
        \Tr\, \big({\rm e}^{-tH} - {\rm e}^{-tH_0} \big) \,.
\end{equation*}
La fonction complexe ainsi d\'efinie
poss\`ede un prolongement analytique en la variable $\sigma$ sur $\C\setminus\{1,0,-1,-2,\cdots\}$.

En effectuant le m\^eme d\'eveloppement que pr\'ec\'edemment, on obtient
pour la contribution correspondant \`a la fonction \`a deux points
\begin{align*}
\Gamma_{1l,2P}^{\sigma}[\vf] &=  \frac{\,\la\,\mu^{2\sigma}\,}{12}
\int_0^\infty \! dt\, t^{\sigma}\,{\rm e}^{-tm^2}\,\Tr\,
\big[\,\bigl( L_{\vf\star\vf} + R_{\vf\star\vf}
+ L_{\vf}\,R_{\vf}\bigr) \,{\rm e}^{-t\tri}\,\big]\\
&=\frac{\la m^2}{\,96\pi^2\,}
\bigg(\frac{\mu^2}{m^2}\bigg)^{\sigma}\Gamma(\sigma-1)
\int d^4x\,\vf^2(x)\\
 &\quad+ \frac{\la}{\,192\pi^3\,\theta^2}
 \bigg(\frac{\mu^2}{m^2}\bigg)^{ \sigma}\Gamma(\sigma)
 \int\! d^4\!x \int \! d^2\!\bar{u}\,
 \vf(\tilde{x},\bar{x}) \vf(\tilde{x},\bar{x}+\bar{u}) \bigg(1
 + \frac{\bar{ u}^2}{m^2\theta^2}\bigg)^{-\,\sigma}.
\end{align*}
On peut alors extraire les fonctions de Green associ\'ees.
On obtient pour la partie planaire,
\begin{equation*}
     \Gamma_{1l,2P,P}^{\sigma}(x_1-x_2) =
     \frac{\la m^2}{\,48\pi^2\,}~
     \bigg(\frac{\mu^2}{m^2}\bigg)^{ \sigma}~\Gamma(\sigma-1)~
     \delta^{(4)}(x_1-x_2)\,,
\end{equation*}
et pour celle non planaire
\begin{equation*}
     \Gamma_{1l,2P,NP}^{\sigma}(x_1-x_2) =
     \frac{\la}{\,96\pi^3\theta^2\,}~
     \bigg(\frac{\mu^2}{m^2}\bigg)^{ \sigma}~\Gamma(\sigma)~
     \delta^{(2)}(\tilde{x}_1-\tilde{x}_2)~ \bigg[\, 1
      + \frac{(\bar{x}_1-\bar{x}_2)^2}{m^2\theta^2}\,\bigg]^{-\sigma}\,.
\end{equation*}
{\it La contribution non planaire est alors ni finie ni locale lorsque $\sigma$
tend vers $0$.
Cette non-localit\'e rend alors la th\'eorie \`a une boucle non-renormalisable par
adjonction de contre-termes locaux},  au sens o\`u de tels termes ne
sont pas pr\'esents dans l'action classique.

Pour comparer ce r\'esultat avec ceux de la litt\'erature, il
est commode de calculer la transform\'ee de Fourier de cette
fonction de Green:
\begin{equation*}
    G_{1l,2P,NP}^{^\sigma}(\tilde{p},\bar{p})
         := \int d^4\!z~ {\rm e}^{-ipz}\,
           \Gamma_{2P,NP}^{\sigma}(z)
         = \frac{\la m^2}{\,24\pi^2\,}~
           \bigg(\frac{\mu^2}{2m^2}\bigg)^{\!\!\sigma}~
           \frac{\,K_{1-\sigma}(\theta m |\bar{p}|)\,}
                {(\theta m |\bar{p}|)^{1-\sigma}}\,.
\end{equation*}
Apparemment, la limite $\sigma\to0$ de cette derni\`ere expression
existe et on est tent\'e d'\'ecrire
\begin{equation*}
    \lim_{\sigma\to0}G^\sigma_{1l,2P,NP}(\tilde{p},\bar{p}) =
    \frac{\,\la m^2}{24\pi^2}~
        \frac{K_1(\theta m |\bar{p}|)}{\theta m |\bar{p}|}~.
\end{equation*}
Bien que finie, cette limite n'existe pas au sens des distributions, car encore
une fois la singularit\'e infrarouge n'est pas int\'egrable.

\subsection{Les divergences dans le cas g\'en\'eral non p\'eriodique}
\label{gnp}

Dans cette partie, nous supposerons que $\vf\in\Coo_c(M)$
et que l'une des hypoth\`eses $(I)$ ou $(II)$ du
paragraphe \ref{section26} soit satisfaite, de mani\`ere
\`a ce que l'une des estimations (\ref{eq:k1}) ou (\ref{eq:k2})
sur le noyau de la chaleur soit valable.

Dans le cas g\'en\'eral (p\'eriodique ou non), nous avons vu que
$\Ga^{\ep}_{1l,NP}[\vf]$ est donn\'ee par
\begin{align*}
\Ga_{1l,NP}^\ep[\vf]&=\frac{1}{2(2\pi)^l}\int_\ep^\infty dt\,e^{-tm^2}
\int_M\mu_g(p)\,\int_{\R^{2l}}\,d^ly\,d^lz\,
e^{-i<y,z>}\,K_t(-\Th y.p,p)\\
&\quad\times\Big\lbrace\frac{\lambda}{3!}\vf(p)\vf(z.p)
-\frac{t}{2}\frac{\lambda^2}{(3!)^2}\big(3\vf\Mop\vf(p)\vf\Mop\vf(z.p)
+4\vf(p)\vf^{\Mop 3}(z.p)
\Big)\Big\rbrace+O(\ep^0).
\end{align*}
Nous allons montrer que cette expression ne peut pas produire
de divergences plus importantes que celles du secteur planaire.
Encore une fois, la r\'egularit\'e de ces int\'egrales d\'epend uniquement
des donn\'ees g\'eom\'etriques, \`a savoir du rang $l$ de la matrice
de d\'eformation (que l'on pourrait
appeler la dimension non commutative effective) et de la m\'etrique
par l'interm\'ediaire de la distance Riemannienne.

Avant d'estimer la partie \`a deux points de $\Ga_{1l,NP}^\ep[\vf]$,
qui est le propos principal de ce paragraphe, il est int\'eressant de remarquer
que la fonction de Green \`a deux points non planaire, a
la forme suivante
$$
G^\ep_{1l,NP,2P}(p,p')=\frac{\lambda}{6(2\pi)^l}\int_{\R^{2l}}d^ly\,d^lz\,
e^{-i<y,z>}\int_\ep^\infty dt\,e^{-tm^2}\,K_t(-\Th y.p,p)\,\delta^g_{z.p}(p').
$$
Dans cette expression distributionnelle, on peut alors lire qualitativement
le ph\'enom\`ene de m\'elange: en utilisant les estimations du noyau
de la chaleur hors diagonale \cite{Davies, Chavel}, valables lorsque la courbure
de Ricci est born\'ee inf\'erieurement et lorsque soit l'inverse du volume soit
de la constante isop\'erim\'etrique d'une
boule de rayon fix\'e est uniform\'ement born\'ee:
\begin{equation}
\label{heatkernelestimate}
(4\pi t)^{-2}e^{-d_g^2(p,p')/4t}\leq K_t(p,p')
\leq C(4\pi t)^{-2}e^{-d_g^2(p,p')/4(1+c)t}.
\end{equation}
On obtient alors l'in\'egalit\'e
\begin{align*}
\int_0^\infty dt\,e^{-tm^2}\,K_t(\Th y.p,p)&\leq
C\int_0^\infty dt\,\frac{e^{-tm^2}}{(4\pi t)^2}e^{-d_g^2(\Th y.p,p)/4(1+c)t}\\
&=\frac{C}{16\pi^2}\,\frac{4m\sqrt{1+c}}{d_g(\Th y.p,p)}\,
K_1\big(\frac{m\,d_g(\Th y.p,p)}{\sqrt{1+c}}\big)\\
&\sim C'\,d_g^{-2}(\Th y.p,p),
\hspace{0.5cm} y\to 0,
\end{align*}
ainsi que celle inverse:
$$
\int_0^\infty dt\,e^{-tm^2}\,K_t(\Th y.p,p)\geq  C''\,d_g^{-2}(\Th y.p,p).
$$
Ces estimations mettent en lumi\`ere le m\'elange UV/IR, car
$y\in\widehat{\R^l}$ doit \^etre interpr\'et\'e comme un moment.

Pour la partie \`a deux points de $\Ga_{1l,NP}^\ep[\vf]$, on a
\begin{align*}
\left|\Ga_{1l,NP,2P}^\ep[\vf]\right|&\leq
\frac{C\,\lambda}{12(2\pi)^l}\,\sup_{p\in M}\left\{\int_{\R^l} \,
d^lz\left|\vf(z.p)\right|\right\}\\
&\hspace{2cm}\times\int_\ep^\infty dt\,\frac{e^{-tm^2}}{(4\pi t)^2}\,
\int_M\mu_g(p)\,\left|\vf(p)\right|\int_{\R^l}\,d^ly\,
e^{-d_g^2(-\Th y.p,p)/4(1+c)t}\\
&\leq
\frac{C\,\lambda}{12(2\pi)^l}\,\sup_{p\in M}\left\{\int_{\R^l} \,
d^lz\left|\vf(z.p)\right|\right\}\|\vf\|_1\\
&\hspace{2cm}\times\sup_{p\in \supp(\vf)}
\left\{\int_\ep^\infty dt\,\frac{e^{-tm^2}}{(4\pi t)^2}\,
\int_{\R^l}\,d^ly\,
e^{-d_g^2(-\Th y.p,p)/4(1+c)t}\right\}.
\end{align*}
D'apr\`es le lemme \ref{lm:sup}, $\sup_{p\in
M}\left\{\int_{\R^l} \, d^lz\left|\vf(z.p)\right|\right\}<\infty$.
De plus, $\a$ agissant isom\'etriquement, la
m\'etrique induite sur les orbites est constante, orbites qui sont des sous-vari\'et\'es
ferm\'ees car l'action est propre \cite{Michor}. Par cons\'equent,
$$
d_g^2(y.p,p)=\sum_{i,j=1}^l\tilde{g}_{ij}(p)y^iy^j,
$$
o\`u les fonctions $\tilde{g}_{ij}(p)$ ne d\'ependent que de l'orbite du point $p$ et
sont strictement positives et continues; dans le cas
non p\'eriodique, l'action est libre et donc $\{(0,p)\in \R^l\times M\}$
est le seul ensemble pour lequel $F(y,p):=d_g( y.p,p)$ s'annule.
Notons aussi que l'on peut trouver un syst\`eme de coordonn\'ees
globales sur  chaque orbite, en choisissant une base de $\R^l$
rendant la m\'etrique induite $\tilde{g}_{ij}(p)$ diagonale. Ainsi, avec
$\th:=(\det\Th)^{1/l}$, on a
$$
\int_{\R^l}\,d^ly\,
e^{-d_g^2(-\Th y.p,p)/4(1+c)t}=
\left(\frac{4\pi(1+c) t}{\th^2}\right)^{l/2}(\det\tilde{g}(p))^{-1/2}.
$$
On obtient alors
$$
\left|\Ga_{1l,NP,2P}^\ep[\vf]\right|\leq
\frac{\lambda}{6}C(l,\tilde{g},\vf,\vf)\,\th^{-l}
\int_\ep^\infty dt\, t^{l/2-2}e^{-tm^2},
$$
avec
\begin{align*}
C(l,\tilde{g},\vf_1,\vf_2)&:=\\
&\frac{C(4\pi)^{l/2-2}(1+c)^{l/2}}{2(2\pi)^l}\|\vf_1\|_1
\sup_{p\in M}\left\{
\int_{\R^l}d^lz\left|\vf_2(z.p)\right|\right\}
\sup_{p\in\supp(\vf_1)}\left\{(\det\tilde{g}(p))^{-1/2}\right\}.
\end{align*}
Des estimations similaires sont valables pour la fonction \`a quatre points:
$$
\left|\Ga_{1l,NP,4P}^\ep[\vf]\right|\leq
\frac{\lambda^2}{72}\,\th^{-l}\,\Big(3C(l,\tilde{g},\vf\Mop\vf,\vf\Mop\vf)
+4C(l,\tilde{g},\vf,\vf\Mop\vf\Mop\vf)\Big)
\int_\ep^\infty dt\, t^{l/2-1}e^{-tm^2}.
$$

Nous avons alors d\'emontr\'e le th\'eor\`eme suivant:
\begin{thm}
Lorsque $M$ est non compacte, satisfait \`a une des hypoth\`eses
$(I)$ ou $(II)$, et munie d'une action isom\'etrique lisse et propre du
groupe $\R^l$, alors
pour tout $\vf\in\Coo_c(M)$ on a:
\item{i)}
$$
\left|\Ga_{1l,NP,2P}^\ep[\vf]\right|\leq\begin{cases}
   C_1(\vf,\Th)  &\text{pour $l=4$}, \\
   C_2(\vf,\Th)|\ln\ep| &\text{pour $l=2$}, \end{cases}
$$
\item{ii)}
$$
\left|\Ga_{1l,NP,4P}^\ep[\vf]\right|\leq
   C_3(\vf,\Th) \, \,\, \,\text{pour $l=4$ ou $l=2$}.
$$
\end{thm}
\noindent
Les possibles divergences restantes  r\'ef\`erent au fait que la singularit\'e IR peut ne pas
\^etre localement int\'egrable, comme illustr\'e pr\'ec\'edemment.

\begin{cly}
Sous les hypoth\`eses du th\'eor\`eme pr\'ec\'edent,
lorsque $l=2$, la th\'eorie quantique \`a une boucle
n'est pas renormalisable par adjonction de contre-termes `locaux'.
\end{cly}

\section{D\'eformations p\'eriodiques}

Les th\'eories quantiques des champs sur d\'eformations p\'eriodiques
se comportent diff\'eremment des non p\'eriodiques.
La raison vient de la diff\'erence structurelle de l'espace des `moments',
d\'efini via l'action $\a$. En effet, le groupe dual du groupe agissant
sur la vari\'et\'e \'etant discret pour les d\'eformations p\'eriodiques (compactes
ou non), les divergences dans l'infrarouge du secteur non planaire n'apparaissent
que pour certaines valeurs discr\`etes des impulsions. Le ph\'enom\`ene du m\'elange
UV/IR est alors bien moins probl\'ematique. Nous verrons, dans les cas favorables
en utilisant la d\'ecomposition de Peter--Weyl du champ de `background', que l'on peut
extraire les modes divergents des fonctions de Green non planaires, pour les renormaliser
ensemble avec le secteur planaire.

Bien que la d\'ecomposition en sous-espaces spectraux (par rapport \`a l'action
de $\T^l$ sur l'espace de Hilbert) existe dans tous les cas
p\'eriodiques, nous ne serons capable de d\'ecrire le comportement individuel
des `graphes de Feynman' (d\'efinis via cette d\'ecomposition) que dans
le cas compact. Cette obstruction d'ordre technique provient du fait que le noyau de la
chaleur ne peut s'\'ecrire comme une somme discr\`ete (d'\'el\'ements
de  l'espace de Hilbert $\H\ox\H$) que lorsque la vari\'et\'e est
compacte. Cependant, nous verrons que les aspects analytiques fins de la th\'eorie
sont justement li\'es aux propri\'et\'es collectives des graphes de Feynman. Les
cons\'equences de cette obstruction sur notre analyse ne seront alors que minimes.
Ainsi, nous verrons comment
intervient la nature arithm\'etique des param\`etres de d\'eformation sur le
comportement analytique de l'action effective.

Nous supposons dans ce paragraphe que le noyau de l'action $\a$ soit un
r\'eseau entier, $\ker\a=\beta \Z^l$ avec $\beta\in M_l(\Z)$ et que son rang
soit maximal (i.e. $\R^l/\beta\Z^l=:\T_\beta^l$ est compact). Pour simplifier
les notations, nous supprimerons l'indice $\b$.

\subsection{Comportement individuel des graphes non planaires}

Revisitons tout d'abord le cas du tore non commutatif.
Soit $M=\T^4$ munie de la m\'etrique plate,
avec une action de $\R^4$ par rotation. En utilisant la base orthonorm\'ee
$\left\{\frac{e^{i<k,x>}}{(2\pi)^2}\right\}_{k\in\Z^4}$ de $L^2(\T^4,d^4x)$,
on peut \'ecrire le noyau de la chaleur comme
$$
K_t(x,y)=(2\pi)^{-4}\sum_{k\in\Z^4}e^{-t|k|^2}e^{i<k,x-y>}.
$$
Rappelons aussi que le produit d\'eform\'e
devient sur les modes de Fourier
$$
e^{i<k,x>}\Mop e^{i<q,x>}=e^{-\sihalf \Th(k,q)}\,e^{i<k+q,x>},
$$
avec $\Th(k,q):=\langle k,\Th q\rangle$. En d\'eveloppant le champ de
`background' $\vf$ dans la base des modes de Fourier
$\vf=\sum_{k\in\Z^4}c_k\,e^{i<k,x>}$, avec
$\{c_k\}_{k\in\Z^4}\in\SS(\Z^4)$ lorsque $\vf\in\Coo(\T^4)$, on obtient
\begin{align*}
\Ga_{NP}^\ep[\vf]&=\frac{1}{2}\sum_k
\frac{e^{-\ep(m^2+|k|^2)}}{m^2+|k|^2}\Big\lbrace
\frac{\lambda}{3!}\sum_r \,c_r\,c_{-r}\,e^{i\Th(k,r)}
   -\frac{\lambda^2}{2(3!)^2}
\frac{1}{m^2+|k|^2}\\
&\quad\times\sum_{r,s,u}\,c_r\,c_s\,c_{u-s}\,c_{-r-u}
   e^{-\sihalf\Th(r+s,u)}\Big(3\,e^{i\Th(k,r+s)}
+4\,e^{i\Th(k,r+u)}\Big)\Big\rbrace+O(\ep^0).
\end{align*}
On peut alors analyser le comportement individuel des diagrammes de
Feynman non planaires. La pr\'esence des phases dans l'expression pr\'ec\'edente
rend  finie la somme sur les impulsions, lorsque les entr\'ees de $(2\pi)^{-1}\Th$
sont irrationnelles et lorsque $r\ne 0$ pour la fonction \`a deux points, $r+s\ne 0$
et $r+u\ne 0$ pour celle \`a quatre points. En effet, en retournant \`a la
param\'etrisation de Schwinger, qui \'echange les divergences aux grands moments $k$
contre des divergences aux petits temps $t$, on obtient en appliquant la formule de sommation
de Poisson par rapport \`a la somme sur $k$:
$$
\sum_{k\in\Z^4}\frac{e^{i \Th(k,r)}}{m^2+|k|^2}=
\sum_{k\in\Z^4} \int_0^\infty dt\,
\frac{e^{-tm^2}}{(4\pi t)^{2}}\,e^{-|2\pi k-\Th r|^2/4t}.
$$
L'int\'egrale sur $t$ est alors finie lorsque $r\ne 0$ et $\frac{\Th
r}{2\pi}\notin \Q^l$. A partir de consid\'erations similaires,
on obtient de semblables conclusions pour la fonction \`a quatre points.

Dans le cas g\'en\'erique p\'eriodique et compact, pour pouvoir
faire des calculs quasiment aussi explicites que dans le cas plat, nous allons tirer profit de
l'invariance du noyau de la chaleur sous $\a$.
D\'ecomposons $\H=L^2(M,\mu_g)$ en sous-espaces spectraux par
rapport \`a l'action induite de $\T^l$ par les op\'erateurs unitaires $V_z$:
$$
\H=\bigoplus_{k\in\Z^l}\H_k.
$$
Chaque $\H_k$ est stable sous $V_z$ pour tout $z\in\R^l$ et tout
$\psi\in\H_k$ satisfait \`a $V_z\psi=e^{-i<z,k>}\psi$. Notons aussi que si
$\psi\in\H_k$ alors $|\psi|\in\H_0$. Soit $P_k$ le projecteur
orthogonal sur $\H_k$. Le Laplacien commutant avec $V_z$,
le semi-groupe de la chaleur commute avec $P_k$; il est alors
diagonalisable en bloc par rapport \`a la d\'ecomposition
$\H=\bigoplus_{k\in\Z^l}\H_k$:
$$
e^{-t\tri}=\sum_{k\in\Z^l}\,P_k\,e^{-t\tri}\,P_k.
$$
L'op\'erateur $0\leq P_k\,e^{-t\tri}\,P_k$ \'etant \`a trace,
il peut \^etre d\'ecompos\'e dans chaque $\H_k$ comme
$$
P_k\,e^{-t\tri}\,P_k=\sum_{n\in\N}
e^{-t\lambda_{k,n}}|\psi_{k,n}\rangle\langle\psi_{k,n}|,
$$
o\`u $\{\psi_{k,n}\}_{n\in\N}$ est une base orthonorm\'ee de $\H_k$
consistant de vecteurs propres de $P_k\tri P_k$ avec valeurs propres
$\lambda_{k,n}$. Le semi-groupe de la chaleur \'etant aussi
un op\'erateur de Hilbert-Schmidt,
son noyau peut s'\'ecrire comme la s\'erie (convergente pour la
topologie normique de l'espace de Hilbert
$L^2(M\times M,\mu_g\times\mu_g)$):
\begin{equation}
\label{eq:hk}
K_t(p,p')=\sum_{k\in\Z^l}\sum_{n\in\N}
e^{-t\lambda_{k,n}}\psi_{k,n}(p)\overline{\psi_{k,n}}(p').
\end{equation}
Puisque chaque $\psi_{k,n}(p)$ appartient \`a $\H_k$, la propri\'et\'e
d'invariance du noyau de la chaleur $K_t(z.p,z.p')=K_t(p,p')$
est explicite.

Tout $\vf\in C^\infty(M)$ poss\`ede une unique
d\'ecomposition en `mode de Fourier' $\vf=
\sum_{r\in\Z^l}\vf_r$, telle que $\{\|\vf_r\|_\infty\}\in\SS(\Z^l)$
et $\a_z(\vf_r) =e^{-i<z,r>}\vf_r$.  Cette d\'ecomposition d\'efinit aussi une notion
de diagrammes de Feynman, c'est-\`a-dire l'amplitude associ\'ee \`a une configuration
de champ fix\'ee.\\
On obtient alors pour l'action effective (non planaire, r\'egularis\'ee et \`a une boucle):
\begin{align*}
\Ga_{1l,NP}^\ep[\vf]&=\frac{1}{2}\int_M\mu_g(p)\sum_{k\in\Z^l}\sum_{n\in\N}
\frac{e^{-\ep(m^2+\lambda_{k,n})}}
{m^2+\lambda_{k,n}}\, |\psi_{k,n}|^2(p)\Big\lbrace
\frac{\lambda}{3!}\sum_{r,s\in\Z^l} \,
\vf_r(p)\,\vf_s(p)\,e^{-i\Th(k,s)} \\
&\quad-\frac{\lambda^2}{2(3!)^2}
\frac{1}{m^2+\lambda_{k,n}}\sum_{r,s,u,v\in\Z^l}\,
\vf_r(p)\,\vf_s(p)\,\vf_u(p)\,\vf_v(p)\\
&\quad\hspace{1cm}\times\Big(3\,e^{-\sihalf(\Th(r,s)+\Th(u,v))}
e^{-i\Th(k,u+v)}
+4\,e^{-\sihalf \Th(r+s,u+s)}
e^{-i\Th(k,v)}\Big)\Big\rbrace+O(\ep^0).
\end{align*}
Bien que ne connaissant pas l'expression
explicite des vecteurs propres $\psi_{k,n}$, les sommes pr\'ec\'e\-den\-tes se
r\'eduisent par conservation des moments, comme pour le tore non commutatif.
\begin{lem}[Conservation des moments]
\label{lem:u}
Soient $\psi_i\in\H_{k_i}\cap L^q(M,\mu_g)$ pour $i=1,\dots,q$. Alors
$$
\int_M\mu_g\,\psi_1\cdots\psi_q=
C(\psi_1,\cdots,\psi_q)\,\delta_{k_1+\cdots +k_q,0}.
$$
\end{lem}
\begin{proof}[Preuve]
En utilisant l'invariance de la forme volume sous $\a$
et la relation
$\a_z(\psi_i)=e^{-i<z,k_i>}\psi_i$, on obtient
$$
\int_M\mu_g\,\psi_1\cdots\psi_q=
e^{i<z,k_1+\cdots+k_q>}\int_M\mu_g\,\psi_1\cdots\psi_q,
$$
pour tout $z\in\T^l$ et le r\'esultat s'ensuit.
\end{proof}

Vu que $|\psi_{k,n}|^2(p)$ est constant sur les orbites de
l'action et que $\vf_r\in\Coo(M)\subset L^q(M,\mu_g)$
pour tout $q\in\N^*$, le lemme \ref{lem:u} donne
\begin{align}
\label{trs}
\Ga_{1l,NP}^\ep[\vf]&=\frac{1}{2}\int_M\mu_g(p)\sum_{k\in\Z^l}\sum_{n\in\N}
\frac{e^{-\ep(m^2+\lambda_{k,n})}}{m^2+\lambda_{k,n}}\,
|\psi_{k,n}|^2(p)\Big\lbrace
\frac{\lambda}{3!}\sum_{r\in\Z^l} \,\vf_r(p)\,
\vf_{-r}(p)\,e^{i<k,\Th r>} \nonumber\\
&\quad-\frac{\lambda^2}{2(3!)^2}
\frac{1}{m^2+\lambda_{k,n}}\sum_{r,s,u\in\Z^l}\,
\vf_r(p)\,\vf_s(p)\,\vf_{u-s}(p)\,\vf_{-r-u}(p)\nonumber\\
&\quad\hspace{2cm}\times e^{-\sihalf\Th(r+s,u)}\Big(3\,e^{i\Th(k,r+s)}
+4\,e^{i\Th(k,r+u)}\Big)\Big\rbrace+O(\ep^0).
\end{align}
Pour analyser les divergences d'une configuration de
champ fix\'ee lorsque $\ep\rightarrow 0$, il suffit de remarquer en
renum\'erotant de mani\`ere usuelle (par
ordre croissant en en comptant les multiplicit\'es) les valeurs propres du Laplacien
$\lambda_{k,n}$ que
$$
\sum_{k\in\Z^l}\sum_{n\in\N}\frac{|\psi_{k,n}|^2(p)}{(m^2+\lambda_{k,n})^N}
=\sum_{n\in\N}\frac{|\psi_n(p)|^2}{(m^2+\lambda_n)^N}
=K_{(m^2+\tri)^{-N}}(p,p)
$$
qui est fini si $N>2$ car
le Th\'eor\`eme de Weyl affirme que $\lambda_n\sim n^{1/2}$.
Ainsi, les sommes sur $n$ et sur $k$ dans~(\ref{trs})
divergent dans la limite $\ep\rightarrow 0$ pour certaines valeurs des moments
($r=0$ pour la fonction \`a deux points, $r+s=0$ et
$r+u=0$ pour celle \`a quatre points) si les entr\'ees de $(2\pi)^{-1}\Th$ sont irrationnelles.
Lorsqu'elles sont rationnelles, il y a un nombre infini de configurations
divergentes, car $e^{-i<k,\Th r>}=1$ pour une infinit\'e de $k$ lorsque
$\frac{\Th r}{2\pi}\in\Q^l$. Nous verrons dans le paragraphe suivant
que la convergence est garantie pour les autres configurations
par les estimations~(\ref{eq:k1}), (\ref{eq:k2}).

{\it En r\'esum\'e, nous avons montr\'e que pour
toutes les d\'eformations isospectrales p\'eriodiques compactes, les
configurations individuelles du secteur non planaire
reproduisent les principales particularit\'es du tore non commutatif.}

Dans le paragraphe suivant, nous verrons l'importance de la nature arithm\'etique
des entr\'ees de la matrice de d\'eformation ainsi que de la structure des
points fixes pour l'action.

\subsection{Condition Diophantienne et points fixes}

Nous supposons toujours l'action p\'eriodique, mais maintenant $M$ peut
\^etre compact ou non. Lorsque que $M$ ne l'est pas, nous supposerons aussi
qu'une des hypoth\`eses $(I)$ ou $(II)$ sur le comportement de la
g\'eom\'etrie \`a l'infini (voir paragraphe \ref{section26}) soit satisfaite.
Sous ces hypoth\`eses, la d\'ecomposition de Peter-Weyl existe toujours,
mais le semi-groupe de la chaleur n'est plus a priori un op\'erateur compact;
l'\'ecriture (\ref{eq:hk}) n'est alors plus valable.  Nous retournons donc aux
estimations du noyau de la chaleur hors diagonale. En utilisant la
conservation des moments (lemme \ref{lem:u}) ainsi que l'invariance
de  $K_t$ sous $\a$, on obtient
\begin{align*}
\Ga_{1l,NP}^\ep[\vf]&=\frac{1}{2}
\int_\ep^\infty\,dt\,e^{-tm^2}\int_M\mu_g(p)\Big\lbrace
\frac{\lambda}{3!}\sum_{r\in\Z^l}K_t(\Th r.p,p)\vf_r(p) \,\vf_{-r}(p)\\
&\quad-\frac{t\,\lambda^2}{2(3!)^2}
\sum_{r,s,u\in\Z^l}\,\vf_r(p)\,\vf_s(p)\,\vf_{u-s}(p)\,\vf_{-r-u}(p)
\,\,e^{-\sihalf \Th(r+s,u)}\\
&\quad\hspace{3cm}\times \Big(
3\,K_t(\Th(r+s).p,p)
+4\,K_t(\Th (r+u).p,p)\Big)\Big\rbrace
+O(\ep^0).
\end{align*}
Le cas rationnel ayant d\'ej\`a \'et\'e discut\'e au paragraphe pr\'ec\'edent,
nous nous restreindrons ici au cas irrationnel. Le m\'elange UV/IR
n'apparaissant que lorsque $r=0$ pour la fonction \`a deux points et lorsque $r+s=0$,
$r+u=0$ pour celle \`a quatre points, il est naturel de d\'efinir \emph{l'action effective non-planaire
r\'eduite} $\Ga_{1l,NP}^{\ep,red}[\vf]$, en soustrayant les configurations
divergentes. Ces derni\`eres doivent \^etre trait\'ees pour la renormalisation avec le
secteur planaire.
\begin{align*}
\Ga_{1l,NP}^{\ep,red}[\vf]&:=\frac{1}{2}
\int_\ep^\infty\,dt\,e^{-tm^2}\int_M\mu_g(p)\Big\lbrace
\frac{\lambda}{3!}\tsum' K_t(\Th r.p,p)\vf_r(p) \,\vf_{-r}(p)\\
&\quad\hspace{3cm}-\frac{t\,\lambda^2}{2(3!)^2}
\tsum'\,\vf_r(p)\,\vf_s(p)\,\vf_{u-s}(p)\,\vf_{-r-u}(p)
e^{-\sihalf \Th(r+s,u)}\\
&\quad\hspace{5cm}\times\Big(
3\,K_t(\Th(r+s).p,p)+4\,K_t(\Th (r+u).p,p)
\Big)\Big\rbrace.
\end{align*}
Ici $\tsum'$ est une notation qui signifie $\sum_{r\in\Z^l,\,r\ne 0}$ dans la partie
\`a deux points, $\sum_{r,s,u\in\Z^l,\,r+s\ne 0}$ et
$\sum_{r,s,u\in\Z^l,\,r+u\ne 0}$ dans respectivement le premier et
le second terme de la partie \`a quatre points.
En utilisant les estimations du noyau de la chaleur et en effectuant l'int\'egrale
sur $t$, on obtient
\begin{align}
\label{belll}
\hspace{-0.5cm}\lim_{\ep\to 0}\left|\Ga_{1l,NP}^{\ep,red}[\vf]\right|
&\leq \frac{C}{32\pi^2}\int_M\mu_g(p)\Big\lbrace
\frac{\lambda}{3!}\tsum'|\vf_r(p) |\,|\vf_{-r}(p)|
\frac{4m\sqrt{1+c}}{d_g(\Th r.p,p)} K_1\big(\frac{m\,d_g(\Th
r.p,p)}{\sqrt{1+c}}\big)\nonumber\\
&\quad+\frac{\lambda^2}{2(3!)^2}\tsum'\,
|\vf_r(p)|\,|\vf_s(p)|\,|\vf_{u-s}(p)|\,|\vf_{-r-u}(p)|\nonumber\\
&\quad\x\Big(3K_0\big(\frac{m\,d_g(\Th(r+s).p,p)}{\sqrt{1+c}}\big)+
4K_0\big(\frac{m\,d_g(\Th(r+u).p,p)}{\sqrt{1+c}}\big)\Big)\Big\rbrace.
\end{align}

\begin{defn}
$\th\in\R^l\setminus\mathbb{Q}^l$
satisfait \`a une condition Diophantienne s'il existe deux constantes $C>0$, $\beta\geq0$
qui soient telles que pour tout $n\in\Z^l\setminus\{0\}$, on ait:
$$
\|\th n\|_{\T^l}:=\inf_{k\in\Z^l}|\th n+k|\geq C|n|^{-(l+\beta)}.
$$
\end{defn}
Cette condition caract\'erise
les nombres irrationnels qui sont ``loin des rationnels'', car
mal approxim\'es par les rationnels. Notons que l'ensemble des
nombres satisfaisant \`a une condition Diophantienne est de mesure
de Lebesgue pleine \cite{Oxtoby}.\\
Cette notion de condition Diophantienne s'\'etend naturellement
aux matrices \`a coefficients ir\-rationels.
\begin{defn}
Une Matrice $\Th\in M_l(\R^l\setminus\mathbb{Q}^l)$, de rang $l$,
satisfait \`a une condition Diophantienne s'il existe deux constantes $C>0$, $\beta\geq0$
qui soient telles que pour tout $n\in\Z^l\setminus\{0\}$, on ait:
$$
\|\Th n\|_{\T^l}:=\inf_{k\in\Z^l}|\Th n+k|\geq C|n|^{-(l+\beta)}.
$$
\end{defn}
La m\'etrique induite sur les orbites est constante comme dans le paragraphe \ref{gnp},
et la distance g\'eod\'esique sur chaque orbite est par cons\'equent donn\'ee par
$$
d_g^2(y.p,p)=\inf_{k \in\Z^l}\left(\sum_{i,j=1}^l
\tilde{g}_{ij}(p)(y^i+k^i)(y^j+k^j)\right).
$$
Nous allons avoir besoin du comportement au voisinage de l'origine des fonctions
de Bessel modifi\'ees
$$
K_1(x)=\frac{1}{x}+O(x^0),\hspace{1cm}K_0(x)=
-\gamma+\ln(2)-\ln(x)+O(x),
$$
o\`u $\gamma$ est la constante d'Euler.  \\
Utilisons le fait que $\{\|\vf_r\|_\infty\}\in\SS(\Z^l)$,
avec $\vf_r\in\Coo_c(M)$ o\`u $\supp(\vf_r)\subset \supp(\vf)$: si
la fonction $d_g^{-2}(\a_y(.),.)$ est localement int\'egrable
pour tout $0\ne y\in\T^l$,
par rapport \`a la mesure donn\'ee par la forme volume Riemannienne,
{\it on a la convergence des sommes dans~(\ref{belll}) si et seulement si
$d_g^{-2}(\Th r.p,p)\in\SS'(\Z^l)$,
c'est-\`a-dire, si et seulement si la matrice $\Th$ satisfait \`a
une condition Diophantienne relativement \`a $(2\pi)$. }\\
Ce r\'esultat semble nouveau, bien que
conjectur\'e par Connes dans le cas du tore non commutatif il y a quelques
ann\'ees. Cette condition joue aussi un r\^ole primordial dans le calcul
de la cohomologie de Hochschild du tore non commutatif \cite{ConnesBook}
et en th\'eorie des champs conforme,
dans les mod\`eles de Melvin \`a param\`etre
de twist irrationnel~\cite{KMM}.

\smallskip

Cependant, {\it l'int\'egrabilit\'e locale de la fonction $d_g^{-2}(\a_y(.),.)$ pour un
$y\in\T^l$ non nul n'est en aucun cas garantie.}
Les probl\`emes sont \`a attendre dans un voisinage de l'ensemble des
points ayant un groupe d'isotropie non trivial, o\`u $d_g^{-2}(\a_y(.),.)$
n'est pas born\'ee. En fait, par une simple analyse
dimensionnelle, nous allons voir que des troubles s\'erieux devraient n'appara\^itre que lorsque les
stabilisateurs sont uni-dimensionnels.\\
Soit $H_p$ le groupe d'isotropie de $p\in M$ et soit
$M_{sing}:=\{p\in M:H_p\ne\{0\}\}$.  Rappelons que $M_{sing}$ est ferm\'e
(pour la topologie m\'etrique) et est de mesure nulle dans $M$ car l'action est propre
(voir \cite[th\'eor\`eme 6.15]{Michor}). Notons aussi que pour $0\ne y\in\T^l$, $d_g(y.p,p)=0$
si et seulement si $p\in M_{sing}$ et $y\in H_p$.  Sur $M_{reg}:=M\setminus M_{sing}$
(l'ensemble des orbites de type principal), l'action \'etant libre, on peut
d\'efinir des coordonn\'ees normales dans un voisinage tubulaire d'une orbite
$\T^l.p$ \`a un point $p\in M$. Soit $(\hat{x}^\mu,\tilde{x}^i)$,
$\mu=1,\cdots,n-l$, $i=1,\cdots,l$ respectivement les coordonn\'ees transverses
et celles donn\'ees par le groupe, d'un point $p\in M_{reg}$.
L'action \'etant isom\'etrique, la m\'etrique s'\'ecrit dans ce syst\`eme de coordonn\'ees
$$
g(\hat{x},\tilde{x})= \begin{pmatrix} h(\hat{x}) & l(\hat{x})\\
l(\hat{x}) & \tilde{g}(\hat{x}) \end{pmatrix},
$$
o\`u $\tilde{g}$ est la m\'etrique (constante) induite sur les orbites.
Un tel syst\`eme de coordonn\'ees est singulier, avec des singularit\'es
localis\'ees en chaque point de $M_{sing}$. Lorsque $x\equiv (\hat{x}^\mu,\tilde{x}^i)$
approche un point $p_0$ de $ M_{sing}$, $\tilde{g}(\hat{x})$ se r\'eduit \`a une matrice de rang
$l-dim(H_{p_0})$.  Dans ce syst\`eme de coordonn\'ees
$\mu_g(p)\,d^{-2}_g(y.p,p)$ s'\'ecrit
$$
\frac{\sqrt{\det
g(\hat{x},\tilde{x})}}{\sum_{i,j=1}^l\tilde{g}_{ij}(\hat{x})y^iy^j}
d^l\tilde{x}\,d^{n-l}\hat{x},
$$
lorsque $dim(H_{p_0})=1$ la singularit\'e de $d^{-2}_g(y.p,p)$ pour $p\to p_0$
ne peut \^etre compens\'ee par le facteur $\sqrt{\det g}$.\\
{\it Nous venons de mettre en lumi\`ere un
nouvel aspect du ph\'enom\`ene de m\'elange UV/IR pour les d\'eformations
isospectrales p\'eriodiques g\'en\'eriques}, qui n\'ecessite d'\^etre analys\'e
dans chaque mod\`ele sp\'ecifique. Comme nous le verrons au paragraphe suivant,
ces divergences apparaissent d'ores et d\'ej\`a dans les sph\`eres de Connes--Landi
et leurs espaces ambiants, mais pas dans les tores non commutatifs pour lesquels
l'action est libre.\\
{\it Notons finalement que l'int\'egrabilit\'e
de l'application $p\to d^{-2}_g(y.p,p)$, pour tout $0\ne y\in\T^l$, est une condition
n\'ecessaire pour que la fonction \`a deux points non planaire puisse d\'efinir une
distribution et donc pour que la th\'eorie soit renormalisable.}

\begin{thm}
Pour $M$ compacte ou non (avec les hypoth\`eses du paragraphe \ref{section26} lorsque
$M$ n'est pas compacte), munie d'une action lisse et isom\'etrique de
$\T^l$, $l=2$ ou $l=4$ et avec une matrice de d\'eformation
satisfaisant \`a une condition Diophantienne (relativement \`a $2\pi$),
alors pour tout $\vf\in\Coo_c(M)$ s'annulant dans un voisinage de $M_{sing}$,
l'action effective \`a une boucle non planaire r\'eduite est finie.
\end{thm}

\begin{cly}
Sous les hypoth\`eses du th\'eor\`eme pr\'ec\'edent, la th\'eorie quantique \`a une
boucle n'est pas renormalisable par adjonction de contre-termes locaux si la
matrice $(2\pi)^{-1}\Th$ ne satisfait pas \`a une condition
Diophantienne, ou si la fonction $d_g^{-2}(\a_y(.),.)$, pour un
$0\ne y\in\T^l$, n'est pas localement int\'egrable par
rapport \`a la mesure donn\'ee par la forme volume Riemannienne.
\end{cly}

\section{Points fixes et divergences: un exemple}

Nous allons illustrer la discussion du paragraphe pr\'ec\'edent
sur les cons\'equences de la non int\'egrabilit\'e de la fonction
$p\mapsto d^{-2}_g(y.p,p)$, en regardant le cas de l'espace ambiant
de la trois-sph\`ere de Connes--Landi $\Sf_\th^3$.

En param\'etrant $\R^4$ en coordonn\'ees sph\'eriques
\begin{align*}
x_1&=R\,\cos\phi\,\cos\kappa\\
x_2&=R\,\sin\phi\,\cos\rho\\
x_3&=R\,\cos\phi\,\sin\kappa\\
x_4&=R\,\sin\phi\,\sin\rho,
\end{align*}
avec $R\in[0,\infty[$, $\phi\in[o,\pi/2[$, $\kappa,\rho\in[0,2\pi[$,
on obtient l'espace non commutatif ambiant des sph\`eres de
Connes--Landi cf. paragraphe \ref{sec:CLDV-basics}
(pour lequel la condition d'homog\'en\'eit\'e, $\sum_i x_i^2=1$
a \'et\'e relax\'ee)
en twistant le produit usuel par l'action de $\T^2$.
Cette action est donn\'ee dans ce
syst\`eme de coordonn\'ees par
$$
\T^2\ni y:(R,\phi,\kappa,\rho)\mapsto y.(R,\phi,\kappa,\rho):=
(R,\phi,\kappa+y_1 \mood 2\pi,\rho+y_2 \mood 2\pi).
$$
Pour cette action, l'ensemble des points singuliers se d\'ecompose
en:
$$
\R^4_{sing}=\{(0,\phi,\kappa,\rho)\}\cup
\{(R,0,\kappa,\rho)\}\cup\{(R,\pi/2,\kappa,\rho)\},
$$
et correspond a l'union des plans $x_2=x_4=0$ et $x_1=x_3=0$.
Le premier sous-ensemble consiste en un seul point:  c'est le seul point fixe pour l'action proprement dit.
Il ne posera aucun probl\`eme
pour l'int\'egrabilit\'e, car son groupe d'isotropie est
bi-dimensionnel. Les stabilisateurs des points des deux derniers ensembles, qui sont des plans priv\'es
de l'origine, sont
uni-dimensionnels et l'application $p\mapsto d^{-2}_g(y.p,p)$ n'est pas
int\'egrable dans leur voisinage. En effet, dans ce syst\`eme de
coordonn\'ees, la distance g\'eod\'esique entre un point $(R,\phi,\kappa,\rho)$
et son translat\'e $y.(R,\phi,\kappa,\rho)$, s'\'ecrit
$$
d^2_g\big((R,\phi,\kappa,\rho);y.(R,\phi,\kappa,\rho)\big)=
2R^2(1-\cos^2\phi\,\cos y_1-\sin^2\phi\, \cos y_2).
$$
Pour $y_1=0$, $d^{-2}_g(p,y.p)$ devient alors
$$
\frac{1}{2R^2\,\sin^2\phi(1-\cos y_2)},
$$
qui n'est pas int\'egrable en $\phi=0$ par
rapport \`a la mesure donn\'ee par
$R^3\sin\phi\,\cos\phi\,dR\,d\phi\,d\kappa\,d\rho$.
Par contre, la singularit\'e en $R=0$
de l'application $p\mapsto d^{-2}_g(y.p,p)$ est int\'egrable.

{\it En conclusion, la fonction de Green non planaire \`a deux points de ce
mod\`ele ne d\'efinit pas une distribution. La th\'eorie \`a une boucle
n'est alors pas renormalisable.}

{\renewcommand{\thechapter}{}\renewcommand{\chaptername}{}
\addtocounter{chapter}{-1}
\chapter{Conclusion et perspectives}\markboth{\sl CONCLUSION ET PERSPECTIVES}
{\sl CONCLUSION ET PERSPECTIVES}}

Dans cette th\`ese, nous avons commenc\'e par \'etudier les plans de Moyal,
les plus anciens espaces non commutatifs connus, dans le cadre axiomatique
d'Alain Connes de la g\'eom\'etrie non com\-mutative. Nous avons montr\'e
que l'on peut construire des triplets spectraux, objets centraux dans
l'interface entre la GNC et la physique des interactions fondamentales, \`a
partir d'alg\`ebres construites avec le produit de Moyal. Le d\'efi \'etait important
car ce type de d\'eformations, d'espaces non commutatifs, a longtemps souffert de
critiques, souvent heuristiques ou inexactes (par exemple la soi-disant
trivialit\'e de leurs dimensions spectrale et homologique), visant \`a les d\'econsid\'erer.

Pour ce faire, nous avons tout d'abord \'etendu la notion de triplet
spectral aux d'alg\`ebres sans unit\'e, i.e. espaces Riemannien
quantiques non compacts. Les premi\`eres pierres, les plus naturelles,
avaient d\'ej\`a \'et\'e pos\'ees par Connes. Un
grand nombre de d\'etails techniques devaient cependant \^etre
ajust\'es. En particulier, il r\'esulte que non pas une mais trois
alg\`ebres doivent \^etre consid\'er\'ees.
En effet, en plus d'une alg\`ebre sans unit\'e, codant la topologie d'un
espace non commutatif non compact, il faut pour des raisons d'ordre aussi bien
analytique qu'homologique, invoquer un plongement dans une alg\`ebre
unif\`ere, i.e. une compactification de l'espace quantique.
Les aspects analytiques, en particulier spectraux,
\'etant plus subtiles dans le cas non compact, une troisi\`eme alg\`ebre (dense dans la
premi\`ere) doit aussi \^etre consid\'er\'ee pour formuler l'axiome de dimension spectrale.

Pour v\'erifier ces axiomes modifi\'es,
de nombreux outils d'analyse fonctionnelle ont \'et\'e
d\'eve\-lop\-p\'es. Ce fut principalement
des extensions, dans le contexte du produit de Moyal, des outils de base de la
th\'eorie de la diffusion et de la th\'eorie des semi-groupes.
L'ensemble des outils math\'ematiques d\'evelopp\'es dans la premi\`ere moiti\'e
du si\`ecle dernier en connection avec la formulation de la m\'ecanique quantique
sur espace de phase, dont est originaire le produit de Moyal, a aussi
\'et\'e largement utilis\'e.

\bigskip

Dans une premi\`ere partie qui concerne l'interface entre
les aspects math\'ematiques de la GNC et ses applications en physique
fondamentale, deux types de fonctionnelles d'action pour champs de jauge
non commutatifs ont \'et\'e calcul\'es. Les r\'esultats obtenus ne sont pas
r\'evolutionnaires; ils reproduisent les actions usuellement prises comme point
de d\'epart dans les programmes d'\'etude des th\'eories de jauge sur plans de Moyal. Cependant,
ils d\'emontrent la puissance et la robustesse des concepts de la GNC ainsi
que la faisabilit\'e de ces programmes dans des situations compl\`etement
non commutatives (par opposition aux situations presque commutatives).

L'action de Connes--Lott est une g\'en\'eralisation directe dans le langage
de la GNC de celle de Yang--Mills, cette derni\`ere donnant une interpr\'etation
g\'eom\'etrique du m\'ecanisme de Higgs lorsque l'on consid\`ere le triplet
spectral presque commutatif du mod\`ele standard de la physique des particules (MS).
Pour les plans de Moyal symplectiques, i.e. avec des matrices de d\'eformation inversible,
cette action reproduit l'action de Yang--Mills pour laquelle le produit de Moyal
remplace celui ordinaire. Les techniques que nous avons utilis\'ees ne sont
cependant pas directement applicable aux plans de Moyal g\'en\'eriques
et/ou pour les d\'eformations courbes.

Le principe d'action spectrale est un principe unificateur en GNC. En d\'efinissant
une fonctionnelle d'action $\Tr\big(\chi(\Dslash^2/\Lambda^2)\big)$
\`a partir du spectre de l'op\'erateur de Dirac,
on dispose directement d'une action
invariante sous le groupe de jauge du triplet (le groupe des automorphismes
de l'alg\`ebre), qui g\'en\`ere par fluctuation de la m\'etrique le Lagrangien du MS
coupl\'e avec la gravitation d'Einstein--Weyl (dans le cas du triplet
presque commutatif du MS). Pour calculer cette action dans le cas des plans de Moyal
g\'en\'eriques, il a fallut \'etendre sa d\'efinition aux triplets
spectraux sans unit\'e. Ce programme n'a \'et\'e que partiellement r\'ealis\'e,
mais les techniques d\'evelopp\'ees, modulo une complexit\'e calculatoire
grandissante, sont directement applicables aux d\'eformations isospectrales courbes.
L'obstruction \`a laquelle nous avons \'et\'e confront\'e est due, d'une part \`a la
r\'egularisation suppl\'ementaire invoqu\'ee, une forme de r\'egularisation
spatiale, mais surtout \`a la manifestation d'un ph\'enom\`ene caract\'eristique
\`a ce type d'espaces quantiques: le m\'elange UV/IR.
Ce point est intrigant car ce ph\'enom\`ene se manifeste habituellement
au niveau quantique, or l'action spectrale est une action classique!

\bigskip

Dans la deuxi\`eme partie de ce travail de recherche, non pas celle de ce m\'emoire
car la pr\'esentation ne suit pas l'ordre chronologique, nous avons \'etudi\'e
les aspects principaux pour la construction de triplets spectraux \`a partir des d\'eformations
isospectrales courbes non compactes. Ces vari\'et\'es non commutatives,
g\'en\'eralisant les plan de Moyal et les tores NC en espace courbe, s'inscrivent dans
une th\'eorie des d\'eformations bien plus g\'en\'erale due \`a Rieffel:
la d\'eformation des alg\`ebres de Fr\'echet munies d'une action (isom\'etrique) de $\R^l$.

A partir de techniques d'analyse sur les vari\'et\'es Riemannienne non compactes,
principalement une estimation du noyau du semi-groupe de la chaleur,
nous avons obtenu des r\'esultats concernant l'appartenance
aux classes de Schatten (ordinaires et faibles) pour les op\'erateurs du type
$L^\Th_f(1+\tri)^{-k}$. Ces estimations ont permis de montrer que la
dimension spectrale de ces d\'eformations
(p\'eriodiques ou non), est \'egale \`a la dimension classique, i.e. la dimension
de la vari\'et\'e non d\'eform\'ee. C'est le point central dans la construction de
triplets spectraux sans unit\'e.

\bigskip

Dans la derni\`ere partie de cette th\`ese, nous nous sommes int\'eress\'e
aux th\'eories quantiques des champs sur les d\'eformations
isospectrales courbes. Au travers d'une th\'eorie scalaire, nous avons d\'emontr\'e
l'existence et le caract\`ere intrins\`eque du ph\'enom\`ene de m\'elange UV/IR.\\
Pour les d\'eformations p\'eriodiques, le m\'elange ne concerne (au niveau de
la fonction \`a deux points) que le mode z\'ero du champ, dans sa d\'ecomposition
en sous-espaces spectraux. Dans ce cas, le m\'elange UV/IR n'est pas tellement probl\'ematique;
les modes affect\'es par le m\'elange UV/IR peuvent \^etre trait\'ees pour la renormalisation avec
le secteur planaire.\\
Dans le cas non p\'eriodique, on obtient des fonctions de Green non planaires qui
pr\'esentent le m\'elange sous une forme tr\`es proche de celle du paradigme
qu'est le plan de Moyal.\\
De plus, notre approche donne une interpr\'etation alg\'ebrique de la pr\'esence
du secteur non planaire de ces th\'eories: il vient des produits d'op\'erateurs de
repr\'esentation r\'eguli\`ere droite et gauche.
Nous avons aussi obtenu que le meilleur comportement du secteur non planaire
est une cons\'equence de la pr\'esence du noyau de la chaleur hors diagonale
dans les int\'egrales de Feynman.

Cependant, son caract\`ere r\'egularisant d\'epend fortement des donn\'ees
g\'eom\'etriques. Pour les d\'eformations non p\'eriodiques, la conclusion est que
lorsque le rang de la matrice de d\'eformation est \'egal \`a deux, la singularit\'e
infrarouge (du m\'elange UV/IR) n'est pas localement int\'egrable. Il s'ensuit
que la fonction de Green 1PI non planaire \`a deux points ne d\'efinit pas une
distribution; la th\'eorie n'est alors pas renormalisable. Seules les actions de groupe de
rang quatre donnent lieu \`a un secteur non planaire sans divergence.\\
Lorsque l'action est p\'eriodique, nous avons montr\'e qu'il est n\'ecessaire
que la matrice de d\'eformation satisfasse \`a une condition Diophantienne
(relativement \`a $2\pi$) pour que les fonctions de Green non planaires d\'efinissent
des distributions. Des divergences additionnelles peuvent exister \`a cause des possibles
points fixes de l'action du groupe.

\bigskip

Plusieurs autres directions de recherches sont envisageables.

\medskip

\begin{itemize}

\item[$\bullet$] {\it Construction de triplets spectraux pour les
d\'eformations isospectrales non compactes}\\
A partir de nos r\'esultats sur la Dixmier-tra\c cabilit\'e (et sur la valeur
de leurs traces de Dixmier) des op\'erateurs $L^\Th(f)(1+\Dslash)^{-n/2}$
(corollaire \ref{cl:periodic}, th\'eor\`eme \ref{th:Dix-tr}) nous avons d\'ej\`a
une v\'erification de l'axiome de dimension. Pour terminer la construction,
il reste \`a d\'efinir des `alg\`ebres lisses' (pr\'e-$C^*$-alg\`ebres)
\`a partir du produit twist\'e. Le point de passage oblig\'e est la construction
de semi-normes, isom\'etrique pour l'action du groupe. L'arsenal d\'evelopp\'e par Rieffel dans
\cite{RieffelDefQ} pourra alors \^etre pleinement employ\'e.

\medskip

\item[$\bullet$] {\it Etude du m\'elange UV/IR pour les th\'eories de jauge}\\
Notre traitement du comportement g\'en\'erique du m\'elange UV/IR
peut \^etre g\'en\'eralis\'e pour les d\'eformations de vari\'et\'es de dimension sup\'erieure
et/ou pour les th\'eories de jauge. Nous nous sommes restreint au cas
quadri-dimensionnel par souci de simplicit\'e et pour son int\'er\^et physique.
Il est cependant clair que les techniques du noyau de la chaleur s'appliquent
aux th\'eories scalaires de dimension sup\'erieure.

Pour les th\'eories de jauges sur d\'eformations isospectrales
de dimensions quelconques, il exis\-te un moyen intrins\`eque,
ne passant ni par la construction de Connes--Lott ni par celle de
Chamseddine--Connes, de
d\'efinir des fonctionnelles d'action de type Yang--Mills:\\
Pour toutes formes diff\'erentielles $\omega\in\Omega^p(M)$,
$\eta\in\Omega^q(M)$,
\`a supports compacts, on peut d\'efinir un produit ext\'erieur twist\'e par
$$
\omega\wedge_\Th\eta:=(2\pi)^{-l}\int_{\R^{2l}}\,d^ly\,d^lz\;e^{-i<y,z>}\;
(\a^*_{-\thalf\Th y}\omega)\wedge(\a^*_{z}\eta),
$$
o\`u $\a^*_z$ d\'esigne dor\'enavant le pull-back de l'action.
En se donnant ensuite un fibr\'e vectoriel associ\'e $\pi:E\rightarrow M$
de groupe de structure $G\subset U(N)$, ainsi qu'une
forme de connection $A\in\Omega^1(M,\Lie(G))$, on peut
d\'efinir l'analogue non commutatif de l'action de Yang--Mills
$$
S_{YM}(A):=\int_M\tr(F_\Th\wedge_\Th \ast_{_H} F_\Th),
$$
o\`u $F_\Th:=dA+A\wedge_\Th A$ et $\ast_{_H}$ est l'\'etoile de Hodge.
Dans ce contexte, on peut aussi \'etablir une identit\'e de `trace':
$$
\int_M\omega\wedge_\Th\ast_{_H}\eta=
\int_M\omega\wedge\ast_{_H}\eta,\hspace{1cm}\forall \omega,\eta\in \Omega^p(M).
$$
Ainsi, $S_{YM}(A)$ est \'egale \`a $\int_M\tr(F_\Th\wedge \ast_{_H} F_\Th)$.
Pour entreprendre la quantification, on peut une fois encore utiliser la m\'ethode
du champ de `background', dans la jauge de `background'. Si
on ignore l'ambigu\"it\'e de Gribov,
l'action effective \`a une boucle se r\'eduit au calcul de d\'eterminants
d'op\'erateurs  (partie quadratique en $A$ de $S_{YM}+S_{gf}$
et d\'eterminant de Faddeev--Popov) qui peuvent \^etre exprim\'es localement par
$$
(\nabla_\mu+L_{A_\mu}-R_{A_\mu})(\nabla^\mu+L_{A^\mu}-R_{A^\mu})+B,
$$
o\`u $B$ est born\'e et contient des sommes et produits d'op\'erateurs de
multiplication twist\'ee \`a gauche et \`a droite.
Il est alors clair que le ph\'enom\`ene de m\'elange UV/IR se
manifestera identiquement aux situations plates
(voir par exemple \cite{KW, MR, MRS}) et que les ph\'enom\`enes li\'es
au propri\'et\'es arithm\'etiques, au rang de la matrice de d\'eformation ainsi
qu'aux points fixes de l'action seront eux aussi pr\'esents.

\medskip

\item[$\bullet$] {\it Calcul de l'action spectrale pour les d\'eformations
isospectrales p\'eriodiques compactes}\\
Pour les d\'eformations isospectrales p\'eriodiques compactes,
les obstructions auquel\-les nous nous sommes heurt\'es lors du calcul de
l'action spectrale pour les plans de Moyal, sont facilement surmontables.
En effet, il n'y a d'une part plus lieu d'introduire une r\'egularisa\-tion `spatiale';
le spectre de l'op\'erateur de Dirac covariant
$\Dslash_A=\Dslash +L^\Th(A)+\epsilon JL^\Th(A)J^{-1}$
est discret et de multiplicit\'e finie pour les vari\'et\'es Riemanniennes
compactes. Les techniques
de d\'eveloppement asymptotique, via la formule BCH, sont aussi directement
applicables dans ces cas. D'autres d\'eveloppements sont envisageables,
en particulier celui de Duhamel (voir \cite{Parthenope, Melpomene} pour des
applications de la formule de Duhamel).
D'autre part, le ph\'enom\`ene de m\'elange
UV/IR est bien moins probl\'ematique pour les d\'eformations p\'eriodiques:
la d\'ecomposition en sous-espaces spectraux des composantes
du champs de jauge $A\in\Omega^1_\Dslash$, la relation
$$
\Tr\left(L_{f_r}\,R_{g_s}\,e^{-t\tri}\right)=
Cte\,\delta_{r,-s}\,e^{-\tihalf r\Th s}\int_M\mu_g(p)\,
f_r(p)\,h_s(p)\,K_t(\Th r.p,p),
$$
ainsi que la majoration du noyau de la chaleur hors diagonale \eqref{eq:k1},
\eqref{eq:k2}, montrent que (si $\Th$ satisfait \`a une condition
Diophantienne et si $d^{-2}_g\big(\a_y(.),.\big)\in L^1_{loc}(M,\mu_g)$)
seule certaines composantes du champs de jauge contribueront \`a
l'action spectrale. En d'autres termes, les valeurs des r\'esidus des p\^oles de la fonction Z\'eta
de l'op\'erateur $\Dslash_A^2$ sont hautement modifi\'ees par la pr\'esence du secteur
non planaire dans l'action spectrale; pr\'esence venant de la repr\'esentation adjointe
$\ad^\Th(.):=L^\Th(.)-R^\Th(.)$.

\medskip

\item[$\bullet$] {\it G\'en\'eralisation des r\'esultats de Grosse et Wulkenhaar}\\
Ce projet consiste \`a appliquer nos techniques au mod\`ele de
Grosse et Wulkenhaar et d'essayer de g\'en\'eraliser leurs r\'esultats pour les
d\'eformations isospectrales courbes non p\'eriodiques.

Dans la s\'erie d'articles \cite{GW, GW1, GW2}, il est d\'emontr\'e
que si on ajoute un couplage avec un potentiel confinant (oscillateur
harmonique dans leur travail) \`a la th\'eorie $\vf^{\mop 4}$ sur le plan
de Moyal quadri-dimensionnel, i.e. si l'on consid\`ere l'action de
Grosse et Wulkenhaar
\begin{align*} S_{GW}[\vf]&:=\int
d^4x\,\Bigl[\thalf(\pa_\mu\vf\mop\pa^\mu\vf)(x)
+2\frac{\Omega^2}{\th^2}(x_\mu\vf)\mop(x^\mu\vf)\nonumber\\
&\quad\hspace{4,4cm}+\frac{m^2}{2}\vf\mop\vf(x)
+\frac{\lambda}{4!}\vf\mop\vf\mop\vf\mop\vf(x)\Bigr],
\end{align*}
alors la th\'eorie est perturbativement renormalisable \`a tout les ordres en $\lambda$.
Si la si\-gni\-fication profonde de ce r\'esultat n'est pas encore enti\`erement comprise,
quelques explications peuvent toutefois \^etre mentionn\'ees. Premi\`erement,
ajouter un potentiel effectif est \'equivalent \`a une compactification du plan de Moyal;
on passe de spectre essentiel \`a  spectre purement ponctuel. Aussi, le choix
particulier du potentiel correspond \`a une `Moyal-d\'eformation' de l'espace des moments
en m\^eme temps que celui de configuration. Ceci peut \^etre vu \`a partir de
l'invariance (\`a des changements d'\'echelle pr\`es) de cette action sous
transform\'ee de Fourier:
$$
p_\mu\longleftrightarrow 2(\Th^{-1})_{\mu\nu}x^\nu,
\sepword{et} \widehat{\vf}(p)\longleftrightarrow (\pi\th)^2\,\vf(x).
$$

On pourrait alors refaire notre analyse du comportement des fonctions de Green
non planaires de cette th\'eorie, en utilisant \`a la place du noyau de la chaleur
le noyau de Mehler, i.e. celui du semi-groupe de l'oscillateur harmonique:
$$
K_M(t,x,y)=(2\pi)^{-n/2}\frac{1}{\big(\sinh(4t)\big)^{n/2}}\exp\left[
\frac{1}{\sinh(4t)}x.y-\thalf\frac{1}{\tanh(4t)}(|x|^2+|y|^2)\right].
$$

La derni\`ere \'etape consiste \`a comprendre le degr\'e de libert\'e existant
dans le choix du potentiel (tout en conservant la renormalisabilit\'e),
afin de voir si ce r\'esultat peut \^etre g\'en\'eralis\'e aux d\'eformations isospectrales
non p\'eriodiques courbes.

\medskip

\item[$\bullet$] {\it Sph\`eres de Connes--Dubois-Violette}\\
Il serait int\'eressant de comprendre si le ph\'enom\`ene UV/IR (ou un analogue) est
une ca\-ract\'e\-ristique g\'en\'erale des th\'eories des champs sur espace non commutatif
(nous avons d\'ej\`a montr\'e au paragraphe \ref{para} que
deux secteurs diff\'erents existent forc\'ement)
ou si c'est une particularit\'e des d\'eformations isospectrales.
Les sph\`eres de Connes--Dubois-Violette \cite{CDV}, une famille \`a trois param\`etres de
vari\'et\'es non commutatives sph\'eriques ne rentrant pas dans le cadre des d\'eformations de
Rieffel, sont de bons exemples pour tester ce point.
Pour ce faire, il faudrait trouver une base particuli\`ere de leurs alg\`ebres,
permettant de d\'eriver les r\`egles de
Feynman. Cette question est compl\`etement ouverte.
\end{itemize}
\appendix
\chapter{Traces de Dixmier}
\label{Dixmier}

J. Dixmier a r\'eussi \`a disjoindre les notions de trace et de normalit\'e dans
les applications sur les op\'erateurs d'un espace de Hilbert en construisant
des traces non normales \cite{Dix}.
Ces traces \'eponymes, $\Tr_{\a,\omega}$, sont des fonctionnelles
lin\'eaires positives, invariantes sous conjugaison par des unitaires,
d\'efinies sur un domaine et s'annulant sur l'id\'eal des op\'erateurs
de rang fini.

L'id\'ee est de d\'efinir une trace via le coefficient d'un certain type de
divergence, en l'occurrence logarithmique, de la trace usuelle:
se donnant une suite $\{\a_n\}_{n\in\N}$, on veut d\'efinir une trace
sur un id\'eal du c\^one positif des op\'erateurs compacts $\K^+(\H)$
d'un espace de Hilbert s\'eparable $\H$ par
$$
\lim_{N\to\infty}\frac{\sigma_N(T)}{\a_N},\hspace{1cm} T\in\K^+(\H),
$$
o\`u les sommes partielles des valeurs singuli\`eres (qui sont \'egales aux valeurs
propres car $T\geq0$) sont d\'efinies par
$$
\sigma_N(T):=\sum_{n=0}^{N-1}\mu_n(T).
$$
En se restreignant aux op\'erateurs pour lesquels les sommes
partielles r\'egularis\'ees forment des suites born\'ees,
$\{\sigma_N(T)/\a_N\}\in l^\infty(\N)$,
deux probl\`emes sont \`a r\'esoudre: la lin\'earit\'e et la convergence.\\
L'invariance sous conjugaison par des unitaires est automatique, \'etant
directement impl\'e\-men\-t\'ee au niveau des valeurs singuli\`eres:
$$
\mu_n(uTu^*)=\mu_n(T),\hspace{0.5cm}\forall u\in\mathcal{U}(\H).
$$

Pour obtenir la lin\'earit\'e, il va falloir
d\'efinir un processus limite, $\lim_\omega$, qui ait du sens lorsque
$\{\beta_n\}\in l^\infty(\N)$ n'est pas convergente. Une condition n\'ecessaire est
qu'il soit invariant d'\'echelle: $\lim_\omega(\tilde \beta)=\lim_\omega(\beta)$
avec $\tilde \beta:=(\beta_1,\beta_1,\cdots,\beta_n,\beta_n,\cdots)$. On demandera aussi qu'il
co\"incide avec la limite usuelle lorsque la suite est convergente. En particulier,
ceci impliquera que $\lim_\omega(\beta)=0$, lorsque $\beta_n\to 0$, $n\to \infty$.
Ces conditions pourront \^etre satisfaites seulement pour des
divergences v\'erifiant les propri\'et\'es suivantes:

Soit $\a$ une suite croissante positive, telle que
\begin{align*}
i)&\, \lim_{n\to \infty}\a_n=\infty,\\
ii)&\, \a_0>\a_1-\a_2,\,\,\,\a_{n+1}-\a_n>\a_{n+2}-\a_{n+1},\,\,\forall n\in\N,\\
iii)& \,\lim_{n\to\infty}\a_n^{-1}\a_{2n}=1.
\end{align*}
Soit ensuite $I_\a(\H)$ l'id\'eal des op\'erateurs compacts d\'efini par:
$$
I_\a(\H):=\left\{T\in\K(\H):\frac{\sigma_N(T)}{\a_N}\in l^\infty(\N)\right\}.
$$
La propri\'et\'e d'id\'eal de cette famille d'op\'erateurs compacts est une
cons\'equence de la sous-multiplicativit\'e des valeurs
singuli\`eres (voir par exemple \cite[th\'eor\`eme 1.5]{SimonTrace}):
$$
\mu_{m+n}(TS)\leq\mu_m(T)\mu_n(S).
$$
Cela implique alors
$$
\mu_n(TS)\leq\|T\|\,\mu_n(S),
$$
et donc
$$
\sigma_N(TS)\leq\|T\|\,\sigma_N(S).
$$
L'id\'eal $I_\a(\H)$ poss\`ede une structure naturelle d'alg\`ebre de Banach,
pour la norme
$$
\|T\|_\a:=\sup_{N\in\N}\left\{\frac{\sigma_N(T)}{\a_N}\right\}.
$$

Pour $T$ appartenant au c\^one positif de $I_\a(\H)$,
on d\'efinit sa trace de Dixmier $\Tr_{\a,\omega}(T)$ par
$$
\Tr_{\a,\omega}(T):=\lim_\omega\Big(\frac{\sigma_N(T)}{\a_N}\Big),
$$
o\`u le processus de limite $\lim_\omega$ est donn\'e par
$$
\lim_\omega(\beta):=\omega\big(M(f_\beta)\big), \hspace{0.5cm}
\beta\in l^\infty(\N).
$$
Ici, $0\leq\omega\in\Big(C_b(\R^*_+)\Big)^*$ est une forme lin\'eaire positive
sur l'espace des fonctions continues et born\'ees sur la demi-droite $\R^*_+$,
dont le noyau contient $C_0(\R^*_+)$, l'id\'eal des fonctions continues s'annulant \`a
l'infini, i.e. $\omega(f)=0$ lorsque $\lim_{x\to\infty}f(x)=0$.
L'application $M$ d\'esigne la moyenne de Ces\`aro associ\'ee au groupe (multiplicatif)
$\R^*_+$ avec mesure de Haar $x^{-1}dx$:
$$
\big(Mg\big)(x):=\frac{1}{\log x}\int_1^x g(t)\frac{dt}{t},
$$
et $f_\a$ est la fonction simple (constante par morceaux) associ\'ee
\`a la suite $\a$:
$$
f:l^\infty(\N)\to L^\infty(\R^*_+),\,\,
f_\a(x):=\a_n, \sepword{pour} x\in]n,n+1],\,\, n\in\N.
$$

\begin{thm}[Dixmier 1966 \cite{Dix}]
$\Tr_{\a,\omega}$ est une forme lin\'eaire positive sur $I_\a(\H)$
qui est invariante sous conjugaison par les unitaires de $\L(\H)$.
\end{thm}

\begin{proof}[Esquisse de d\'emonstration]
Que l'application $I^+_\a(\H)\ni T\mapsto \Tr_{\a,\omega}(T)$,
soit invariante par conjugaison sous unitaires, est une cons\'equence de
l'invariance des valeurs singuli\`eres. La positivit\'e vient de la positivit\'e de
la fonctionnelle $\omega$ ainsi que de celle de la moyenne de Ces\`aro.

Nous allons montrer la lin\'earit\'e. A cette fin,
notons tout d'abord que $\lim_\omega$ est invariante d'\'echelle:
$$
\lim_\omega(\beta)=\lim_\omega(\tilde\beta),
\sepword{avec}
\tilde \beta:=(\beta_1,\beta_1,\cdots,\beta_n,\beta_n,\cdots).
$$
En effet, avec $\theta_a:C(\R_+^*)\to C(\R_+^*)$ l'op\'erateur d'\'echelle
$\big(\theta_ag\big)(x):=g(ax)$, d\'efini pour tout $a>0$,
on a:
$$
f_{\tilde\beta}(x)=f_\beta(\thalf x)=\big(\theta_{1/2}f_\beta\big)(x).
$$
Puisque la moyenne de Ces\`aro est asymptotiquement
invariante d'\'echelle, i.e. pour tout $g\in L^\infty(\R_+^*)$, tout $a>0$
$$
\lim_{x\to\infty}\left|M\big(\th_a g\big)(x)-M\big(g\big)(x)\right|=0,
$$
on obtient en utilisant le fait que $\omega(g)=0$ lorsque $\lim_{x\to\infty}g=0$:
$$
\lim_\omega(\tilde\beta)=\omega\big(M(f_{\tilde\beta})\big)=
\omega\big(M(\th_{1/2}f_\beta)\big)=\omega\big(M(f_\beta)\big)=
\lim_\omega(\beta).
$$

Posons pour $T,S\in I^+_\a(\H)$:
\begin{equation*}
\gamma_N:=\frac{1}{\a_N}\sigma_N(T+S),\,\,
\beta_N:=\frac{1}{\a_N}\sigma_N(T),\,\,
\delta_N:=\frac{1}{\a_N}\sigma_N(S).
\end{equation*}
On remarque que les sommes partielles $\sigma_N(.)$
sont aussi des normes, car pouvant aussi \^etre exprim\'ees en termes de suprema
sur des normes traces:
\begin{equation}
\label{nor}
\sigma_N(T)=\sup_E \|TE\|_1,
\end{equation}
o\`u $E$ varie dans l'ensemble des op\'erateurs auto-adjoints de
rang $N$. On obtient alors
$$
\sigma_N(T+S)\leq\sigma_N(T)+\sigma_N(S),
$$
et donc $\gamma_N\leq\beta_N+\delta_N$.
Ainsi, on a
$$
\Tr_{\omega,\a}(T+S)\leq \Tr_{\omega,\a}(T)+\Tr_{\omega,\a}(S).
$$
Pour obtenir l'in\'egalit\'e inverse, nous allons nous servir
de la relation
$$
\sigma_N(T)+\sigma_M(S)\leq\sigma_{N+M}(T+S).
$$
Cette relation s'obtient en consid\'erant $E$, $F$ et $G$, projecteurs de rang $n$,
$m$ et $n+m$, qui soient tels que l'image de $E+F$ soit contenue dans celle de $G$.
On a alors pour $0\leq T,S\in\K(\H)$
$$
\Tr(ET)+\Tr(FS)\leq \Tr(GT)+\Tr(GS)=\Tr(G(T+S)),
$$
qui donne le r\'esultat d'apr\`es la caract\'erisation
(\ref{nor}) des sommes partielles. Il en r\'esulte que
$$
\sigma_N(T)+\sigma_N(S)\leq\sigma_{2N}(T+S),
$$
et donc $\beta_N+\delta_N\leq\frac{\a_{2N}}{\a_N}\,\gamma_{2N}$.
Ceci conclut la preuve, car par hyphoth\`ese
$$
\lim_{N\to\infty}\frac{\a_{2N}}{\a_N}=1,
$$
et d'apr\`es l'invariance d'\'echelle du processus de limite,
 on a $\lim_\omega(\gamma_{2N})=
\lim_\omega(\gamma_{N})$.
\end{proof}

Les fonctionnelles $\Tr_{\omega,\a}$ s'\'etendent ensuite par
lin\'earit\'e \`a tout $I_\a(\H)$, en utilisant la d\'ecomposition
de tout op\'erateur born\'e
en une somme \`a coefficients complexe de (quatre) \'el\'ements
positifs.

La propri\'et\'e remarquable des traces de Dixmier, h\'erit\'ee
de celle de la fonctionnelle $\omega$ d\'efinissant le processus de limite
($\omega(g)=0$ lorsque $\lim_{x\to\infty}g=0$), est de poss\'eder
un noyau consistant en la fermeture en norme $\|.\|_\a$ de l'id\'eal des
op\'erateurs de rang fini, dans lequel est inclus l'id\'eal des op\'erateurs \`a trace:
$$
\Ker(\Tr_{\omega,\a})=\overline{O.R.F^{\|.\|_\a}}\supset \L^1(\H).
$$

Les suites donnant lieu \`a d'int\'eressantes traces de Dixmier sont
typiquement les polyloga\-rithmes et les puissances de logarithme (voir
par exemple \cite{Nicola}). Ces suites satisfont \'evidemment aux propri\'et\'es
requises. \\
En g\'eom\'etrie non commutative, c'est la divergence logarithmique
qui est la plus pertinente: $\a_n=\log n$, $\Tr_{\omega}:=\Tr_{\omega,log}$.
En particulier, l'id\'eal $I_{log}(\H)$ co\"incide avec la premi\`ere classe de
Schatten faible $\L^{(1,\infty)}(\H)$. Notons aussi que les classes
de Schatten faibles ainsi que leurs duaux
$$
\L^{(\infty,k)}(\H):=\{T\in\K(\H);\mu_m(T)=o(m^{-1/k}),\forall m\in\N\},\hspace{1cm}k\geq 1,
$$
peuvent \^etre alternativement d\'efinis \`a partir des classes de Schatten ordinaires
par interpolation r\'eelle \cite[IV.2.$\a$]{ConnesBook}. Aussi, il existe
une in\'egalit\'e de H\"older pour les classes de Schatten faibles avec la `trace
de Dixmier logarithmique' \cite[proposition 7.16]{Polaris}.

Les traces des Dixmier sont \'evidemment tr\`es difficiles
\`a calculer lorsque, comme dans la majeure partie des cas, on ne conna"t
pas explicitement les valeurs propres d'un op\'erateur.
Heureuse\-ment, les r\'esidus des fonctions Z\'etas donnent, sous certaines
conditions, un acc\`es rapide aux valeurs des `traces de Dixmier logarithmiques'.

\begin{thm}[Connes\cite{ConnesBook}]
Pour $0\leq T\in \L^{(1,\infty)}(\H)$, soit
$\zeta_T(s):=\Tr(T^s)$ sa fonction Z\'eta. Si
$\lim_{s\to1^+}(s-1)\zeta_T(s)=L<\infty$ alors
$\Tr_\omega(T)=L$,
ind\'ependamment de $\omega$.
\end{thm}
Ce r\'esultat caract\'erise le fait que si la suite $\sigma_N(T)/\log N$
converge, alors $\lim_\omega$ co\"incide avec la limite ordinaire.
Notons finalement que ce r\'esultat a \'et\'e g\'en\'eralis\'e comme suit:

\begin{thm}[ \cite{CareyPS, LSS}]
Pour $0\leq T\in \L^{(1,\infty)}(\H)$ et $S\in\L(\H)$, soit
$\tilde\zeta_T(s):=\Tr(ST^s)$. Si
$\lim_{s\to1^+}(s-1)\tilde\zeta_T(s)=L<\infty$
alors $\Tr_\omega(ST)=L$,
ind\'ependamment de $\omega$.
\end{thm}

\newpage

\chapter{Base de Wigner}
\label{fmn}

\section{Limite $\th\to 0$}
Rappelons que les \'el\'ements de la base de Wigner de l'oscillateur harmonique
ont \'et\'e d\'efinis par
\begin{equation}
\label{chien}
f_{mn}
:= \frac{1}{\sqrt{\th^{|m|+|n|}\,m!n!}}\,(a^*)^m \mop f_{00} \mop a^n,
\end{equation}
o\`u $f_{00}$ est la Gaussienne
$$
f_{00}(x) := 2^N \exp\left(-\frac{1}{\th}\sum_{i=1}^{2N}x_{_i}^2\right),
$$
et les fonctions de cr\'eation et d'annihilation sont
$$
a_l := \frac{1}{\sqrt{2}} (x_{_l} + i\,x_{_{l+N}})  \sepword{et}
a_l^* := \frac{1}{\sqrt{2}} (x_{_l} - i\,x_{_{l+N}}).
$$
Les produits $\mop$ et ponctuel sont reli\'es par
\begin{align*}
a_l\mop f&=a_l.f+\frac{\th}{2}\frac{\pa f}{\pa a_l^*}\hspace{0.5cm}
f\mop a_l=a_l.f-\frac{\th}{2}\frac{\pa f}{\pa a_l^*}\\
a^*_l\mop f&=a^*_l.f-\frac{\th}{2}\frac{\pa f}{\pa a_l}\hspace{0.5cm}
f\mop a^*_l=a^*_l.f+\frac{\th}{2}\frac{\pa f}{\pa a_l},
\end{align*}
o\`u
$$
\frac{\pa}{\pa a_l}:=\frac{1}{\sqrt{2}} \Big(\frac{\pa}{\pa x_{_l}} -i\,\frac{\pa}{\pa  x_{_{l+N}}}\Big)  \sepword{et}
\frac{\pa}{\pa a_l^*} := \frac{1}{\sqrt{2}} \Big(\frac{\pa}{\pa x_{_l}}+ i\,\frac{\pa}{x_{_{l+N}}}\Big).
$$
Ainsi
$$
(a_l^*)^{\mop m} \mop f_{00}=2^m {a_l^*}^mf_{00}\sepword{et}
f_{00} \mop a_l^{\mop m}=2^ma_l^mf_{00}.
$$
En dimension $2N$, avec d'\'evidentes notations multi-indicielle, il en r\'esulte que
\begin{align}
\label{oi}
f_{mn}
&= \frac{1}{\sqrt{\th^{|m|+|n|}\,m!n!}}\,\big({a^*}^m \mop f_{00}\big) \mop a^n\nonumber\\
&= \frac{1}{\sqrt{\th^{|m|+|n|}\,m!n!}}\,\Big(a-\frac{\th}{2}\frac{\pa }{\pa a^*}\Big)^n
\big({a^*}^m \mop f_{00}\big)\nonumber\\
&= \frac{1}{\sqrt{\th^{|m|+|n|}\,m!n!}}\,\Big(a-\frac{\th}{2}\frac{\pa }{\pa a^*}\Big)^n
\big(2^m{a^*}^m  f_{00}\big)\nonumber\\
&= \frac{1}{\sqrt{\th^{|m|+|n|}\,m!n!}}\,\Big(\Big(2a-\frac{\th}{2}\frac{\pa }{\pa a^*}\Big)^n
2^m{a^*}^m \Big) f_{00}\nonumber\\
&= \frac{1}{\sqrt{\th^{|m|+|n|}\,m!n!}}\,\sum_{\N^N\ni p=0}^{\min(m,n)}(-1)^{|p|}
\left(\!\!\!
\begin{array}{c}
m\\
p
\end{array}\!\!\!\right)
\left(\!\!\!
\begin{array}{c}
n\\
p
\end{array}\!\!\!\right)
\,p!\,2^{|n|+|m|-2|p|}\,\th^{|p|} \,{a^*}^{m-p}\,a^{n-p}\,f_{00}.
\end{align}
En utilisant
$$
\lim_{\th\to 0}\,(\pi\th)^{-N}\,\exp\Big(-\frac{1}{\th}\sum_{\mu=1}^{2N}x_{_\mu}^2\Big)
=\delta^{2N}(x),
$$
on obtient que pour $m$ et $n$ fix\'es, seuls les
termes de la somme \eqref{oi} pour lesquels $|p|=\frac{1}{2}(|n|+|m|) -N$
donnent des contributions finies et non identiquement nulles
dans la limite $\th\to 0$ (au sens des distributions).
Lorsque $|p|>\frac{1}{2}(|n|+|m|) -N$ les contributions sont nulles et
lorsque $|p|<\frac{1}{2}(|n|+|m|) -N$ elles sont infinies.

\section{$f_{mn}$ et op\'erateur d'Euler}

Nous allons voir un \'etrange lien entre l'op\'erateur d'Euler
$$
\widehat{\E}_c:=\th\pa_\th+\sum_{i=1}^N\,
c\, x_{_i}\pa_{x_{_i}}+ (1-c)\, x_{_{i+N}}\pa_{x_{_{i+N}}},
\sepword{avec} c\in[0,1],
$$
agissant sur le champs continu de $C^*$-alg\`ebre
$\prod_{\th\in\R^+}A_\th^0$ ($A_\th^0$ est la $C^*$-compl\'etion
de $\A_\th$), et l'exponentielle imaginaire quadratique
$$
h_\beta(x):=\exp\Big(i\beta \sum_{i=1}^{N}x_{_i}x_{_{i+N}}\Big).
$$
Nous avons vu au paragraphe 2.2.1.3, que cette exponentielle
appartient \`a $\M^\th$ (l'alg\`ebre des multiplicateurs bilat\`eres de $\A_\th$)
si est seulement si $|a|\ne2/\th$. Nous allons
d\'emontrer que $\widehat{\E}_c$ est une d\'erivation sur $\prod_{\th\in\R^+}A_\th^0$ et
qu'elle est ``int\'erieure'' sur une de ses sous-alg\`ebres.
L'automorphisme correspondant sera justement donn\'e par
l'\'el\'ement `pathologique' $h_{2/\th}$.

Montrons tout d'abord que $\widehat{\E}_c$ satisfait \`a la r\`egle de Leibniz.
Par d\'efinition, la $C^*$-alg\`ebre $\prod_{\th\in\R^+}A_\th^0$ consiste en
les fonctions sur $\R^+$ \`a valeur dans $A_\th^0$, pour lesquelles
la norme
$$
\|f\|:=\sup_{\th\in\R^+}\|f(\th)\|_{A_\th^0},
$$
est finie. Ici $\|.\|_{A_\th^0}$ d\'esigne la norme op\'eratorielle de $A_\th^0$. Il suffit
alors de v\'erifier que $\widehat{\E}_c$ satisfait \`a la r\`egle de Leibniz
pour le produit de Moyal de deux fonctions $f$ et $g$ d\'ependant
a priori de $\th$.

D'une part nous avons
\begin{align*}
\th\pa_\th\big(f\mop g\big)(x)&=
\th\pa_\th\int\frac{d^{2N}y}{(\pi\th)^N}\frac{d^{2N}z}{(\pi\th)^N}
\,f(y)\,g(z)\,e^{\frac{2i}{\th}(x-y).S(x-z)}\\
&=-2N \,\big(f\mop g\big)(x)+\big(\th\pa_\th f\mop g\big)(x)
+\big(f\mop \th\pa_\th g\big)(x)\\
&+\frac{2i}{\th}\sum_{\mu=1}^{2N}\big((x^\mu.f)\mop g\big)(x).(Sx)^\mu
-\big(f\mop (x^\mu.g)\big)(x).(Sx)^\mu-\big((x^\mu.f)\mop((Sx)^\mu.g)\big)(x).
\end{align*}
Finalement, en utilisant la r\`egle de `produit mixte' \eqref{pro4} ainsi
que le fait $x.(Sx)=0$, on obtient
$$
\th\pa_\th\big(f\mop g\big)(x)=\big(\th\pa_\th f\mop g\big)(x)
+\big(f\mop \th\pa_\th g\big)(x)-\frac{2i}{\th}\sum_{\mu=1}^{2N}
\big(\pa_{(S x)^\mu}f\big)\mop\big(\pa_{x^\mu}g\big)(x).
$$
D'apr\`es la r\`egle de Leibniz pour le produit de Moyal \eqref{pro3} et encore
\eqref{pro4}, on a d'autre part
\begin{align*}
\sum_{i=1}^N x_{_i}.\pa_{x_{_i}}\big(f\mop g\big)(x)&=
\sum_{i=1}^N x_{_i}.\Big(\big(\pa_{x_{_i}}f\big)\mop g(x)+
f\mop\big( \pa_{x_{_i}}g\big)(x)\Big)\\
&=\sum_{i=1}^N\Big( \big(x_{_i}.\pa_{x_{_i}}f\big)\mop g(x)
+f\mop \big(x_{_i}.\pa_{x_{_i}}g\big)(x)  \Big)
+\frac{2i}{\th}\sum_{\mu=1}^{2N}
\big(\pa_{(S x)^\mu}f\big)\mop\big(\pa_{x^\mu}g\big)(x).
\end{align*}
Par un calcul similaire
\begin{align*}
\sum_{i=1}^N x_{_{i+N}}.\pa_{x_{_{i+N}}}\big(f\mop g\big)(x)&=
\sum_{i=1}^N\Big( \big(x_{_{i+N}}.\pa_{x_{_{i+N}}}f\big)\mop g(x)
+f\mop\big( x_{_{i+N}}.\pa_{x_{_{i+N}}}g\big)(x)  \Big)\\
&+\frac{2i}{\th}\sum_{\mu=1}^{2N}
\big(\pa_{(S x)^\mu}f\big)\mop\big(\pa_{x^\mu}g\big)(x).
\end{align*}
Ainsi, pour tout $c\in[0,1]$, $\widehat{\E}_c$ satisfait \`a la r\`egle de Leibniz:
\begin{equation}
\label{eq:euler}
\widehat{\E}_c\big(f\mop g)=\big(\widehat{\E}_c f\big)\mop g
+f\mop\big(\widehat{\E}_c g\big).
\end{equation}

Nous allons voir dans quel sens, et pour quelle sous-alg\`ebre de
$\prod_{\th\in\R^+}A_\th^0$, cette d\'erivation est int\'erieure.
En utilisant l'\'ecriture \eqref{chien} des $f_{mn}$ ainsi que la propri\'et\'e de
d\'erivation, on va \'etablir
l'action de l'op\'erateur d'Euler sur les $f_{mn}$ (qui d\'ependent de $\th$):
\begin{align*}
\widehat{\E}_c\big(f_{mn}\big)&=-\frac{|n|+|m|}{2}f_{mn}+(\th^{|n|+|m|}m!n!)^{-1/2}
\Big\{(a^*)^m\mop\widehat{\E}_c\big(f_{00}\big)\mop a^n\\
&\quad+\sum_{j=1}^m(a^*)^{m-j}\mop\widehat{\E}_c(a^*)\mop(a^*)^{j-1}\mop f_{00}
\mop a^n\\
&\quad+\sum_{j=1}^m(a^*)^m\mop f_{00}\mop a^{m-j}\mop\widehat{\E}_c(a)\mop a^{j-1}
\Big\}.
\end{align*}
Or,
\begin{align*}
\widehat{\E}_c\big(f_{00}\big)&=\frac{1-2c}{4\th}\sum_{i=1}^N(a_i^*)^2\mop f_{00}+
f_{00}\mop a_i^2,\\
\widehat{\E}_c\big(a_i\big)&=\frac{1}{2}\big(a_i-(1-2c)a_i^*\big),\\
\widehat{\E}_c\big(a_i^*\big)&=\frac{1}{2}\big(a_i^*-(1-2c)a_i\big).
\end{align*}
Ainsi,
\begin{align*}
\widehat{\E}_c\big(f_{mn}\big)&=\frac{1-2c}{4}\sum_{i=1}^N\left\{
\sqrt{(m_i+1)(m_i+2)}f_{m+2u_i,n}+\sqrt{(n_i+1)(n_i+2)}f_{m,n+2u_i}\right.\\
&\hspace{2cm}\left.-\sqrt{m_i(m_i-1)}f_{m-2u_i,n}-\sqrt{n_i(n_i-1)}f_{m,n-2u_i}\right\},
\end{align*}
o\`u $u_i$ d\'esigne le $i$-i\`eme vecteur de base de $\N^N$. En utilisant
la d\'efinition des $f_{mn}$ en termes de produit de Moyal it\'er\'e des
fonctions de cr\'eation et d'annihilation,  cf. \'equations
\eqref{riw} et \eqref{riv}, on identifie la derni\`ere expression \`a
une action adjointe:
$$
\widehat{\E}_c\big(f_{mn}\big)=\frac{1-2c}{4\th}\sum_{i=1}^N\big(
L(a^*_i\mop a^*_i -a_i\mop a_i)-R(a^*_i\mop a^*_i -a_i\mop a_i)\big)f_{mn}.
$$
Ainsi, sur la sous-alg\`ebre de $\prod_{\th\in\R^+}A_\th^0$ engendr\'ee par les
combinaisons de $f_{mn}$ \`a coefficients constants (ind\'ependants de $\th$),
$\widehat{\E}_c$ est une d\'erivation int\'erieure:
$$
\widehat{\E}_c=\ad_b,
$$
avec
$$
b=\frac{1-2c}{4\th}\sum_{i=1}^Na^*_i\mop a^*_i -a_i\mop a_i=
\frac{-2i(1-2c)}{\th}\sum_{i=1}^N\,x_{_i}\,x_{_{i+N}},
$$
qui est justement l'argument de l'exponentielle `pathologique' $h_{\pm 2/\th}$,
pour $c=0$ ou $c=1$. \\
Pour $c=1/2$, les $f_{mn}$ sont dans le noyau de $\widehat{\E}_c$ et sont des points fixes
pour l'automorphisme associ\'e.

\end{document}